\documentclass[11pt]{report} 
\pdfoutput=1
\usepackage[T1]{fontenc}
\usepackage{varwidth}
\usepackage{rotating}
\usepackage{las-master}
\usepackage{epsfig}
\usepackage{amsmath}
\usepackage{subfigure}
\usepackage{multirow}
\usepackage{verbatim}
\usepackage{url}

\usepackage{graphics}
\usepackage{graphicx}
\usepackage{xcolor}
\usepackage{pstricks}
\usepackage{eepic}
\usepackage{color}
\usepackage{float}
\newfloat{source}{htbp}{}
\floatstyle{ruled}

\makeatletter

\newcommand{\Rmnum}[1]{\expandafter\@slowromancap\romannumeral #1 @}
\makeatother

\usepackage{float}
\usepackage{natbib}

\usepackage{setspace} 
\newcommand{\solarmass}{$M_{\odot}$}
\newcommand{\deriv}{\rm d}
\bibpunct{(}{)}{;}{a}{,}{,}

\title{Non-LTE Luminosity and Abundance Diagnostics of 
Classical Novae in X-rays}
\thesistitle{Non-LTE Luminosity and Abundance Diagnostics of 
Classical Novae in X-rays}
\student{Peter Nemeth}
\studenttitle{Master of Science in Astronomy\\
University of Szeged\\
Hungary, 2004}

\begin{document}
\pagenumbering{roman}
\titlepage

\newpage
\pagenumbering{roman}
\thispagestyle{empty}
\begin{center} 
\vspace*{0.5in}
\copyright~Copyright \number\year{} \lasstudent\\
All Rights Reserved\\
\vspace{6in}
The author grants permission to make single copies\\
\vspace{0.5in}
\line(1,0){180}
\end{center}

\newpage
\begin{small}
\committeepage
\end{small}

\newpage
\abstract
\setstretch{2.00}
\vspace{5mm}

Classical novae have fundamental importance in astronomy as they are
relevant to both an understanding of individual 
stellar evolution and to taking proper distance
measurements on galactic and cosmological scales. Also, novae are significant
sources of interstellar material, especially carbon, nitrogen, oxygen and 
aluminum.
These standard candles are only behind supernovae and $\gamma$-ray bursts
as the third brightest objects in the sky, and the most probable progenitors
of the brightest, type Ia supernovae. 

Just after 
a nova outburst the system enters into the constant bolometric luminosity
phase and the nova maintains a stable hydrogen burning in the surface layers
of the white dwarf. As
the expanding shell around the nova attenuates,
progressively deeper
and hotter layers become visible. At the end of the constant bolometric
luminosity phase, the hottest layers are exposed and novae radiate X-rays.

This work uses the static, plane-parallel model atmosphere code TLUSTY to
calculate atmospheric structure,
and SYNSPEC to calculate synthetic X-ray spectra of novae. It was necessary to incorporate 
atomic data for the
highest ionization stages for elements ranging from
hydrogen to iron for both programs. 
Atomic data on energy levels, bound-free, bound-bound
transitions and natural broadening were taken from NIST and TOPbase. 

Extensive tests revealed the importance of line opacities on atmospheric
parameters and on the final spectra. A correlation can be defined between
effective temperature and surface gravity. The spectral
appearance is not very sensitive to the joint changes of both. Due to this
effect both parameters might be over-estimated with static models. These tests
also 
showed that N VI and N VII lines are good indicators of effective temperature.
 
Model fitting of V4743 Sgr and V2491 Cyg confirmed the anticipated impact of
modeling geometry and stellar wind. Both novae are close to or over the
Eddington limit. Ionization balance and line profiles also
indicate this. These 
results are consitent with previous studies; further and unambiguous details 
require a comprehensive update of TLUSTY, what is under way.

\setstretch{2}
\newpage
\tableofcontents
\newpage
\listoffigures
\newpage
\listoftables

\newpage
\acknowledgements
In the 
first place I would like to thank to my Parents for their continous support and 
encouragement to follow my hobby, even though they knew this will 
take me away for many years. I am working on to turn 
this childhood hobby into a scientific profession.

I am indebted to my advisor, Dr. Stephane Vennes, for suggesting this hot topic,
teaching me the steps of spectral modeling and helping me throughout my 
research.
I thank Dr. Matthew A. Wood for his professional help during my studies,
especially in the last year when his
supervision helped me to finish this work on time. 
Gratitude goes to Dr. Terry D. Oswalt for his advices, his careful reading and
innumerable suggestions to improve this dissertation, and for 
his financial support
in the last year. I thank to all the members in my doctoral 
committee for their useful comments.

My appreciation goes to Dr. Adela Kawka for her help and to the 
Ond\v{r}ejov Observatory for supporting this project during my summer visits. 
I also thank Nora Reeves for her cheering and support, and Dr. Laszlo Baksay
for drawing my attention to the Ph.D. program at Florida Tech. 

Special thanks go to Dr. Ivan Hubeny and Dr. Thierry Lanz for 
developing, documenting and making TLUSTY and SYNSPEC publicly
available. Similarly, I thank to the TOPbase and NIST groups for their well
organized atomic data.

I thank to the Department of Physics and Space Sciences at Florida
Institute of Technology for supporting me as a teaching assistant for five
years. 

I am grateful 
to all my former teachers and professors for their work. Without their
inspiration, enthusiasm and professionalism I would not be here today. 

Last but not least, I thank all my friends who stood by me and 
brightened my days in ``Boreida''. Thank You!








\newpage
\dedication
\hspace{3in}{\normalsize to Runci $\dots$}
\setstretch{1}
\begin{eqnarray*}
\hspace{1.7in}
{\parbox{71mm}{\small{\it 
"$\dots$ because there, 
where you want to go for further studies, you will learn 
spectroscopy $\dots$"}}}\\ 
{\parbox{71mm}{\small (High school chemistry, $\sim${1996})}}
\end{eqnarray*}

\newpage
\pagenumbering{arabic}
\setstretch{2.}

\chapter{Introduction}\label{chap:introduction}

In ancient times the Latin name {\it nova stella} (new star) 
was applied to the sudden appearance of a previously unrecorded bright
star. The name was coined by 
Tycho Brahe after he discovered a new bright star (SN
1572) in
the constellation Cassiopeia in 1572.  
Reports of such new stars can be traced back to about 1500 BCE in eastern and 
European records. According to \citet{lundmark21} there were six novae 
(namely: Nova Vulpeculae 1670, Nova Sagittae
1783, Nova Ophiuchi 1848, Nova Corona Borealis 1866, Nova Cygni 1876 and Nova
Andromedae 1885) 
discovered in the post-telescopic 
era before the application of wide-field photography in 1887. 
After this date their 
discovery rate quickly rose to the current 4$-$6 per year in the Milky Way. 
From this number 
the Galactic nova rate estimated by \citet{shafter02} is 30$\pm$10 yr$^{-1}$.
Observations of the evolution of Nova Ophiuchi 1848 showed that the star 
returned to its faint, original state at apparent magnitude 13.5 
a few years after the 
outburst, confirming that novae are not new objects, 
just enormous brightenings
of existing stars. 
Also the recurrence of some other novae suggested that the cataclysmic 
process that
produces the sudden outburst does not destroy the star. The 
different nature of novae and supernovae was not realized until the late 
1930's. With respect to novae where there is a thermonuclear explosion on the
surface of the star, in supernovae the entire star explodes, releasing about
$10^3$ times more energy. Without knowing the difference between the two
types of objects Harlow 
Shapley was led to the wrong conclusion regarding the distance to 
spiral nebulae. This error contributed to the Great Debate about the
size of the Universe in 1920. 
Following the detailed work of Dean B. McLaughlin in the early 1940's and 
Robert Kraft in the 1960's, it was realized 
that novae were invariably members of
close binary systems. Observations showed that 
the progenitor is a hot star with a late-type main sequence (MS) companion. 
The mass estimates suggested the eruptive stars were white
dwarf (WD) stars. 
Later the discovery of accretion disks around WDs further supported this idea. 
Recent theoretical work and more advanced multi-wavelength observations
 led to our current, fairly comprehensive picture of classical novae. Thus,
novae are WDs accreting matter from a cool MS 
companion in a close binary system.

The initial mass function for MS stars shows an average stellar mass of 0.6
$M_{\odot}$. \citet{ninkovic06} observationally 
found the same average mass for individual 
MS stars and
about 2 $M_{\odot}$ for binaries in the solar neighborhood up to 10 pcs. 
Consequently, the vast majority 
($>80 \%$) of stellar mass is in low and intermediate mass stars. 
 This ratio is related to their formation scenario and the fragmentation of 
interstellar clouds. \citet{lada06} asserted that about one third of stars 
in our galaxy are in binary or multiple systems, and found the larger the
mass
the more probable the star is not alone. 
On the other hand, low and intermediate mass stars ($< 8$ \solarmass) 
finish their evolution on the Hertzsprung-Russell 
diagram as WDs. When hydrogen, the nuclear fuel of MS 
stars, runs out in the core the radiation and gas pressure can not 
counterbalance the force of gravity. In the collapsing core of the star matter
is 
squeezed so much that electrons are forced to fill higher energy levels as
all low levels are filled and they can not occupy the same states
according to the Pauli exclusion principle. 
Energy is deposited in the kinetic energy of electrons and these free
electrons give rise to the so called degenerate pressure which can stop
contraction. For more massive WDs (over $\sim0.8\ M_{\odot}$) 
electrons become relativistic. The degenerate pressure can support the star  
up to the Chandrasekhar mass limit. This limit is 
between 1.25 and 1.45 M$_{\odot}$ depending on the composition of the core.  
This process is accompanied by large scale mass loss, leaving behind the hot 
core of the star, known as a pre-WD. 
There is no energy generation in WDs, 
only the gradual release of their residual thermal energy. 

The population of binary stars and the fact that WDs are the most common end
products of stellar evolution suggest there are 
a large number of binaries which consist of
WDs and MS companions. The close binaries in which members affect
each others' evolution are called cataclysmic variable stars (CVs). 
When the secondary star 
fills its last stable gravitational equipotential surface (Roche lobe), 
it starts 
losing mass to the primary. 
The mass transfer occurs through the inner Lagrangian point, where the
Roche lobes of the stars are connected. The hydrogen-rich gas ends
 up in an accretion disk around, and finally on the surface of the WD. 
This
 mass transfer and its related eruptive processes cause
cataclysmic variable stars to show sudden light
 variations. 

CVs fall in two 
main classes depending on their magnetic fields. 
Classical novae (CNe), recurrent novae and U Geminorum (or dwarf novae) systems 
lack strong 
magnetic fields and may acquire an extended accretion disk. AM Herculis, 
and 
intermediate polars have strong magnetic fields which prevent 
the formation of a 
disk. In such cases 
the material spirals along the field lines directly to the poles 
of the WD.  

Some classical novae emit hard X-ray radiation that originate in 
the shocked wind 
around the system and/or soft X-rays 
from the atmosphere of the WD. Here we will concentrate on the modeling of
the soft X-ray radiation emerging from the atmospheres 
during the constant bolometric luminosity phase which starts just after a
nova outburst.



\section{The light curve of  CNe}

\begin{figure}[ht]
\begin{center}
\epsfig{file=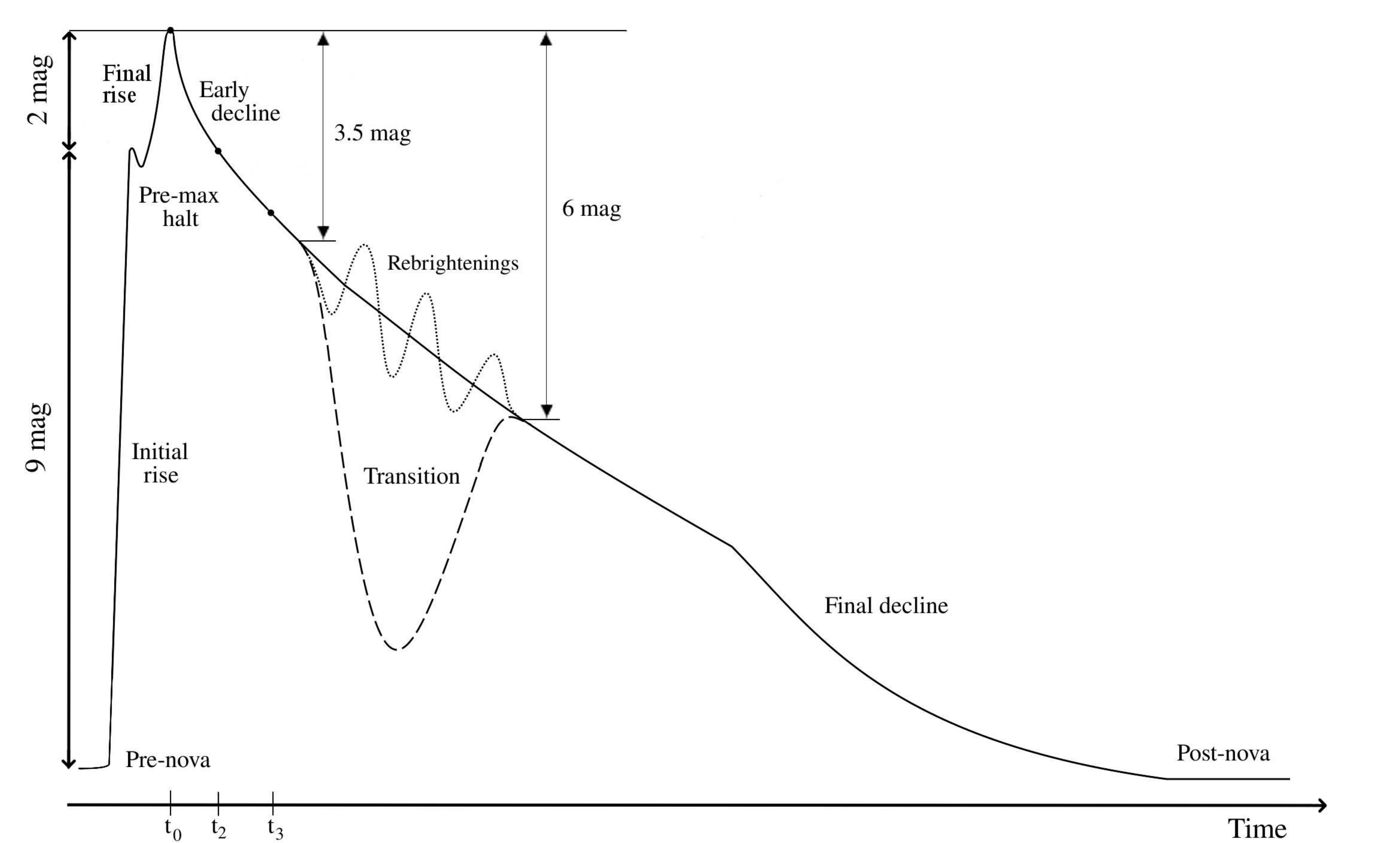, width=14cm}
\parbox{130mm}{\caption[General light curve of CNe.]
{The general light curve of CNe. 
Figure is based on \citet{mclaughlin43}.\label{fig:nlc}}}
\end{center}
\end{figure}
\begin{figure}[!h]
\begin{center}
\epsfig{file=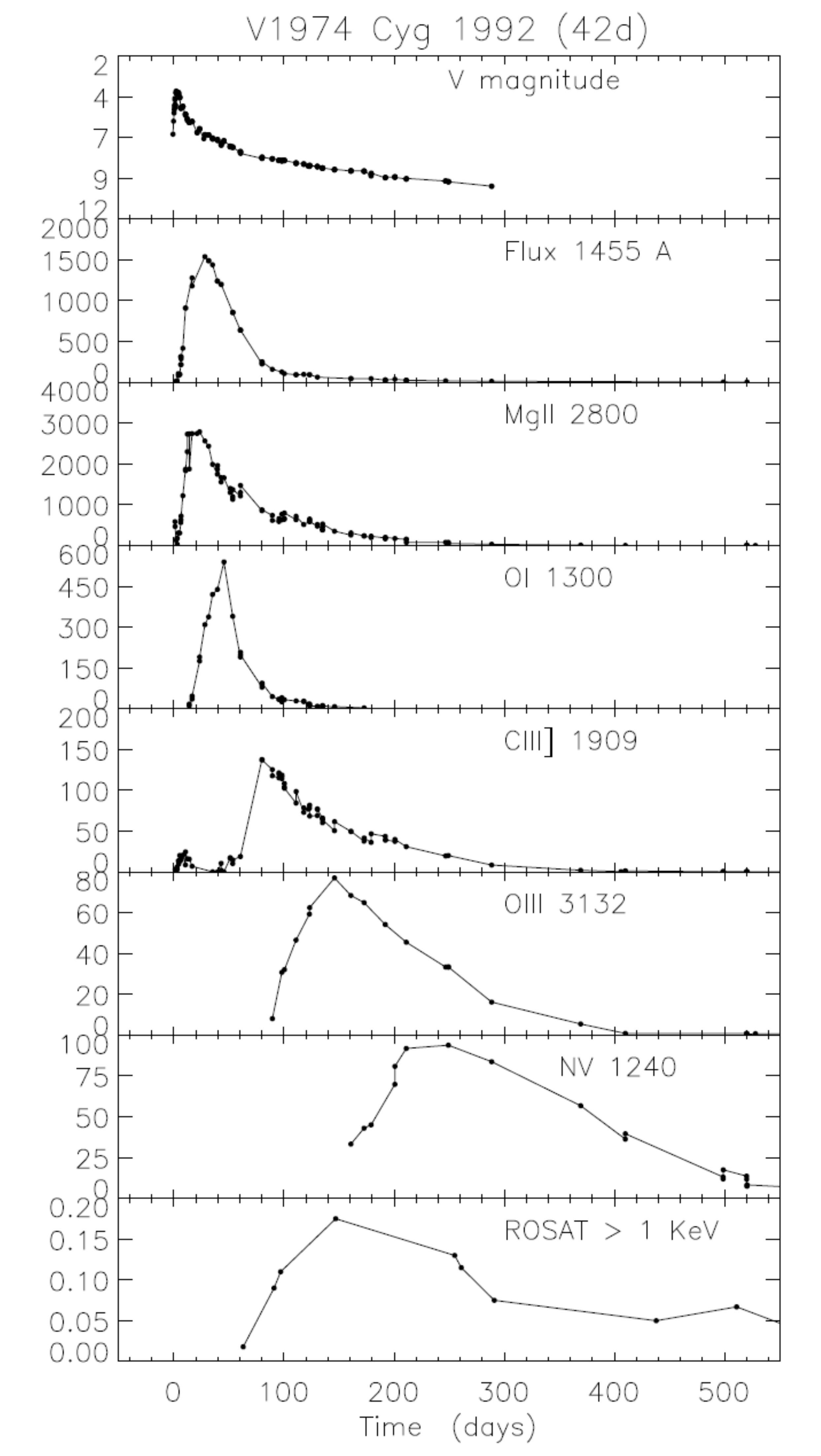, width=10cm}
\parbox{130mm}{\caption[Temporal spectral evolution of V1974 Cyg.]
{Temporal evolution of V1974 Cyg. Figure taken
    from \citet{cassatella04}. Top panel shows optical V band lightcurve, bottom panel
    is ROSAT PSPC X-ray light curve. The rest are IUE lightcurves tracking
    different spectral features of V1974 Cyg. 
\label{fig:v1974}}}
\end{center}
\end{figure}

Figure \ref{fig:nlc} shows the general light-curve of a CNe based on the
original plot published by
\citet{mclaughlin43}. 
The optical 
light variation is characterized by the remarkable initial rise,
typically 6-13 magnitudes in  one or two days, often with a temporary halt
lasting a few hours for fast novae, to a few days for slow novae, before the
final steep rise of about 2 magnitudes to maximum. The early decline is smooth
for all but the slowest novae, where irregular variations are seen. The speed
class of a nova is defined by the rate at which the early decline takes
place. Starting about 3.5 magnitudes below maximum light, through a 2.5
magnitude drop known as the transition region, showing a variety of
behaviors. Faster novae usually continue to decline smoothly, others pass
through a deep, 7-10 magnitude minimum lasting for 2-3 months, after which the
smooth decline resumes. Still others show quasi-periodic brightness variations
with amplitudes 1-1.5 mag, with a period of 5-25 days. 
Infrared
observations show that the flux grows slowly during the initial decline and
increases greatly if there is a deep minimum during the transition
region. This suggests 
dust condensation in the expelled cooling shell causes
the light variation in the transition region. 
The final decline to the
post-nova state 
is usually smooth, with dwarf-nova-like variations in a few cases. 
Ultraviolet and X-ray
observations show that the peak of the flux distribution moves
steadily to shorter wavelengths after maximum light as can be seen for
V1974 Cyg in
Figure \ref{fig:v1974}. 
The early decline in
visual luminosity is due almost entirely to the redistribution of flux.
The total bolometric luminosity remains almost constant at least until the
end of the transition region. 
 
Novae are primarily classified according to the time of their decline from
maximum light. This scheme is also known as the 
MMRD method [maximum magnitude vs.\ rate of
decline, first introduced by \citet{mclaughlin45}]. 
For a review, see \citet{stellarcandles}, and references therein.
The rate of decline is described by the time of fading 2 or 3 magnitudes
following optical maximum luminosity 
over a time interval denoted by $t_2$ and $t_3$, 
\begin{figure}[h]
\begin{center}
  \epsfig{file=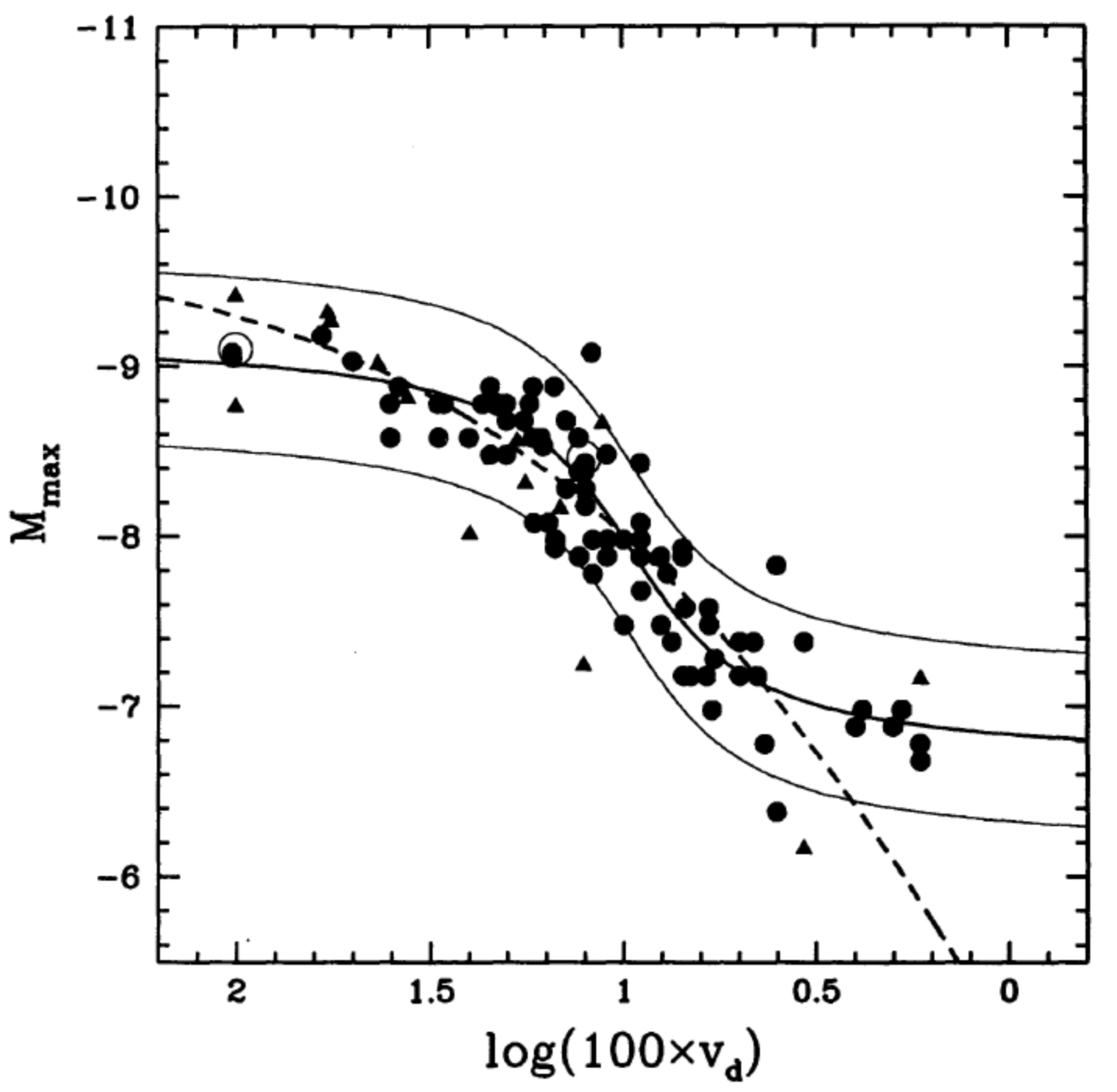, width=8cm}
\parbox{130mm}{\caption[Maximum magnitude versus rate of decline (MMRD) curve
  for CNe.]{The maximum magnitude versus rate of decline (MMRD) curve for
  classical novae. 
The decline rate is expressed as $v_d=2/t_2$ and $v_d=3/t_3$. 
Filled circles represent novae in M31 and triangles in LMC. The solid curves
are the S shaped fit and its $3\sigma$ ranges by \citet{dellavalle95}, dashed
curve represents the theoretical model of \citet{livio92}.\label{fig:mmrd}}}
\end{center}
\end{figure}
respectively. Figure \ref{fig:mmrd} shows the relation between the absolute
magnitude and the rate of decline for novae. 
\begin{table}[h]
\centering
\parbox{130mm}{\caption[Speed classes of classical novae.]{Speed classes of classical novae.\label{tab:speedc}}}
\begin{tabular}{lr@{$-$}l}
\\
\hline\hline
Speed class &\multicolumn{2}{c}{$t_2$ (days)}\\
\hline
Very fast   & \multicolumn{2}{c}{$<10$}\\
Fast        & $11$ &$25$\\
Moderately fast & $26$ &$ 80$\\
Slow        & $81$&$150$\\
Very slow   & $151$&$250$\\
\hline
\end{tabular}
\end{table}
There are 5
speed classes based on the $t_2$ listed in Table \ref{tab:speedc}, although 
these are 
qualitative groups rather than distinct classes because the variation of
the decline rate is continuous. 
The discovery that the rate of decline is
inversely proportional to the absolute maximum brightness made such objects 
valuable in distance measurements. 
Another distance
 measurement technique uses the fact that all novae, almost independent of speed
class, reach $M_V \sim -5.51\pm0.62$ after 15 days of maximum light.
These relationships are useful distance indicators for novae, or
extragalactic systems in which novae occur. The extra value of novae
relative to the bright and well-calibrated Cepheids is that novae, are on
average, brighter by 2 magnitudes than Cepheids of the longest
periods. Moreover, novae can occur in both spiral and elliptical galaxies,
while Cepheids have been observed exclusively in spirals. A further advantage is
that novae have a bimodal luminosity function with peaks at $M_V=-8.8$ and
$M_V=-7.5$ magnitudes. The dip between
 the peaks at $M_V=-8.2\pm0.15$ can be used
for testing the Malmquist-bias\footnote{A selection effect which occurs in flux
  limited samples. The observer sees an increase in luminosity with
  distance due to undetected faint objects. Because in a given intensity
  range it is
  more likely that a faint object remains undetected than a bright one.}, which affects 
the completeness of the novae sample.  
On the other hand, due to
the unpredictable
nature of novae their discovery in other galaxies 
requires a lot of telescope time.   
A new calibration of MMRD with 28 objects using HST data gave the following 
linear fits, 
\citep{dellavalle95}:
\begin{equation}
M_V=(2.54\pm0.35)\times\log{t}_3-(11.99\pm0.56)
\end{equation}
\begin{equation}
M_V=(2.55\pm0.32)\times\log{t}_2-(11.32\pm0.44)
\end{equation}
Although these linear fits proved to be only approximate relations, 
they are still in use because the majority of novae fall close to them. The
real MMRD function is more like an S shaped function, 
as shown in Figure \ref{fig:mmrd}, \citep{dellavalle95}: 
\begin{equation}
M_V=-7.92-0.81\arctan\left(\frac{1.32-\log{t_2}}{0.23}\right)
\end{equation}
A theoretical model calibrated for V1500 Cygni 
by \citet{livio92} relates $t_3$ to the luminosity, or
to the mass of the underlying WD:
\begin{equation}
t_3=51.3\times10^{\frac{M_B+9.76}{10}}\times\left(10^{\frac{M_B+9.76}{15}}-10^{\frac{-M_B-9.76}{15}}\right)^{3/2}
\label{eq:mmrdlum}
\end{equation} 
\begin{equation}
t_3=51.3\left(\frac{M_{WD}}{M_{Ch}}\right)^{-1}\left[\left(\frac{M_{WD}}{M_{Ch}}\right)^{-2/3}-\left(\frac{M_{WD}}{M_{Ch}}\right)^{2/3}\right]^{3/2}
\label{eq:mmrdmass}
\end{equation}
The flattening of the curves at the high luminosity end is real, due to 
super-Eddington luminosity that occurs in 
novae approaching the Chandrasekhar limit. The flattening at
the faint end might
be a selection effect, because 
the magnitude limit was $m_{pg}=19$ for the M31 survey. 
Its physical background is still
under investigation but there is 
some indication that low mass WDs can cause this
effect.     
From equations \ref{eq:mmrdlum} and \ref{eq:mmrdmass} it follows that the 
relationship
between the absolute magnitude and mass of the WD is:
\begin{equation}
M_B\approx-8.3-10\log\frac{M_{WD}}{M_{\odot}}
\end{equation}
Thus, the more massive the WD is, the more luminous the outburst it produces.
The characteristics of a light curve are also reflected in the
spectral evolution.

\section{Spectral evolution}


\begin{figure}[htb]
\begin{center}
  \epsfig{file=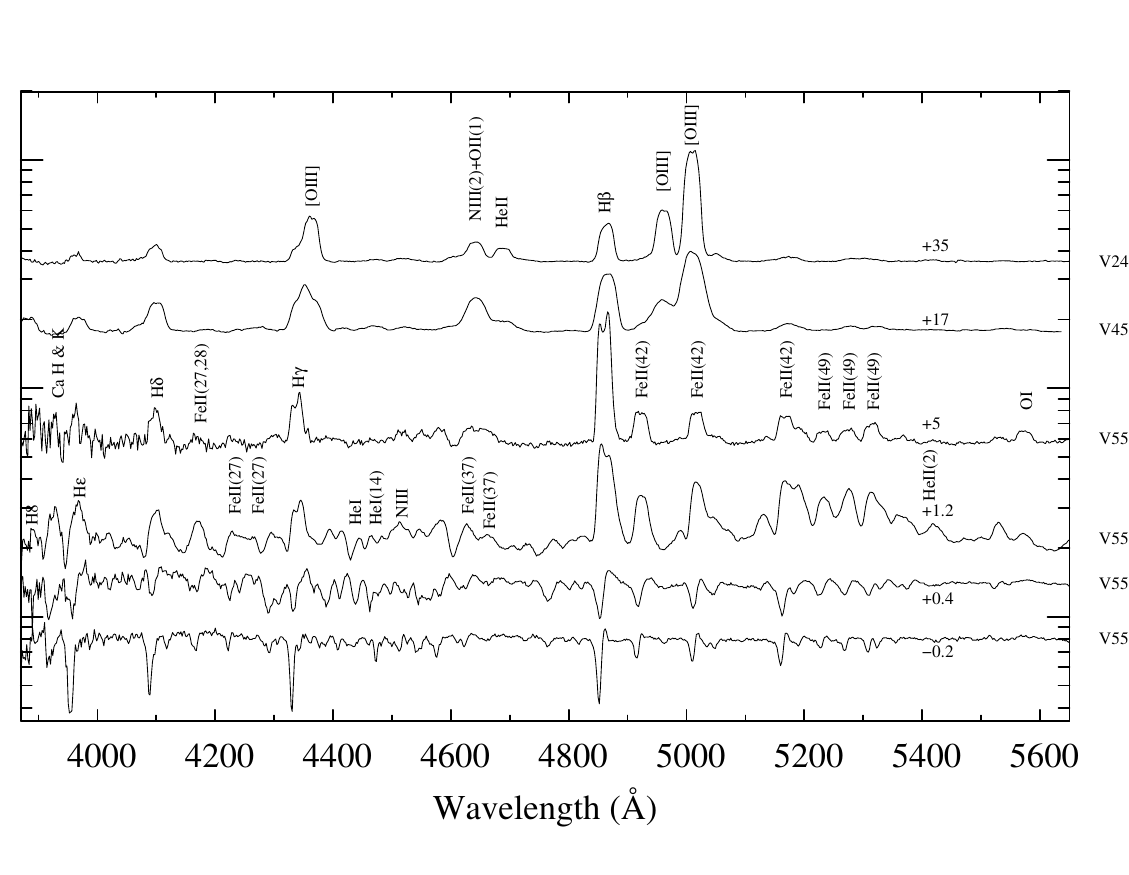, width=11cm}
\parbox{130mm}{\caption[Spectral evolution of the FeII class.]
{The major stages of evolution of some recent FeII class CNe. 
The epoch since
  optical maximum is indicated in days after the names of novae.  
Note the spectra
  are shifted for clarity and plotted on log scale, the offset can be found at
  5400 \AA\ for each novae.  
All spectra were taken with the Ortega 0.8m telescope between April and October
2008. The decline times were: $t_{2, (V459)}=18.3$ d, $t_{2, (V2468)}=10.5$ d and
$t_{2, (V5579)}=6.1$ d.\label{fig:specevol}}}
\end{center}
\end{figure}
The spectral description of classical nova evolution dates back to the
extensive work of
Dean B. McLaughlin in the early 1940's \citep{mclaughlin43}, and was revisited by \citet{williams92}.
The spectral evolution is illustrated from bottom to top in Figure
\ref{fig:specevol}. 
The pre-maximum stage is characterized by the spectrum of a hot star with a
strong continuum, blue-shifted absorption lines and weak or absent emission
lines. The
line shifts indicate lower velocities than after maximum. The
evolution of the blue-shifted absorption line spectrum indicates an expanding
and cooling photosphere. In the final expansion towards maximum, the
photospheric temperature declines rapidly to 9000-4000 K depending on the
speed of the
nova. Spectral types in this stage 
are from B to A, sometimes F. The maximum luminosity stage 
includes the
early decline, with the appearance of the ``principal'' spectrum showing strong
P-Cygni profiles of H and Fe. Blue-shifts range from a few to several
hundred km/s for slow, and around a $10^3$ km/s for fast novae. The principal
spectrum is followed
by the ``diffuse enhanced'' spectrum about 1 or 2 days after maximum with about 
twice as large
blue-shifts and numerous emission lines. 
It is characterized by strong P-Cygni profiles superimposed on a
relatively cool and strong continuum, which starts fading and shifting to the
blue. Strong low-ionization H, CNO and Fe emission lines appear in the
spectrum. The ``Orion''\footnote{Named after the apparent similarity with
  Orion-nebula, which also shows strong emission lines.} 
spectrum appears during the transition phase, characterized
by flare and coronal lines. The continuum is weaker and the shell grows
optically thin as the photosphere recedes towards deeper and hotter
regions. This can be seen in Figure \ref{fig:v1974} as well; 
the spectrophotometric
light curves show increasing ionization stages according to increasing
effective temperature.
\begin{figure}[htb]
\begin{center}
  \epsfig{file=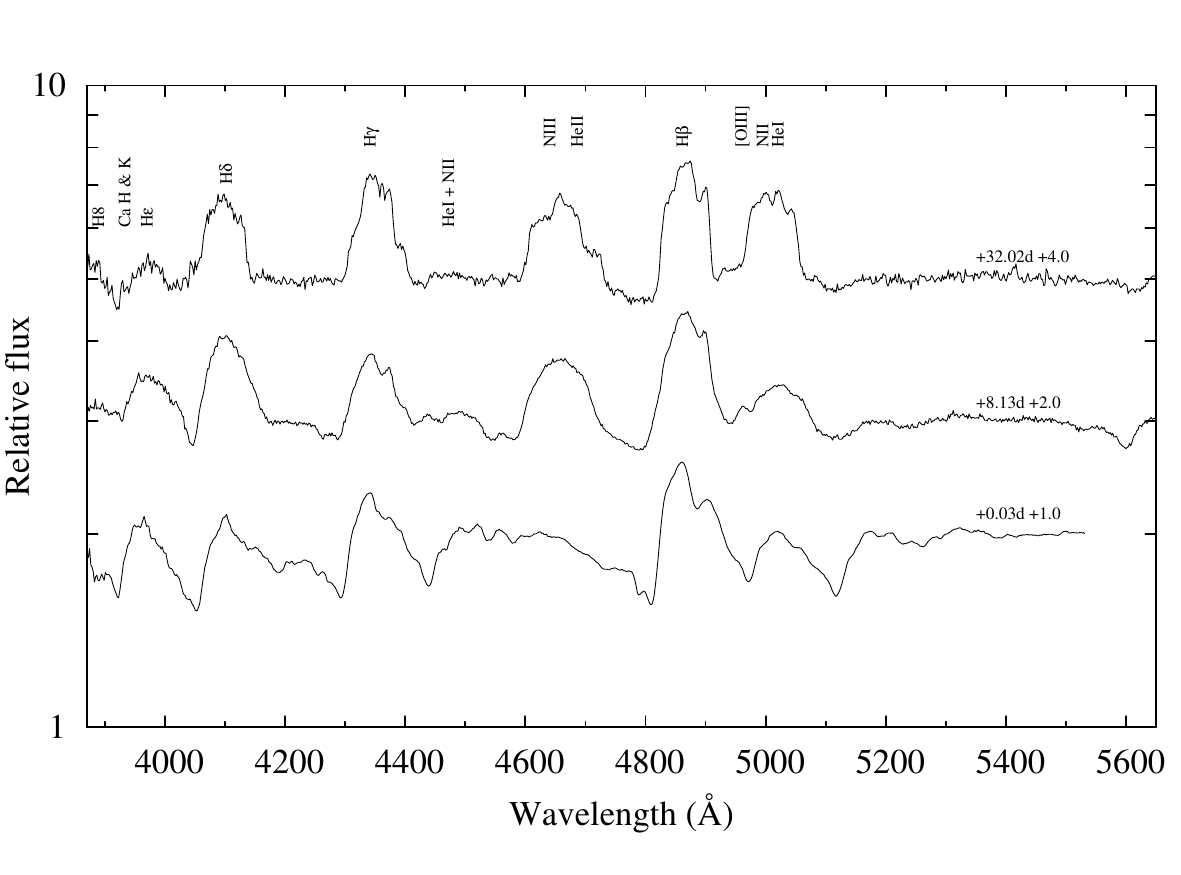, width=11cm}
\parbox{130mm}{\caption[Spectral evolution of the He/N class.]
{Evolution of He/N nova V2491 Cygni in the same spectral range like in 
Figure \ref{fig:specevol}. 
The broad, rounded and jagged line profiles
 of H, He and N typical for these novae are
clearly visible. V2491 Cyg was a very fast nova with $t_2=5.65$ d. Spectra
were taken with the Ortega 0.8m telescope in April and May
2008.\label{fig:specevol2}}}
\end{center}
\end{figure}
The broad emission lines show velocities higher than ever before and 
forbidden lines appear, confirming the thinning atmosphere. 
The photosphere is small and hot, emitting in UV and later in X-rays 
for novae having enough residual envelope mass or high enough
accretion rate to maintain hydrogen burning.
In the last, so called
nebular phase, the forbidden lines strengthen and eventually dominate the
spectrum, which resembles those of planetary nebulae. 
After years the envelope runs out of fuel, the soft X-ray radiation turns off,
the star returns to its pre-nova state and the cycle
starts over again. 

The bimodal behavior that is seen in light curves
is also present in the spectral evolution. \citet{williams92} classified novae
an either FeII or He/N classes (Cerro-Tololo system) 
based on the line properties in their permitted line spectra. 
About 60\% of novae belong to the FeII class. The common feature of
these objects are the strong FeII lines, pronounced P-Cygni profiles, slow
evolution and narrow emission lines showing expansion velocity less than 2500
km/s. Novae shown in Figure \ref{fig:specevol} belong to the FeII
type. The remaining 40\% belong to the 
He/N class. These usually show higher excitation levels, flat, saddle
shaped or jagged line emission line peaks and rapid expansion with velocity
extending to 5000 km/s and fast evolution like in Figure \ref{fig:specevol2}. 
The classification is more complex
in the nebular phase. Some FeII and He/N novae can evolve to Ne novae, forming
a hybrid class. In general the main difference during the nebular phase 
is that FeII novae tend to have strong auroral transitions of O and N, 
while He/N novae either show very weak, quickly fading forbidden lines, 
or [FeX], or [NeIII] - [NeV] forbidden lines. He/N spectra are probably formed
in discrete shells ejected at the maximum of the outburst. The
narrower FeII spectra are likely formed in the subsequent wind.

\section{Degeneracy}

Because WDs are supported by electron degeneracy and this has a major role
in nova explosions as well, a brief introduction to its physics is appropriate.
Under the influence of gravity 
all celestial objects would be in contraction, unless
some other force can counterbalance it. In MS stars the total pressure
consists of the gas
and the radiation pressure which comes 
from the liberated energy by hydrogen
burning. By neglecting lower order effects like rotational and magnetic
pressure the total pressure can be written as:
\begin{equation}
  P_{\rm total} = P_{\rm gas} + P_{\rm rad}
\label{eq1} 
\end{equation}
\begin{equation}
 P_{\rm total} = NkT + \frac{4\sigma}{c}\frac{T^4}{3}
\label{eq2} 
\end{equation}
Where $\sigma$ is the Stephan-Boltzmann, $k$ is the Boltzmann
constant, $N$ is the total number density and $c$ is the speed of
light. $T$ is the temperature of the star and $P$ is the pressure.

\begin{figure}[!h]
\begin{center}
\epsfig{file=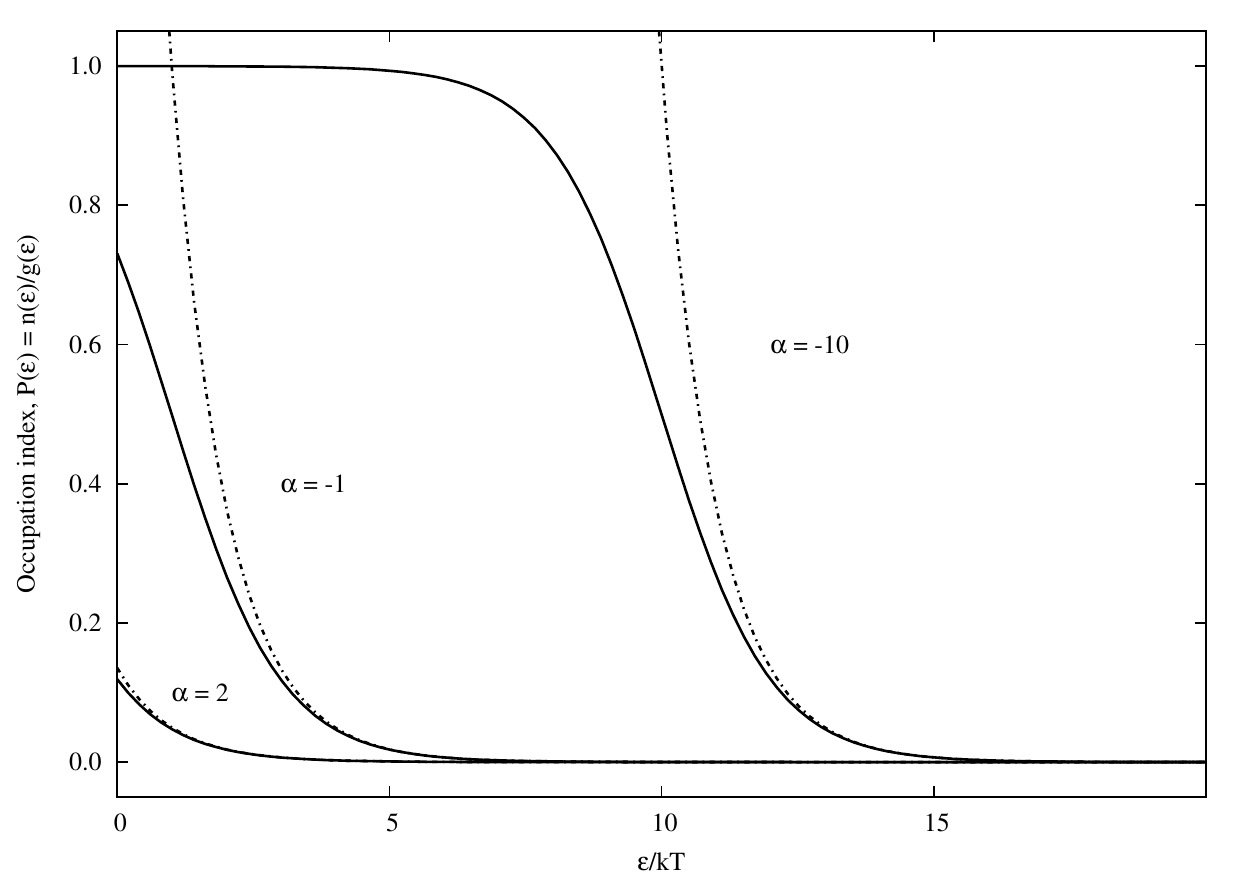, width=10cm}
\parbox{130mm}{\caption[Fermi-Dirac distribution.]
{Fermi-Dirac distributions for different degeneracy 
parameters
  ($\alpha$) are showed by solid lines. Maxvell-Boltzmann distributions shown
  by dashed-dotted lines, for low occupation indices the two statistics are
  identical.\label{fig:degeneracy}}}
\end{center}
\end{figure}

When hydrogen is depleted in the core, the balance between
gas, radiation pressure and gravity ends. 
The core starts to collapse, releasing
part of its gravitational energy which initiates hydrogen shell burning around
the core. The increased energy production inflates the star to a red giant
and initiates large scale mass loss. 
Meanwhile the core is still unbalanced, contracting,
heating up and getting denser. The particle energy distribution in the core is:
\begin{equation}
  n(\epsilon)=\frac{g(\epsilon)}{e^{\alpha+\epsilon/kT}+\beta} 
 \label{eq1}
 \end{equation}
\begin{center}
\[ \alpha = \left\{ 
\begin{array}{cl}
    0 & $Bose-Einstein statistics, integral spin;$\\
    -\frac{h^2}{2mkT}\left(\frac{3N_e}{8\pi}\right)^{\frac{2}{3}}  & 
$Fermi-Dirac statistics, half-integer spin.$
\end{array} \right. \] 
\end{center}
\begin{center}
\[ \beta = \left\{
\begin{array}{rl}
    -1 & $Bose-Einstein statistics, integral spin;$\\
    1  & $Fermi-Dirac statistics, half-integer spin.$
  \end{array} \right. \]
\end{center}
Where $g(\epsilon)$ is the number density of possible states at energy
$\epsilon$, $\alpha$ is the degeneracy parameter $\beta$ is the
statistical parameter and $N_e$ is the electron density. 
In Figure \ref{fig:degeneracy} the occupation 
index $n(\epsilon)/g(\epsilon)$ is
plotted against electron energy at different degeneracy parameters. 
For large electron 
energies the \mbox{Fermi-Dirac distribution} shows a maximum 
that is a
consequence of the Pauli exclusion principle. The low energy states are all
occupied, electrons require more and more energy to be raised to empty 
high energy
states and this introduces a new, electron degeneracy pressure. 
Also shown in Figure \ref{fig:degeneracy} are
the classical Maxwell-Boltzmann statistics for each
degeneracy parameter for low occupation numbers matching with
Fermi-Dirac statistics. This is analogous to the Newtonian and
the relativistic theory of gravity. The Maxwell-Boltzmann statistics describe
the system well for low occupation numbers 
and only a limiting case of the more general
Fermi-Dirac distribution.
At the density:
\begin{equation}
 \frac{\rho}{\mu_e}=\frac{8\pi{M_{\mu}}}{3}\frac{(5m_e{k})^{3/2}}{h^3}T^{3/2}
 \label{eq:degdensity}
\end{equation}
the pressure of degenerate electrons is:
\begin{equation}
  P_{e}=\frac{h^2}{5m}\left(\frac{3}{8\pi}\right)^{\frac{2}{3}}N_e^{\frac{5}{3}}
  \hspace{5mm}\left(=P_{total}\right)
 \label{eq:degpress}.
 \end{equation}
This can stop the contraction of the core. Thus, WDs are the 
exhausted stellar cores 
supported by the pressure of degenerate electrons. 
Note the degenerate pressure does not depend on temperature, just on electron
density. As a consequence, in degenerate stars the radius is largely 
independent of temperature. Temperature affects the outer, normal gas
envelope of these stars. 
\begin{figure}[!htb]
\begin{center}
  \epsfig{file=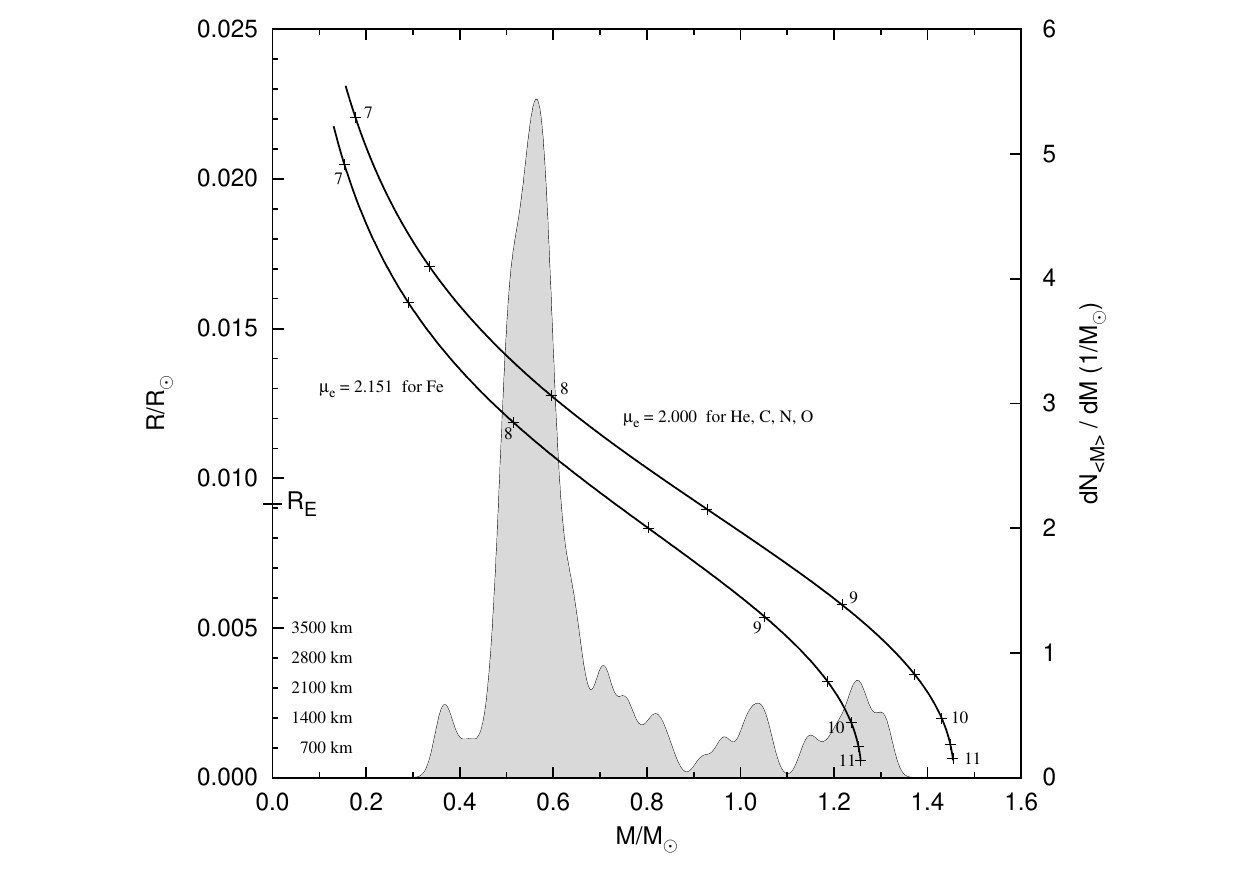, width=13cm}
\parbox{130mm}{\caption[The mass--radius relationship and mass distribution of
  WDs]{Zero temperature mass-radius relations for different electron molecular
    weights $\mu_e$. Surface gravities ($\log g$) are labeled along the curves 
in cm/s$^2$ units. The shaded area 
shows the observed mass distribution by \citet{vennes99}. R$_{\rm E}$
indicates the radius of Earth. For the most massive WDs radii are also shown
in km.\label{fig:massradius}}}
\end{center}
\end{figure}

An interesting consequence of degeneracy is the WD mass-radius relation. 
As the degenerate pressure is a function of the density, the higher the
mass of a white dwarf the smaller its radius. In Figure \ref{fig:massradius} 
the boundaries of stable regions were calculated with a 4th order Runge-Kutta 
numerical integration for relativistic cores consisting of 
light metals and iron. Also shown is the observed
mass distribution by \citet{vennes99} based on 141 EUV/soft X-ray selected
white dwarfs. The mass limit of WDs depends on the core composition, in solar
units: 
\begin{equation}
M=\frac{5.87}{\mu_e^2}M_\odot
\label{eq:chandrasekhar}
\end{equation}
where $\mu_e$ is the is the mean molecular mass per electron. This maximum mass
was first realized by Subrahmanyan Chandrasekhar in the early 1930's; this
work earned him the Nobel Prize in physics in 1983 and in his
honor was named the
 Chandrasekhar limit, \citep{chandrasekhar35}.   

\section{The outburst}

The requirements for a pre-nova are: a close binary system in
which the secondary fills its Roche lobe and mass transfer rates 
between $10^{-8} - 10^{-11}$ \solarmass$/yr$ to the WD. 
At this accretion rate the 
hydrogen-rich material from the secondary accumulates in a layer on the WD.
As the mass of the WD increases, so does the central 
density and the density 
at the boundary of the core and envelope. Eventually this
reaches the critical density for relativistic 
degeneracy. Theoretical studies showed that
the accreted layer grows until a temperature of $>10^7$ K and a pressure of
$>10^{19}$ dyne/cm$^2$ is reached. 
More massive WDs require less material in their envelope to reach
this critical density. Thus, massive WDs are expected to produce more
frequent nova outbursts than their lower mass counterparts. The transferred
material also carries extra heat to the envelope from released potential
energy. 
With this extra heat and the hydrogen-rich degenerated layer in the envelope,
the explosive is loaded in the gun. 
The core is nearly isothermal due to its high conductivity and the excess
heat is expected to be distributed evenly throughout the surface layers.   

When the temperature reaches the point that hydrogen ignites, the envelope 
suffers a thermonuclear
runaway. Because the temperature goes up exponentially, the reaction rate
increases suddenly and causes a chain reaction. 
Although the degenerate equation of state (Equation \ref{eq:degpress}) does not
depend on temperature, the degeneracy parameter does. At very high
temperatures $\alpha$ reaches small negative values, degeneracy is lifted and 
the material goes back 
to normal gas. Its temperature dependence turns on and the outer envelope
expands adiabatically
blowing off a significant part of the previously accreted material.  

During the outburst the atmosphere expands and cools adiabatically. The
nova reaches its optical maximum at this phase. The WD gains an atmosphere
like a red giant, 
reaching out nearly 100 R$_\odot$ and forming a common envelope
around the stars, 
but it can not support this for long. 
Due to the strong radiation and dynamical friction with 
the secondary the atmosphere loses mass and the photosphere recedes. As we can
see
deeper and deeper into the atmosphere where temperature is higher, 
the peak of
the energy distribution gradually shifts towards shorter wavelengths. 
Meanwhile there is stable
hydrogen fusion at a constant rate (constant bolometric luminosity phase) 
on the surface of the WD which can be
supported by further accretion from the secondary. The spectral
analysis of this static hot WD atmosphere is the scope of this work. 

\section{Supersoft X-ray Sources}\label{SSS}

Supersoft X-ray 
sources are astronomical sources of low energy X-rays. The peak of
their spectral energy distribution (SED) 
is between 5 and 120 \AA. Normal stars burn hydrogen in their
cores producing X-rays and gamma rays. By the time this radiation reaches
the surface of the star, it has become 
visible light. However, in the mid eighties
a new class of object was discovered with the Einstein Observatory: 
WDs which have hydrogen-burning
close to their surfaces. These hot and dense stars in cataclysmic binaries can
get hydrogen-rich material from their companion as a result of stellar
evolution. Burning of this layer keeps the surface temperature at a few
hundred thousand Kelvins and thus is the source of the soft X-rays. 
Depending on the mass transfer rate there are two main classes of
supersoft sources. If the mass transfer ($\dot{M}$) rate is around 
$10^{-7}\ {\rm
  M}_{\odot}\ yr^{-1}$, it is possible
that the ongoing burning is continuously supported 
by fresh material. These stars are referred as Close Binary Supersoft Sources
(CBSS). In this case the
hydrogen-burning roughly constant. The mass transfer rate must not be too
hig; 
at $\dot{M} \ge 4\times10^{-7}\ {\rm M}_{\odot}\ yr^{-1}$ 
the released gravitational energy can stop further accretion, regulating the
mass transfer and luminosity. For even higher accretion rates a common
envelope forms around the stars.

If the accretion rate is low, on the order of 
$10^{-8} - 10^{-11}\ {\rm M}_{\odot}\ yr^{-1}$, 
the burning is episodic. 
In this case, a hydrogen-rich envelope builds up on the surface of
the WD and it becomes a classical nova system. The recurrence time of the nova
depends on the mass transfer rate and the WD's mass. The supersoft stage in
this case starts with the nova event and its duration strongly depends on the
WD mass and the chemical composition of the accreted material.

\begin{figure}[!h]
\begin{center}
\epsfig{file=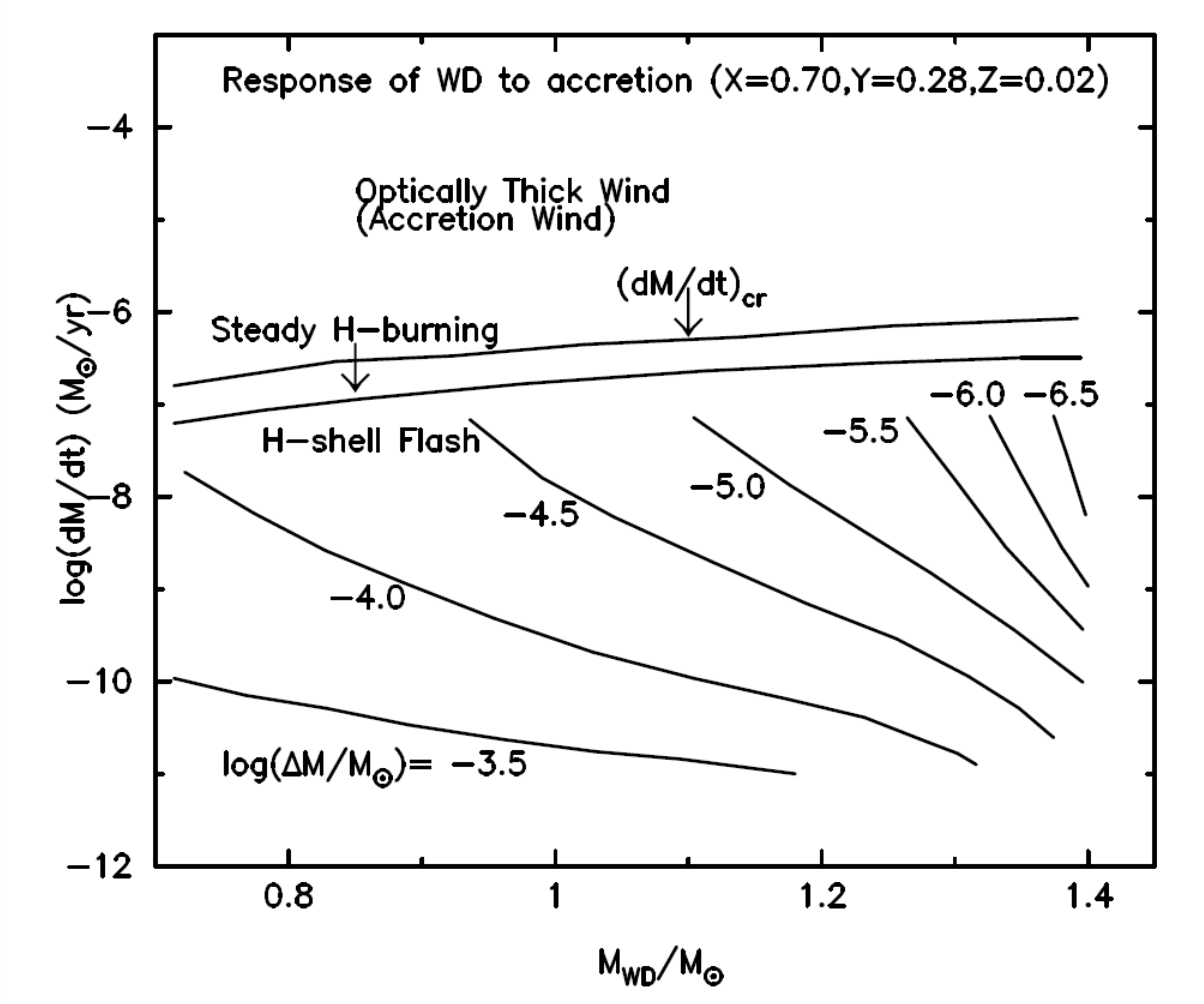, width=12cm}
\parbox{130mm}{\caption[Response of WDs to mass accretion.]{Response of WDs
    for mass accretion in the WD mass and mass-accretion rate plane. Figure
    is taken from \citet{kato10}.\label{fig:kato}}}
\end{center}
\end{figure}
Figure \ref{fig:kato} summarizes the response of WDs for accretion as
function of WD mass and accretion rate, \citep{kato10}.  
The abscissa shows the WD mass in solar mass and the ordinate is the 
logarithm of the accretion rate ($\dot{M}$) in M$_\odot$ yr$^{-1}$. Next to
the curves are the envelope masses necessary for a nova outburst 
(hydrogen-shell flash). 
It can be seen
that more massive WDs require less material for an outburst. Also
shown that there is a critical rate ($\dot{M}$) almost independent from WD mass
at which steady hydrogen-burning occurs. Above this rate an optically thick
wind forms.
 
\citet{ogelman83} discovered that soft X-ray radiation came from the 
classical
nova GQ Mus (Nova Mus 1983) in April 1984, only two months before
\citet{macdonald85} published the theoretical background and 
detectability of such radiation by EXOSAT, the only X-ray satellite that
time.

The fate of these objects is also interesting. If a ONeMg WD gains enough mass
it can collapse to a neutron star. If the secondary star stops donating
material, 
hydrogen-burning turns off and the WD returns to the cooling sequence. 
When a CO WD reaches the Chandrasekhar limit the star explodes
completely in a type Ia supernova (SNIa). The carbon core of the WD converts to
nickel instantaneously, which decays to cobalt and iron in a few hundred days.  

There is no clear distinction between CBSS and classical novae that emit
supersoft radiation. It is possible that they are similar systems and 
only appear
to be in a different evolutionary stage. However, while CBSS show constant, 
long term hydrogen-burning, CNe burn hydrogen for only a few months, and in
exceptional cases, a few years (GQ Mus, V723 Cas). 

\chapter{Modeling Hot Atmospheres}\label{chap:atmosphere}

\section{Model Atmospheres}

The calculation of model atmospheres 
involves the solution of the equations
governing the stellar structure and the flow of radiation. 
One must make some simplifying assumptions to pursue
self-consistent solutions.
For example, we assume that the stellar photosphere is thin compared
to the radius of the star (WD). In plane-parallel atmospheres this
approximation is valid and the surface gravity
can be considered constant. 
We assume the atmosphere is in steady-state,
neglecting the effects of pulsation, 
shock, stellar wind, magnetic and tidal heating. It is 
in hydrostatic equilibrium, such that the pressure balances the gravity
in every layer of the star. Also, the atmosphere is in radiative equilibrium,
energy transportation is by convection and conduction or other means can be 
neglected,
\citep{mihalas70}.

Computationally the problem is addressed in two 
basic iterative steps. First, one finds the radiation field, determining 
the flow of radiation from the interior. Second,
 keeping the energy conserved (radiative equilibrium), one determines
the structure and
finds the corrections to the radiation field ($\rho$, $T$, $P$). Repeating
these steps iteratively until self-consistency is achieved gives a solution. 
The model atmosphere is given by the surface 
gravity $g$, the effective temperature
$T_{\rm eff}$ and the chemical composition of the star.
Assuming local thermodynamic equilibrium (LTE) makes 
the calculation much easier,
because the total line opacity depends only on the local
temperature, electron density and abundance. 
However, in the hot atmospheres of supersoft
sources the photon mean free path is large in the extended low density 
atmosphere and this introduces non-LTE effects. 
The hot deep interior of the atmosphere
is coupled with outer layers, changing their 
atomic level populations and ionization
balance from the LTE values, \citep{hl95}.   
 
\begin{figure}[!h]
\begin{center}
\epsfig{file=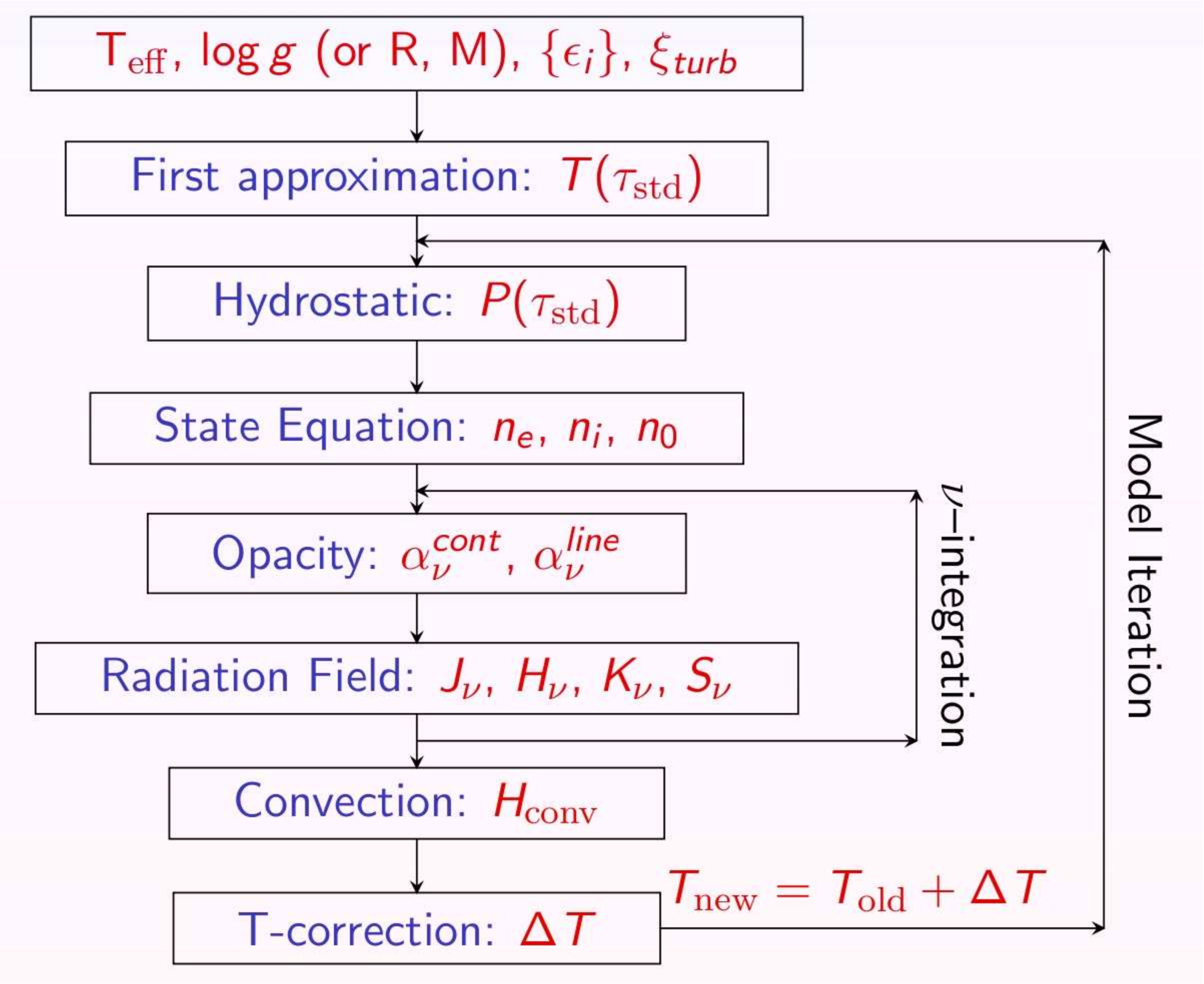, width=14.2cm}
\parbox{130mm}{\caption{The main steps of model atmosphere
    calculation.\label{fig:flowchart}}}
\end{center}
\end{figure} 
In general, a model calculation proceeds as shown in
Figure \ref{fig:flowchart}. 
The 
effective temperature ($T_{\rm eff}$), the surface
gravity ($\log g$) and the atomic input data defines the atmosphere. In Figure
\ref{fig:flowchart} at the first step
$\epsilon_i$ means not only the atomic energy levels, but also the data for 
bound-free and bound-bound transitions and abundances of all species. 
The microturbulent velocity is $\xi_{turb}$. The first step is to estimate
the temperature at the standard depth that is usually related to the effective
temperature ($T_{\rm std}=0.75T_{\rm eff}$). 
One then solves the hydrostatic equilibrium
equation:
\begin{equation}
\frac{{\deriv}P}{{\deriv}m}=g
\label{eq:hydro}
\end{equation} 
where $P$ is the total pressure, involving the ideal 
gas, radiation and microturbulent
pressures, d$m=-{\rho}{\deriv}r$ is the Lagrangian mass,
$\rho$ is the density and ${\deriv}r$ is the geometrical vertical distance
in the atmosphere. The gravitational acceleration, $g=GM_{\star}/R_{\star}^2$ 
is assumed to
be constant throughout the atmosphere, 
$M_{\star}$ and $R_{\star}$ are the mass and radius of the star,
respectively. $G$ is the gravitational constant. Level populations are
determined by the statistical equilibrium equation, also called rate equation:
\begin{equation}
n_i\sum_{j{\ne}i}(R_{ij}+C_{ij})=\sum_{j{\ne}i}n_j(R_{ji}+C_{ji})
\label{eq:stat}
\end{equation}
where $n_i$ is the population of level $i$, and $R_{ij}$ and $C_{ij}$ are the
radiative and collisional rate, respectively, for transitions from level $i$ to
level $j$. Radiative rates depend on radiation intensity and collisional rates
are functions of local temperature and density. The set of rate equations for
all levels of an atom would be a linearly dependent system. Therefore one
equation of each set has to be replaced by the total particle 
number conservation
equation or abundance definition equation $\sum_in_i=N_{atom}$, summed over
all levels of all ions of a given element. The radiative equilibrium equation 
expresses energy
conservation, i.e. 
the absorbed and emitted energy must be the same everywhere in 
the atmosphere. 
In its most useful form this is, \citet{hl95}:
\begin{equation}
\alpha\left[\int_0^{\infty}(\kappa_{\nu}J_{\nu}-\eta_{\nu}){\deriv}\nu\right]+\beta\left[\int_0^{\infty}\frac{{\deriv}(f_{\nu}J_{\nu})}{{\deriv}\tau_{\nu}}{\deriv}{\nu}-\frac{\sigma}{4{\pi}}T_{\rm
  eff}^{4}\right]=0
\label{eq:radeq} 
\end{equation}
where $\sigma$ is the Stefan-Boltzmann constant, and 
$\kappa_{\nu}$ and $\eta_{\nu}$ are the thermal absorption and emission
coefficients, respectively. Scattering coefficients cancel out when coherent
scattering may be assumed. 
The two terms of the equation are equivalent; the first
is the integral form and the second is the differential. At large optical
depths the differential form shows better numerical stability and higher
accuracy, while at small optical depths the integral form is preferable. The
two empirical parameters $\alpha$ and $\beta$ can have values between 0 and 1,
and this 
sets the linear combination of the two forms. The flux is expressed by
the variable Eddington factor $f_{\nu}=K_{\nu}/J_{\nu}$, where $K_{\nu}$ is
the flux, or second moment of the specific intensity and $J_{\nu}$ is the mean intensity. For
grey atmospheres the
radiative equilibrium equation is
solved with frequency
independent opacities. 
The propagation of energy in the form of photons is expressed by the radiative
transfer equation:
\begin{equation}
\frac{{\deriv}^2(f_{\nu}J_{\nu})}{{\deriv}\tau_{\nu}^2}=J_{\nu}-S_{\nu}   
\label{eq:transfer}
\end{equation}
where $\tau_{\nu}$ is the monochromatic optical depth. The Eddington
factor $f_{\nu}$ is not known but 
must be calculated by a separate set of formal
solutions for the specific intensity, and iteratively updated using the current
optical depth $T_\nu$ and source function, 
$S_{\nu}\equiv\eta_{\nu}/\kappa_{\nu}$.   
The global electric neutrality of the medium is expressed by the charge
conservation equation:
\begin{equation}
\sum_in_iZ_i-n_e=0
\label{eq:chargeconserv}
\end{equation}
where $Z_i$ is the charge of level $i$ (0 for neutrals, 1 for singly ionized,
etc.) and $n_e$ is the electron density. 

The structural equations must be solved simulaneously at different layers of
the atmosphere. This is done by the Hybrid CL/ALI method in TLUSTY. Before
explaining these numerical methods, we first introduce the equations for
LTE. Assuming LTE gives the starting values for non-LTE calculations. LTE is
valid deep in the atmosphere, where due to high pressure, the free optical
path of photons is small. In such an environment the level populations are
determined by local conditions. 
The distribution of atoms among their bound
states depends only on the local 
temperature $T$. This is expressed by the Boltzmann-equation:
\begin{equation}
\frac{N_{i,j}}{N_{0,j}}=\frac{g_i}{g_j}e^{-\chi/kT}
\label{eq:boltzmann}
\end{equation}
The distribution of ions depends on the temperature $T$
and electron density $n_e$. 
This is expressed by the Saha-Equation. In general the
relative ionization fraction of any two states is:
\begin{equation}
\frac{N_{j+1}n_e}{N_j}=\left(\frac{2\pi{m}kT}{h^2}\right)^{3/2}\frac{2U_{j+1}(T)}{U_j(T)}e^{-\chi_I/kT}
\label{eq:saha}
\end{equation}
where 
\begin{equation}
U_j(T)=\sum_ig_ie^{\chi_I/kT}
\label{eq:partition}
\end{equation}
is the partition function, and $j$ is the degree of ionization. 
These equations can replace the rate equations
(\ref{eq:stat}) and the Planck function can be used for the source function
in the transfer equation (\ref{eq:transfer}):
\begin{equation}
S_{\nu}=B_{\nu, T}
=\frac{2 h\nu^{3}}{c^2}\frac{1}{ e^{\frac{h\nu}{kT}}-1}\hspace{5mm}
\label{eq:ltetransfer}
\end{equation}

In general, $T(\tau)$ does not satisfy the equations and one needs a correction
term, $\Delta{T}$ for the temperature. 
In this way, 
the whole problem is reduced to finding the
correct temperature structure. The correction 
term $\Delta{T}$ makes use of the difference between the 
computed values and the integrated flux $F$, as well as 
the change of total flux with
optical depth ${\rm d}F/{\rm d}\tau$ to force radiative equilibrium. Writing
the temperature correction as: 
\begin{equation}
T(\tau) = T_0(\tau)+\Delta{T}(\tau)
\label{eq:tempcorr1}
\end{equation}
and using it in the radiative equilibrium:
\begin{equation}
\int^{\infty}_0k_{\nu}J_{\nu}{\rm d}\nu=\int^{\infty}_0k_{\nu}S_{\nu}(T_0+\Delta{T}){\rm d}\nu
\label{eq:radiativeeq}
\end{equation}
the temperature correction has the form:
\begin{equation}
\Delta{T} =
\frac{\int^{\infty}_0\kappa_{\nu}[J_{\nu}-B_{\nu}(T_0)]{\rm
    d}\nu}{\int^{\infty}_0\kappa_{\nu}\frac{\partial{B_{\nu}}}{\partial{T}}{\rm
  d}\nu}
\label{eq:tempcorr2}
\end{equation}
With the altered temperature in the next iteration, 
and repeating the correction
until the difference is small enough one can acheive a self-consistent
solution. 


\section{TLUSTY and SYNSPEC}

TLUSTY\footnote{\protect\url{http://nova.astro.umd.edu/index.html}}
\citep{tlusty200} is a program for calculating plane-parallel, 
horizontally homogeneous model stellar atmospheres in radiative and 
hydrostatic equilibrium following the steps in Figure \ref{fig:flowchart}. 
For a detailed description of the code see 
\citet{hubeny88} and \citet{hl95}. It solves the radiative transfer,
hydrostatic, radiative and statistical equilibrium, charge and particle
conservation equations using the Hybrid CL/ALI methods, 
a combination of the complete linearization (CL) and
the Accelerated Lambda Iteration (ALI) methods.
Beyond LTE, TLUSTY is also capable of calculating 
non-LTE models for a set of occupation numbers of selected atomic species and
energy 
levels, and allows for convection. Starting with version 200, TLUSTY treats 
NLTE metal line blanketing with full consistency either using Opacity
Distribution Functions (ODF) or Opacity Sampling (OS), (see
\citet{hl95}).
\begin{singlespace}
\begin{table}[!h]
\begin{center}
\caption{First and second blocks of TLUSTY input.\label{tab:block1}}
\begin{small}
\begin{tabular}{l}
\\
\hline
\verb,700000.0 8.6       ! TEFF; GRAV,\\
\verb, F  F              ! LTE;  LTGRAY,\\
\verb, 'flags'           ! name of file containing non-standard flags,\\
\verb,*,\\ 
\verb,* frequencies,\\
\verb,*,\\
\verb, 500               ! NFREAD,\\
\verb,*,\\
\hline
\end{tabular}
\end{small}
\end{center}
\end{table}
\end{singlespace}
An example TLUSTY input can be seen in Tables \ref{tab:block1}, \ref{tab:block2}
and \ref{tab:block3}. This file
can be separated into four blocks. The first block 
defines $T_{\rm eff}$ and $\log{g}$ of the atmosphere along with two 
switches that control whether LTE or non-LTE is preferred, and either grey or 
frequency-dependent model calculation is performed. 
In the third line one can pass an optional flag
file to the program by which numerous (140) 
non-standard parameters can be changed. These parameters control additional
physics and set different treatments of various processes and numerical
methods in the calculation. The second input block is only one number, NFREAD.
The program sets up continuum frequencies automatically and by default, two
frequencies near discontinuities corresponding to the bound-free transitions
from all explicit levels, plus NFREAD frequency points in between. 
\begin{singlespace}
\begin{table}[!h]
\begin{center}
\caption{Third block of TLUSTY input: explicit atoms.\label{tab:block2}}
\begin{small}
\begin{tabular}{l}
\\
\hline
\verb,* data for atoms   ,\\
\verb,*,\\
\verb, 10                ! NATOMS,\\
\verb,* mode abn	 modpf,\\
\verb,    2   1.	   1    ! H,\\
\verb,    2   1.e-1 1    ! He,\\
\verb,    0   0	    0 	  ! Li,\\
\verb,    0   0	    0 	  ! Be,\\
\verb,    0   0	    0 	  ! B,\\
\verb,    2   1.e-2 1    ! C,\\
\verb,    2   1.e-2 1    ! N,\\
\verb,    2   1.e-2 1    ! O,\\
\verb,    0	  0	    0 	  ! F,\\
\verb,    2   1.e-3 1    ! Ne,\\
\verb,*    ,\\
\verb,*,\\
\hline
\end{tabular}
\end{small}
\end{center}
\end{table}
\end{singlespace}
The third block is shown in Table \ref{tab:block2}. 
Here one can pass the list of atomic species included in the
model. NATOMS is the highest atomic number considered in the model. MODE 
can have integer values between 0 and 2. If 0, the given atom is not
considered at all. 
By setting it equal to 1, the element is taken into account in the
particle and charge conservation equations, but not in line transitions. For
MODE=2, the element is fully considered. ABN is the abundance. In
this form it corresponds to the number ratio to the reference element,
$N_{\rm elem}/N_{\rm H}$. 
If ABN is coded as negative, it corresponds to the relative
solar abundance. MODPF is the mode of the evaluation of the partition
function: 1 means to use Opacity Project ionization fraction tables, 0 is
for standard evaluation (see Traving et al. 1966).

\begin{singlespace}
\begin{table}[!h]
\begin{center}
\caption{Fourth block of TLUSTY input: explicit ions.\label{tab:block3}}
\begin{small}
\begin{tabular}{l}
\\
\hline
\verb,* data for ions,\\
\verb,*,\\
\verb,* iat  iz   nlevs  ilast ilvlin  nonstd typion  filei,\\
\verb,*,\\
\verb,   1    0     9      0     90      0    ' H 1' 'data/h1x.dat',\\
\verb,   1    1     1      1      0      0    ' H 2' ' ',\\
\verb,   2    1    20      0     90      0    'He 2' 'data/he2x.dat',\\
\verb,   2    2     1      1      0      0    'He 3' ' ',\\
\verb,   6    4     5      0     90      0    ' C 5' 'data/c5.dat',\\
\verb,   6    5    10      0     90      0    ' C 6' 'data/c6.dat',\\
\verb,   6    6     1      1      0      0    ' C 7' ' ',\\
\verb,   7    4     2      0     90      0    ' N 5' 'data/n5.dat',\\
\verb,   7    5     9      0     90      0    ' N 6' 'data/n6.dat',\\
\verb,   7    6    10      0     90      0    ' N 7' 'data/n7.dat',\\
\verb,   7    7     1      1      0      0    ' N 8' ' ',\\
\verb,   8    5     5      0     90      0    ' O 6' 'data/o6.dat',\\
\verb,   8    6     9      0     90      0    ' O 7' 'data/o7.dat',\\
\verb,   8    7    10      0     90      0    ' O 8' 'data/o8.dat',\\
\verb,   8    8     1      1      0      0    ' O 9' ' ',\\
\verb,  10    6     1      0      0      0    'Ne 7' 'data/ne7.dat',\\
\verb,  10    7     5      0     90      0    'Ne 8' 'data/ne8.dat',\\
\verb,  10    8     9      0     90      0    'Ne 9' 'data/ne9.dat',\\
\verb,  10    9    10      0     90      0    'Ne10' 'data/ne10.dat',\\
\verb,  10   10     1      1      0      0    'Ne11' ' ',\\
\verb,   0    0     0     -1      0      0    '    ' ' ',\\
\verb,*,\\
\verb,* end,\\
\hline
\end{tabular}
\end{small}
\end{center}
\end{table}
\end{singlespace}

In the last block (\ref{tab:block3}) 
one can set the data for ions. One line for each
ionization level is considered in the model. IAT and IZ are the atomic 
number of the parent element and the charge of the ion, respectively. 
NLEVS is the
number of atomic energy levels to be 
considered explicitly. The number of explicit
levels can be lower than the energy levels in 
the model atom file. In this case only the lowest NLEVS levels will be
included in the models. ILAST indicates
if the given ion is the highest considered ionization degree of the given
element. ILVIN
changes the line treatment globally, 
all transitions related to energy levels with 
relative index
 smaller than ILVIN will be calculated in detailed radiative
 balance. This is used to calculate continuum-only models and neglects
 spectral lines. 
NONSTD is a flag, defining non-standard parameters for 
individual
 ions. These parameters must follow immediately after the block of the given 
ion. TYPION
 is a 10-character constant for ion identification. FILEI is the path to
 model atom 
data files. These data files were built in this work and will be described in
more detail in
Chapter \ref{chap:modatom}.

\section*{Complete linearization}

Equations \ref{eq:hydro} -- \ref{eq:chargeconserv} are solved at
discrete depths in the atmosphere and in numerous frequency points. These
frequency points are carefully 
selected to represent the most important parts of the
spectrum such as ionization edges, line centers and the wings of strong
lines. A certain number of frequency points are also added equidistantly 
between these points. 
This yields a set of highly-coupled, non-linear equations. Complete
linearization or CL, (Auer and Mihalas, 1969), 
was the first efficient and powerful
method used 
to solve the structural equations simultaneously. The physical state of
the atmosphere is described by a set of vectors $\psi_d$ at each depth point
$d$: 
\begin{equation}
\psi_d=\{J_1,\dots,J_{NF},N,T,N_e,n_1,\dots,n_{NL}\}
\label{eq:vector}
\end{equation}
where $J_i$ is the mean intensity of radiation at the $i$-th frequency
point. $N$, $T$ and $N_e$ are the number 
density, temperature and electron density at
depth point $d$, respectively. The population of energy level $i$ is $n_i$. In
a model represented by $NF$ number of frequency points and having $NL$ atomic
energy levels, the dimension of $\psi_d$ is $NN = NF + NL + 3$. The number $3$
stands for the three variables: number and electron density, and the
temperature.  
The set of structural equations can be formally written as
\begin{equation}
P({\bf x})=0
\label{eq:formal}
\end{equation}
where ${\bf x}$ is the matrix formed from all state vectors, ${\bf
  x}={\psi_1,\dots,\psi_{ND}}$. Equation \ref{eq:formal} can be solved by
the Newton-Raphson method:
\begin{equation}
{\bf x}^{(n+1)}={\bf x}^{(n)}-J\left({\bf x}^{(n)}\right)^{-1}P\left({\bf x}^{(n)}\right)
\label{eq:nrm}
\end{equation}
$J$ is the Jacobi matrix, $J_{ij}=\partial{P_i}/\partial{x_j}$. This
represents a finite difference solution and Equation \ref{eq:formal} reduces
to:
\begin{equation}
-{\bf A}_d\delta{\psi_{d-1}}+{\bf B}_d\delta\psi_d-{\bf
  C}_d\delta\psi_{d+1}={\bf L}_d
\label{eq:reduced}
\end{equation}
Here ${\bf A}, {\bf B}$ and ${\bf C}$ are $NN{\times}NN$ matrices, and ${\bf
  L}_d=P_d\left({\bf x}^{(n)}\right)$ is defined as 
the residuum vector. Equation
\ref{eq:reduced} is solved in place as a block-tridiagonal system. Only one
$NN{\times}NN$ matrix needs to be inverted at each depth point. Therefore,
computing time scales as:
\begin{equation}
(NF + NL + 3)^3 \times ND \times N_{\rm iter}
\label{eq:comptime}
\end{equation}
where $N_{\rm iter}$ is the number of iterations to solution.
To accurately describe the radiation field it is necessary to work with $10^4$
-- $10^5$ frequency points. From the computer time requirement 
it is clear that the complete 
linearization method in its 
original form is not practical on a single computer. 

To optimize the CL method, TLUSTY utilizes some improvements over its original
form. One of these improvements reduces the number of frequency points
without compromising model accuracy. It is done by linearizing the mean
intensities in the most 
essential frequency points while keeping others fixed and
updating them during a subsequent formal solution for one equation at a
time for the current values of other variables. 
Another approach uses the
multi-frequency/multi-gray method.
In this method the frequency
points for which the radiation forms in a similar way are grouped together,
rearranging the $NF$ frequency points into $NB$ blocks, where $NB \ll
NF$. $\bar{J}_i$ 
represents the mean intensity typical for the $i$-th frequency block. 
These blocks do not need to be composed of continuous frequency intervals. For
instance, one block can have all frequency points of the wings of weak lines,
another the cores, and so on. 
The Accelerated Lambda Iteration is the most significant method, as it allows
all the frequency points to be completely eliminated; more on this in the next
section. 

From Equation 
\ref{eq:comptime} it can bee seen that $NL$ also increases the cubic
term. Significant improvement in computing speed 
can be achieved by reducing the
number of energy levels, especially for heavy elements. Iron-peak elements can
have thousands of levels per ion, and so the total number of levels can be
comparable to the number of frequency points. The number of energy levels can
be reduced by using simpler model atoms, including only the most important
low-lying levels explicitly. The remaining levels can be taken in LTE through
the partition function \citep{hubeny88} or neglected, although the latter 
decreases
accuracy. 
Another way to reduce the number of
levels is to average individual levels into groups by their spectroscopic
terms. 
Levels belonging to the same term share similar properties and can be
averaged in multiplets. To further decrease the number of variables, individual
levels or multiplets can be assigned to level groups. In this simplification, 
levels, in which populations can be
assumed to change in a coordinated way during linearization are grouped
together. Then, instead of linearizing the individual 
level populations one needs to
linearize only the entire population of the group. 
Level averaging and grouping was used in this work and is outlined in Chapter
\ref{chap:modatom}. Level zeroing is a numerical trick to decrease the
sizes of matrices. If a level
population falls to a negligible value, the population is set exactly to zero 
and
the level is taken out of the calculations. This is also practical when one
needs to include multiple ionization degrees in a model, the program will work
only with those levels that contribute.

A 
much better approach to decrease the number of levels for heavy elements is to 
consider superlevels, \citep{hl95}. Here many levels  (50 -- 100) 
are assigned
to a superlevel assuming that each is in thermodynamic
equilibrium and share the same non-LTE
departure coefficient. There are two requirements for this: all  
levels
belonging to a superlevel must have similar energies and the same parity. These
two criteria assure that individual levels have indeed similar departure
coefficients, and the same parity prevents transitions in a superlevel. This
option was used only for iron in my work, but is available for any other species
as well. 

Further computational speed-up can be achieved 
by decreasing the number of depth points $ND$
or the number of iterations $N_{\rm iter}$. 
However, as timing is a linear function
of these parameters, such an improvement is moderate. 
In hot atmospheres with line-blanketing many
ionization degrees are present simultaneously and require many depth
points making this option not feasible. 
To decrease the number of iterations TLUSTY uses Ng \citep{ng74} and
Kantorovich \citep{kantorovich49} 
accelerations. Both acceleration methods are described in \citet{hubeny92}. 
These mathematical accelerations of convergence
can significantly reduce the number of iterations, particularly for ALI. 
The Ng method uses information from earlier steps of the iteration. The
Kantorovich method keeps the Jacobian in Equation \ref{eq:nrm} fixed after a 
certian number of iterations and re-evaluates only the residuum vector, 
${\bf L}_d$ in Equation \ref{eq:reduced}. The number of iterations after which
the Ng or Kantorovich acceleration is applied can be
adjusted to the given model atmosphere by inspecting the convergence
properties of the calculation.
By default, TLUSTY
uses an adjustable combination of the two methods. The Kantorovich
acceleration starts after the 3rd iteration and refreshes the Jacobian after 3
iterations. Ng accelerations starts at the 7th iteration and is repeated after
every 4 iterations. 

\section*{Accelerated Lambda Iteration}\label{sec:ali}

In Lambda Iteration the mean intensity is expressed as the product of 
the source function
and the so-called Lambda operator:
\begin{equation}
J_{\nu}=\Lambda_{\nu}S_{\nu}
\label{eq:lambda}
\end{equation}
The iteration proceeds by consecutively updating the mean intensity using
the previous source function and updating the new source function with the new
mean intensity. In LTE this is:
\begin{eqnarray}
J^{(n)}_{\nu}&=&\Lambda_{\nu}S_{\nu}^{(n-1)}\\
S_{\nu}^{(n)}&=&(1-\epsilon)J_{\nu}^{(n)}+\epsilon B_{\nu}
\label{eq:ize}
\end{eqnarray}
where $\epsilon$ is the thermal coupling coefficient and $B_{\nu}$ is the
Planck function in LTE, and $n$ is the iteration number. 
This procedure has slow convergence at large optical
depths where $\epsilon\approx1$ and at high opacities. To avoid this, ALI uses
operator splitting. The Lambda operator is written as:
\begin{equation}
\Lambda=\Lambda^*+(\Lambda-\Lambda^*)
\label{eq:opsplit}
\end{equation}
where $\Lambda$ and $\Lambda^*$ are the exact and approximate Lambda
operators, respectively. Both Lambda operators are valid for a specific
frequency. With this, the mean intensity and source function 
have the form:
\begin{eqnarray}
J^{(n)}_{\nu}&=&\Lambda_{\nu}^*S_{\nu}^{(n)}+(\Lambda_{\nu}-\Lambda_{\nu}^*)S_{\nu}^{(n-1)}\\
S_{\nu}^{(n)}&=&(1-\epsilon)\Lambda_{\nu}^*S_{\nu}^{(n)}+(1-\epsilon)(\Lambda_{\nu}-\Lambda_{\nu}^*)S_{\nu}^{(n-1)}+\epsilon{B_{\nu}}
\label{eq:meanali}
\end{eqnarray}
The second correction term of the mean intensity is known from previous
iterations. 
The first term represents an action of the approximate operator on
the source function which is a function of temperature, density and level
populations. 
In this way the radiative transfer equations are separated from the
structural equations. 
The new problem is to find a suitable approximate Lambda
operator. Its form must give rise to a simpler matrix inversion while
incorporating the essential properties of the original Lambda operator in
order to reduce the necessary iterations. 
TLUSTY uses either the Rybicki-Hummer algorithm \citep{rybicki91}, 
or the Olson-Kunasz operator
\citep{kunasz88} to determine $\Lambda^*$.

\section*{Output}

The main output of TLUSTY is the structure of the atmosphere, 
stored in an unformatted ASCII file. 
The first line
contains the number of depth points and number of model parameters for a depth
point. The following block is the depth grid, the column mass (in
g/cm$^2$) at each depth point. This is followed by the effective
temperature, electron density, mass density and population numbers for all
levels at each
depth points. This output model can be used as an input for the next run of
TLUSTY, or for calculating a synthetic spectrum with SYNSPEC. Also from this
file it is possible to extract data for ionization balance and the structure
of the atmosphere. In non-LTE mode the main output 
file is accompanied by an other
file having the exact same structure, but listing the departure
coefficients instead of occupation numbers.  

\section*{SYNSPEC}

SYNSPEC\footnote{\protect\url{http://nova.asto.umd.edu/Synspec43/synspec.html}}
\citep{synspec43} can work with a given model atmosphere, which 
either can be calculated by TLUSTY or a Kurucz model (ATLAS)
taken from the literature. 
The program reads a line list (see Appendix for O VI line list) 
and selects lines which contribute to the total opacity of 
the given atmosphere. Then it solves the
radiative transfer equation (Equation \ref{eq:transfer}),
in a user-specified wavelength range and resolution and selects frequency
points automatically. The program calculates the flux
at the centers of two neighboring lines and in the midpoint between the
two. 
Then it adds a certain number of frequency points equidistantly spaced
between each pair to achieve the required resolution. 
In this way both the line
centers and the continuum are sampled well. 
SYNSPEC is built in a similar way to
TLUSTY and uses the same subroutines, but while TLUSTY calculates the model
atmosphere self consistently and returns the structure and occupation numbers,
SYNSPEC only recalculates the transfer equation and gives the detailed
spectrum while considering natural, Stark and Van der Waals line broadening.    

The calculated spectrum is the emergent flux versus wavelength at the top
of the atmosphere ($\tau_{\rm Ross}=10^{-7}$). 
SYNSPEC also provides the list of contributing spectral
lines and their approximate equivalent widths. 
To compare this to observations one needs to take into
account the effects of stellar rotation, interstellar absorption
and convolve the data with the instrumental profile of the observing detector.

\chapter{Model Atoms}\label{chap:modatom}
\setstretch{2.}

Model atoms are the primary data input for TLUSTY. These are the physical
models of chemical species and their ions. This chapter describes the
steps of constructing model atoms.

In order to model hot atmospheres one must include model atoms for all
astrophysically important species and all of their ionization stages relevant
to the work. TLUSTY itself is data independent, it has hard-coded data only for
hydrogen and helium. Data for 
other elements must be communicated with model atom
files. 
These files can be split into three main sections: data for energy
levels, data for bound-free and bound-bound transitions.    

In this work I closely followed the works of Ivan Hubeny and Thierry Lanz 
for sdB \citep{lanz07} and sdO \citep{lanz03}
stars, the atomic data update of PHOENIX \citep{apetz05} and supersoft
modeling of CAL 83 by \citet{lanz05}. However, my approach differs
from theirs in several ways. 
To keep TLUSTY and SYNSPEC standard I did not modify the code, but extended the
built-in
atomic data arrays for higher ionization stages of all included
metals. SYNSPEC version 48 has highest ionization degree of VII for all
elements, except for Fe and Ni (for them it is X). I increased this limit to 
XXVI to consider all ionization degrees of all elements up to iron. 

The challenge of modeling novae 
comes from the extreme conditions throughout the
atmosphere. There is a large temperature gradient below the photosphere and
outer layers of the atmosphere are radiatively heated. For this reason many
ionization degrees are present, which increases the complexity and 
calculation time
considerably. Another difficulty arises from the lack of reliable atomic data
for high ionizations. The available data in the supersoft range 
is pretty much limited to 
TOPBASE\footnote{\protect
\url{http://cdsweb.u-strasbg.fr/topbase/topbase.html}} \citep{cunto93},
NIST\footnote{\protect\url{http://www.nist.gov/physlab/data/asd.cfm}}, 
APED\footnote{\protect
\url{http://cxc.harvard.edu/atomdb/index.html}} and 
Chianti\footnote{\protect
\url{http://www.ukssdc.ac.uk/solar/chianti/}} 
(\citet{dere97} and \citet{dere09}). 

In the following I describe TOPBASE and NIST, from
which I collected atomic data and recalculated them to model
atom files that are readable by TLUSTY.

\section{TOPbase}

TOPbase 
is the atomic database of Opacity Project (OP), an international
collaboration to calculate stellar envelope opacities by the {\it ab initio} 
methods of
\citet{seaton87}. The prime initiative of the project was to provide
atomic data to better understand pulsating stars. The aim was to tabulate
accurate data
on energy levels, oscillator strengths, photoionization cross-sections and
line-broadening parameters for all contributing processes and astrophysically
abundant ions. OP calculations were based on the close-coupling method of
scattering theory. An N-electron system is described in terms of an N-1-electron
target system and an active electron. The wave function of such a system is:
\begin{equation}
\Psi^{SL\pi} = \mathcal{A}\sum_i\chi_i\theta_i+\sum_ic_j\Phi_j
\label{wavefunction}
\end{equation}
where $\mathcal{A}$ is the antisymmetrisation operator, $\chi_i$ is the
target wave function and $\theta_i$ is the active electron function. The
$\Psi_j$ are bound-state type functions for the total system, introduced to
compensate for orthogonality conditions imposed on $\theta_i$ and to render
short-range correlations. Applying the Kohn variational principle
\citep{kohn96} with the
functions $\theta_i$ and coefficients $c_j$ as variational parameters leads to
a set of integro-differential equations which were solved
numerically. In the calculations relativistic 
effects were neglected and LS-coupling was assumed. 
The close-coupling approximation addresses both electron correlations and
threshold effects in the same framework, which allowed calculations of
both bound-free and bound-bound transitions. 
This unique feature made OP data
state-of-the-art for its time and it is still the most complete atomic data set
for the supersoft range. 
The integro-differential equations were solved with
the R-matrix method of \citet{burke71} and asymptotic techniques developed by
\citet{seaton85}. 

TOPbase has atomic data from Hydrogen to Silicon, for 
Sulfur, Argon, Calcium
and Iron. The full content and data tables are 
available on-line at the Centre de Donne\'{e}s Astronomiques de
Strasbourg (CDS) under
TIPTOPbase.
The principal and
orbital angular momentum quantum numbers are limited to $N\le10$ 
and $l\le4$ respectively, 
except for hydrogenic ions $N\le10$ and $l\le9$. The total data
file is about 500 Mb. Table \ref{tab:topbasel} shows a small subset of energy
levels listed in TOPbase; here are shown 
the first 10 levels of N VI. These levels are
actually spectroscopic terms, as TOPbase neglects hyperfine structure. The
first column {\it i} is the ID number of the level in the current file, 
{\it NZ} and
{\it NE} 
are the atomic number and the number of electrons. The spectroscopic term
is expressed with the spin quantum number ({\it S}), 
the orbital angular momentum
quantum number ({\it L}) and parity ({\it P}) for level i:
\begin{equation}
SLP_i=(2S_i+1)\cdot100+L_i\cdot10+P_i
\label{eq:slpi}
\end{equation}
\begin{table}[!h]\tiny
\setstretch{1}
\begin{center}
\parbox{130mm}{\caption{First ten energy levels of N VI in TOPbase.} \label{tab:topbasel}}
\begin{tabular}{l}
\\
\hline
\verb,==================================================================================================,\\
\verb,      i NZ NE iSLP iLV iCONF                 E(RYD)      TE(RYD)   gi         QD     EQN    RL(NS),\\
\verb,==================================================================================================,\\
\verb,...,\\
\verb,   1817  7  2  100   1 1s2             -4.05249E+01  0.00000E+00  1.0  0.000E+00  0.9425  1.00E+30,\\
\verb,   1818  7  2  100   2 1s 2s           -9.22280E+00  3.13021E+01  1.0  2.431E-02  1.9757  1.00E+30,\\
\verb,   1819  7  2  100   3 1s 3s           -4.06733E+00  3.64576E+01  1.0  2.493E-02  2.9751  8.91E-02,\\
\verb,   1820  7  2  100   4 1s 4s           -2.27854E+00  3.82464E+01  1.0  2.513E-02  3.9749  1.31E-01,\\
\verb,   1821  7  2  100   5 1s 5s           -1.45466E+00  3.90702E+01  1.0  2.526E-02  4.9747  2.08E-01,\\
\verb,   1822  7  2  100   6 1s 6s           -1.00849E+00  3.95164E+01  1.0  2.531E-02  5.9747  3.19E-01,\\
\verb,   1823  7  2  100   7 1s 7s           -7.40041E-01  3.97849E+01  1.0  2.533E-02  6.9747  4.71E-01,\\
\verb,   1824  7  2  100   8 1s 8s           -5.66081E-01  3.99588E+01  1.0  2.534E-02  7.9747  6.69E-01,\\
\verb,   1825  7  2  100   9 1s 9s           -4.46959E-01  4.00779E+01  1.0  2.535E-02  8.9746  9.20E-01,\\
\verb,   1826  7  2  100  10 1s10s           -3.61833E-01  4.01631E+01  1.0  2.536E-02  9.9746  1.23E+00,\\
\verb,...,\\
\hline
\end{tabular}
\setstretch{2}
\end{center}
\end{table}

The next column, {\it iLV}, is the level index for a spectroscopic term,
followed by its configuration. Columns seven and eight list the energy of the
level in Rydbergs with respect to the ionization energy and to the ground
level, respectively. The last columns are the statistical weight of the term
($g_i$),
quantum defect ({\it QD}), effective quantum number ({\it EQN}) and radiative 
lifetime ({\it RL}) in nanoseconds.  

Table \ref{tab:topbasex} illustrates the photoionizaton data. Here, 
only five cross section
values are shown for the first two levels. For each level the first number
({\it I}) is the level index in the given file, {\it NZ, NE, ISLP, ILV} and 
{\it E} has the same meaning as before. For each level the first line is
followed by {\it NP} number of data pairs of wavenumber (in cm$^{-1}$) and cross
section (in Mb). 

\begin{table}
\setstretch{1}
\begin{center}
\parbox{130mm}{\caption[Photoionization cross-sections in TOPbase.]
{First two entries of photoionization cross-sections for N VI. Only the
   first five data points are listed for each level.} \label{tab:topbasex}}
\begin{scriptsize}
\begin{tabular}{l}
\\
\hline
\verb,=================================================,\\
\verb,       I  NZ  NE  ISLP  ILV        E(RYD)      NP,\\
\verb,=================================================,\\
\verb,      56   7   2   100    1  -4.05249E+01      62,\\
\verb,  4.022729E+01 3.490E-01,\\
\verb,  4.030170E+01 3.475E-01,\\
\verb,  4.037610E+01 3.460E-01,\\
\verb,  4.045039E+01 3.445E-01,\\
\verb,  4.052451E+01 3.431E-01,\\
\verb,...,\\
\verb,      57   7   2   100    2  -9.22280E+00      62,\\
\verb,  8.925290E+00 4.156E-01,\\
\verb,  8.999670E+00 4.089E-01,\\
\verb,  9.074050E+00 4.024E-01,\\
\verb,  9.148430E+00 3.960E-01,\\
\verb,  9.222450E+00 3.898E-01,\\
\verb,...,\\
\hline
\end{tabular}
\setstretch{2}
\end{scriptsize}
\end{center}
\end{table}

\begin{table}[!h]\tiny
\begin{center}
\parbox{130mm}{\caption{First ten lines of $f$-values data for N VI.} \label{tab:topbasef}}
\setstretch{1}
\begin{tabular}{l}
\\
\hline
\verb,=======================================================================================================,\\
\verb,      i NZ NE iSLP jSLP iLV jLV iCONF           jCONF                  gF   gA(S-1)     WL(A)   gi   gj,\\
\verb,=======================================================================================================,\\
\verb,    331  7  2  111  100   1   1 1s 2p           1s2              6.74E-01  5.41E+12 2.883E+01  3.0  1.0,\\
\verb,    332  7  2  111  100   1   2 1s 2p           1s 2s            7.69E-02  5.79E+07 2.977E+03  3.0  1.0,\\
\verb,    333  7  2  111  100   1   3 1s 2p           1s 3s           -5.94E-02 -1.12E+10 1.879E+02  3.0  1.0,\\
\verb,    334  7  2  111  100   1   4 1s 2p           1s 4s           -1.28E-02 -4.53E+09 1.373E+02  3.0  1.0,\\
\verb,    335  7  2  111  100   1   5 1s 2p           1s 5s           -5.03E-03 -2.25E+09 1.221E+02  3.0  1.0,\\
\verb,    336  7  2  111  100   1   6 1s 2p           1s 6s           -2.55E-03 -1.28E+09 1.152E+02  3.0  1.0,\\
\verb,    337  7  2  111  100   1   7 1s 2p           1s 7s           -1.49E-03 -8.00E+08 1.115E+02  3.0  1.0,\\
\verb,    338  7  2  111  100   1   8 1s 2p           1s 8s           -9.50E-04 -5.32E+08 1.091E+02  3.0  1.0,\\
\verb,    339  7  2  111  100   1   9 1s 2p           1s 9s           -6.46E-04 -3.72E+08 1.076E+02  3.0  1.0,\\
\verb,    340  7  2  111  100   1  10 1s 2p           1s10s           -4.60E-04 -2.70E+08 1.065E+02  3.0  1.0,\\
\verb,...,\\
\hline
\end{tabular}
\setstretch{2}
\end{center}
\end{table}

Table \ref{tab:topbasef} contains radiative data for NVI. The first nine columns
identify the transition using the same entries that are also listed in Table 
\ref{tab:topbasel}. Oscillator strength ($gF_{ij}$) is multiplied by $g_i$, the
statistical weight of level {\it i}. Transitions from higher to lower
levels 
(emission) are
taken as positive. Simlarly, Einstein coefficients ($A_{ij}$) are also
multiplied by the statistical weight and emission lines are taken as
positive. 
Wavelength calculated from the
energy difference of the levels involved ({\it WL}) is
listed in {\AA}ngstr\"{o}ms.

\section{NIST/ASD}

National Institute of Standards and Technology (NIST) 
is a measurement standards laboratory in
Gaithersburg, Maryland. NIST has been collecting 
data, maintains and publishes
scientific
and technical databases. One of these databases is the Atomic Spectra
Database (ASD), which contains data for radiative transitions and
energy levels in atoms and atomic ions. Data are included for observed
transitions of 99 elements and energy levels of 52 elements. ASD contains data
for about 900 spectra from 1 {\AA}ngstr\"{o}m to 200 $\mu$m, with about 70,000
energy levels and 91,000 lines, 40,000 of which have transition probabilities
listed. 
Its convenient, html interface allows for interactive data collection
and download as ASCII tables, as shown 
in Table \ref{tab:nistl}. Levels are listed with increasing energy with 
respect to
the ground level and identified by their configurations, spectroscopic terms
and total electronic angular-momentum quantum number ({\it J}). Level energies
and splittings are given in cm$^{-1}$ units. 

\begin{table}[!h]
\begin{center}
\parbox{130mm}{\caption{First 10 energy levels of NVI in NIST/ASD.} \label{tab:nistl}}
\begin{scriptsize}
\setstretch{1}
\begin{tabular}{l}
\\
\hline
\verb,----------------------------------------------------------------,\\
\verb,Configuration   | Term   |   J |      Level     |    Splitting |,\\
\verb,                |        |     |     (cm-1)     |       (cm-1) |,\\
\verb,----------------------------------------------------------------,\\
\verb,1s2             | 1S     |   0 |          0     |              |,\\
\verb,                |        |     |                |              |,\\
\verb,1s.2s           | 3S     |   1 |    3385890     |    3385890   |,\\
\verb,                |        |     |                |              |,\\
\verb,1s.2p           | 3P*    |   0 |    3438304     |      52414   |,\\
\verb,                |        |   1 |    3438320     |         16   |,\\
\verb,                |        |   2 |    3438610     |        290   |,\\
\verb,                |        |     |                |              |,\\
\verb,1s.2s           | 1S     |   0 |    3439274     |        664   |,\\
\verb,                |        |     |                |              |,\\
\verb,1s.2p           | 1P*    |   1 |    3473790     |      34516   |,\\
\verb,                |        |     |                |              |,\\
\verb,1s.3s           | 3S     |   1 |    3991860     |     518070   |,\\
\verb,                |        |     |                |              |,\\
\verb,1s.3s           | 1S     |   0 |    4006000     |      14140   |,\\
\verb,                |        |     |                |              |,\\
\verb,1s.3p           | 3P*    |   0 |    4006160     |        160   |,\\
\verb,                |        |   2 |    4006160     |          0   |,\\
\verb,                |        |   1 |    4006160     |          0   |,\\
\verb,                |        |     |                |              |,\\
\verb,1s.3d           | 3D     |   3 |    4013460     |       7300   |,\\
\verb,                |        |   2 |    4013460     |          0   |,\\
\verb,                |        |   1 |    4013460     |          0   |,\\
\verb,                |        |     |                |              |,\\
\verb,1s.3d           | 1D     |   2 |    4013770     |        310   |,\\
\verb,                |        |     |                |              |,\\
\verb,...,\\
\hline
\end{tabular}
\setstretch{2}
\end{scriptsize}
\end{center}
\end{table}

Table \ref{tab:nistline} is an example of line data in NIST/ASD. Observed
wavelengths are tabulated if either or both energy levels of the transition
are unavailable. In most cases wavelength is calculated from the energy
difference of the two levels (these are the Ritz wavelengths). All wavelengths
are given for vacuum. Relative intensities are tentative and serve as
identifications only. Einstein coefficients $A_{ki}$ are
derived from lifetime measurements. Oscillator strengths are calculated:
\begin{equation}
f_{ik}=\frac{m_ec\epsilon_0\lambda^2}{2{\pi}e^2}A_{ki}\frac{g_k}{g_i}
\label{eq:aki}
\end{equation} 
where $g_x=2J_x+1$. The transition strength is also function of $A_{ki}$:
\begin{equation}
S_{ik}=\frac{3h\epsilon_0\lambda^3g_k}{16\pi^3}A_{ki}
\label{eq:ski}
\end{equation}

\section{TOPAtom}

Building model atoms from TOPbase and NIST/ASD data tables is done through
several steps. Each of these is written in short modules of Python
scripts, which are called by a top-level program, TOPAtom. The simple command
line program can be
invoked with flags setting various treatments, or formats of the output.  

The first task is to extract level configurations, terms, total electronic
angular-momentum quantum numbers and energies 
from both NIST/ASD and TOPbase. These levels are then cross-correlated. Because
calculated and observed data are different, levels are identified according to
their configuration and term, instead of their energies or wavelengths. 
Both datasets provide only the significant
part of the configuration. To convert them to the same format I rebuilt the
full configuration and stored it in a 36 character string variable. 
For example, for a level of 
Ca XII NIST provides the configuration $2s2p^5(^3P_{\circ})3s$; in my level 
list this level has the configuration:
\verb,1s2.2s.2p5.(3PO).3s_________________,. In the 
correlation work
all NIST levels are kept and the list is completed with TOPbase data. With
the level list generated, the next step is to generate a new dataset for
transitions. The levels can be combined with selection rules considered,
although NIST has data for forbidden transitions as well, so currently 
all linear
combinations of the levels are considered. This increases the size
of the transition file and the computational time, but does not skip any
transitions. The transition list contains level energies, configurations,
terms, $J$s, level database indices (NIST or TOPbase) and the wavelengths of the
transitions. This is the input for the next step, where the program 
cross-correlates the databases and assigns transition strengths and transition
probabilities to level pairs.

There are three data sets for the transitions: the previously described 
transition
list, the $f$-values data from TOPbase, and data for 
lines from NIST. The latter two
are converted to the same data format with the first. 
Correlation is done by
taking each data entry in the transition file and assigning oscillator strengths
and transition probablilities to them from NIST lines. 
Upon finding missing data in NIST, it is taken from TOPbase. 
If the given entry is missing from both databases the transition is dropped. 
This way
it is possible to assign TOPbase oscillator strengths and transition
probabilities to the more accurate NIST levels. 
Correlated results are stored
in transition 
files \verb,f77united_list.ion, for each ion. An example for N VI is
shown in Table \ref{tab:f77united}. Each line of the file 
describes one transition including oscillator strengths and transition
probabilities.

\begin{sidewaystable}[!htbp]
\begin{center}
\parbox{130mm}{\caption{Example for NIST lines data for N VI.} \label{tab:nistline}}
\tiny
\begin{tiny}
\setstretch{1}
\begin{tabular}{l}
\\
\hline
\verb,-------------------------------------------------------------------------------------------------------------------------------------------------------------------------------,\\
\verb,    Observed  |      Ritz    |  Rel. |    Aki    |    fik    |     S     |  log_gf | Acc. |     Ei           Ek     | Configurations  |    Terms   |  Ji   Jk  | gi   gk |Type|,\\
\verb,   Wavelength |   Wavelength |  Int. |    s^-1   |           |           |         |      |   (cm-1)       (cm-1)   |                 |            |           |         |    |,\\
\verb,    Vac  (A)  |    Vac  (A)  |  (?)  |           |           |           |         |      |                         |                 |            |           |         |    |,\\
\verb,-------------------------------------------------------------------------------------------------------------------------------------------------------------------------------,\\
\verb,              |              |       |           |           |           |         |      |                         |                 |            |           |         |    |,\\
\verb,              |      23.0240 |       | 6.290e+10 | 1.500e-02 |  1.137e-03|  -1.824 | AA   |       0    -  4343290   |    1s2 - 1s.6p  |  1S  - 1P* |   0 - 1   |  1 - 3  |    |,\\
\verb,              |      23.2770 |       | 1.091e+11 | 2.659e-02 |  2.037e-03|  -1.575 | AA   |       0    -  4296090   |    1s2 - 1s.5p  |  1S  - 1P* |   0 - 1   |  1 - 3  |    |,\\
\verb,              |      23.7710 |       | 2.145e+11 | 5.451e-02 |  4.266e-03|  -1.263 | AA   |       0    -  4206810   |    1s2 - 1s.4p  |  1S  - 1P* |   0 - 1   |  1 - 3  |    |,\\
\verb,              |      23.8277 |       | 4.59e+01  | 1.17e-11  |  6.91e-08 | -10.931 | C    |       0    -  4196800   |    1s2 - 1s.4s  |  1S  - 3S  |   0 - 1   |  1 - 3  | M1 |,\\
\verb,              |      24.8980 |       | 5.158e+11 | 1.438e-01 |  1.179e-02|  -0.842 | AA   |       0    -  4016390   |    1s2 - 1s.3p  |  1S  - 1P* |   0 - 1   |  1 - 3  |    |,\\
\verb,              |              |       |           |           |           |         |      |                         |                 |            |           |         |    |,\\
\verb,              |      24.9142 |       | 3.847e+07 | 1.790e-05 |  1.649e-03|  -4.747 | AA   |       0    -  4013770   |    1s2 - 1s.3d  |  1S  - 1D  |   0 - 2   |  1 - 5  | E2 |,\\
\verb,              |      25.0510 |       | 8.36e+01  | 2.36e-11  |  1.46e-07 | -10.627 | C    |       0    -  3991860   |    1s2 - 1s.3s  |  1S  - 3S  |   0 - 1   |  1 - 3  | M1 |,\\
\verb,              |      28.7870 |       | 1.809e+12 | 6.742e-01 |  6.390e-02|  -0.171 | AA   |       0    -  3473790   |    1s2 - 1s.2p  |  1S  - 1P* |   0 - 1   |  1 - 3  |    |,\\
\verb,              |      29.0815 |       | 1.03e+05  | 6.53e-08  |  1.62e+00 |  -7.185 | A    |       0    -  3438610   |    1s2 - 1s.2p  |  1S  - 3P* |   0 - 2   |  1 - 5  | M2 |,\\
\verb,              |      29.5343 |       | 2.55e+02  | 1.00e-10  |  7.31e-07 | -10.000 | A+   |       0    -  3385890   |    1s2 - 1s.2s  |  1S  - 3S  |   0 - 1   |  1 - 3  | M1 |,\\
\verb,              |              |       |           |           |           |         |      |                         |                 |            |           |         |    | ,\\
\verb,              |     104.884  |       | 1.58e+09  | 7.82e-03  |  2.70e-03 |  -2.107 | A    | 3439274    -  4392710   |  1s.2s - 1s.8p  |  1S  - 1P* |   0 - 1   |  1 - 3  |    | ,\\
\verb,              |     107.151  |       | 2.31e+09  | 1.19e-02  |  4.21e-03 |  -1.923 | A    | 3439274    -  4372540   |  1s.2s - 1s.7p  |  1S  - 1P* |   0 - 1   |  1 - 3  |    | ,\\
\verb,              |     110.618  |       | 3.642e+09 | 2.004e-02 |  7.299e-03|  -1.698 | AA   | 3439274    -  4343290   |  1s.2s - 1s.6p  |  1S  - 1P* |   0 - 1   |  1 - 3  |    |,\\
\verb,              |     116.711  |       | 6.275e+09 | 3.844e-02 |  1.477e-02|  -1.415 | AA   | 3439274    -  4296090   |  1s.2s - 1s.5p  |  1S  - 1P* |   0 - 1   |  1 - 3  |    |,\\
\verb,              |     116.786  |       | 7.460e+09 | 4.576e-02 |  1.759e-02|  -1.339 | AA   | 3438304    -  4294570   |  1s.2p - 1s.5d  |  3P* - 3D  |   0 - 1   |  1 - 3  |    |,\\
\hline
\end{tabular}
\end{tiny}
\setstretch{2}
\end{center}
\vspace{.5cm}
\begin{center}
\parbox{130mm}{\caption[Cross correlated TOPbase and NIST line lists.]{Example for transitions in {\tt f77united\_list.n6},
    for N VI. Level configurations are stored in a 36 character string
    variable, here it is trimmed to 20.} \label{tab:f77united}}
\tiny
\begin{tiny}
\setstretch{1}
\begin{tabular}{l}
\\
\hline
\verb,  NZ  NE  Fij                   Level i (m-1)   Level j (m-1)      Configuration i       Configuration j          Ji   Jj   ISLP JSLP Level i   Level j      Aji              Tr. id,\\
\verb,  7   2   0.150000E-01          0.0000000000    4343290.0000000000 1s2_______________... 1s.6p_______________...  0.0  1.0  100  111 NIST______ NIST______   0.629000E+11     NISTtr 1,\\
\verb,  7   2   0.265900E-01          0.0000000000    4296090.0000000000 1s2_______________... 1s.5p_______________...  0.0  1.0  100  111 NIST______ NIST______   0.109100E+12     NISTtr 1,\\
\verb,  7   2   0.545100E-01          0.0000000000    4206810.0000000000 1s2_______________... 1s.4p_______________...  0.0  1.0  100  111 NIST______ NIST______   0.214500E+12     NISTtr 1,\\
\verb,  7   2   0.117000E-10          0.0000000000    4196800.0000000000 1s2_______________... 1s.4s_______________...  0.0  1.0  100  300 NIST______ NIST______    45.9000         NISTtr 1,\\
\verb,  7   2   0.143800              0.0000000000    4016390.0000000000 1s2_______________... 1s.3p_______________...  0.0  1.0  100  111 NIST______ NIST______   0.515800E+12     NISTtr 1,\\
\verb,  7   2   0.179000E-04          0.0000000000    4013770.0000000000 1s2_______________... 1s.3d_______________...  0.0  2.0  100  120 NIST______ NIST______   0.384700E+08     NISTtr 1,\\
\verb,  7   2   0.236000E-10          0.0000000000    3991860.0000000000 1s2_______________... 1s.3s_______________...  0.0  1.0  100  300 NIST______ NIST______    83.6000         NISTtr 1,\\
\verb,  7   2   0.674200              0.0000000000    3473790.0000000000 1s2_______________... 1s.2p_______________...  0.0  1.0  100  111 NIST______ NIST______   0.180900E+13     NISTtr 1,\\
\verb,  7   2   0.653000E-07          0.0000000000    3438610.0000000000 1s2_______________... 1s.2p_______________...  0.0  2.0  100  311 NIST______ NIST______    103000.         NISTtr 1,\\
\verb,  7   2   0.100000E-09          0.0000000000    3385890.0000000000 1s2_______________... 1s.2s_______________...  0.0  1.0  100  300 NIST______ NIST______    255.000         NISTtr 1,\\
\verb,  7   2   0.674000              0.0000000000    3473790.0000000000 1s2_______________... 1s.2p_______________...  0.0  1.0  100  111 NIST______ NIST______   0.541000E+13       TBtr 2,\\
\verb,  7   2   0.144000              0.0000000000    4016390.0000000000 1s2_______________... 1s.3p_______________...  0.0  1.0  100  111 NIST______ NIST______   0.154000E+13       TBtr 2,\\
\verb,  7   2   0.545000E-01          0.0000000000    4206810.0000000000 1s2_______________... 1s.4p_______________...  0.0  1.0  100  111 NIST______ NIST______   0.642000E+12       TBtr 2,\\
\verb,  7   2   0.265000E-01          0.0000000000    4296090.0000000000 1s2_______________... 1s.5p_______________...  0.0  1.0  100  111 NIST______ NIST______   0.325000E+12       TBtr 2,\\
\verb,  7   2   0.150000E-01          0.0000000000    4343290.0000000000 1s2_______________... 1s.6p_______________...  0.0  1.0  100  111 NIST______ NIST______   0.188000E+12       TBtr 2,\\
\hline
\end{tabular}
\end{tiny}\setstretch{2}\end{center}\end{sidewaystable}

\subsubsection{Line list}

The transition files are actually line lists, they 
can be easily reformatted for SYNSPEC. The files are read in line
by line and written to \verb,gfXRA.ion, files like in Table
\ref{tab:linelist} for 
N VI. Then these files are merged and sorted by wavelength.  
Currently the line list contains 380,092 lines without iron, 
in the supersoft
range (10 \AA\ -- 100 \AA) there are 
77,299 lines with oscillator strengths and natural
broadening coefficients. 
The structure of the file follows those of the original SYNSPEC 
line lists
for ultraviolet and optical. The
first column is the wavelength in nm. The code of the ion is listed in the
second column in the form 
$(NZ).(NZ-NE)$, where $NZ$ is the atomic number and $NE$ is the
number of electrons. The third column lists $\log({g_if_{ij}})$, followed by
the excitation potential (in cm$^{-1}$) 
and the $J$ quantum number of the lower and the upper
level. The eighth column has data for natural broadening $\log(A_{ji})$,
followed by coefficients for Stark and Van der Waals broadening which are in
this case set to taken with their classical expression. 
The eleventh column is a flag indicating that
there are no more records for the given line. The last column is a comment
indicating which database the oscillator strength and natural broadening were
taken from.     

\begin{table}[!h]
\begin{center}
\parbox{130mm}{\caption{N VI line list up to 100 \AA.}\label{tab:linelist}}
\begin{scriptsize}
\setstretch{1}
\begin{tabular}{l}
\\
\hline
\verb,2.2688 07.05  -2.983          0.000 0.0    4407643.300 1.0 10.13 0.00 0.00 0 !TBtr,\\
\verb,2.2736 07.05  -2.844          0.000 0.0    4398403.418 1.0 10.27 0.00 0.00 0 !TBtr,\\
\verb,2.2765 07.05  -2.688          0.000 0.0    4392710.000 1.0 10.42 0.00 0.00 0 !TBtr,\\
\verb,2.2870 07.05  -2.510          0.000 0.0    4372540.000 1.0 10.59 0.00 0.00 0 !TBtr,\\
\verb,2.3024 07.05  -1.824          0.000 0.0    4343290.000 1.0 10.80 0.00 0.00 0 !NISTtr,\\
\verb,2.3277 07.05  -1.575          0.000 0.0    4296090.000 1.0 11.04 0.00 0.00 0 !NISTtr,\\
\verb,2.3771 07.05  -1.264          0.000 0.0    4206810.000 1.0 11.33 0.00 0.00 0 !NISTtr,\\
\verb,2.3828 07.05 -10.932          0.000 0.0    4196800.000 1.0 1.66  0.00 0.00 0 !NISTtr,\\
\verb,2.4898 07.05  -0.842          0.000 0.0    4016390.000 1.0 11.71 0.00 0.00 0 !NISTtr,\\
\verb,2.4914 07.05  -4.747          0.000 0.0    4013770.000 2.0 7.59  0.00 0.00 0 !NISTtr,\\
\verb,2.5051 07.05 -10.627          0.000 0.0    3991860.000 1.0 1.92  0.00 0.00 0 !NISTtr,\\
\verb,2.8787 07.05  -0.171          0.000 0.0    3473790.000 1.0 12.26 0.00 0.00 0 !NISTtr,\\
\verb,2.9082 07.05  -7.185          0.000 0.0    3438610.000 2.0 5.01  0.00 0.00 0 !NISTtr,\\
\verb,2.9534 07.05 -10.000          0.000 0.0    3385890.000 1.0 2.41  0.00 0.00 0 !NISTtr,\\
\verb,9.7408 07.05  -2.444    3385890.000 1.0    4412500.000 2.0 8.93  0.00 0.00 0 !TBtr,\\
\verb,9.8308 07.05  -2.301    3385890.000 1.0    4403100.000 2.0 9.06  0.00 0.00 0 !TBtr,\\
\verb,9.9501 07.05  -2.135    3385890.000 1.0    4390900.000 2.0 9.22  0.00 0.00 0 !TBtr,\\
\verb,...,\\
\hline
\end{tabular}
\end{scriptsize}
\end{center}
\setstretch{1}
\end{table}

\subsubsection{Atoms -- Levels}

Model atoms are derived from the same transition files. At the first step all
levels with configurations and term designations were extracted. These levels
then were listed according to TLUSTY input format 
shown in Table \ref{tab:man6}. The principal quantum number (N) of each level 
is taken as the principal quantum number of the highest orbit in the
configuration. Level grouping
is automatic for $N\ge3$. Each level is appended by a level index, 
spectroscopic term, level energy in 1000 cm$^{-1}$ and database ID as comments
at the end of the lines. 

\begin{table}[!h]
\begin{center}
\parbox{130mm}{\caption{First ten energy levels of N VI.}\label{tab:man6}}
\begin{scriptsize}
\setstretch{1}
\begin{tabular}{l}
\\
\hline
\verb,****** Levels,\\
\verb,1.334899368397e+17      1.      1     '_______1s2'    0     0.         0 ! 1 1Se 0.0 NIST,\\
\verb,3.198350827796e+16      3.      2     '_____1s.2s'    0     0.         0 ! 2 3Se 3385.89 NIST,\\
\verb,3.041217608860e+16      1.      2     '_____1s.2p'    0     0.         0 ! 3 3Po 3438.304 NIST,\\
\verb,3.041169642066e+16      3.      2     '_____1s.2p'    0     0.         0 ! 4 3Po 3438.32 NIST,\\
\verb,3.040300243938e+16      5.      2     '_____1s.2p'    0     0.         0 ! 5 3Po 3438.61 NIST,\\
\verb,3.038309622017e+16      1.      2     '_____1s.2s'    0     0.         0 ! 6 1Se 3439.274 NIST,\\
\verb,2.934833257214e+16      3.      2     '_____1s.2p'    0     0.         0 ! 7 1Po 3473.79 NIST,\\
\verb,1.381698470053e+16      3.      3     '_____1s.3s'    0     0.      -103 ! 8 3Se 3991.86 NIST,\\
\verb,1.339307816492e+16      1.      3     '_____1s.3s'    0     0.      -103 ! 9 1Se 4006.0 NIST,\\
\verb,1.338828148559e+16      1.      3     '_____1s.3p'    0     0.      -103 ! 10 3Po 4006.16 NIST,\\
\verb,...,\\
\hline
\end{tabular}
\end{scriptsize}
\end{center}
\setstretch{1}
\end{table}

\subsubsection{Atoms -- Bound-free transitions}

Bound-free transitions are from the 
TOPbase $x$-data. Table \ref{tab:topbasex} shows
the content of these files. Levels can be identified by level indices, which
can be
matched with indices in the TOPbase level data. 
But, instead of tracking these back to find the configuration, 
I chose to find levels based on energy
and terms. The previous level list is the basis for this search. For each
level the program seeks levels in the cross section data with the same
term and within an energy difference of 1\%, 5\%, 25\% and 100\%. This way the
search returns cross section data for all TOPbase levels and accuracy can be
tracked for NIST levels. Levels not having cross section data in TOPbase or
outside acceptable limits are taken in hydrogenic approximation. 

\begin{figure}[!h]
\begin{center}
\epsfig{file=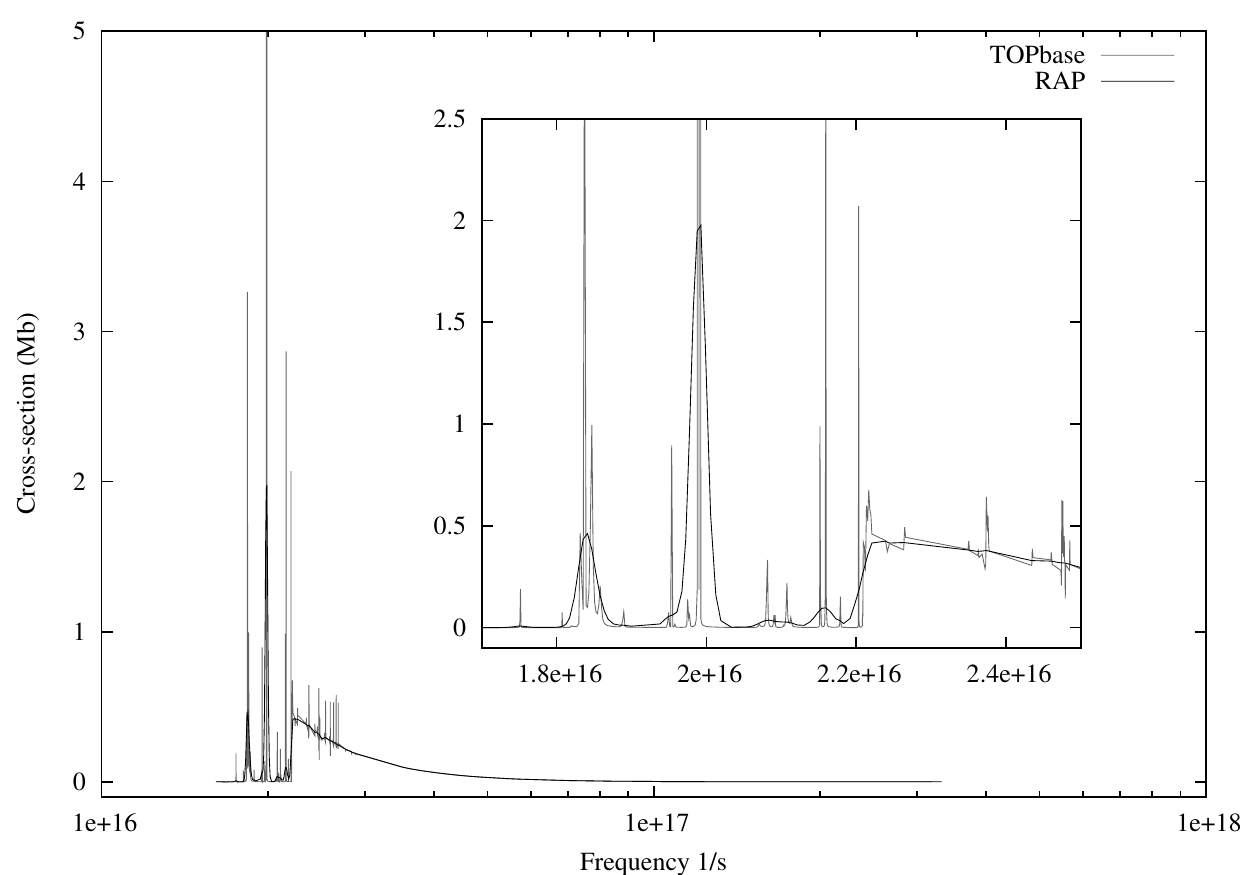, width=12cm}
\parbox{130mm}{
\caption[Example for photoionization cross-sections for Mg.]{Photoionization cross-section of Mg VIII $1s^22s2p(^1\mathrm P_{\circ})3d$
  level (grey) and its resonance averaged representation (black).  
\label{fig:mgviii}}}
\end{center}
\end{figure}

\begin{figure}[!h]
\begin{center}
\epsfig{file=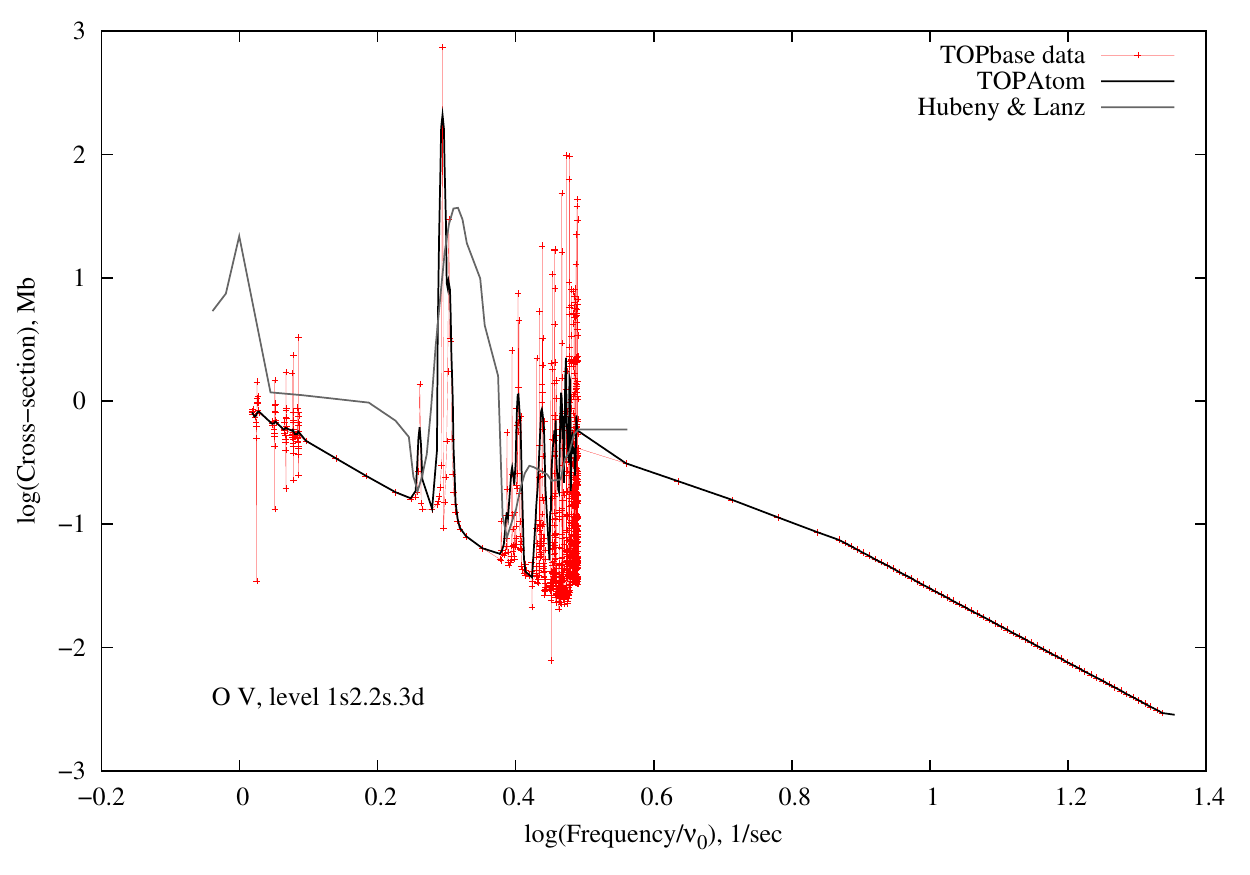, width=12cm}
\parbox{130mm}{\caption[Example for photoionization cross-sections for O.]
{Photoionization cross-section of the O V, {\tt
      1s2.2s.3d} level. $\nu_0$ is the ionization edge frequency. The
    original data consists of 967 data points and are shown here in red. The
    TOPAtom RAP smoothed data is black, it has 192 data points. For
    comparison the cross-section from TLUSTY's web site is plotted with grey.
\label{fig:ov}}}
\end{center}
\end{figure}

Once photoionization data is assigned to a given level, the frequency and
cross section columns are written into separate files. This data set can be
very extensive for TLUSTY. Thus, they need to be resampled, smoothed or
fitted. TOPAtom has three modes to do this. The first one uses the analytic
functions and coefficients for ground states from \citet{verner96}. These 
are adequate only for the highest ionization states. With more electrons
and more complex resonance structure these analytic functions 
do not fit the data. 
The second treatment is the Resonance Averaged Photoionization
Cross-section (RAP) mode as prescribed by \citet{bautista98}. The cross
section is convolved with a Gaussian at constant resolution $\delta{E}/E$:
\begin{equation}
\sigma_A(E) =
C\int_{E-5\delta{E}}^{E+5\delta{E}}\sigma(x)e^{\frac{-(x-E)^2}{2(\delta{E})^2}}{\deriv}x 
\label{eq:bautista}
\end{equation}
where $\sigma$ and $\sigma_A$ are the detailed and averaged photoionization 
cross sections, respectively, and $C$ is a normalizing constant.  
At high
energies the resolution $\delta{E}/E=0.01$ gives reasonable results, both
the general run of the function as well as resonances are represented well, as
shown in Figure \ref{fig:mgviii} for Mg VIII and Figure \ref{fig:ov} for O V. 

It is also possible to manually select representative data points. In this
mode TOPAtom plots the graph of the cross-section and the user can select
data points 
by mouse clicks. This mode allows full control by the user, but would
make fitting hundreds of levels a tedious work, and so was not used here.
Once the resampled cross section is available it is written to the model atom
file.

\subsubsection{Atoms -- Bound-bound transitions}

Bound-bound transitions are built based on the level list. Taking the
first level and forming level pairs with each higher level gives the
transitions. If the level pair can be found in the transition file its
oscillator strength is taken in a primarily ALI transition with Doppler
profile. The collisional rate is evaluated by Van Regemorter's formula 
(Equation \ref{eq:regemorter}) with
$\bar{g} = 0.2$ for transitions $nl \rightarrow n'l'$ (where $n{\ne}n'$) and
$\bar{g}=0.7$ if $nl \rightarrow nl'$. 
\begin{equation}
\Omega=19.7363f_{ij}T^{-3/2}U_0^{-1}\exp(-U_0){\rm max}[\bar{g},0.276\exp(U_0)E_1(U_0)]
\label{eq:regemorter}
\end{equation}
where $U_0=(E_j-E_i)/(kT)$, $E_i$ and $E_j$ are the energies of the lower and
upper levels, $T$ is the temperature and $f_{ij}$ is the oscillator
strength. $E_1(U_0)$ is the first exponential integral function:
\begin{equation}
E_1(U_0)=\int_{U_0}^\infty\frac{e^{-t}}{t}{\rm d}t
\label{eq:elliptical}
\end{equation}
If the transition is neither in NIST nor
in TOPbase its oscillator strength is set to ``scaled hydrogenic''. These lines
are then 
treated in detailed radiative balance with Doppler profiles and 
collisional rates
are evaluated by the Eissner-Seaton formula (Equation \ref{eq:eissner}) 
with $c_0=0.05$:  
\begin{equation}
\Omega=8.631\cdot10^{-6}g_i^{-1}T^{-1/2}{\rm exp}(-U_0)c_0
\label{eq:eissner}
\end{equation}
where $g_i$ is the statistical weight of level $i$.

\begin{table}[!h]\small
\begin{center}
\caption{First ten transitions of N VI.\label{tab:matn6}}
\setstretch{1}
\begin{tabular}{l}
\\
\hline
\verb,*** Line transitions,\\
\verb,1 2             -1 0 1 0 0       1.0000e-10 2.000e-01,\\
\verb,   F  1  7  0.  0.,\\
\verb,1 3              0 0 4 0 0       0.0000e+00 5.000e-02,\\
\verb,1 4              0 0 4 0 0       0.0000e+00 5.000e-02,\\
\verb,1 5             -1 0 1 0 0       6.5300e-08 2.000e-01,\\
\verb,   F  1  7  0.  0.,\\
\verb,1 6              0 0 4 0 0       0.0000e+00 5.000e-02,\\
\verb,1 7             -1 0 1 0 0       6.7420e-01 2.000e-01,\\
\verb,   F  1  7  0.  0.,\\
\verb,1 8             -1 0 1 0 0       2.3600e-11 2.000e-01,\\
\verb,   F  1  7  0.  0.,\\
\verb,1 9              0 0 4 0 0       0.0000e+00 5.000e-02,\\
\verb,1 10             0 0 4 0 0       0.0000e+00 5.000e-02,\\
\verb,...,\\
\hline
\end{tabular}

\end{center}
\setstretch{1}
\end{table}

\subsection*{Averaged model atoms}

The detailed model atom output of TOPAtom is readable by TLUSTY in theory, 
although very extensive. The few
hundred energy levels and thousands of transitions would make model
calculations very ineffective and impractical. 
For this reason, it is necessary to reduce the
number of levels to a few dozen. This is also done with TOPAtom. Levels can
be grouped according to their principal quantum  number, 
their spectroscopic terms or any other user-defined way. TOPAtom
needs only the level indices rearranged into groups and does the averaging
automatically. In the first step it calculates the weighted average of all
levels in a group using the energies and statistical weights of the individual
levels. Autoionizing levels included in TOPbase are also incorporated in model
atom files; they are listed separately, as negative levels.

Bound-free transitions require a bit more attention. These are given with a
few hundred data points of the RAP cross-sections for each individual
transition. First the frequencies are collected from all the cross-sections
belonging to a new level group. This gives the frequency points of the
averaged cross-section. This scale is trimmed to the shortest data set to
avoid spurious extrapolations. Following this, each cross-section data set is
interpolated to the new frequency points and their weighted average is
calculated. This way every feature of the cross-section is
considered. The procedure returns a large number of close lying
data points as well. To sort out unnecessary data points a filtering is
applied for each cross-section. First, the total frequency range is
divided into 1000 bins. Then the relative change is calculated
between adjacent data points; if this is below 0.001 the data point is 
considered
unnecessary and gets deleted. The first and last data points are always kept.
This way the data is sampled with
many data points where the cross section changes 
fast and only by a few in
the smooth parts, such as the high energy tails. 

Transitions are averaged according to Equation \ref{eq:multiplet}, where
$\bar\lambda_{ik}$ is the wavelength of the averaged transition:   
 \begin{equation}
 f_{ik}^{\rm multiplet} = \frac{\sum\limits_{J_i,J_k} (2J_i+1) \lambda(J_i,J_k)
   f(J_i,J_k)}{\bar{\lambda}_{ik} \sum\limits_{J_i}(2J_i+1)} = \frac{
   \sum\limits_{i,k}g_i \lambda_{ik}f_{ik}}{\bar{\lambda}_{ik}\sum\limits_ig_i}
 \label{eq:multiplet}
 \end{equation}

\subsection*{Hydrogen and Helium}

Hydrogen and helium have no spectral lines in the soft X-rays and the energies
considered in my work exceeds their highest ionization degree in most of the
atmosphere. They are important because they are abundant and provide 
reference species. 
Hydrogen and helium
are well-described and incorporated in TLUSTY and SYNSPEC by default. For
this reason I used the hard-coded atomic data and model atoms from TLUSTY's
web page.   

The hydrogen 
model atom is adopted from \citet{lanz03} sdO grid
 (OSTAR2002). It consists of the eight 
lowest levels and one superlevel merging higher excitation
levels from $N = 9$ to $80$, in this model atom 
Lyman and Balmer series are included with approximate Stark profiles and 
all other lines have Doppler profiles. Hydrogen is considered in detailed
balance in my NLTE-C and NLTE-L models.

Similarly to hydrogen, helium is also taken from OSTAR2002 model grid. He I
consists of 19 individual levels up to $N=4$ and five averaged levels up to $N =
8$. Although He I is somewhat populated in hot models, to avoid convergence
problems it was not always included. He II has $20$ levels up to $N=20$. 

I updated my line list for SYNSPEC 
and included all hydrogen and helium lines from $880$
\AA\ to $7500$ \AA\ from the
line lists provided on TLUSTY's page. These $1715$ lines have no significance in
this work, but they are included to keep the line list consistent with the model
atoms.

\section*{Carbon, Nitrogen and Oxygen }

\begin{figure}
\begin{center}
\includegraphics[width=12.5cm]{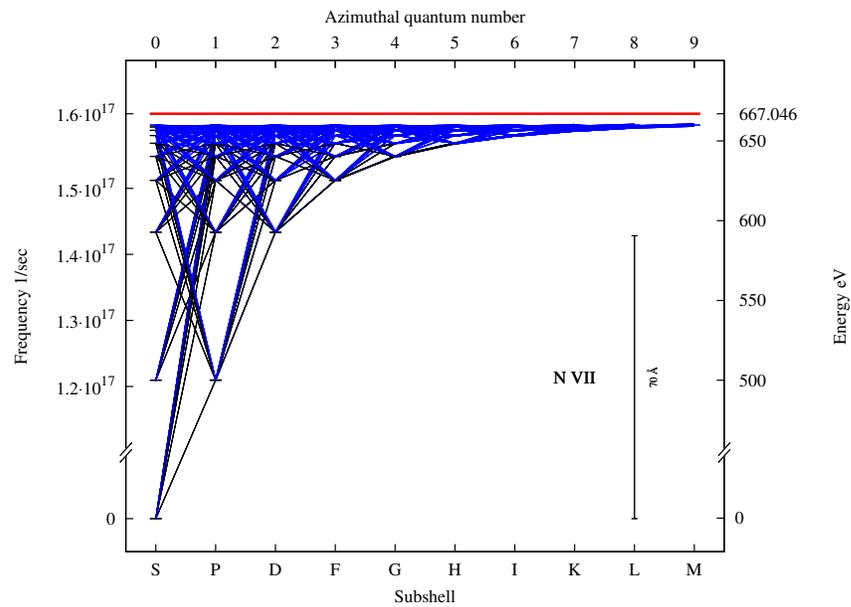}
\parbox{130mm}{
\caption[Grotrian diagram for N VII.]{Grotrian diagram for N VII. Black lines show NIST levels
  and transitions,
  blue ones are from TOPbase. The red horizontal bar is the ionization
  limit. For transitions line widths are proportional
  to oscillator strengths. The {\tt 1s} level is not on the linear scale 
for better clarity of the graph. The $70$ \AA\ bar represents the long
wavelength limit for practical purposes. Levels that are farther apart (Lyman-series) than
this limit have transitions in the supersoft range. 
\label{fig:n7}}}
\end{center}
\end{figure}

NIST has abundant data on energy levels and a significant number of transitions
for C, N and O. Only the highest energy
levels were taken from TOPbase. This means at 
least one level was taken from NIST for
each transition, which improves the accuracy of the wavelengths
of individual lines. Model atoms for these light metals are published from
neutrals up to lithium-like ions. I built helium-like and hydrogenic model
atoms for all three species and lithium-like ions of carbon and oxygen for
verification of my code. The O VI model atom can be found in Appendix 
\ref{appendixa}. 
The hydrogenic ions of C, N and O were fully considered, Figure 
\ref{fig:n7} shows the Grotrian diagram for N VII. 

\begin{table}[!h]
\begin{center}
\parbox{130mm}{\caption[CNO levels and transitions.]
{Number of energy levels and transitions for CNO model
    atoms. $\lambda_e$ is the ionization edge in {\AA}ngstr\"{o}ms.}}
\begin{tabular}{|l|r|rr|rr|rr|rr|}
\hline
&\multicolumn{1}{c|}{$\lambda_e$}&\multicolumn{2}{c|}{TOPbase}&\multicolumn{2}{c|}{NIST}&\multicolumn{2}{c|}{Total}&\multicolumn{2}{c|}{Model
atom}\\
Atom&\multicolumn{1}{c|}{\AA}&Levels&Lines&Levels&Lines&Levels&Lines&Levels&Lines\\
\hline\hline
C V   &31.62&15 &684 &64 &146 &79 &830&22 &151\\
C VI  &25.30&10 &703 &81 &137 &91 &840&20 &115\\
N V   &126.66&0  &72  &45 &402 &45 &474&18 &113\\
N VI  &22.46&17 &483 &51 &90  &68 &573&29 &214\\
N VII &18.59&10 &704 &81 &137 &91 &841&52 &339\\
O VI  &89.77&2  &345 &43 &126 &45 &471&18 &112\\
O VII &16.77&18 &600 &60 &188 &78 &788&29 &203\\
O VIII&14.23&0  &839 &99 &137 &99 &976&49 &295\\
\hline
\end{tabular}
\end{center}
\label{tab:cno}
\end{table}

\section*{Neon and Magnesium}

About one third of classical nova explosions occur
on ONeMg WDs. This core material 
can mix with the
envelope before the nova explosion, causing Ne and Mg enrichment in the
atmosphere and shell. 
The effectiveness of this process is not well understood, but the 
presence of these
elements is observed in both UV and visible regions. 
Unfortunately, neon has very
few lines in the supersoft range. Resonance transitions of Ne IX and X 
ending in the \verb,1s, level are all either shorter than 13.5~\AA\ or longer
than 37~\AA. 

With increasing atomic number the number of NIST transitions reduces 
to the strongest lines of the lower excitation levels. For Ne X and Mg XII 
NIST has no data on lines at all. 

\begin{table}[!h]
\begin{center}
\parbox{130mm}{\caption[Ne and Mg levels and transitions.]
{Number of energy levels and transitions for Ne and Mg
    model atoms. $\lambda_e$ is the ionization edge in
    {\AA}ngstr\"{o}ms.}\label{tab:nemg}}
\begin{tabular}{|l|r|rr|rr|rr|rr|}
\hline
&\multicolumn{1}{c|}{$\lambda_e$}&\multicolumn{2}{c|}{TOPbase}&\multicolumn{2}{c|}{NIST}&\multicolumn{2}{c|}{Total}&\multicolumn{2}{c|}{Model atom}\\
Atom&\multicolumn{1}{c|}{\AA}&Levels&Lines&Levels&Lines&Levels&Lines&Levels&Lines\\
\hline\hline
Ne VII &59.82&80 &4407&148&147 &228 &4554&18 &132\\
Ne VIII&51.85&4  &303 &35 &65  &39  &368 &11 &54\\
Ne IX  &10.37&25 &705 &50 &23  &75  &728 &17 &102\\
Ne X   &9.10&0  &978 &100&0   &100 &978 &25 &145\\
Mg VIII&46.62&225&6649&151&920 &319 &7569&26 &244\\
Mg IX  &37.81&127&2849&71 &406 &198 &3255&23 &196\\
Mg X   &33.73&9  &190 &30 &184 &39  &374 &19 &120\\
Mg XI  &7.04&25 &361 &42 &192 &67  &553 &25 &152\\
Mg XII &6.32&30 &538 &40 &0   &70  &538 &20 &115\\
\hline
\end{tabular}
\end{center}
\end{table}

\section*{Aluminum and Silicon}

CNe are believed to be important sources of galactic Al,
\citet{kolb97}.
To include Al, I built
six ions from Al VIII -- Al XIII. My Al IX contains 339 levels
that is tremendously simplified to the nine lowermost levels and eight
averaged levels for spectral modeling. 
This blurs individual transitions but Al IX is needed to keep the
ionization balance and to set the flux distribution. It does not affect models
over 500,000 K significantly. Al X and higher ions have fewer levels and so my
model atoms have more details on these ions. 

\begin{table}[!h]\begin{center}
\parbox{130mm}{\caption[Al and Si levels and cross sections.]
{Number of energy levels and transitions for Al and Si
    model atoms. $\lambda_e$ is the ionization edge in {\AA}ngstr\"{o}ms}}
\begin{tabular}{|l|r|rr|rr|rr|rr|}
\hline
&\multicolumn{1}{c|}{$\lambda_e$}&\multicolumn{2}{c|}{TOPbase}&\multicolumn{2}{c|}{NIST}&\multicolumn{2}{c|}{Total}&\multicolumn{2}{c|}{Model atom}\\
Atom&\multicolumn{1}{c|}{\AA}&Levels&Lines&Levels&Lines&Levels&Lines&Levels&Lines\\
\hline\hline
Al VIII&43.56&478 &13539&88 &418&566 &13957&   &   \\
Al IX  &37.56&273 &7995 &104&338&377 &8333 &   &   \\
Al X   &31.05&160 &2938 &40 &169&200 &3107 &23 &187\\
Al XI  &28.05&8   &183  &32 &208&40  &391  &25 &158\\
Al XII &5.94 &25  &361  &42 &192&67  &553  &27 &176\\
Al XIII&5.38 &35  &538  &35 &  0&70  &538  &21 &127\\
Si VII &50.50&387 &10998&102&212&489 &11210&25 &219\\
Si VIII&40.85&443 &16332&64 &167&507 &16499&26 &204\\
Si IX  &35.31&517 &14611&63 &270&580 &14881&29 &257\\
Si X   &30.89&297 &8820 &78 &289&375 &9109 &26 &238\\
Si XI  &26.04&150 &2908 &52 &244&202 &3152 &31 &283\\
Si XII &23.68&9   &186  &30 &174&39  &360  &25 &159\\
Si XIII&5.09 &8   &252  &71 &585&79  &837  &27 &173\\
Si XIV &4.64 &30  &538  &40 &  0&70  &538  &21 &127\\
\hline
\end{tabular}\end{center}\label{tab:alsi}\end{table}

\section*{Sulfur, Argon and Calcium}



Sulfur, argon and calcium have numerous lines and provide important line
blanketing. Their complex structure prevents them from being included 
in modeling in detail. 
In my models these ions were extremely simplified, according to their principal
quantum numbers, into 9-10 levels only. In all cases the ground levels were
treated separately and all higher energy levels were grouped together assuming
their population changes in a coordinated way.

\section*{Iron}

\begin{figure}[!h]
\begin{center}
\includegraphics{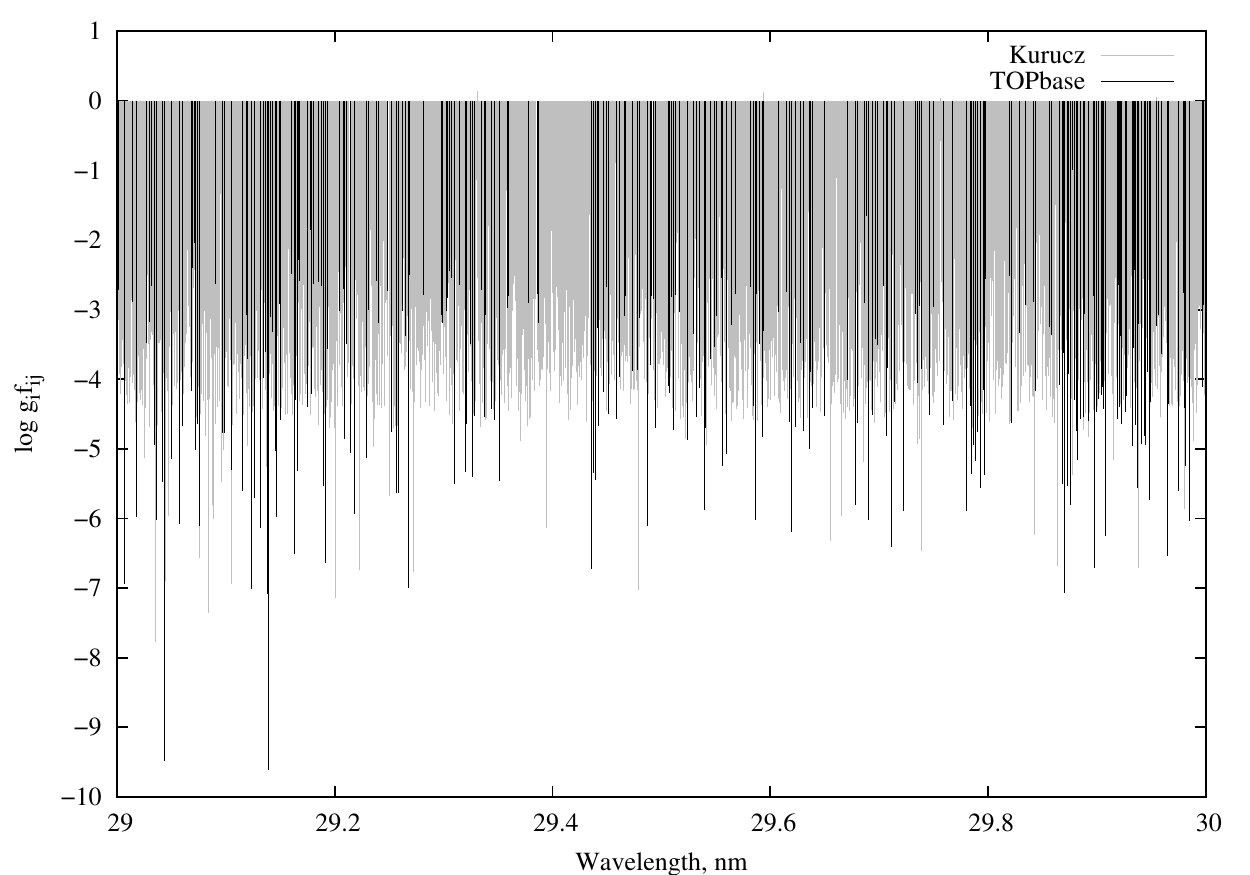}
\parbox{130mm}{\caption{
Comparison of Fe VI lines in Kurucz database and TOPbase. 
}\label{fig:kurucztopbase}}
\end{center}
\end{figure}

Similarly to sulfur, argon and calcium, the 
large number of energy levels and transitions of Iron cannot be treated in
the format of light metals in the calculations. 
For this reason I first combined NIST and TOPbase atomic data with TOPAtom
into the standard format. 
Then I split
these large files into separate data files for energy levels, 
continuum transitions
and line transitions in Kurucz CD-ROM 22 format. \citet{lanz03} and
\citet{lanz07} gave model atoms
for iron from singly ionized to Fe VI. 
To validate my method I built an Fe VI model atom and
compared my result to the published input. The original Kurucz Fe VI has 4740
energy levels and 475,750 transitions. In my model atom I have 761 levels and
37,047 transitions,
although as TOPbase levels are actually multiplets, the
number of individual levels and transitions are somewhat greater. 
I show a comparison between Kurucz and TOPbase transitions in Figure
\ref{fig:kurucztopbase}. 
TOPbase transitions are weaker in general and it is obvious 
that the Kurucz data is superior for low ionizations. 
Low energy 
transitions are not very significant as they do not contribute lines in the
supersoft range and therefore they do not
change the atmospheric structure. 
Similarly, due to the low spectral resolution the short wavelength
transitions form blends and are not distinguishable separately, but they 
affect the
emergent flux through blanketing. 

\begin{figure}[htbp]
\begin{center}
\includegraphics[width=14.2cm]{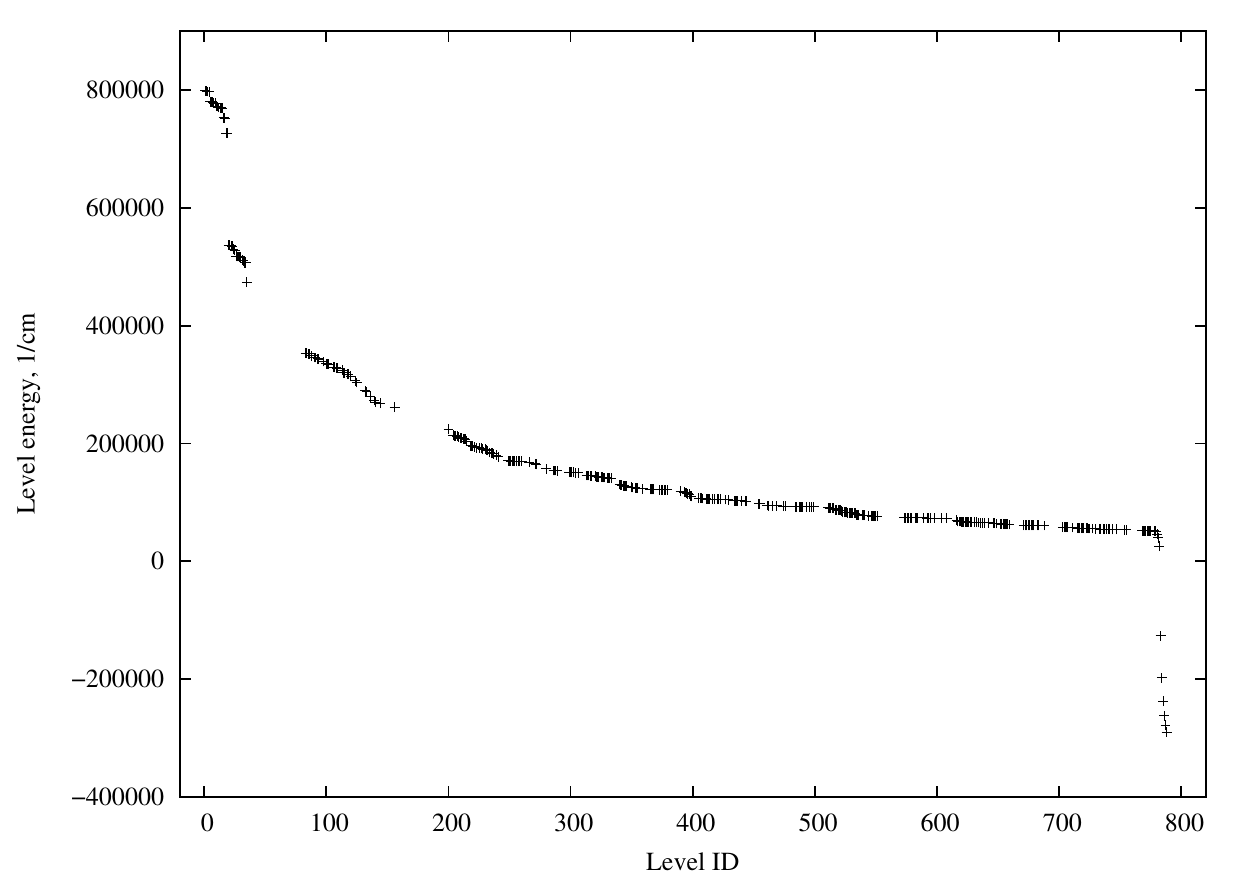}
\parbox{130mm}{\caption{
349 even parity energy levels for Fe VI in TOPbase and NIST. 
}\label{fig:even}}
\end{center}
\end{figure}

I worked out Fe VI in the OS mode and Fe XIV -- Fe XXVI 
model atoms in standard mode.  
In OS mode levels are assigned into superlevels based on their energies and
parity. Figure \ref{fig:even} shows the even parity energy levels for Fe VI.
Energy levels are labeled with integer IDs from the ground level up to the
highest levels. The ionization limit is at zero energy. Doubly excited levels
cause autoionization, in this case an excited atom returns to the ground
state of its next ionization degree. These levels lie lower than the
ionization limit and assigned negative energies and they are
always treated separately from regular levels.
TOPAtom shows this kind of 
graph for even and odd parity levels and one can select the
boundaries of a level group by mouse clicks. The interactive 
graphs allow
for zooming as well. 
Once the superlevels are set up the program calculates their
energies as in the standard mode and constructs the model atom file. 
However, in
this case the model atom file refers to auxiliary data files in which data for
bound-free and bound-bound transitions are stored. Bound-free transitions are
not smoothed, they are represented by the weighted average of all individual
cross sections in the same superlevel. Data on photoionization cross sections 
is stored in \verb,ion.pcs, file. 
For lines and energy levels data is written into
\verb,ion.lin, and \verb,ion.gam, files in Kurucz CD-ROM 22 format, 
respectively.  

\chapter{X-ray Modeling of Classical Novae}

The first detailed 
X-ray spectral modeling of a supersoft source with TLUSTY was done by
\citet{lanz05} on CAL 83. I followed their methodology as well as general
considerations for hot stars from \citet{lanz03}. 
I used my model atoms and carried out 
spectral analysis of two recent novae (V4743 Sgr and V2491 Cyg) in
their supersoft phase to test the new atomic input, and to set 
up boundaries for their luminosities and
abundances. 
In general, to model stellar atmospheres 
one needs to construct a series of spectra in a
possible parameter space for effective 
temperature, gravity and abundance. Then the
elements of this model grid can be compared to observed data 
using $\chi^2$ minimization method to find the best fit model. 
Using the results to update input parameters, the 
fitting process can be repeated until a satisfactory match is achieved.
In this work I used TLUSTY version 200 for the atmospheric 
structure, and SYNSPEC
version 48 to calculate synthetic x-ray spectra.

\begin{figure}[!h]
\begin{center}
\includegraphics[width=13cm]{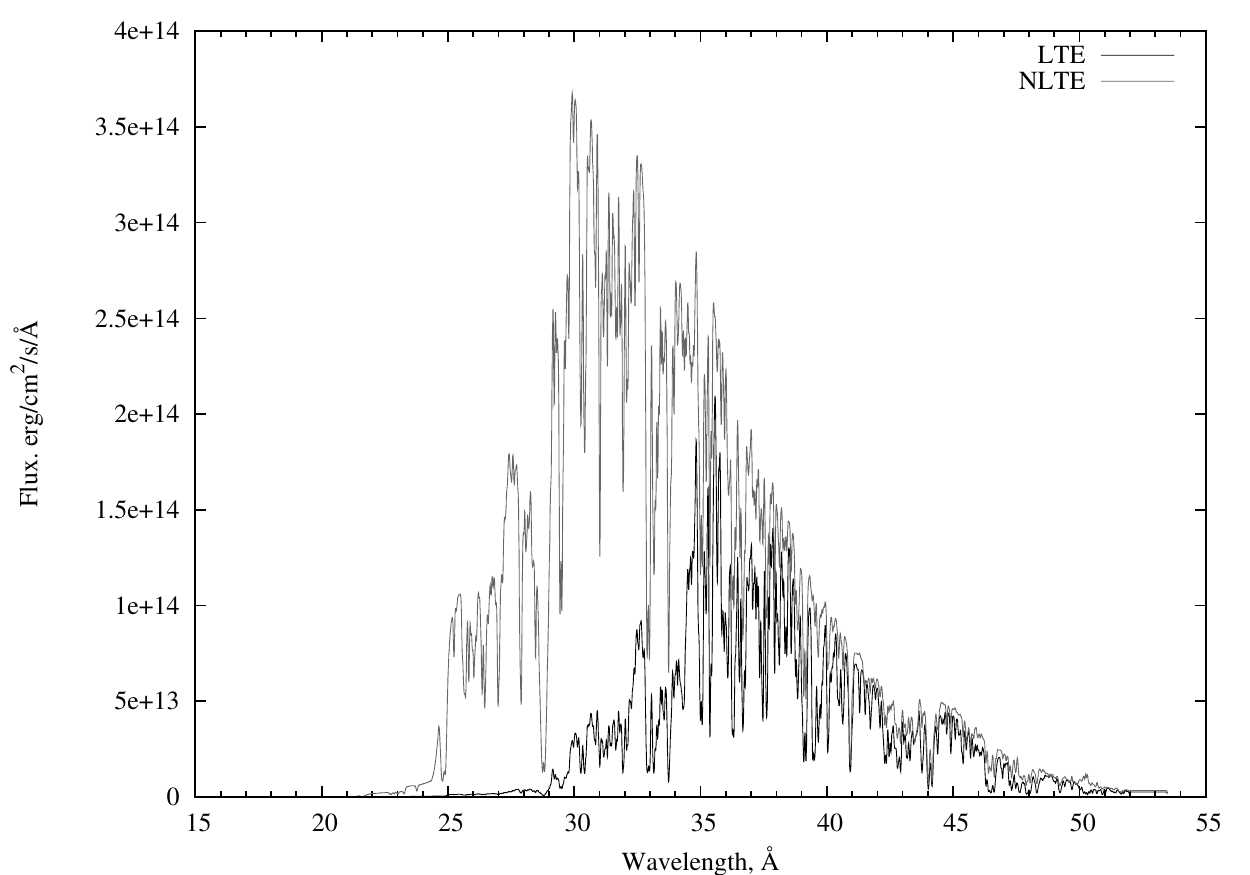}
\parbox{130mm}{\caption[LTE -- NLTE comparison.]{
Comparison of LTE and non-LTE 
synthetic spectra calculated with the same model input at $T_{\rm
  eff}=450,000$ K and $\log g=8.00$ cm/s$^{-2}$.
Spectra are corrected for interstellar absorption with
$N_h=4.1\times10^{-20}$ cm$^{-2}$ and convolved for Chandra resolution (see
Section \ref{sec:res}).
\label{fig:lte_nlte}}}
\end{center}
\end{figure}

There are severe complications to 
modeling supersoft sources. Rough estimates of effective temperatures and 
surface gravities can be derived from
observations. Unfortunately, determination of chemical composition is not
straightforward. The CNO-processed hydrogen-burning envelope has a peculiar
abudance with increased helium, nitrogen, oxygen and depleted carbon. This 
abundance is further complicated 
by both the mixing of WD core material into the envelope 
and the unknown original composition of the
accreted material. Line-blanketing has a crucial role in both the structure of
the atmosphere 
and in the formation of the final spectrum. 
The flux blocked by these
lines emerges at other, higher frequencies. Because the bandwith of the
spectrum in which energy transport occurs is restricted by lines, a steeper
temperature gradient is necessary in underlying layers to drive the flux, 
causing the effect called backwarming \citep{mihalas78}.   
These effects can be seen in Figure
\ref{fig:lte_nlte} where LTE and non-LTE spectra with same composition,
effective temperature and surface gravity were compared. With increasing
surface gravity this effect decreases and spectra calculated in LTE are similar
to non-LTE. 
Ionization balance can change significantly throughout the atmosphere due to
non-LTE effects over a few thousand
Kelvins. The extended and expanding atmosphere complicates the
spectral appearance further. TLUSTY's basic assumptions, that is, the
atmosphere is in hydrostatic equilibrium and can be considered in
plane-parallel geometry break down for novae. To include the effects of an
extended and dynamic atmosphere in modeling
a major upgrade of the code is needed, which is not within the scope of this
work. However, the atomic data input presented in Chapter \ref{chap:modatom}
is necessary for future work with supersoft sources and important conclusions
on atomic data can be drawn from static and plane-parallel tests.  

The final 
spectrum is dominated by both the few 
strong
lines of abundant
light metals and the blanketing of numerous weak lines of heavier
metals. Simple CNO models can be calculated fast based on observed line
strengths, but because of the
strong coupling between species it is necessary to include all heavier 
elements in the
modeling at once. 
With increasing atomic number, intermediate levels become more
significant, neglecting these decrease modeling accuracy. 
By including these levels computer time also increases drastically.
Also, inaccuracies in
atomic data, the blurred energy levels of averaged model atoms and the
overwhelming interstellar absorption altogether prevent
qualitative methods. Model fitting was done by eye, within tolerances chosen
by
changing each model parameter one-by-one and 
inspecting their effects on the final
spectrum.  

Proton collision rates were neglected in modeling as \citet{apetz05} did not
find significant contribution of this process.
A microturbulent velocity of 50 km/s was used, similar to O-star models of
\citet{lanz03}. The frequency range was automatically 
set up for continuum transitions 
by TLUSTY
according to $\nu_{\rm max}=8\times10^{11}T_{\rm eff}$, or
$h\nu_{\rm max}/kT_{\rm eff}\approx 38$. The minimum frequency was $10^{12}$
Hz for all models. 
The same frequency range was applied for line transitions as well. The
Rosseland optical depth was $10^{-7}$ in the first, and increased to $10^4$ in
the last depth point to consider the low opacity of hydrogen. The radiative
transfer equation was solved by an ordinary second-order Feautrier method with
a full ALI scheme, where the diagonal $\Lambda^*$ operator was 
evaluated by the algorithm
of \citet{rybicki91}. Bound-free transitions were taken as primarily
linearized transitions with photoionization cross sections taken from
TOPbase. Electron collisional rates were calculated by Seaton's formula
(Equation \ref{eq:seaton}) for continua: 
\begin{equation}
\Omega=1.55\times10^{13}T^{-1/2}U_0^{-1}\exp(-U_0)\sigma_0\bar{g}
\label{eq:seaton}
\end{equation}
where $U_0=(E_j-E_i)/kT$, $\sigma_0$ is the absorption cross section at the
ionization edge and $\bar{g}=0.3$ for all ions.  
Bound-bound transitions (spectral lines) were treated
in ALI with Doppler profiles. Only the strongest resonance lines of nitrogen
and oxygen were
considered with depth-dependent Voigt profiles. Collisional rates were
calculated by Van Regemorter's formula (Equation \ref{eq:regemorter}). 
Light metals up to silicon
were considered with detailed model atoms. Sulfur, argon and calcium were
included with their simple model atoms having only 9--10 levels per ion. 
Ions of iron were considered
with their 
ground levels only to account for the effects of iron on the
ionization balance. At least
50 depth points were set up, for some models it was increased to 80. With 
less than 50, or more than 80 depth points the convergence properties 
of the models got worse. The maximum relative change of the state vector was
kept at its default value of 
$10^{-3}$, as shown in Figure \ref{fig:convC}. Figure
\ref{fig:convC} 
shows only the relative changes of occupation numbers as temperature,
and pressure convergence were much faster 
for every model. Opacity Sampling was not used as with light metals it did not
change the spectra but increased the computation time. For iron-peak elements
however, the many new frequency points introduced by the OS mode exceeded
computer limits.

\begin{sidewaysfigure}[p] 
\centering
          \includegraphics[width=.6\textwidth]{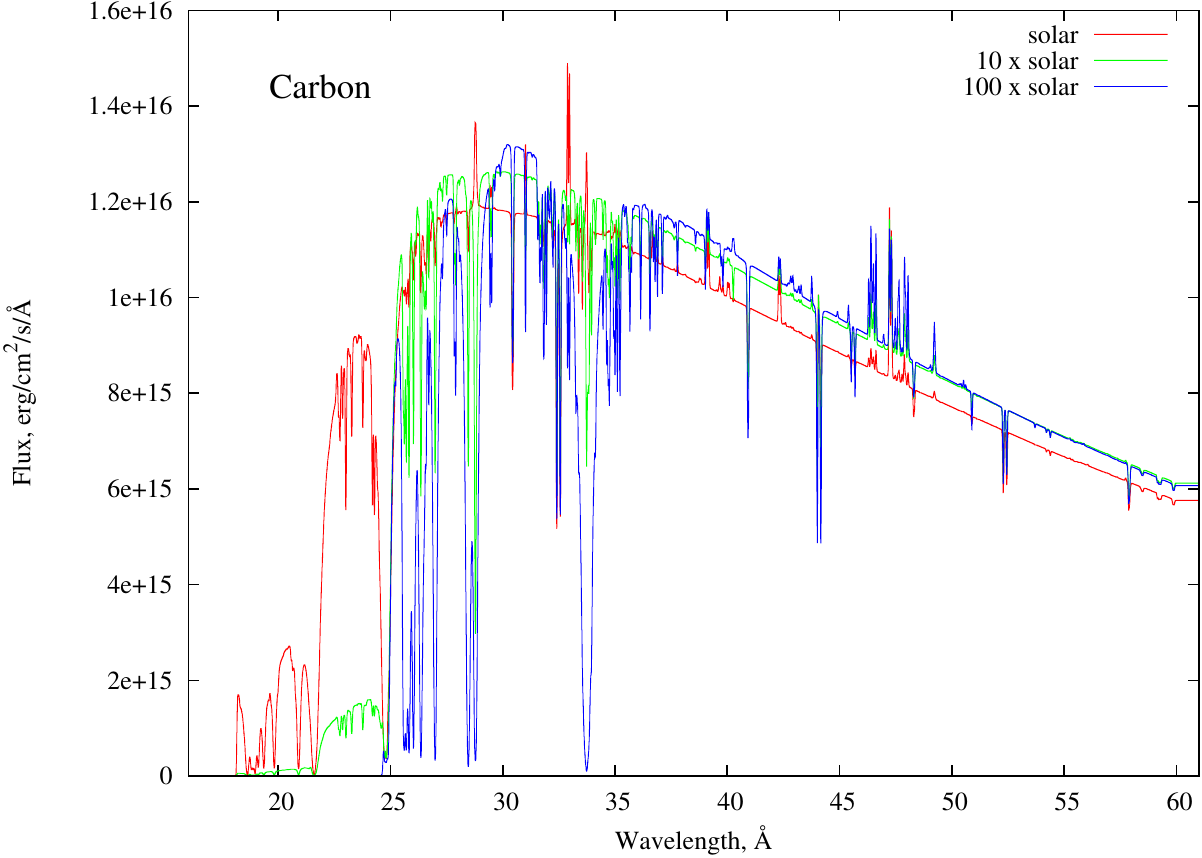}
     \caption{Effects of different abundances of carbon on the
       final spectra.\label{fig:abnch_c}}
\end{sidewaysfigure}

\begin{sidewaysfigure}[p] 
\centering
          \includegraphics[width=.6\textwidth]{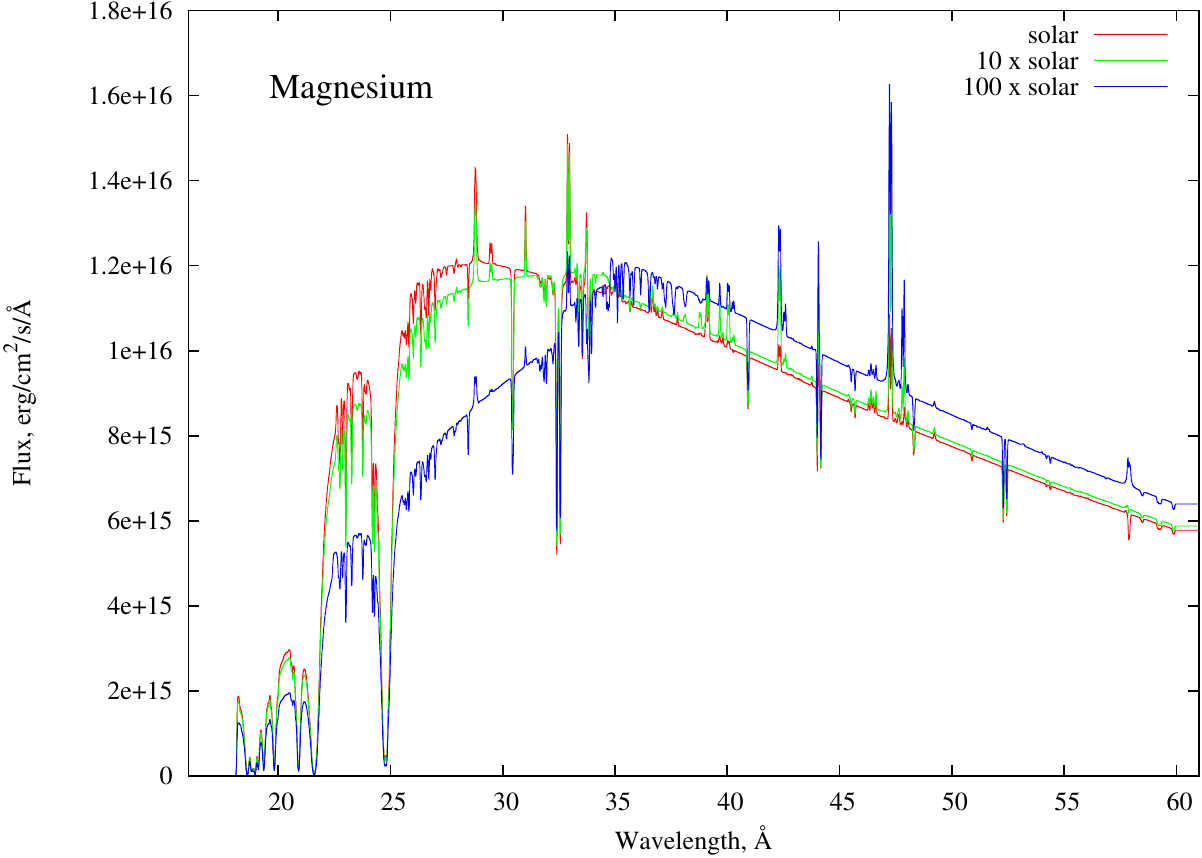}
     \caption{Effects of different abundances of magnesium on the
       final spectra.\label{fig:abnch_mg}}
\end{sidewaysfigure}

\begin{figure}[!h]\begin{center}
\includegraphics{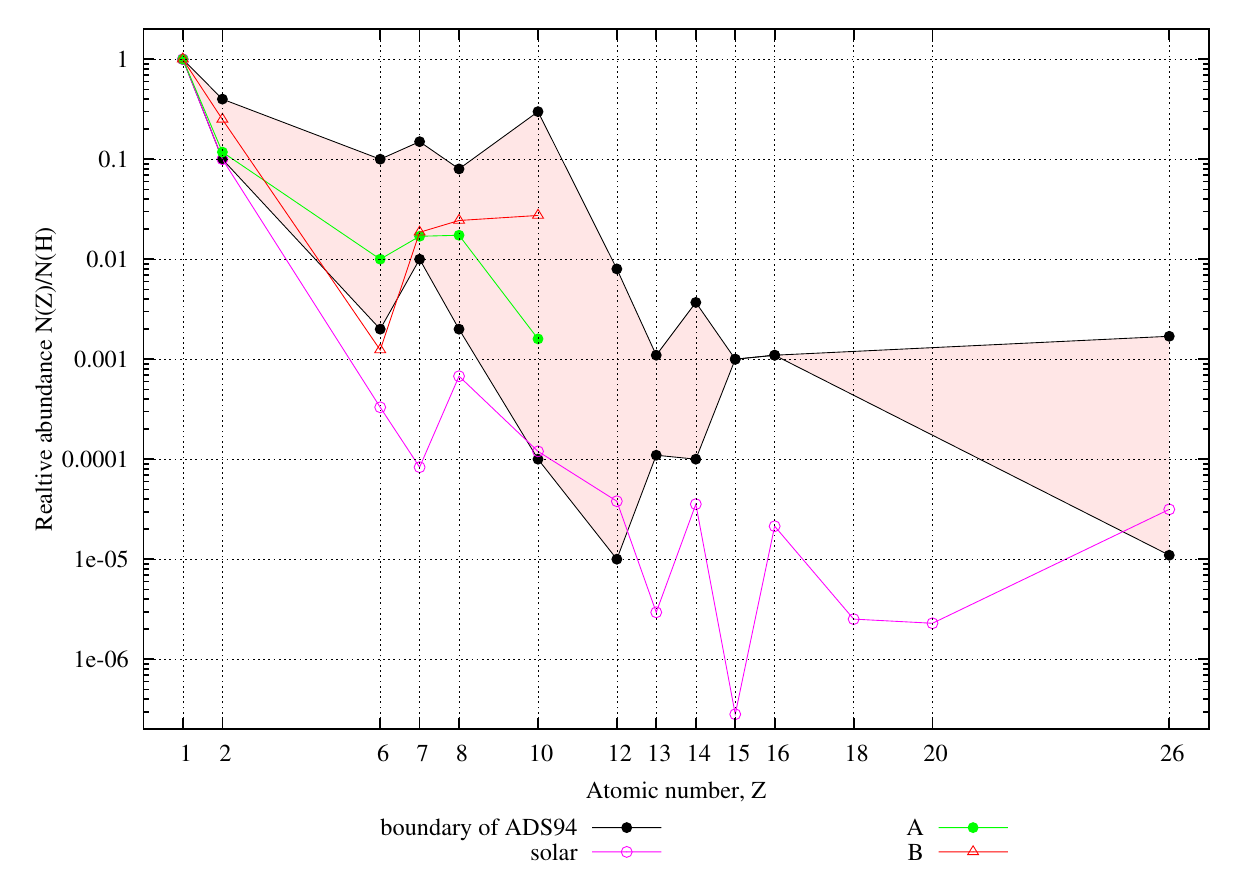}
\parbox{130mm}{\caption[Nova shell abundances.]{Nova abundance limits based on UV and optical observations of nova
  shells, data taken from \citet{ads94} and \citet{gehrz98}. Solar abundance
  is from TLUSTY, \citet{grevesse98}. Lines A and B shows abundance from
  Table \ref{tab:abundance} \citep{macdonald91}.}\label{fig:abund}}
\end{center}\end{figure}

Model calculation started with invoking TLUSTY and SYNSPEC consecutively in a
user defined temperature, surface gravity range and resolution. 
Super-Eddington models were
skipped as they do not converge and are out of TLUSTY's modeling domain. 
Each model atmosphere was calculated in two
steps. First a continuum gray 
model was calculated, keeping all lines in detailed
balance. This was followed by a full non-LTE model calculation with lines and 
frequency dependent
opacities. Optionally, LTE models can be calculated before non-LTE, but this
is not necessary for the convergence of non-LTE models. 
Element abundance can be changed parallel for all elements, 
starting with solar abundance and increasing to $3$, $30$, $100$ and $10^3$ 
times their solar
values. Convergence properties of the solar models are poor and this method 
failed for higher abundances and high energy levels included. 
Also the large number of overlapping lines (up 
to 15) made it hard or impossible to identify individual lines in the spectra.
Abundances were set up individually for each star based on nova shell
abundances in \citet{ads94} and \citet{gehrz98} and by investigating the
effects of different species on the spectra. The shell compositions can
be seen in Figure \ref{fig:abund}. 
To discover the contribution of each element to the spectra, I calculated
models with similar input parameters, only changing the abundance of one
element at a time. Figure \ref{fig:abnch_c} and \ref{fig:abnch_mg} show these
graphs for carbon, nitrogen, neon and magnesium, respectively. For carbon, the 
increasing abundance 
changes the flux significantly. A strong cut-off can be seen below 25~\AA, 
while the the flux increases in the continuum over 25~\AA. An 
increase in line strengths of the Lyman series (33.7, 28.5, 27.0, 26.3
26.0~\AA) can be also seen. Similar effects can be found in the case of
nitrogen only on a moderate level. The high energy cut-off is at 21.5~\AA\ and
the Ly$\alpha$ (24.7~\AA) line is dominant. The soft part of the spectrum is
practically unchanged.      

\begin{sidewaysfigure}[p] 
\centering
          \includegraphics[width=.6\textwidth]{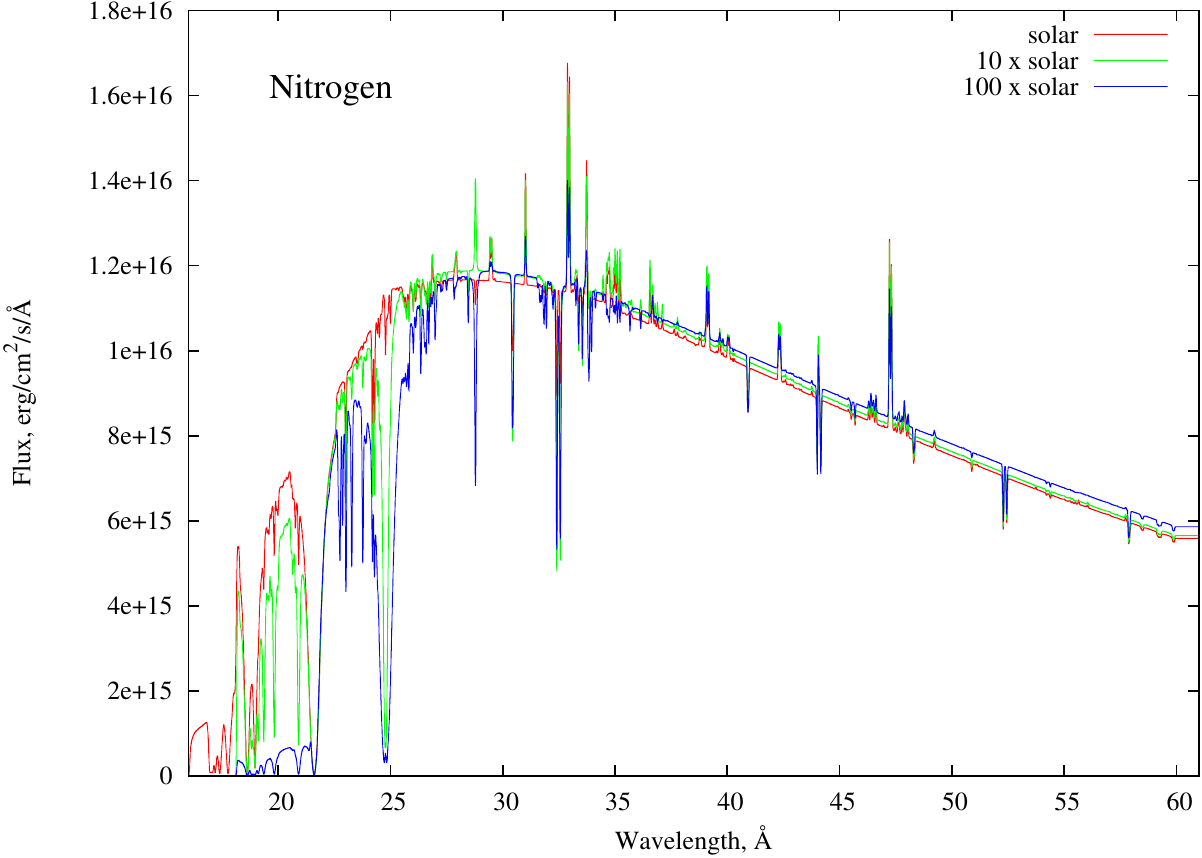}
     \caption{Effects of different abundances of nitrogen on the
       final spectra.\label{fig:abnch_n}}
\end{sidewaysfigure}

\begin{sidewaysfigure}[p] 
\centering
          \includegraphics[width=.6\textwidth]{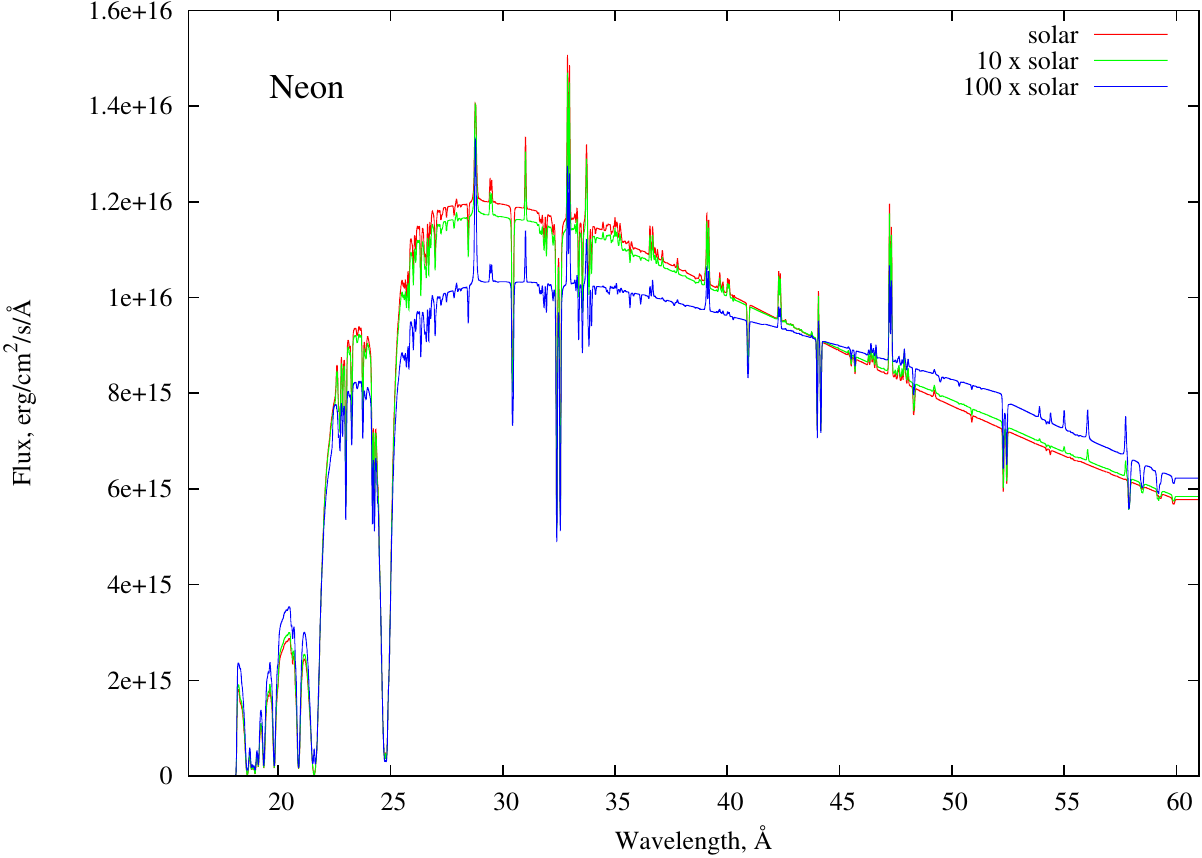}
     \caption{Effects of different abundances of neon on the
       final spectra.\label{fig:abnch_ne}}
\end{sidewaysfigure}

The neon abundance has not much effects on the spectra. 
Unfortunately neon has no
spectral lines in the supersoft range. Transitions to the \verb,1s, level
lie in the hard X-rays, and transitions to \verb,2s, and \verb,2p, 
levels are all softer than 30 \AA. 
Neon decreases the flux between 23 and 43 \AA, while it
does not introduce significant changes below 23 \AA. A small increase can be
seen over 43 \AA. More interesting are the effects of magnesium. 
While increasing magnesium abundance only slightly reduces
the flux below 22 \AA, there is a great impact between 23 and 35 \AA. The
large photoionization cross-section of magnesium redistributes the energy into
the soft tail of the spectrum. Similar graphs were calculated for all included
species and chemical composition was set up with the help of these diagnostics.

Once the grid had been generated all spectra were corrected for
interstellar absorption as described in Section 
\ref{sec:reddening}, and convolved with the
instrumental resolution, as described in Section \ref{sec:res}. 
Then spectra were scaled to observations, which
directly provided $d^2/R^2$ factors, where $R$ is the radius of the star and
$d$ is its distance. The next step would be to select some strong lines and
cross-correlate the model and data to get Doppler-shifts. This can be
done for lines of separate species or ionization degrees. The last step would
be to
select some lines of a given ion and get ${\chi}^2$ for each
spectrum. All these steps can be done by TGRID, a procedure-oriented 
script which generates the grid and carries out the steps of spectral analysis. 
Unfortunately, the large number of overlaping lines in the soft X-rays makes
$\chi^2$ minimization unreliable. Due to the strong line blanketing radial
velocities could be measured only for the strongest lines of 
nitrogen and oxygen.

\begin{table}[!h]
\centering
\parbox{130mm}{\caption[Abundance of CO and ONeMg noave.]
{Abundance for CO and ONeMg novae 
from \citet{macdonald91}.}\label{tab:abundance}}
\vspace{2mm}
\begin{tabular}{llcccccc}
\hline\hline
Abundance&\multirow{2}{*}{WD type}& \multicolumn{6}{c}{Fractions}\\
pattern&  & H    & He   & C     & N    & O    & Ne\\
\hline
A (mass)&CO      & 0.47 & 0.22 & 0.059 & 0.11 & 0.13 & 0.015\\
A (number)&CO      & 1.00 & 0.118 & 0.01 & 0.017 & 0.0174 & 0.0016\\
A (number$_\odot$)&CO      & 1.00 & 1.21 & 29.42 & 150.6 & 20.51 & 13.9\\
B (mass)&ONeMg   & 0.31 & 0.31 & 0.0046& 0.08 & 0.12 & 0.17\\
B (number)&ONeMg   & 1.00 & 0.25 & 0.00124& 0.0186 & 0.0244 & 0.0274\\
B (number$_\odot$)&ONeMg   & 1.00 & 2.58 & 3.47& 166.1 & 28.7 & 238.8\\
\hline
\end{tabular}
\end{table}

\begin{figure}[!h]
\begin{center}
\includegraphics[width=12cm]{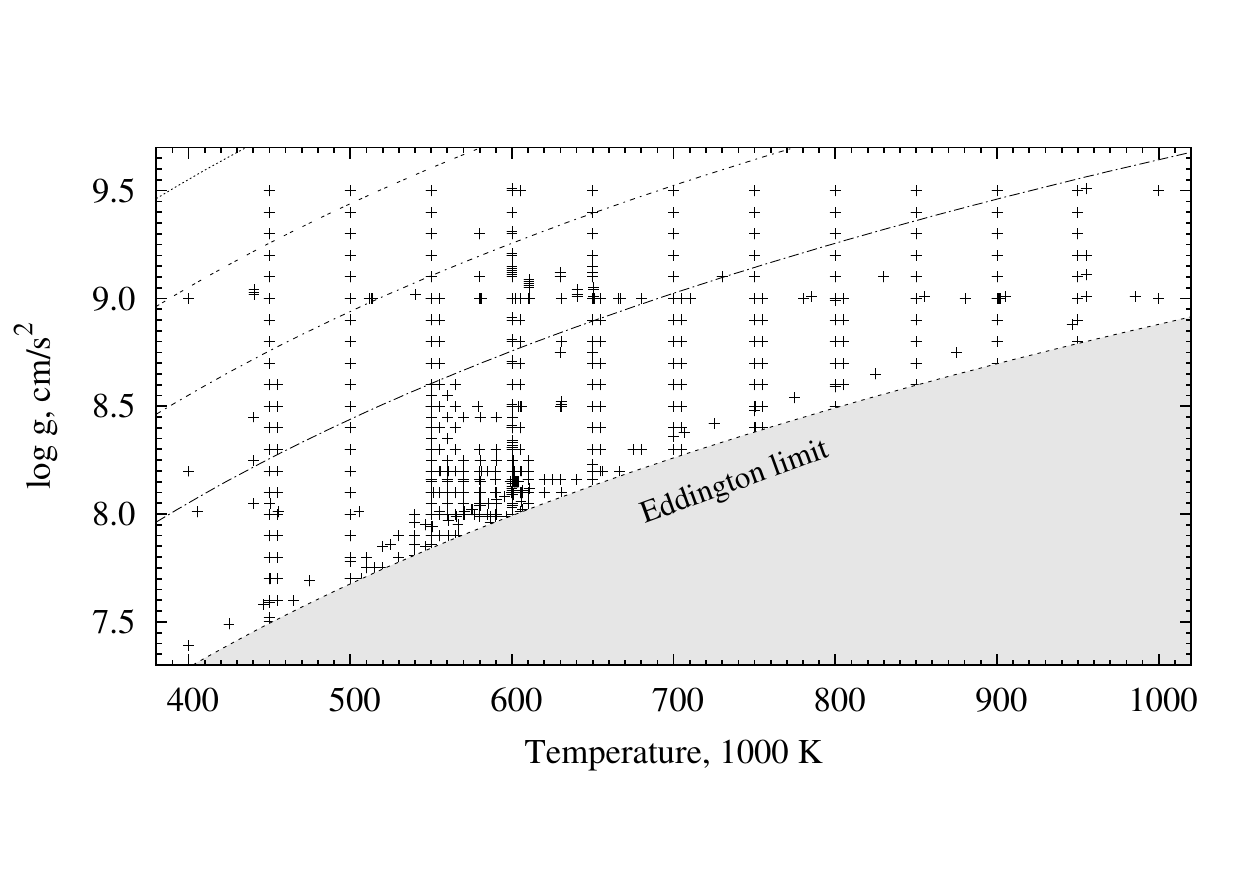}
\parbox{130mm}{\caption[Model grid.]{Model grid. Each point indicates 
a TLUSTY model. The
    Eddington limit and 4 iso-luminosity curves are also shown assuming a given
    stellar mass and black body radiation.\label{fig:grid}}}
\end{center}
\end{figure}

Temperature and gravity ranges of my model grid can be seen on 
Figure \ref{fig:grid}. It is worth comparing surface
gravities for some
grid points with
Figure \ref{fig:massradius} to find mass, radius and
relative numbers of these WD cores. However, this mass-radius relation valid
only for the degenerate cores, as the atmosphere of CNe in the
constant bolometric luminosity phase is significantly extended, up to a few
stellar radii. In Figure \ref{fig:grid} each point represents a TLUSTY
model. With SYNSPEC it is possible to change the composition assuming
that small changes in abundance do not change the atmospheric structure, so
the number of synthetic spectra calculated is larger than shown in Figure
\ref{fig:grid}.   
This $\log g$ vs. $T_{\rm eff}$ diagram also shows the effective
modeling domain of TLUSTY. It can calculate atmospheres
with effective temperature up to $10^9$ K, although
at high gravities the Chandrasekhar-limit sets a
natural boundary, WDs over $\log g=9.5$ were not considered in my work. 
At low gravities and high temperatures typical for
these atmospheres, 
the Eddington-luminosity sets another limit, as a strong 
stellar wind breaks down the basic assumption of TLUSTY, i.e. the atmosphere
is no longer static.
The Eddington-luminosity as function of mass 
can be written as:
\begin{equation}
L_{\rm Edd} = \frac{4{\pi}GMm_{\rm p}c}{\sigma_{\rm T}} \cong 3.3\times10^{4}\left(\frac{M}{M_{\odot}}\right)L_{\odot}
\label{ledd}
\end{equation}
where $m_{\rm p}$ is the proton mass and $\sigma_{\rm T}$ is the Thomson
cross-section for electrons. The black body luminosity is:
\begin{equation}
L = 4R^2{\pi}{\sigma}T_{\rm eff}^{4}
\label{bblum}
\end{equation}
where $\sigma$ is the Stephan-Boltzmann constant and $T_{\rm eff}$ is the
effective temperature of the star. From the two equations above and the
definition of surface gravity: 
\begin{equation}
g = \frac{GM}{R^2}
\label{grav}
\end{equation}
it follows that the temperature 
dependence of the surface gravity at Eddington-luminosity is:
\begin{equation}
\log g = 4\times\log T_{\rm eff} - 15.12
\label{ledd}
\end{equation}
where $\log g$ is in cm/s$^2$ and $T_{\rm eff}$ is in Kelvins. This modeling
limitation is also shown in Figure \ref{fig:grid}

Following \citet{lanz05}, Compton scattering was neglected. Their analysis did
not reveal significant contribution from Compton scattering at $500,000$ K and
$\log g = 8.5$, while extended the computation time. 
Also, their comprehensive study
on line broadening mechanisms concluded that at the high temperature typical
for supersoft sources 
natural broadening
dominates linear Stark broadening. In my models natural broadening was included
for the strongest resonance lines in model atoms and for all the lines in the
line list either from TOPbase or NIST/ASD.

\section{Model Example}

A representation of all models is beyond the scope of this dissertation.
Because
the  
structures of these atmospheres are only slightly different, here only 
one of them
is discussed in detail. This atmosphere was calculated as part of the
grid for V4743 Sgr. The model parameters are presented in Table
\ref{tab:modelp}. 

\begin{table}[!h]
\centering
\parbox{115mm}{\caption[Example model parameters.]
{Model parameters of example atmosphere calculated at $T_{\rm eff}=599,900$ K and
  $\log g=8.54$ cm/s$^2$. Abundances are given in solar units.}\label{tab:modelp}}
\vspace{2mm}
\\
\begin{tabular}{lr@{.}ll}
\hline
Parameter&\multicolumn{2}{c}{Value}&unit\\
\hline
Effective Temperature, $T_{\rm eff}$&599,900&0& K\\
Surface gravity, $\log{g}$         &8&54& cm/s$^2$\\
H/H$_\odot$ &1&0&\\
He/He$_\odot$&3&5&\\
C/C$_\odot$ &4&0&\\
N/N$_\odot$ &27&0&\\
O/O$_\odot$ &12&0&\\
Ne/Ne$_\odot$&50&0&\\
Mg/Mg$_\odot$&30&0&\\
Al/Al$_\odot$&10&0&\\
Si/Si$_\odot$&2&0&\\
S/S$_\odot$ &1&0&\\
Ar/Ar$_\odot$&0&1\\
Ca/Ca$_\odot$&0&05&\\
Fe/Fe$_\odot$&0&01&\\
\hline
\end{tabular}
\end{table}

\begin{figure}[!h]
\begin{center}
\includegraphics[width=13cm]{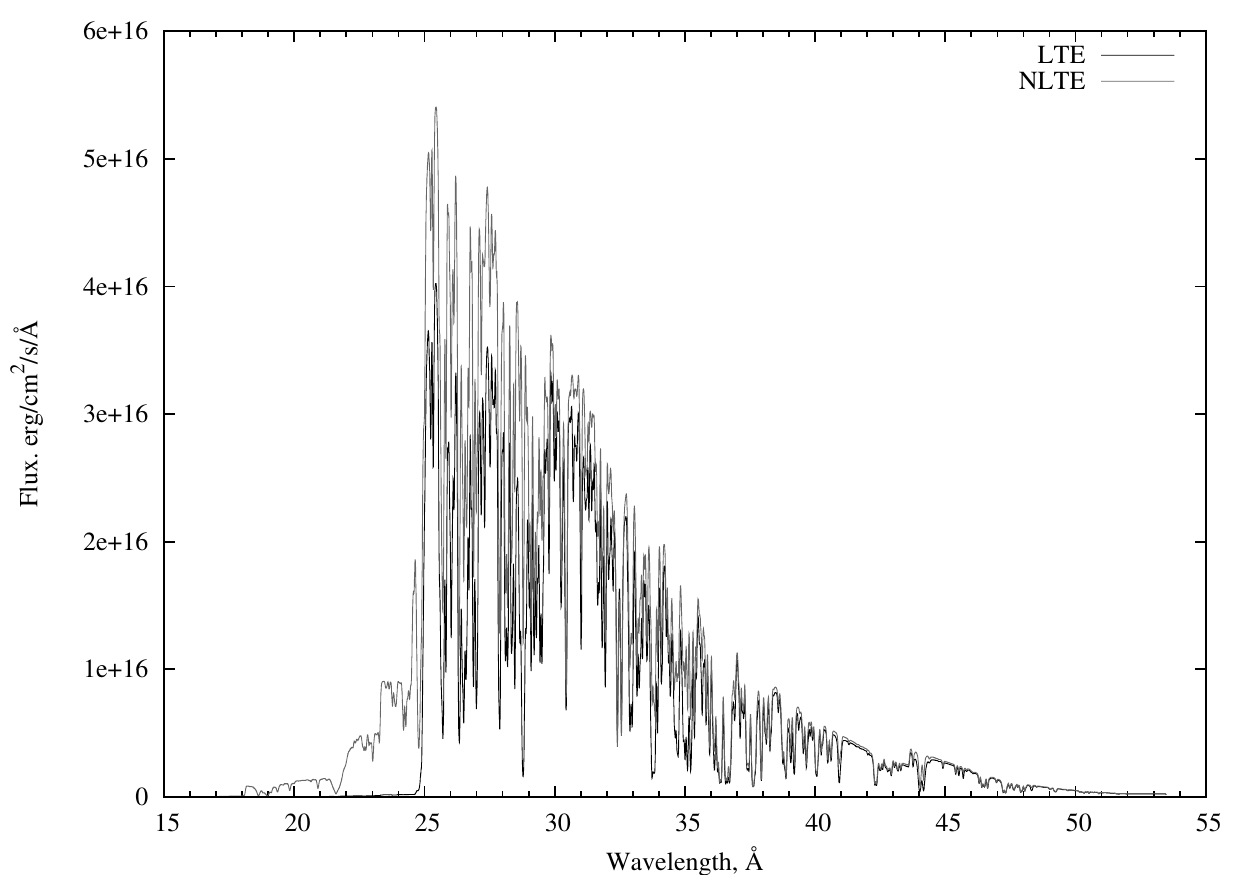}
\parbox{130mm}{\caption[Final spectrum in LTE and NLTE.]{
Comparison of LTE and non-LTE 
synthetic spectra calculated with the same model input.
Spectra are corrected for interstellar absorption with
$N_h=4.1\times10^{-20}$ cm$^{-2}$ and convolved for Chandra resolution (see
Section \ref{sec:res}).
\label{fig:lte_nlte600}}}
\end{center}
\end{figure}

Figure \ref{fig:lte_nlte600} shows the final spectra assuming LTE and non-LTE at
identical input parameters. Strong non-LTE effects clearly 
dominate the spectrum and cause flux redistribution towards high energies as
\citet{hartmann97} pointed out with ground level model atoms. This result also
shows that hot nova atmospheres can not be modeled in LTE.  

\begin{figure}[!h]
\begin{center}
\includegraphics[width=13cm]{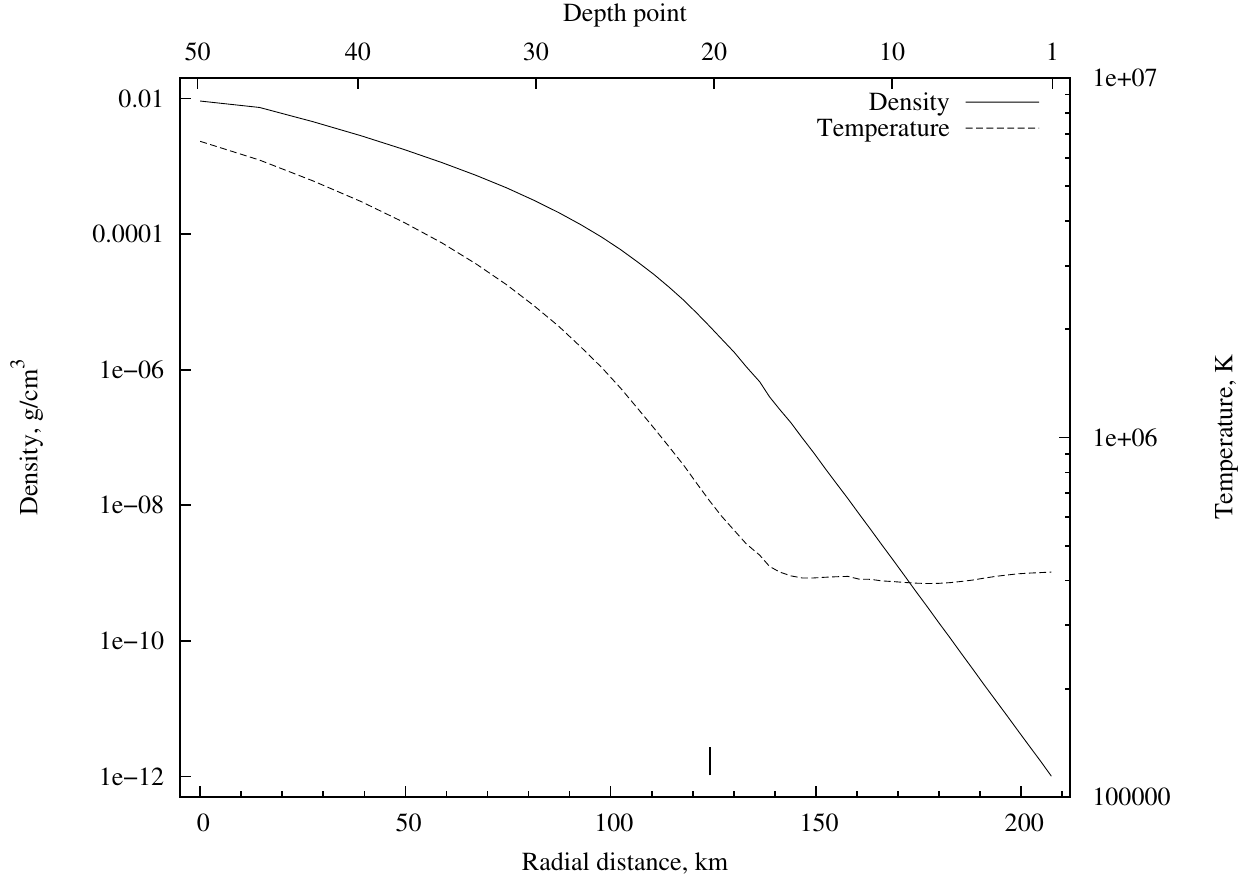}
\parbox{130mm}{\caption[Atmospheric structure.]{Atmospheric structure. The two
    lines represent the temperature and density throughout the atmosphere as
    function of radial distance from the bottom of the
    atmosphere. Corresponding depth points are also indicated. 
\label{fig:structure1}}}
\end{center}
\end{figure}

The temperature and density 
structures of the atmosphere are shown in Figure \ref{fig:structure1}. The
bottom horizontal axis shows the radial distance in the atmosphere calculated
from the column mass and density at each depth point. The horizontal axis on
the top indicates the corresponding depth points to the radial
distance. The short vertical line at 124 km indicates $\tau_{\rm Ross}=1$, the
location of the continuum forming region measured from the bottom of the
atmosphere. 
Temperature decreases sharply with radius 
in the inner parts of the atmosphere, then
above the photosphere it becomes nearly constant. In the outermost layers it
shows a slight increase. Density changes slow at the beginning and over the
photosphere it decreases logarithmically. 

\begin{figure}[!h] 
     \centering
     \subfigure[Carbon.]{
          \label{fig:ionfrac.C}
          \includegraphics[width=.57\textwidth]{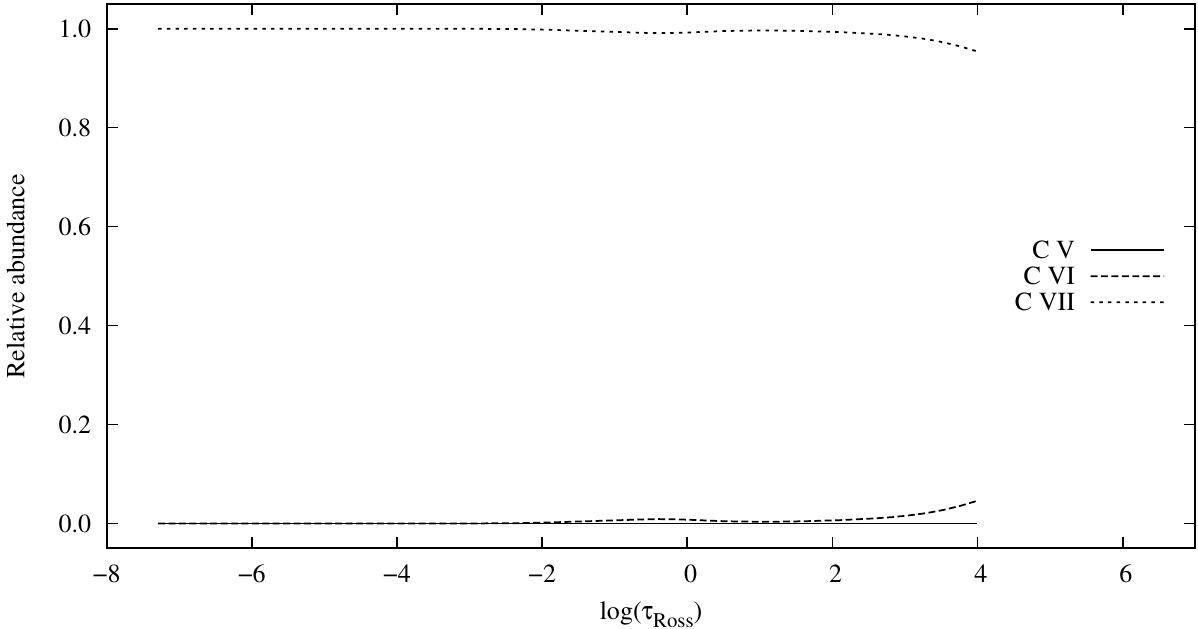}}\\
     \vspace{.3in}
     \subfigure[Nitrogen.]{
          \label{fig:ionfrac.N}
          \includegraphics[width=.57\textwidth]{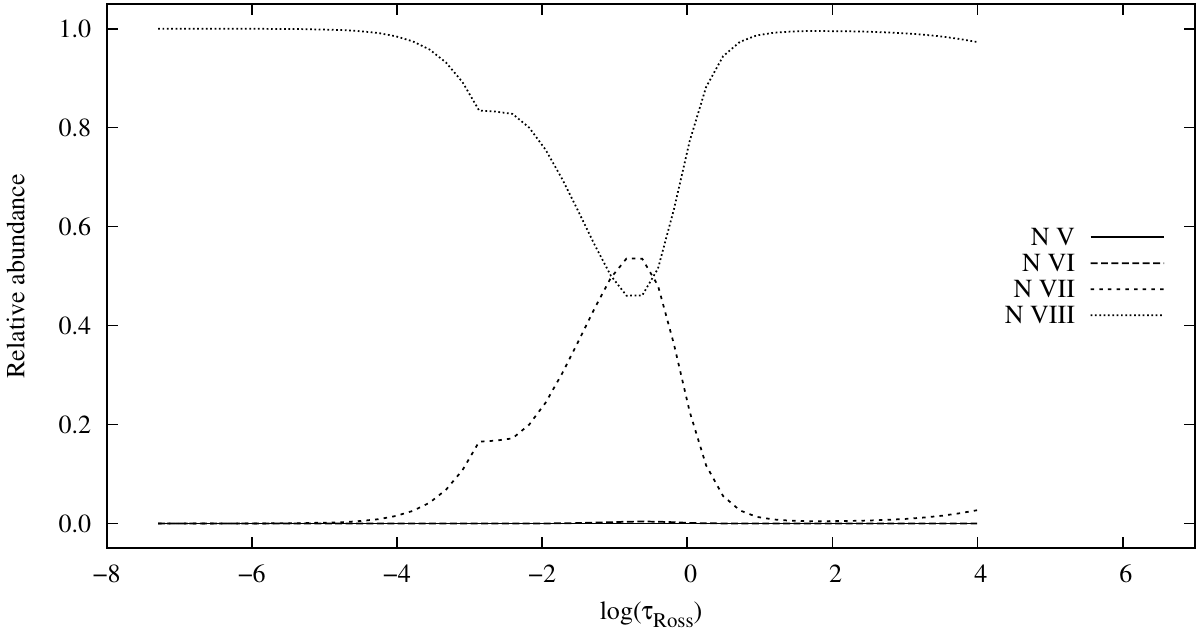}}\\
     \vspace{.3in}
     \caption{Ionization fractions for carbon and nitrogen.}
     \label{fig:ionfracCN}
\end{figure}

\begin{figure}[!p] 
     \centering
     \subfigure[Oxygen.]{
          \label{fig:ionfrac.O}
          \includegraphics[width=.57\textwidth]{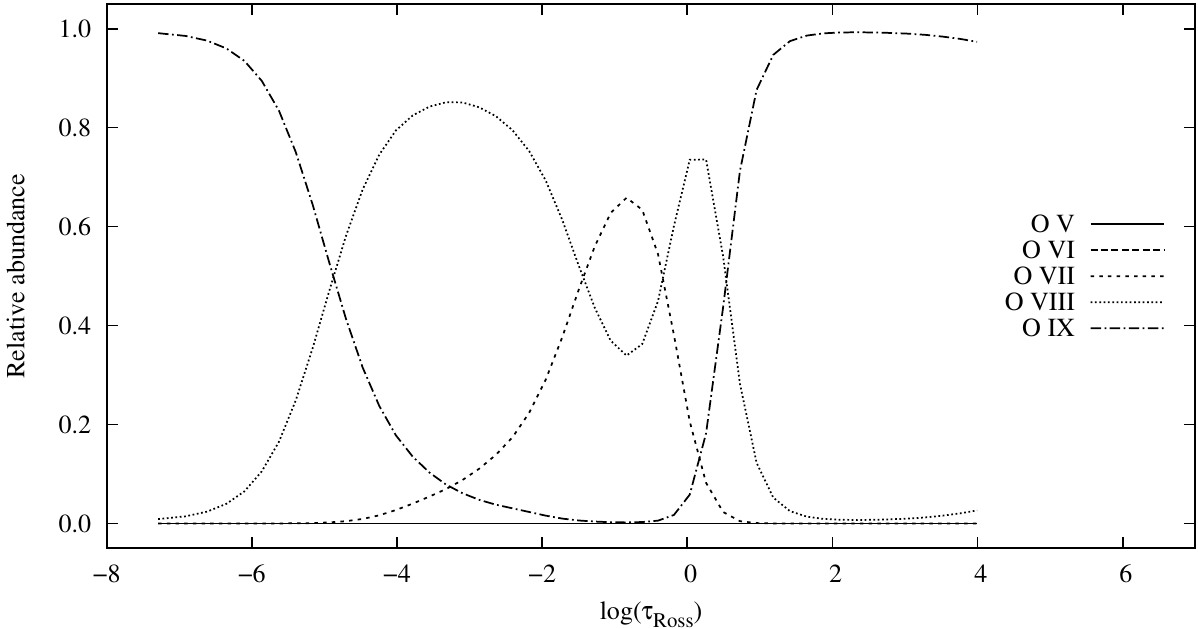}}\\
     \vspace{.3in}
     \subfigure[Neon.]{
          \label{fig:ionfrac.Ne}
          \includegraphics[width=.57\textwidth]{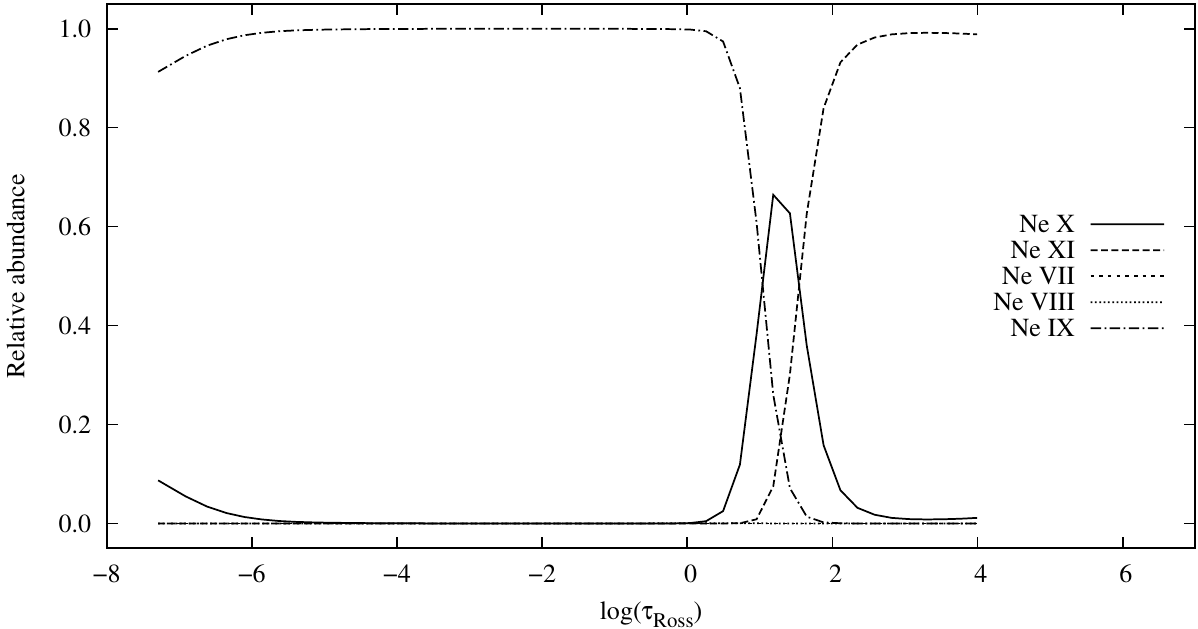}}\\
     \vspace{.3in}
     \subfigure[Magnesium.]{
           \label{fig:ionfrac.Mg}
           \includegraphics[width=.57\textwidth]{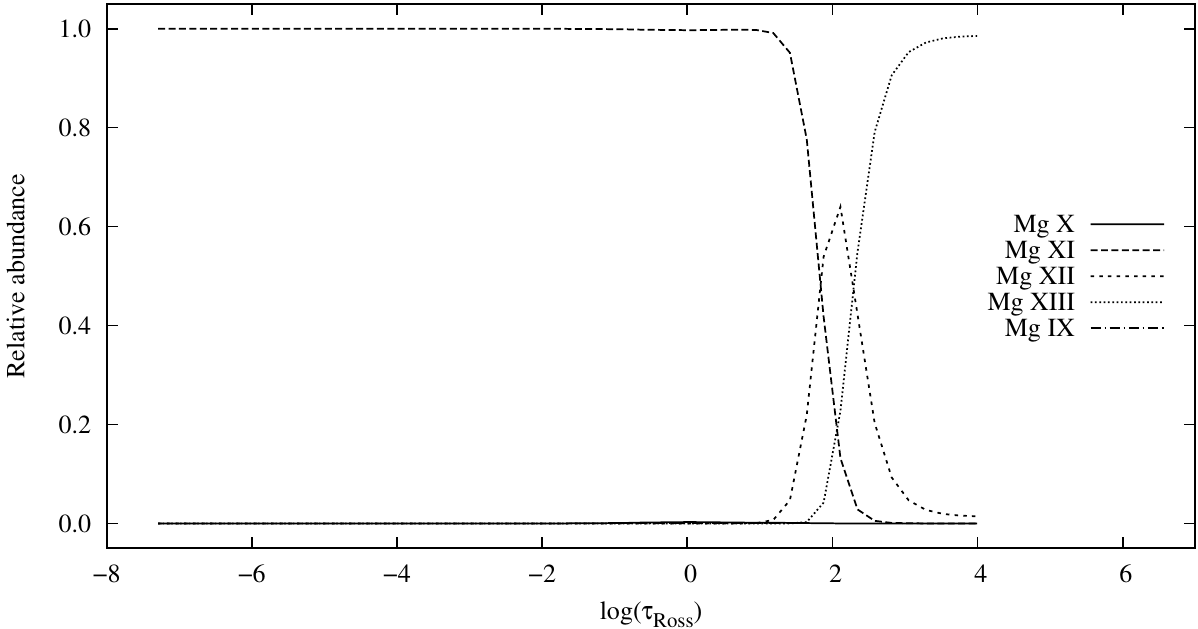}}\\
     \caption{Ionization fractions for oxygen, neon and magnesium.}
     \label{fig:ionfracONeMg}
\end{figure}

\begin{figure}[!p] 
     \centering
     \subfigure[Aluminum.]{
          \label{fig:ionfrac.Al}
          \includegraphics[width=.57\textwidth]{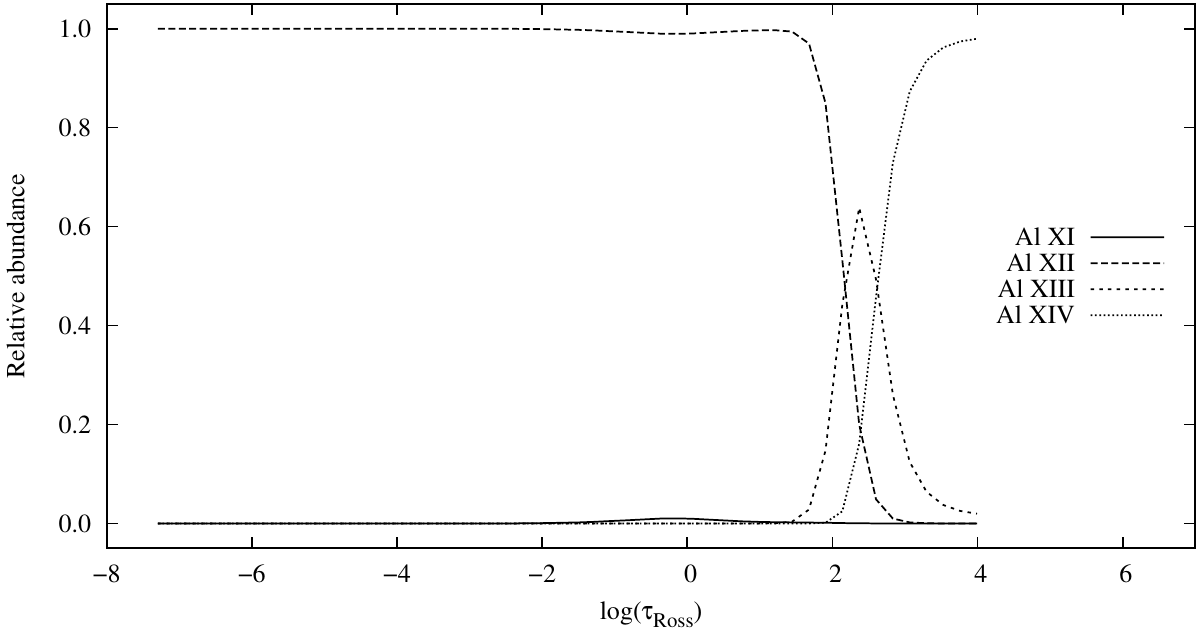}}\\
     \vspace{.3in}
     \subfigure[Silicon.]{
          \label{fig:ionfrac.Si}
          \includegraphics[width=.57\textwidth]{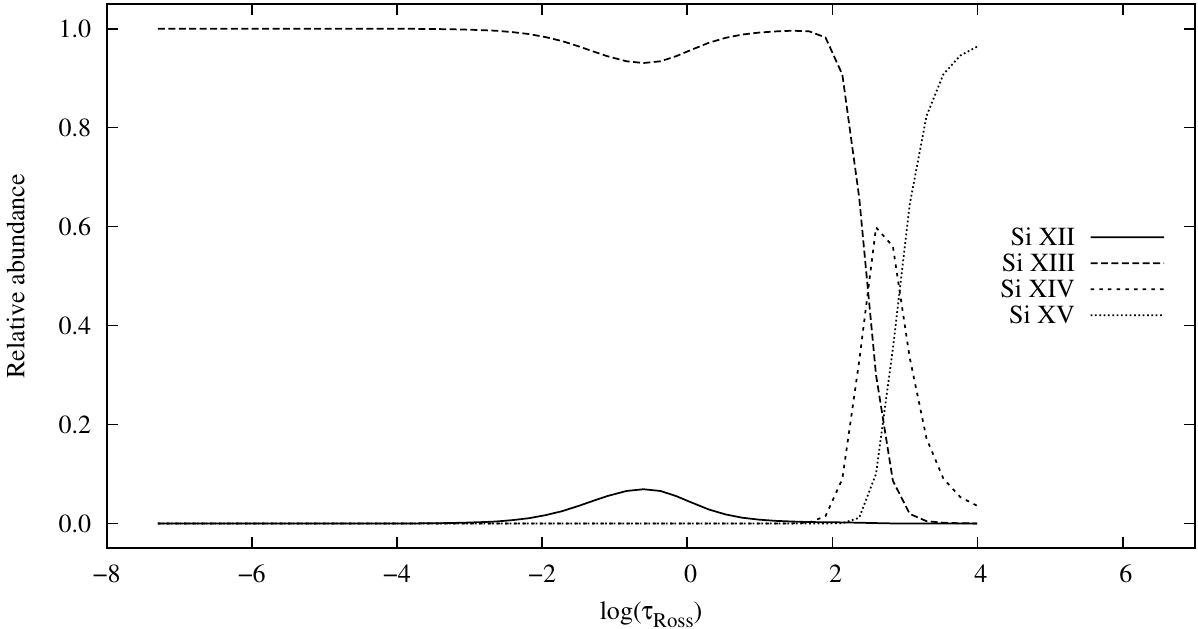}}\\
     \vspace{.3in}
     \subfigure[Sulfur.]{
           \label{fig:ionfrac.S}
           \includegraphics[width=.57\textwidth]{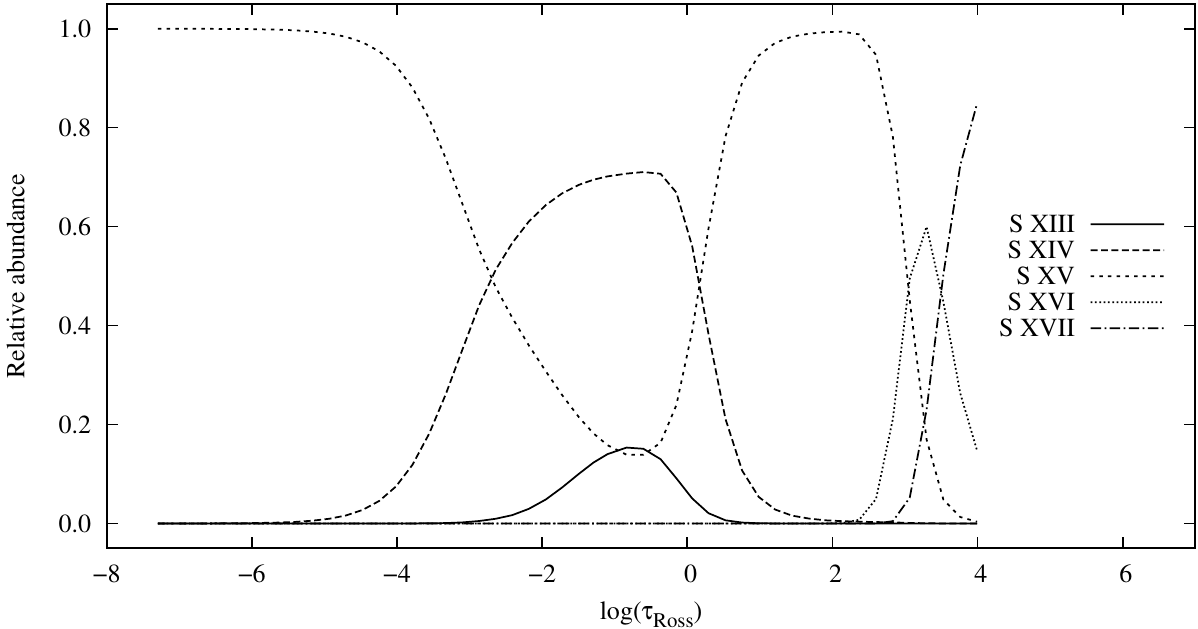}}\\
     \caption{Ionization fractions for aluminum, silicon and sulfur.}
     \label{fig:ionfracAlSiS}
\end{figure}

\begin{figure}[!p] 
     \centering
     \subfigure[Argon.]{
          \label{fig:ionfrac.Ar}
          \includegraphics[width=.57\textwidth]{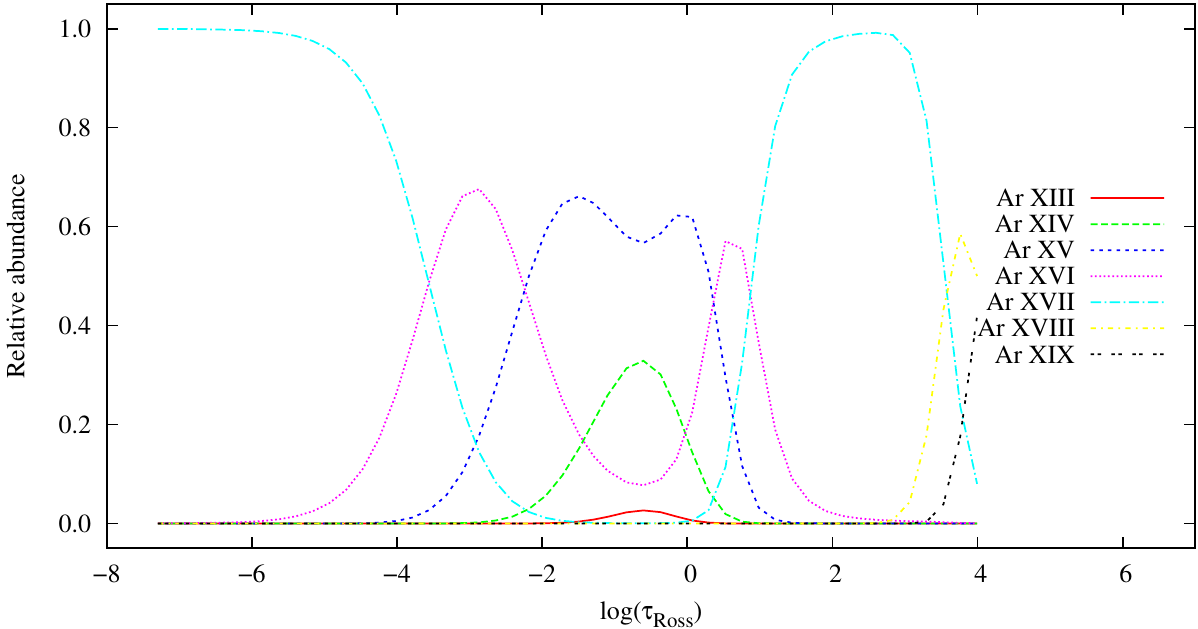}}\\
     \vspace{.3in}
     \subfigure[Calcium.]{
          \label{fig:ionfrac.Ca}
          \includegraphics[width=.57\textwidth]{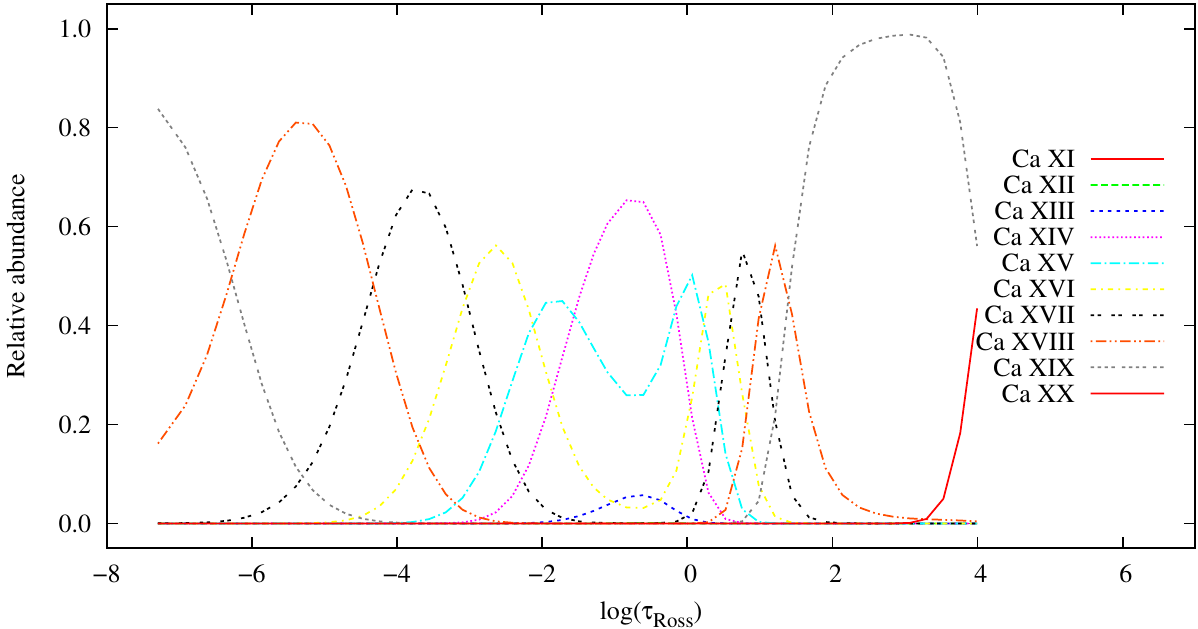}}\\
     \vspace{.3in}
     \subfigure[Iron.]{
           \label{fig:ionfrac.Fe}
           \includegraphics[width=.57\textwidth]{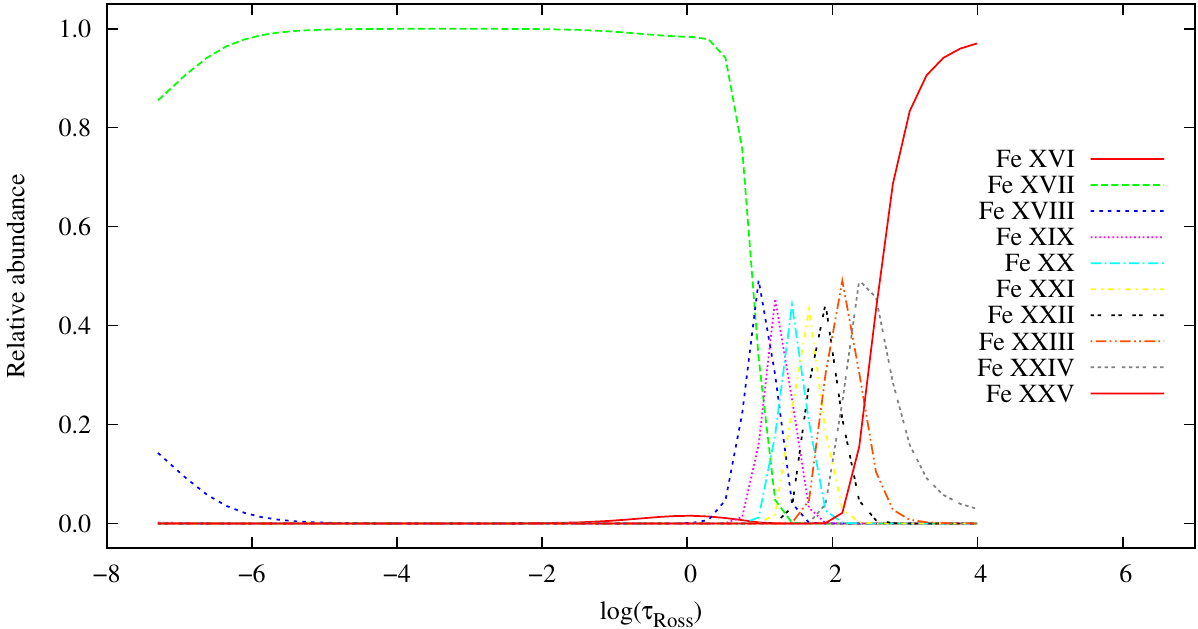}}\\
     \caption{Ionization fractions for argon, calcium and iron.}
     \label{fig:ionfracArCaFe}
\end{figure}

From level populations given by TLUSTY one can 
calculate the ionization balance throughout the
atmosphere. For this, the ratio of each level population and the total
population of the ion must be calculated at each depth point. These graphs are
shown in Figures \ref{fig:ionfracCN} -- \ref{fig:ionfracArCaFe}.
Due to the high temperature, hydrogen and helium are completely ionized. Figure
\ref{fig:ionfrac.C} shows that carbon is also ionized throughout the
atmosphere, with only a very tiny contribution of C VI around $\log\tau_{\rm
  Ross}\approx 0$, what is also a non-LTE effect. The temperature structure is
not monotonic in non-LTE. 
Proceeding outward from the atmosphere the
temperature decreases rapidly, reaching a minimum above the photosphere and
increasing in the outermost layers. This effect is more noticeable for
nitrogen and heavier elements. From neon through silicon the helium-like ions
are dominant in most of the atmosphere. Sulfur, argon, calcium and iron show
complex structures with 3--5 dominant ions simultaneously in the
atmosphere. Iron shows a stratified structure in the inner parts and Fe XVII
alone dominates the atmosphere in most of the line forming region. These
graphs are important diagnostics to improve the model input. By excluding
levels which do not contribute to the ionization balance, both the
computation time and modeling stability improves.

\begin{figure}[p]
     \centering
     \subfigure[For HI.]{
          \label{fig:HI}
          \includegraphics[width=.45\textwidth]{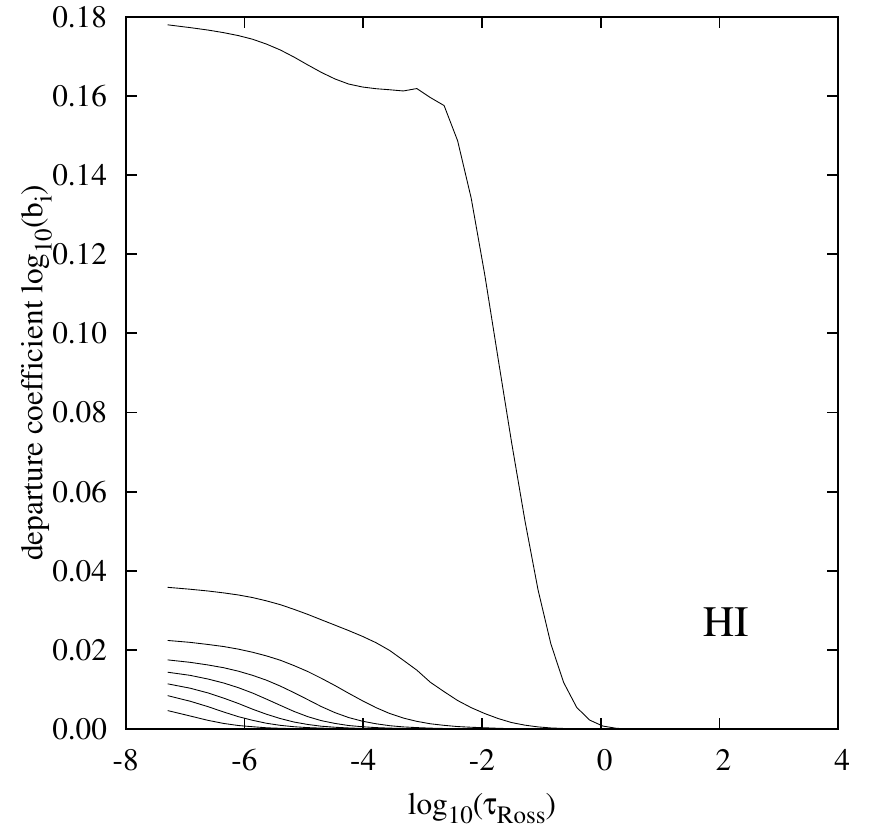}}
     \hspace{.1in}
     \subfigure[For HeII.]{
          \label{fig:HeII}
          \includegraphics[width=.45\textwidth]{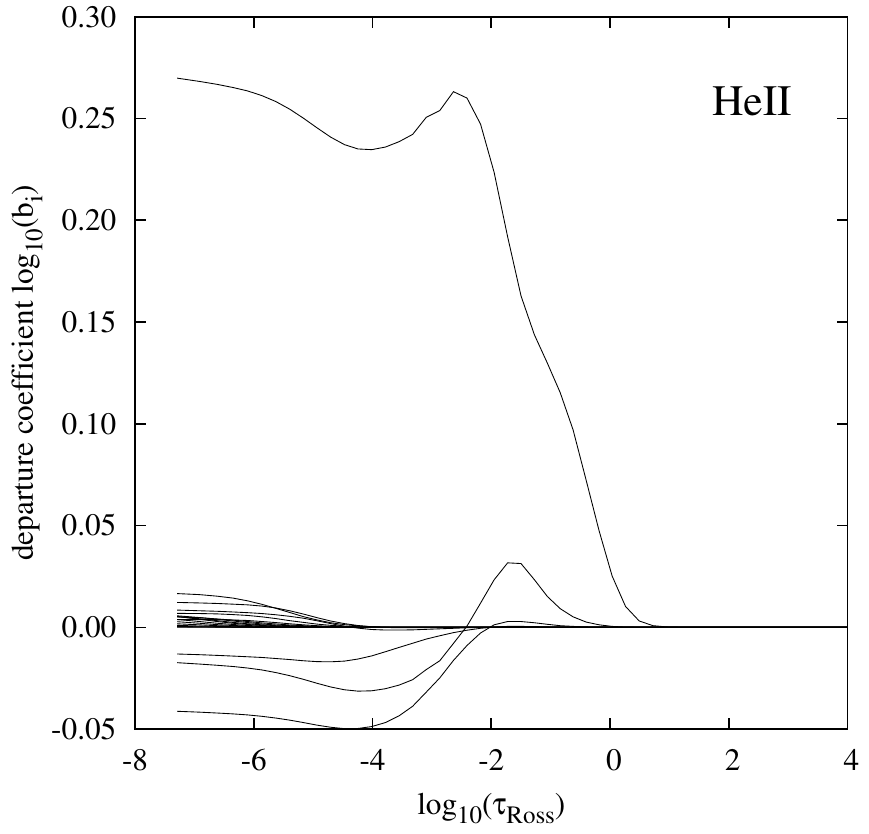}}
     \vspace{.2in}
     \hspace{.0in}
     \subfigure[For Carbon V.]{
           \label{fig:CV}
          \includegraphics[width=.45\textwidth]{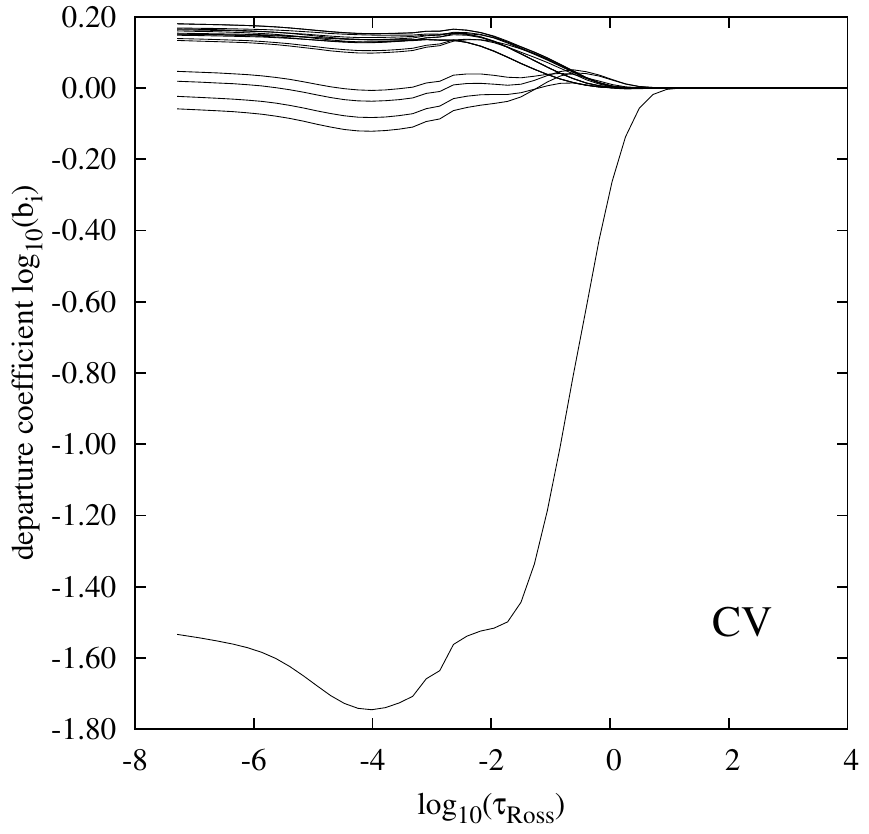}}
     \vspace{.2in}
     \hspace{.0in}
     \subfigure[For Carbon VI.]{
           \label{fig:CVI}
           \includegraphics[width=.45\textwidth]{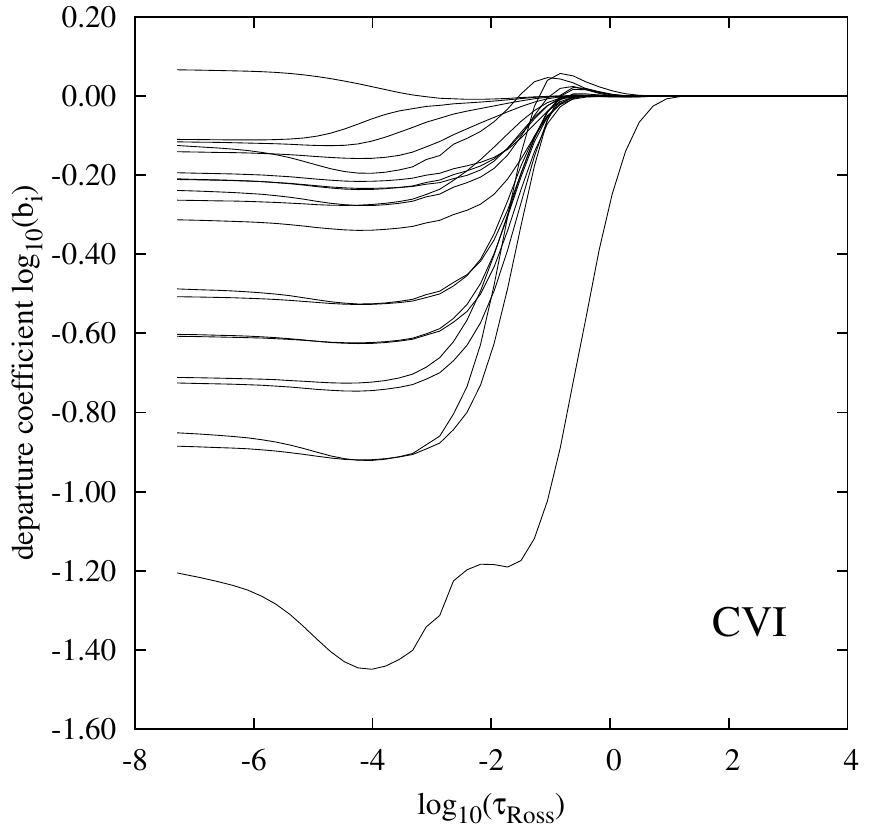}}
     \caption{NLTE departure coefficients of ions from H I -- C VI.}
     \label{fig:Depco1}
\end{figure}

\begin{figure}[p]
     \centering
     \subfigure[For Nitrogen V.]{
          \label{fig:NV}
          \includegraphics[width=.45\textwidth]{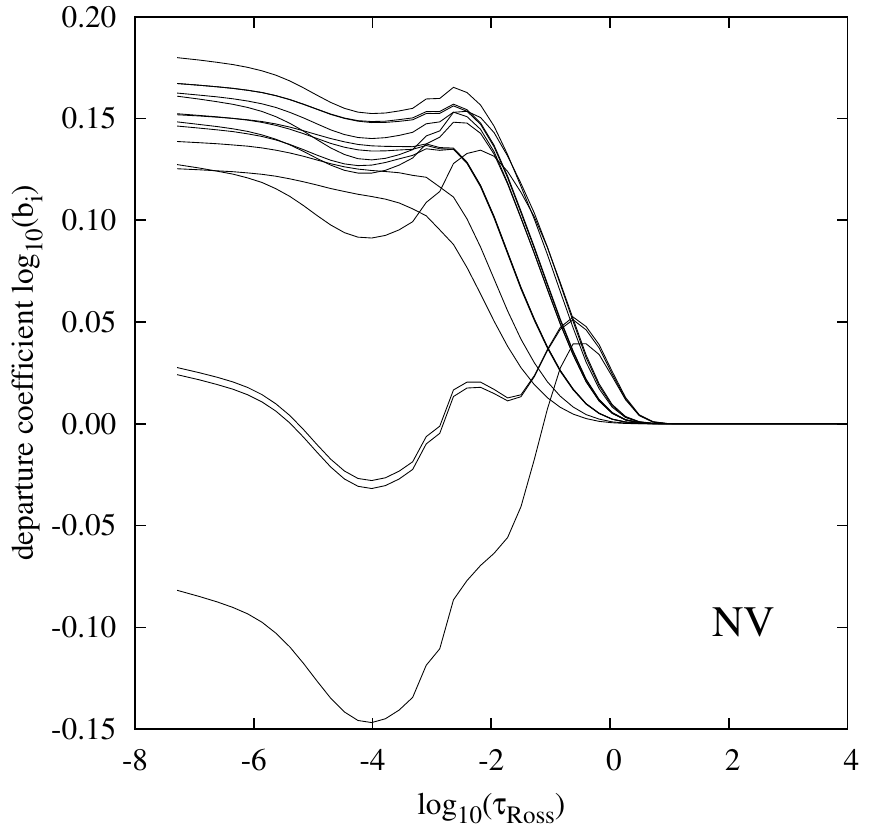}}
     \hspace{.1in}
     \subfigure[For Nitrogen VI.]{
          \label{fig:NVI}
          \includegraphics[width=.45\textwidth]{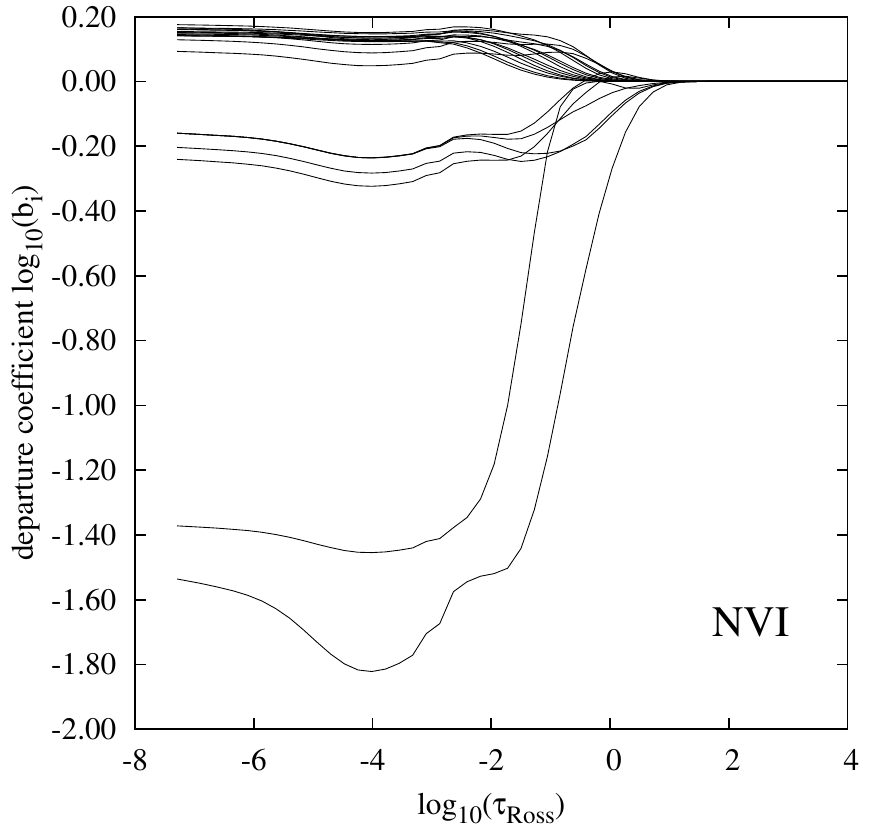}}\\
     \vspace{.2in}
     \subfigure[For Nitrogen VII.]{
           \label{fig:NVII}
           \includegraphics[width=.45\textwidth]{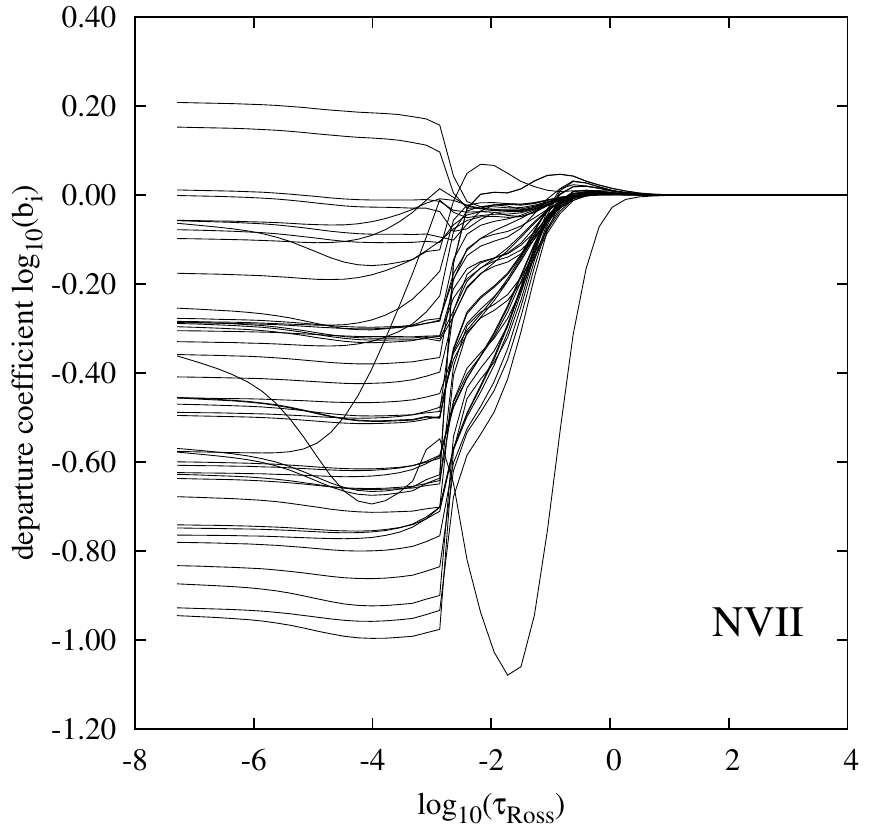}}
     \caption{NLTE departure coefficients of ions from N V -- N VII.}
     \label{fig:Depco2}
\end{figure}

\begin{figure}[p]
     \centering
     \subfigure[For Oxygen V.]{
          \label{fig:OV}
          \includegraphics[width=.45\textwidth]{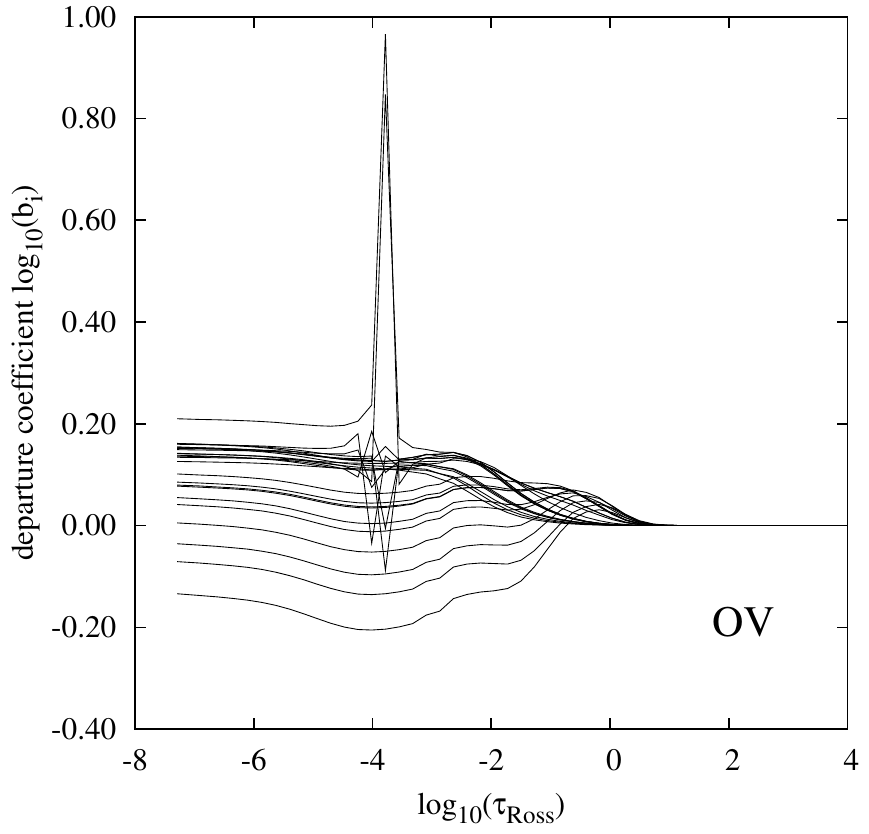}}
     \hspace{.1in}
     \subfigure[For Oxygen VI.]{
          \label{fig:OVI}
          \includegraphics[width=.45\textwidth]{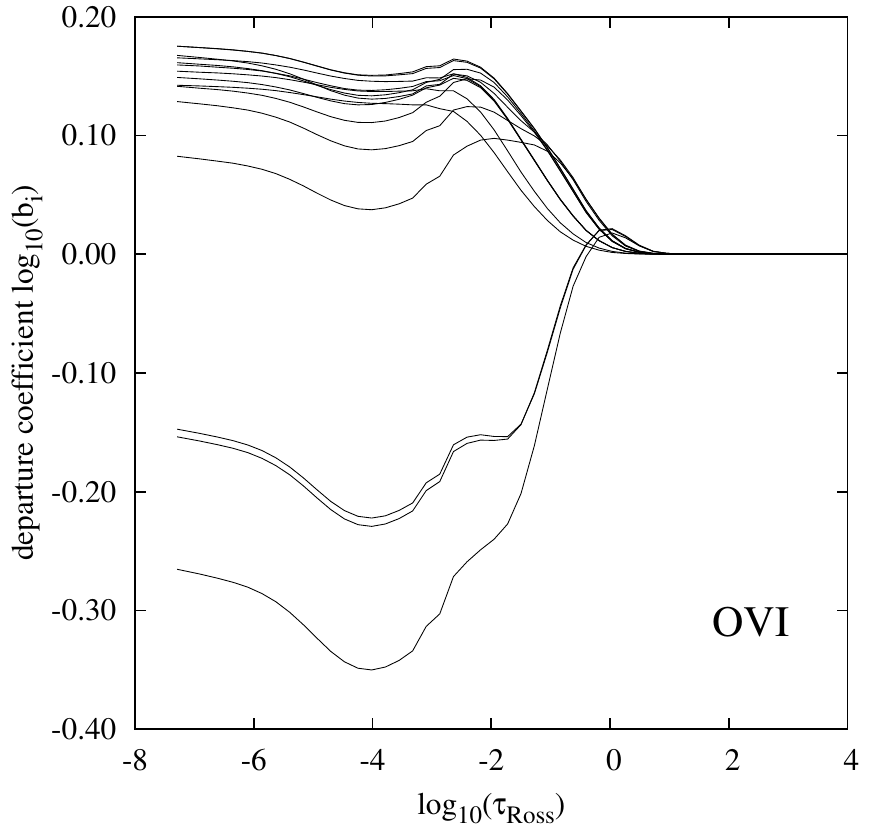}}\\
     \vspace{.2in}
     \hspace{.0in}
     \subfigure[For Oxygen VII.]{
           \label{fig:OVII}
           \includegraphics[width=.45\textwidth]{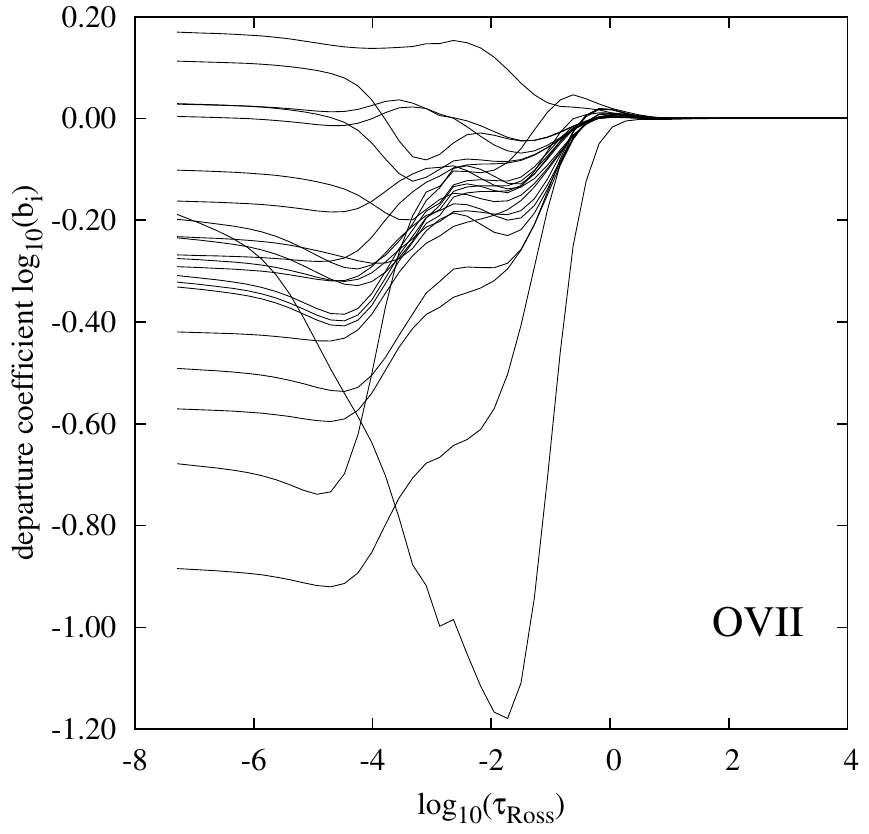}}
     \hspace{.1in}
     \subfigure[For Oxygen VIII.]{
           \label{fig:OVIII}
          \includegraphics[width=.45\textwidth]{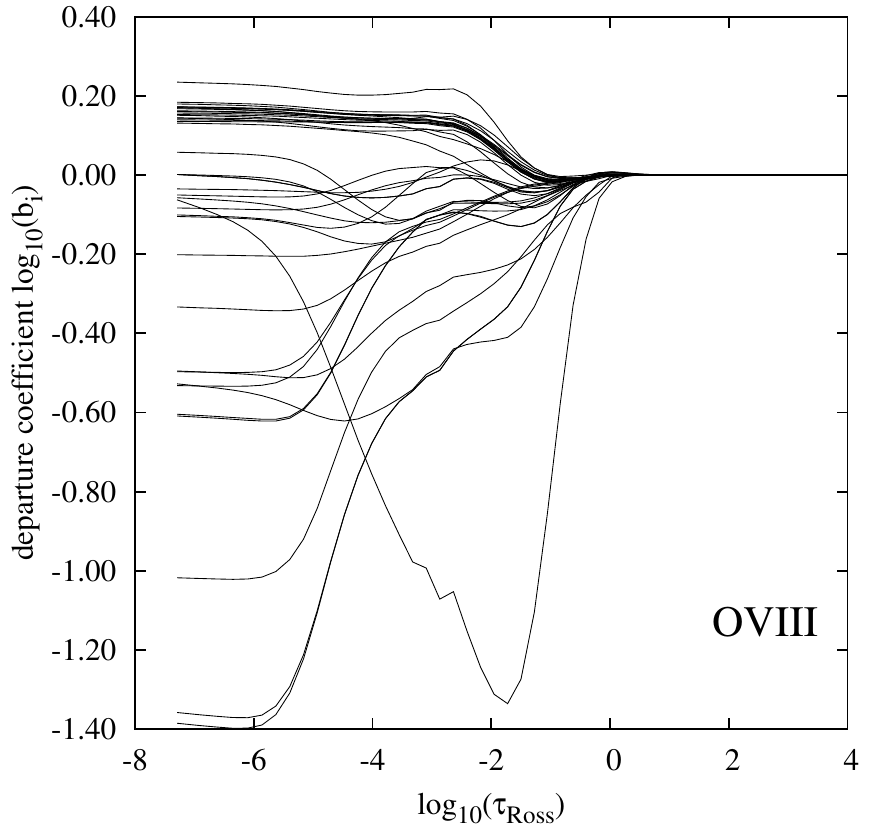}}
     \caption{NLTE departure coefficients of ions from O V -- O VIII.}
     \label{fig:Depco3}
\end{figure}

\begin{figure}[p]
     \centering
     \subfigure[For Neon VII.]{
          \label{fig:OV}
          \includegraphics[width=.45\textwidth]{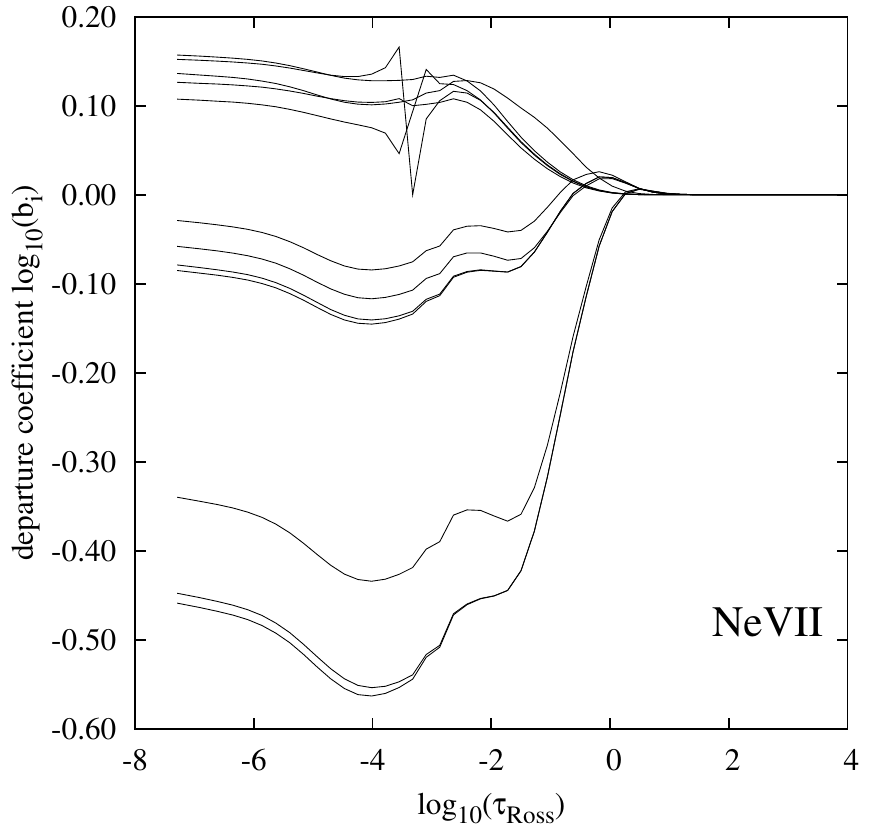}}
     \hspace{.1in}
     \subfigure[For Neon VIII.]{
          \label{fig:OVI}
          \includegraphics[width=.45\textwidth]{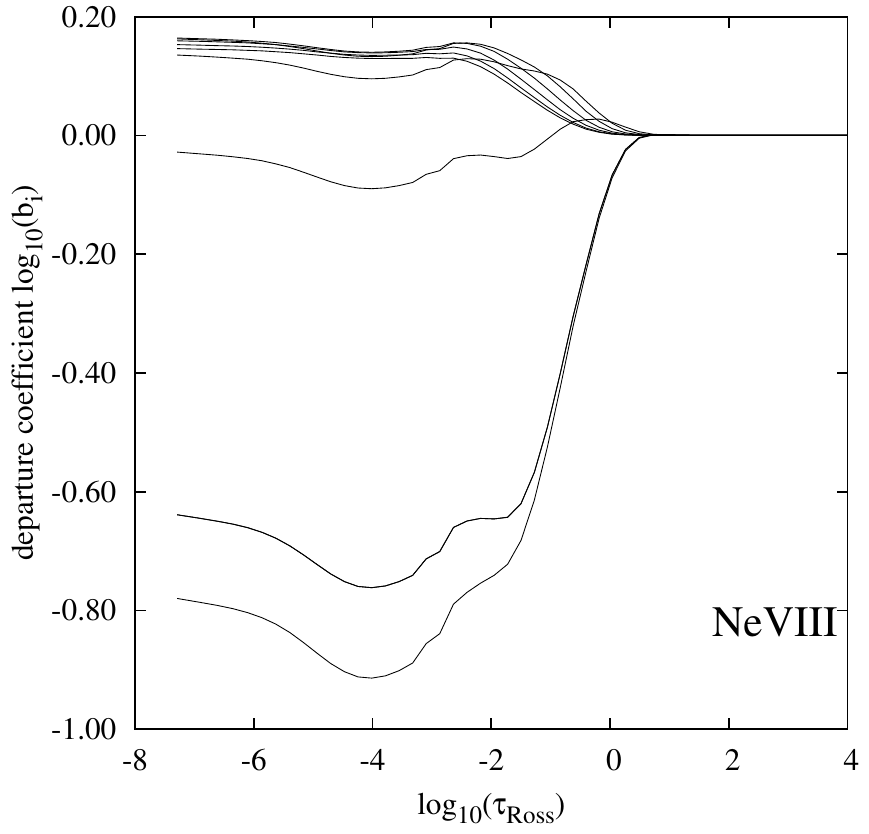}}\\
     \vspace{.2in}
     \hspace{.0in}
     \subfigure[For Neon IX.]{
           \label{fig:OVII}
           \includegraphics[width=.45\textwidth]{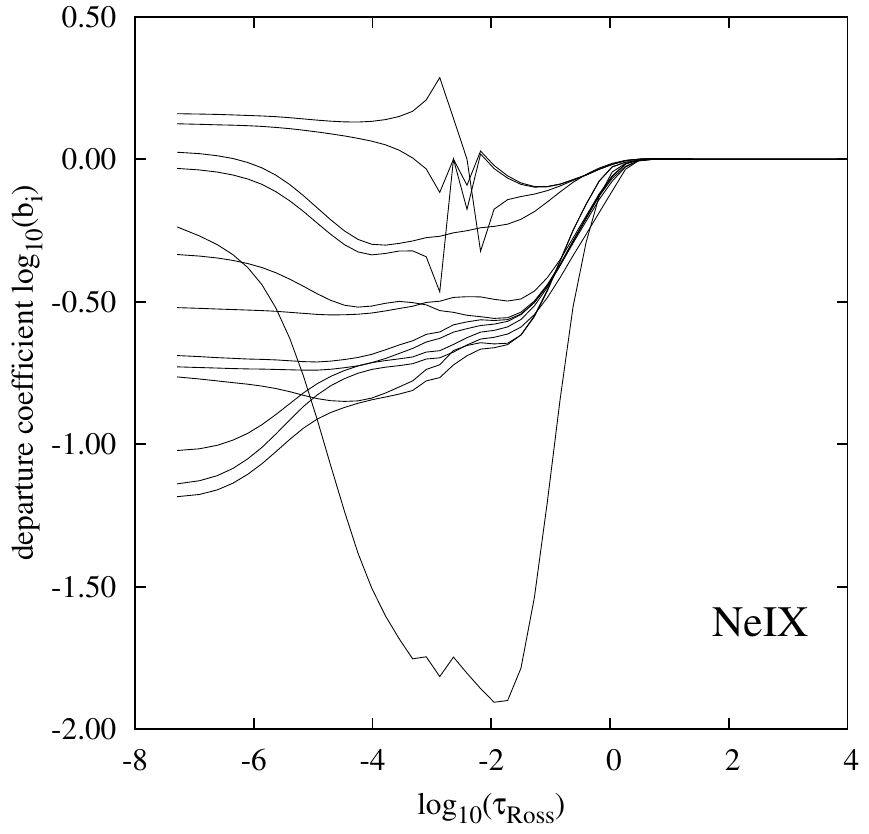}}
     \hspace{.1in}
     \subfigure[For Neon X.]{
           \label{fig:OVIII}
          \includegraphics[width=.45\textwidth]{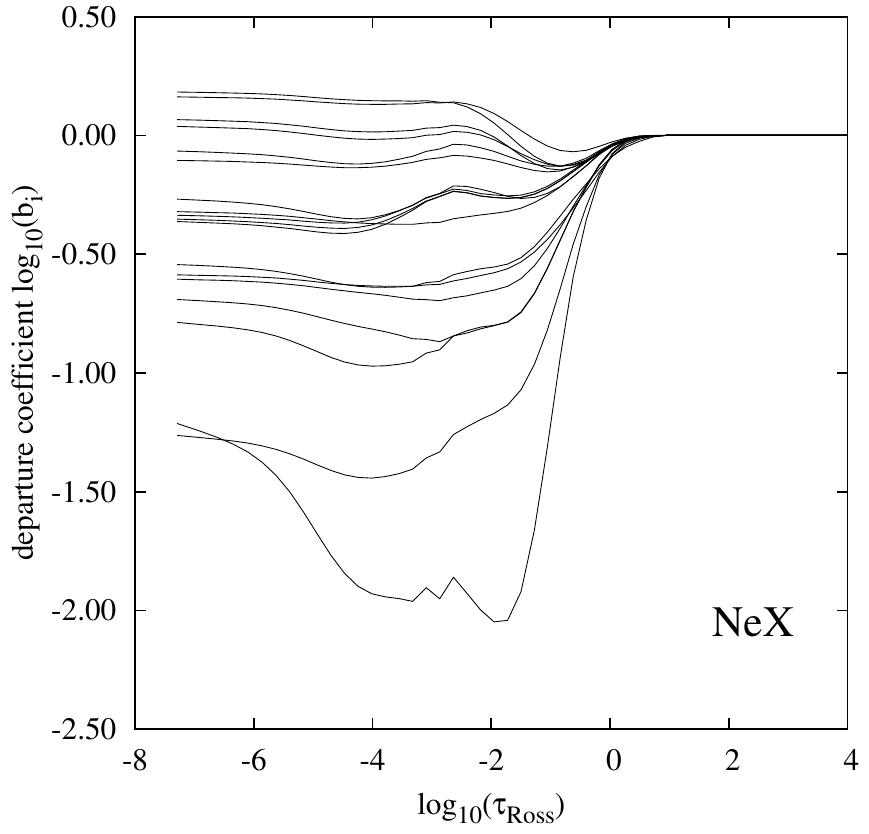}}
     \caption{Departure coefficients of included ions from Ne VII -- Ne X.}
     \label{fig:Depco4}
\end{figure}

\begin{figure}[p]
     \centering
     \subfigure[For Neon VII.]{
          \label{fig:OV}
          \includegraphics[width=.45\textwidth]{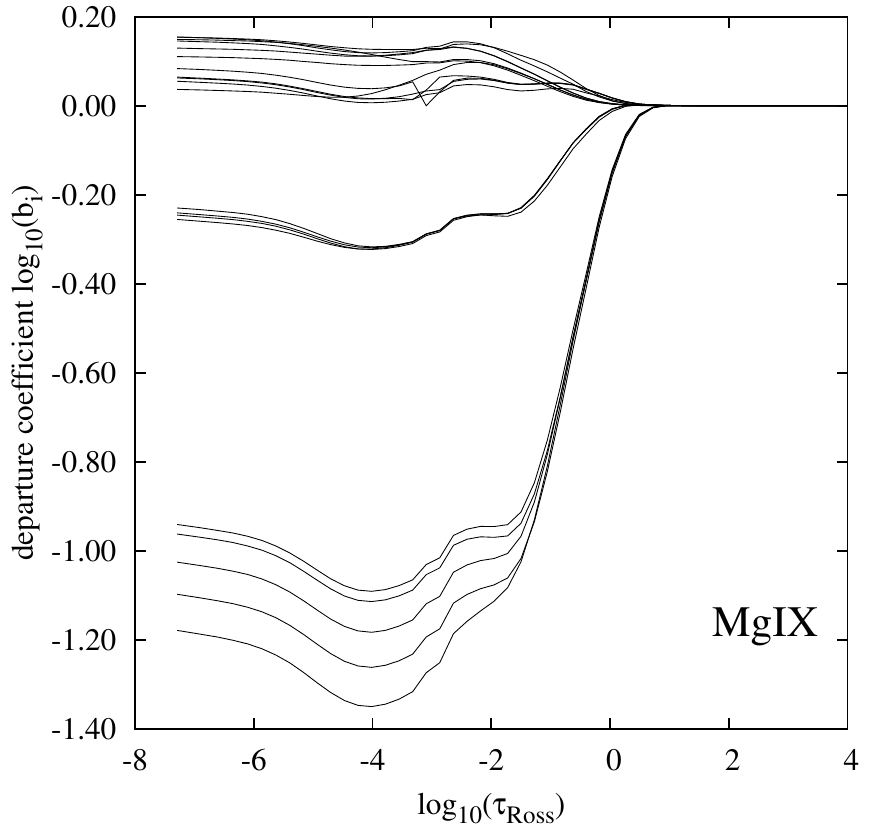}}
     \hspace{.1in}
     \subfigure[For Neon VIII.]{
          \label{fig:OVI}
          \includegraphics[width=.45\textwidth]{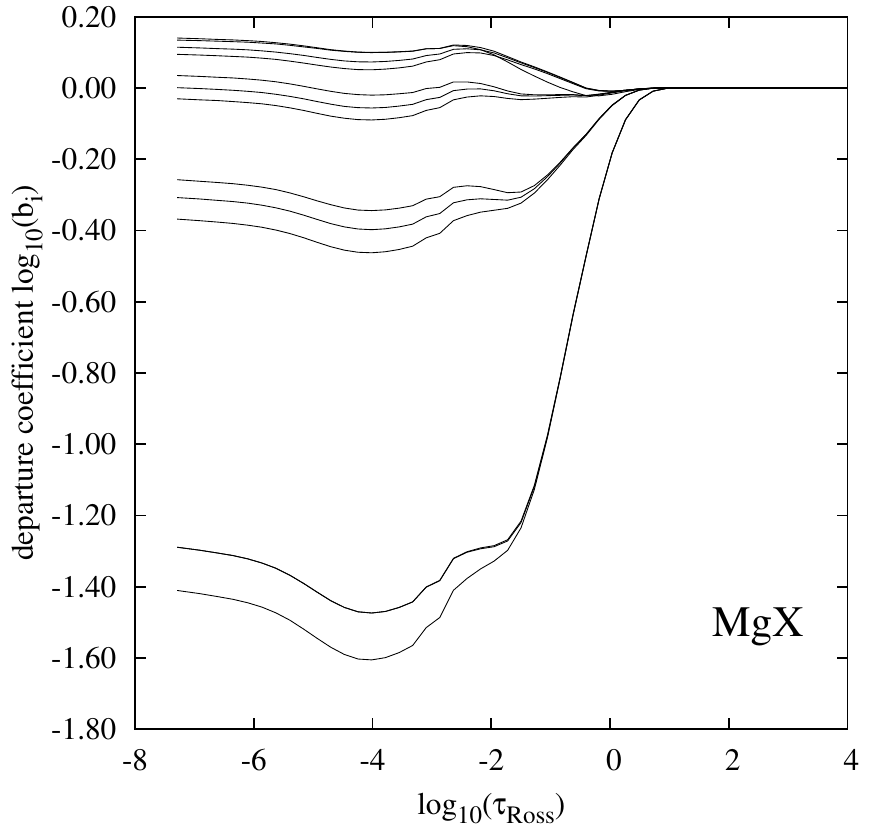}}\\
     \vspace{.2in}
     \hspace{.0in}
     \subfigure[For Neon IX.]{
           \label{fig:OVII}
           \includegraphics[width=.45\textwidth]{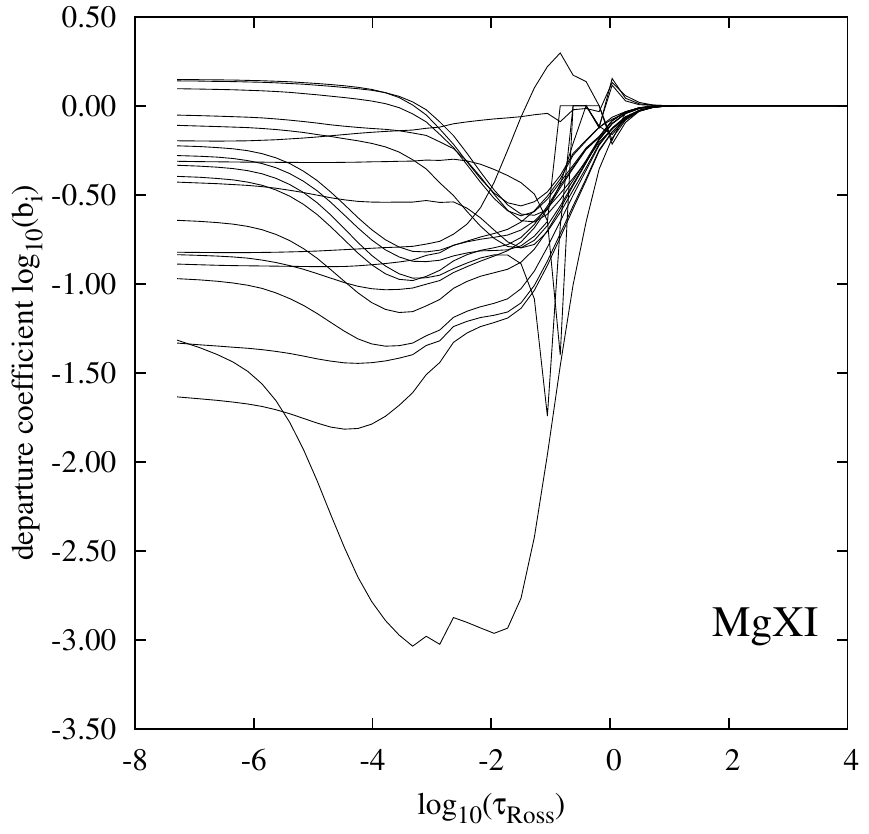}}
     \hspace{.1in}
     \subfigure[For Neon X.]{
           \label{fig:OVIII}
          \includegraphics[width=.45\textwidth]{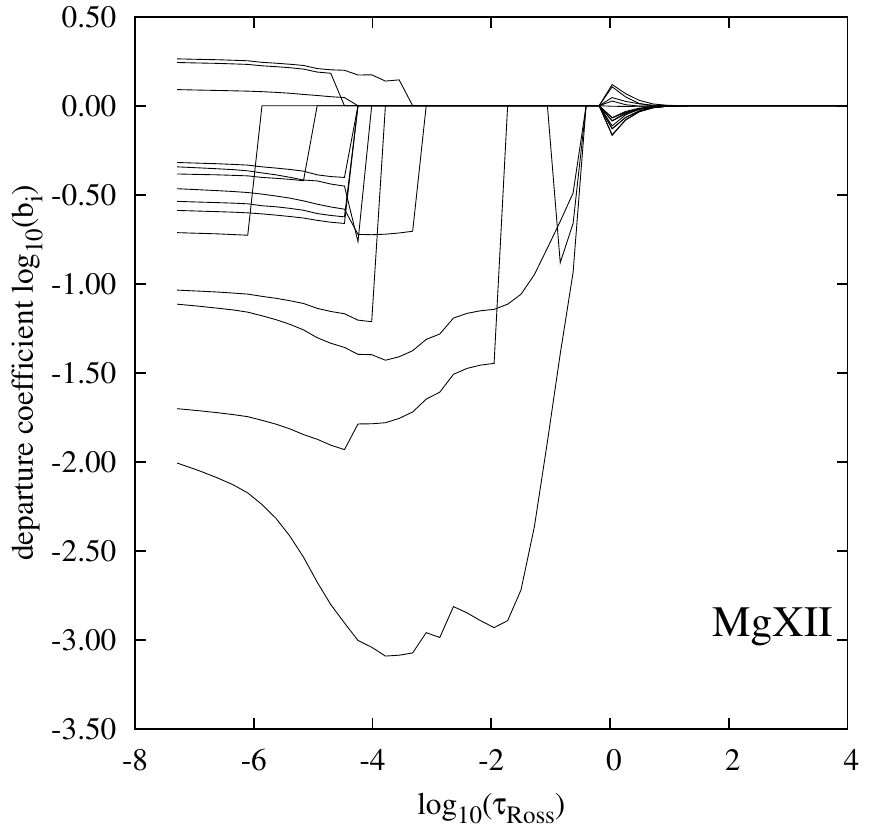}}
     \caption{Departure coefficients of included ions from Mg IX -- Mg XII.}
     \label{fig:Depco5}
\end{figure}

Along with population numbers, in non-LTE TLUSTY also provides departure
coefficients for each depth point and all levels.
Departure coefficients are defined as the relative fraction of the population
of level $i$ calculated in non-LTE to the level population calculated in LTE:
\begin{equation}
b_i=\frac{n_i(\tiny {\rm NLTE} )}{n_i({\rm LTE})}
\label{eq:bi}
\end{equation}
Departure coefficients
 from carbon to magnesium are shown in Figures
\ref{fig:Depco1} -- \ref{fig:Depco5}. These graphs depict the strength of 
non-LTE effects on each level. In general, deep in the atmosphere LTE prevails
due to high pressure and the fact that departure coefficients are close to
one. In the
outer half of the atmosphere population numbers can significantly differ from
their LTE values. Ground and low lying levels are over-populated
while high levels show less deviation.


\begin{figure}[htbp]\begin{center}
\includegraphics[angle=0,width=11cm]{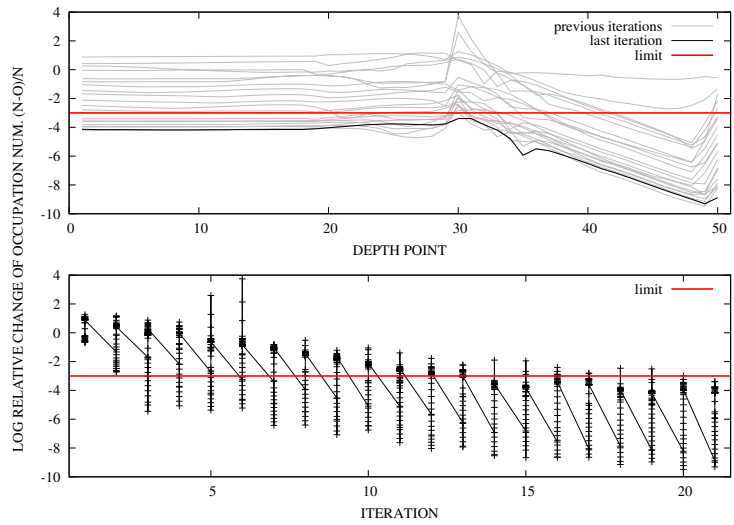}
\parbox{138mm}{\caption[NLTE-C convergence log.]{Convergence log of the non-LTE continuum model. Upper
    panel shows relative change vs.\ depth points. Gray
    lines represent 
    the relative change of population numbers at each depth point,
    the thin black line is the relative change for the last iteration. Thick
    horizontal 
    line shows the convergence limit, $\delta{n_i}/n_i=10^{-3}$. The lower
    panel is the relative change vs.\ iteration number, like looking at the top
    panel from the side.\label{fig:convC}}}
\end{center}\end{figure}

\begin{figure}[htbp]\begin{center}
\includegraphics[angle=0,width=11cm]{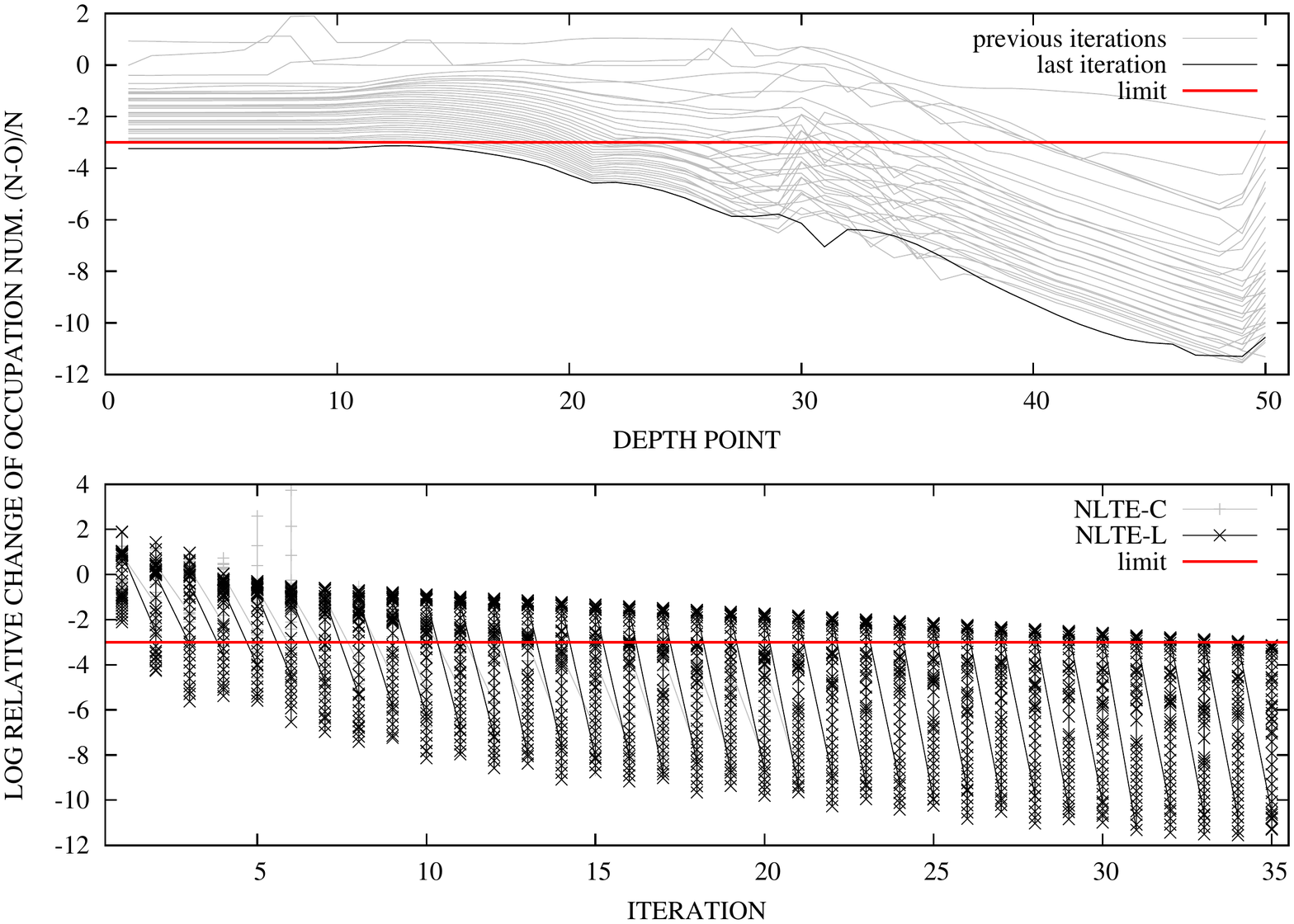}
\parbox{130mm}{\caption[NLTE-L convergence log.]{Convergence of the full non-LTE model.\label{fig:convL}}}
\end{center}\end{figure}

Figures \ref{fig:convC} and \ref{fig:convL} show the convergence log for
continuum and full non-LTE models, respectively. Convergence is very sensitive
for model parameters, and usually about 25--55 iterations are needed. 
Close to the Eddington limit the 
convergence slows down, and shows oscillatory behavior. Model
atoms and abundances also have great impact on the convergence.  
During model calculations TLUSTY prints the relative change of 
effective temperature, electron
density and population numbers into a file for each depth point at each
iteration. By plotting these quantities one can
investigate the convergence properties of the model. In Figure \ref{fig:convC}
the continuum model is presented. The top panel shows the relative-change
vs. depth-point curves. Numbering and calculation starts at the top of the
atmosphere, depth-point zero means the outermost layer 
($\tau_{\rm Ross}=10^{-7}$) 
and 50 is the
innermost layer ($\tau_{\rm Ross}=10^4$) of the atmosphere. 
Thin grey lines represent the relative changes at each
iteration. Usually relative changes are large at the beginning and quickly
fall below the convergence limit (horizontal line at
$\delta{n_i}/n_i=10^{-3}$) in the innermost layers. The rest of the atmosphere
converges more slowly, 
especially at the photosphere around depth-point 30 and in
the outermost layers.
The lower panel in Figure \ref{fig:convC} shows the relative-change vs.\
iteration number relations. It is like looking at the top panel from the
side. 
This graph shows that 21 iterations were required for full
convergence. 
Kantorovich acceleration started at the third iteration and
repeated after every third iteration. 
Ng
acceleration was applied from the seventh iteration and repeated after every
fourth step. 
Figure \ref{fig:convL} has the same structure as Figure \ref{fig:convC} 
only for the NLTE-L model,
which consideres spectral lines as well. Similar convergence properties can be
seen, but non-LTE effects are more prominent. The convergence is very slow in
the outermost layers where non-LTE effects of spectral lines are the
strongest. The model required 35 iterations for a solution. 
Figures \ref{fig:convC}
and \ref{fig:convL} show the normal behavior of convergence. It is also
possible to show the convergence of the effective temperature and electron
density in similar graphs, however as they converge much faster the
convergence of population
numbers alone is adequate. 

Convergence log graphs are continuosly updated and displayed 
during the calculations at a
user-defined frequency. This helps to be aware of convergence properties 
and to improve the parameters
which control the calculation. 
Graphs for both ionization fractions, departure coefficients and convergence
logs are automatically generated by TGRID.



\section{Trends in the model grid}

In order to experience the effects of effective temperature, gravity,
different abundances and other variables 
on the final spectra model sequences were calculated. All models were
constructed with parameters from Table \ref{tab:modelp} unless otherwise
noted.  

\subsection*{Gravity sequence}

\begin{figure}[htb]\begin{center}
\includegraphics[width=14.2cm]{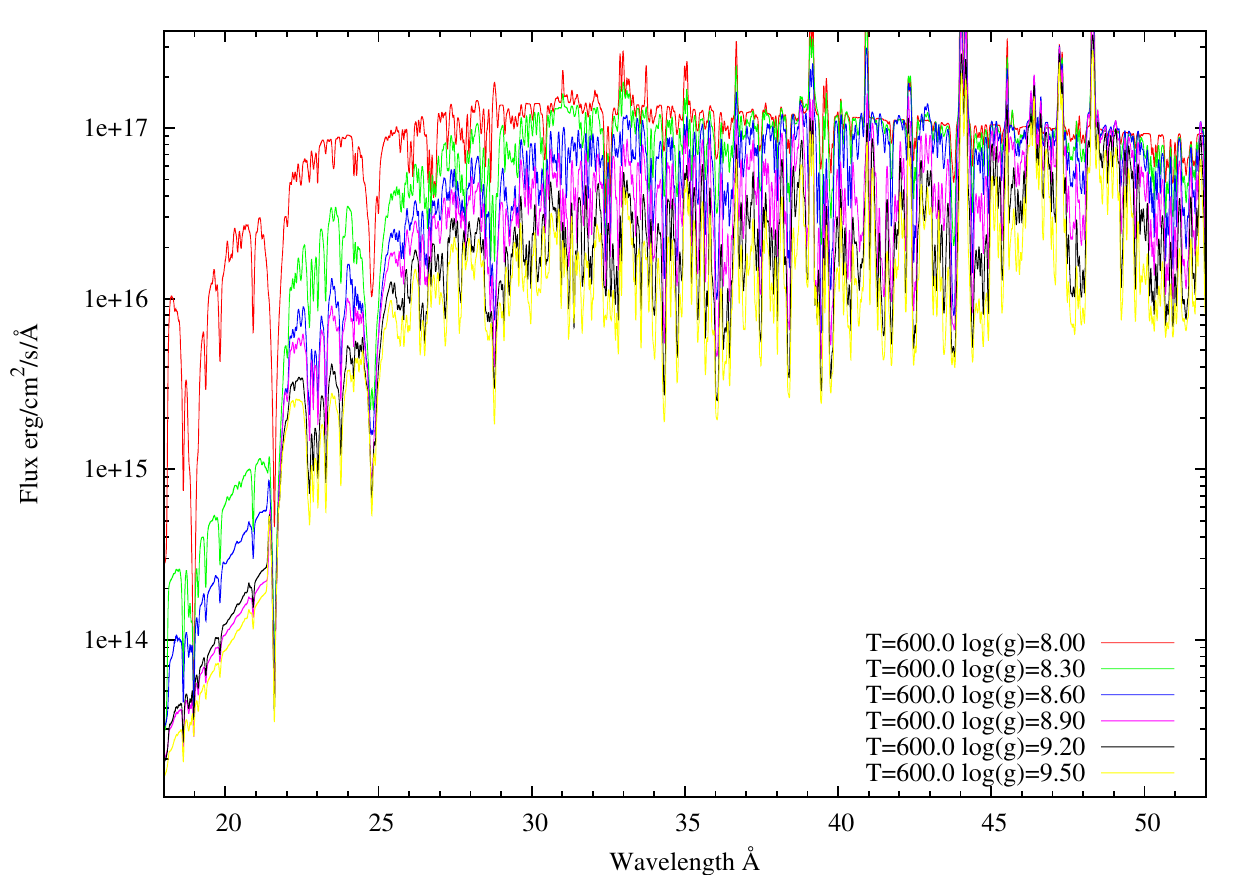}
\parbox{130mm}{\caption[Gravity sequence.]{Effects of surface gravity on the
    spectra. Models were calculated at the same composition, at effective 
temperature of 600,000 K, gravity is indicated in cgs units.  
\label{fig:gseq}}}
\end{center}\end{figure}

\begin{figure}[htbp]\begin{center}
\includegraphics[width=14.2cm]{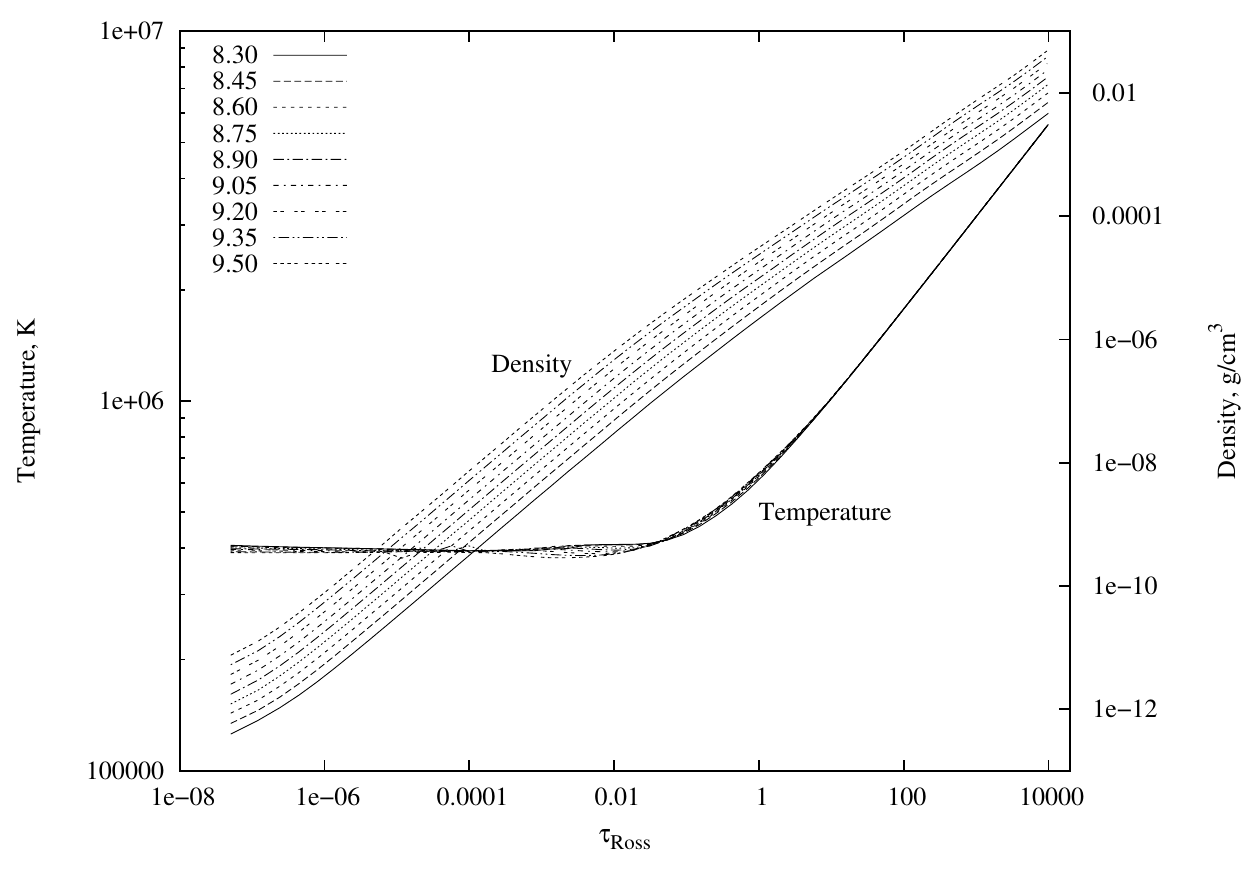}
\parbox{130mm}{\caption[Gravity sequence -- structure.]{Effects of surface
    gravity on the atmospheric structure. 
Models were calculated with same composition, at effective 
temperature of 600,000 K, gravity is indicated in cgs units.  
\label{fig:gseq_td}}}
\end{center}\end{figure}

The
effects of gravity were examined 
in the range of $\log g= 8.00$ -- $9.50$ cm/s$^2$ at $T_{\rm
  eff}=600,000$ K. 
These models are shown in Figure \ref{fig:gseq}. 
The atmospheric
composition was kept unchanged. The most important effect of gravity is on the
flux below 30 \AA. In this region with increasing gravity 
ionization edges become stronger and the energy distribution becomes
softer. Observed continuum emission below 25 \AA\ confirms that novae in the
constant bolometric luminosity phase are close to, or over the Eddington
limit. The atmospheric structure for these models can be seen in Figure
\ref{fig:gseq_td}, where temperature and density are plotted as function of
optical depth between $\tau_{\rm Ross}=10^4$ -- $10^{-7}$. The surface gravity
does not affect the temperature profile of the atmosphere. There are only
small differences above the photosphere between $\tau_{\rm Ross}=1$ and
$10^{-4}$. The density profile shows gradual change with surface gravity (WD
mass) which increases the density throughout the atmosphere. 

\begin{figure}[htbp]\begin{center}
\includegraphics[width=14.2cm]{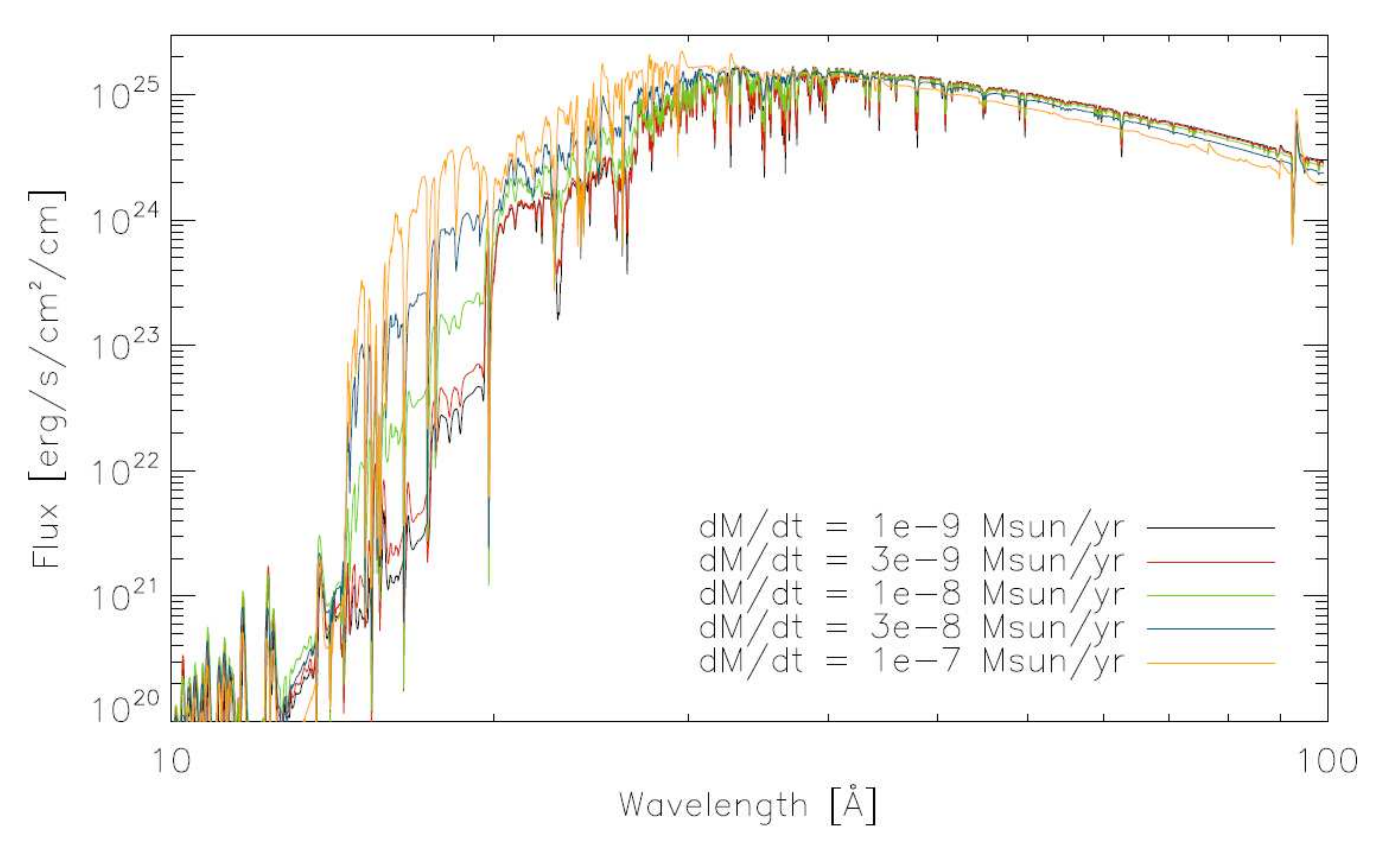}
\parbox{130mm}{\caption[PHOENIX flux vs. mass-loss rate.]{Impact of the mass-loss rate on the spectrum. Five
    spectra are shown at different mass loss rates calculated with PHOENIX
    with $T_{\rm eff}=600,000$ K and $v_{\infty}=2400$ km/s. Figure is taken
    from \citet{rossum10}.
\label{fig:rossum}}}
\end{center}\end{figure}

The effects of surface gravity are in good agreement with the calculations of
\citet{rossum10}. Figure \ref{fig:rossum} shows their models with different
mass-loss rates calculated with
PHOENIX\footnote{\protect\url{http://www.hs.uni-hamburg.de/EN/For/ThA/phoenix/index.html}}. PHOENIX
is a general purpose stellar atmosphere code for calculating hydrodynamic model
atmospheres in spherical geometry. Their models show that the softening of the
energy distribution with increasing gravity is continued over the Eddington
limit as well. 

\subsection*{Temperature sequence}

\begin{figure}[htb]\begin{center}
\includegraphics[width=14.2cm]{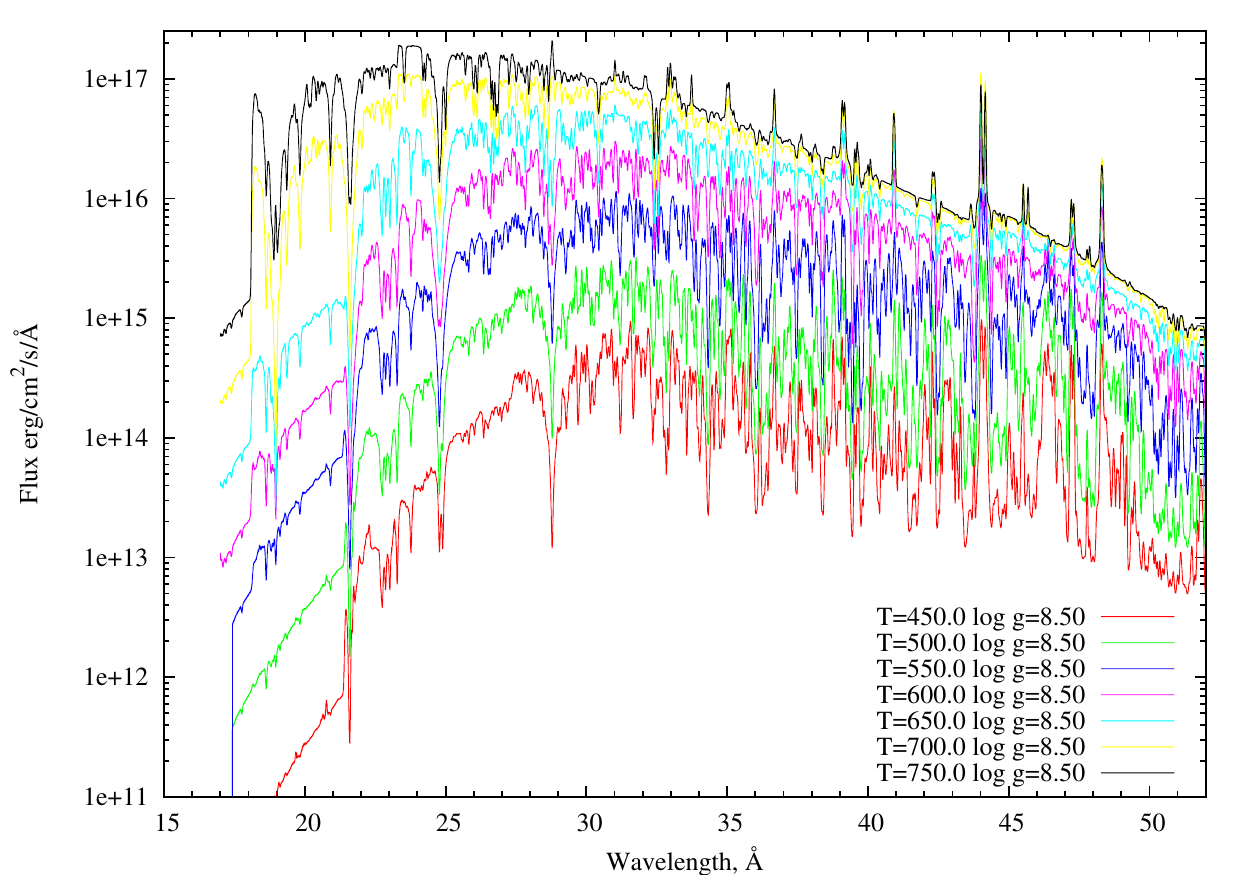}
\parbox{130mm}{\caption[Temperature sequence.]{Temperature sequence. Models
    were calculated from $T_{\rm eff}=450,000$ to $750,000$ K with 
    50,000 K increments at $\log
    g=8.5$ cm/s$^2$ and composition listed in Table \ref{tab:modelp}. 
\label{fig:tempseq}}}
\end{center}\end{figure}

\begin{figure}[htb]\begin{center}
\includegraphics[width=14.2cm]{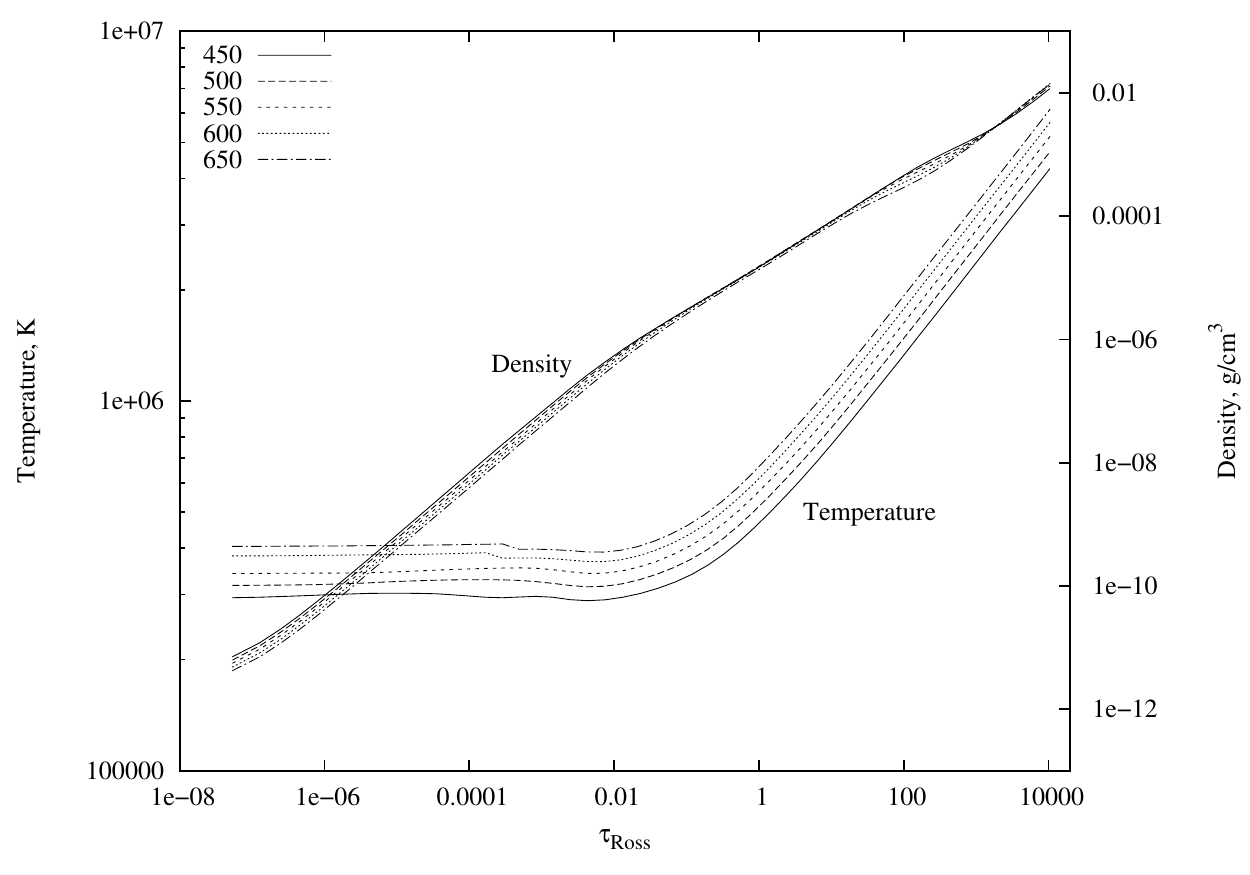}
\parbox{130mm}{\caption[Temperature sequence -- structure.]
{Effects of the effective temperature on the atmospheric structure.
\label{fig:tempseq_td}}}
\end{center}\end{figure}

Other interesting trend is how the effective temperature changes the energy
distribution. This sequence can be seen in Figure \ref{fig:tempseq}. At high
energy an opposite
effect can be seen like at the gravity sequence, 
increasing temperature decreases
the ionization egdes and makes the spectra harder. The change of the
ionization balance can also be noticed when the relative strengths of
Ly$\alpha$ lines of N VI (28.8 \AA) and N VII (24.8 \AA) are compared.
Figure \ref{fig:tempseq_td} shows the structure of model atmospheres when only
temperature was changed. The gravity and temperature profiles are independent
from each other. Only a small change of the density can be seen and variations
around the photosphere are particularly small. The effective temperature
scales the local temperature homogeneously throughout the atmosphere.  

\subsection*{Temperature -- Gravity sequence}

\begin{sidewaysfigure}[!hp]\begin{center}  
\includegraphics[width=20.2cm]{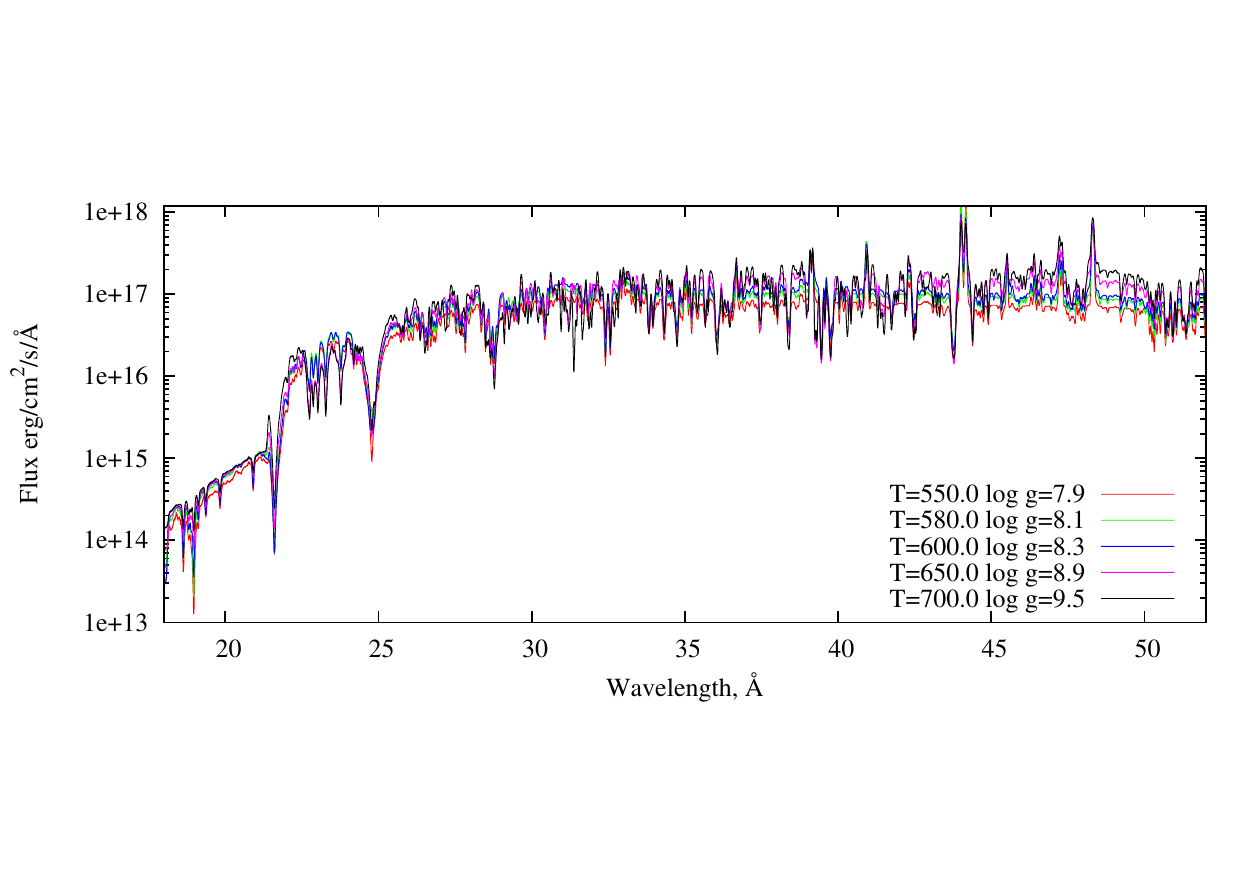}
\parbox{130mm}{\caption[Temperature--gravity sequence.]
{Change of spectral appareance when both
    temperature and gravity are varied.
\label{fig:isospec}}}
\end{center}\end{sidewaysfigure}

From the characteristic changes of Figures \ref{fig:gseq} and 
\ref{fig:tempseq} a more interesting trend can
be outlined. By changing both the gravity and the temperature in a coordinated
way, the two effects nearly 
cancel out and the spectra remain the same as shown in Figure
\ref{fig:isospec}. This approximately linear correlation along $\log
g=(0.011\pm0.0006)T+(1.8\pm0.4)$ at 550,000 K is valid at other
temperatures as well. This correlation causes difficulty in determinig the
temperature and gravity and can lead to overestimation of both. 

\newpage
\subsection*{Close to the Eddington limit}

\begin{figure}[htb]\begin{center}
\includegraphics[width=14.2cm]{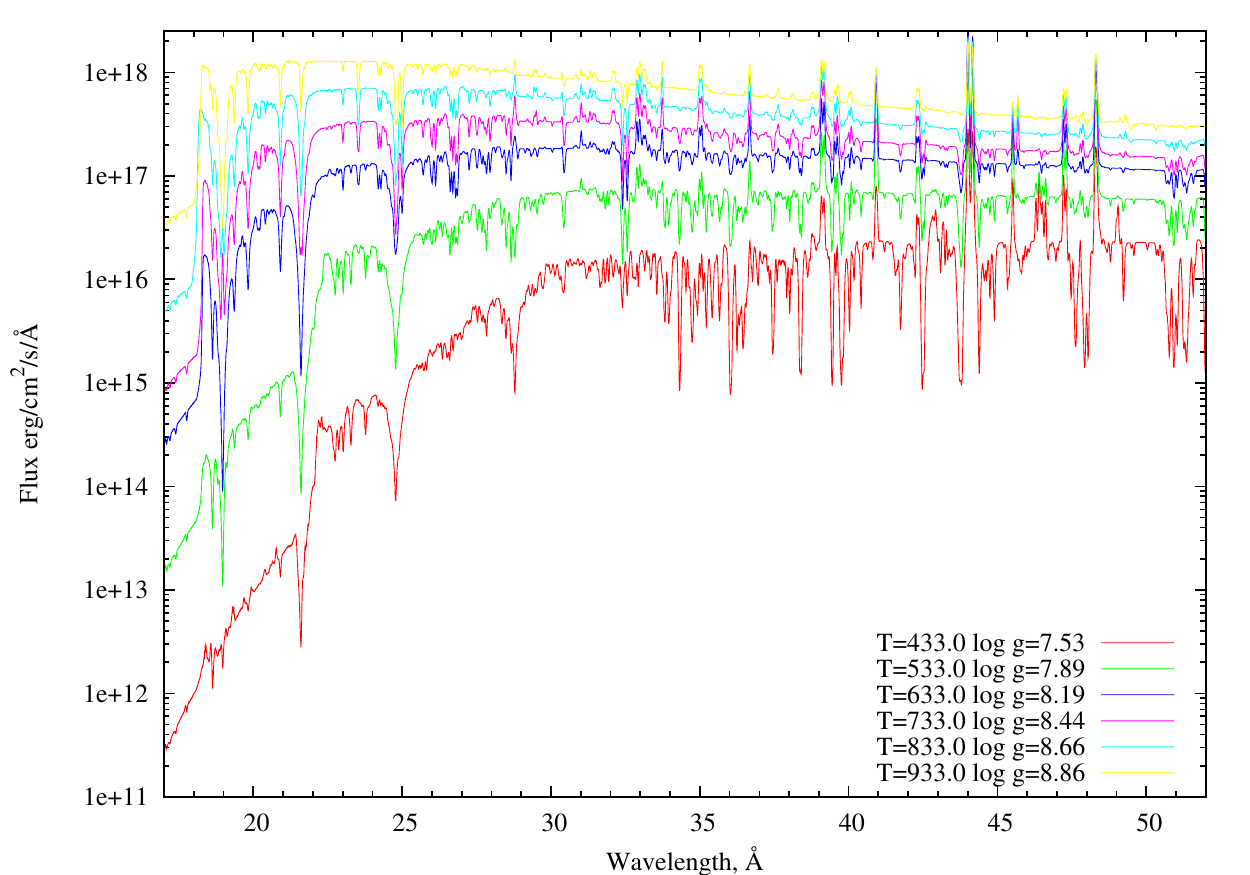}
\parbox{130mm}{\caption[Models close to the Eddington limit.]
{Model sequence close to the Eddington limit.
\label{fig:eddington}}}
\end{center}\end{figure}

The spectral appearance just below the Eddington limit was also investigated. 
Equation \ref{ledd} was used to calculate the surface gravity for a given
temperature at Eddington luminosity. Then 
to avoid convergence problems the surface
gravity was increased by $0.1$ dex for each model from $T_{\rm eff}=433,000$
to $T_{\rm eff}=933,000$~K. The spectral sequence of these models are shown in
Figure \ref{fig:eddington}. All models show a smooth continuum 
energy distribution with relatively weak ionization edges. Flux below 25~\AA\
is higher than for models with higher gravities. These features are also in
accordance with the conclusion of \citet{rossum10} and Figure
\ref{fig:rossum}. 

The strong absorption lines of N VII at $24.8$~\AA\ and N VI at $28.8$~\AA\
can be found in all spectra of the entire sequence. With increasing
temperature and changing ionization balance both lines get weaker and the N VI
$28.8$~\AA\ line goes into emission. Due to the observed high abundance
of nitrogen in novae these spectral lines are good indicators of effective
temperature. Similarly, the lines of O VIII at $18.97$~\AA\ and O VII at
$21.6$~\AA\ can be used for temperature diagnostics. 
However, these lines are present at
higher temperatures where they are accompanied by strong continuum emission.

\newpage
\subsection*{Effects of different model atoms}

\begin{sidewaysfigure}[!hp]\begin{center}  
\includegraphics[width=20.2cm]{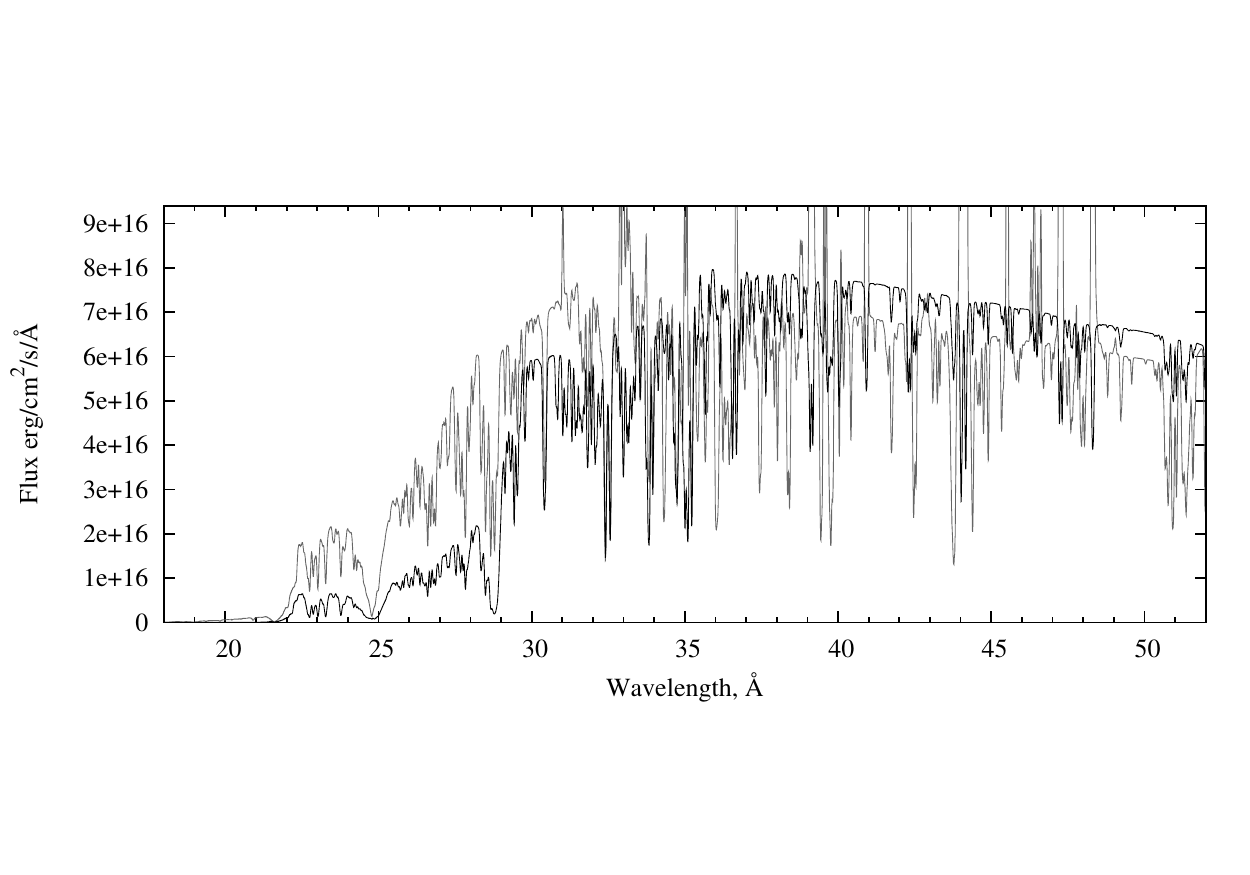}
\parbox{130mm}{\caption[Effects of higher energy levels.]
{Effects of higher energy levels. Models were calculated at $T_{\rm
    eff}=533,000$ K and $\log g=7.9$ cm/s$^{-2}$ with two-level atoms
  (grey) and full model atoms (black).
\label{fig:hilev}}}
\end{center}\end{sidewaysfigure}

Also important is to look at the effects of higher energy levels. This
comparison can be seen in Figure \ref{fig:hilev}, where spectra with two-level 
atoms and full model atoms are compared. These models were calculated at
$T_{\rm eff}=533,000$ K, $\log g=7.89$ cm/s$^2$ with identical abundances. 
The effects of
opacities of the 
high energy levels are obvious. The new opacity sources redistribute
the energy and make the spectrum softer. Some lines appear in emission with
only a few energy levels in the model atoms, these are in absorption if higher
levels are considered. 

\subsection*{Abundance sequence}

\begin{figure}[htbp]\begin{center}  
\includegraphics[width=14.2cm]{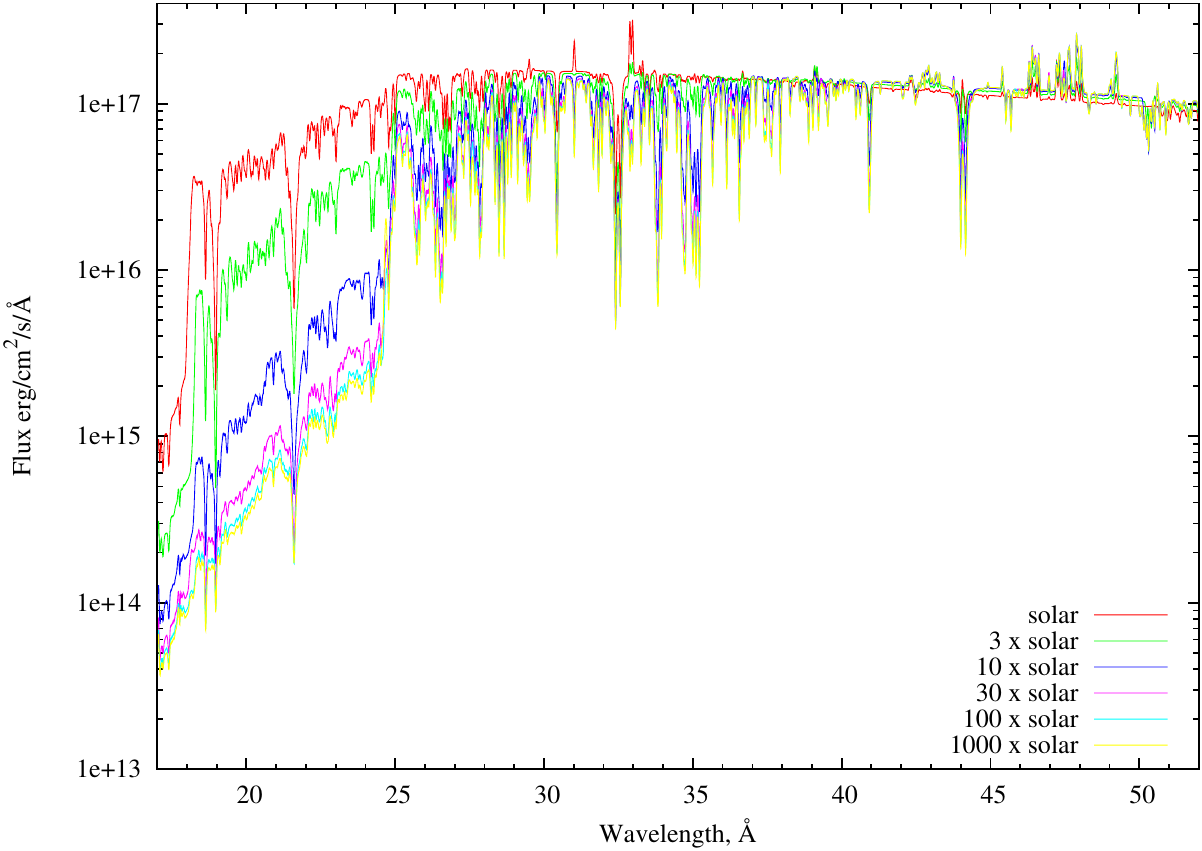}
\parbox{130mm}{\caption[Effects of metallicity.]
{Effects of different metallicity. Models were calculated at $T_{\rm
    eff}=600,500$ K and $\log g=8.23$ cm/s$^{2}$ with solar, 
and $3$, $10$,
  $30$, $100$ and $10^3$ times the solar values.
\label{fig:abnch}}}
\end{center}\end{figure}

The effects of individual abundances has already been shown for some elements 
in Figures
\ref{fig:abnch_c} -- \ref{fig:abnch_mg}, however abundances can be changed for
all elements simultaneously. This was done for $T_{\rm eff}=600,500$ K and
$\log g=8.23$ cm/s$^2$ at $1$, 
$3$, $10$, $30$, $100$ and $10^3$ times solar abundances
and is shown in Figure \ref{fig:abnch}. Model atoms with at most 
five levels were used
to avoid convergence problems. These models show the important opacity of
CNO elements below $25$ \AA, what significantly reduces the flux in this
range. The strong line blanketing is also noticeable between $25$ and $40$ \AA.

\begin{figure}[htbp]\begin{center}  
\includegraphics[width=14.2cm]{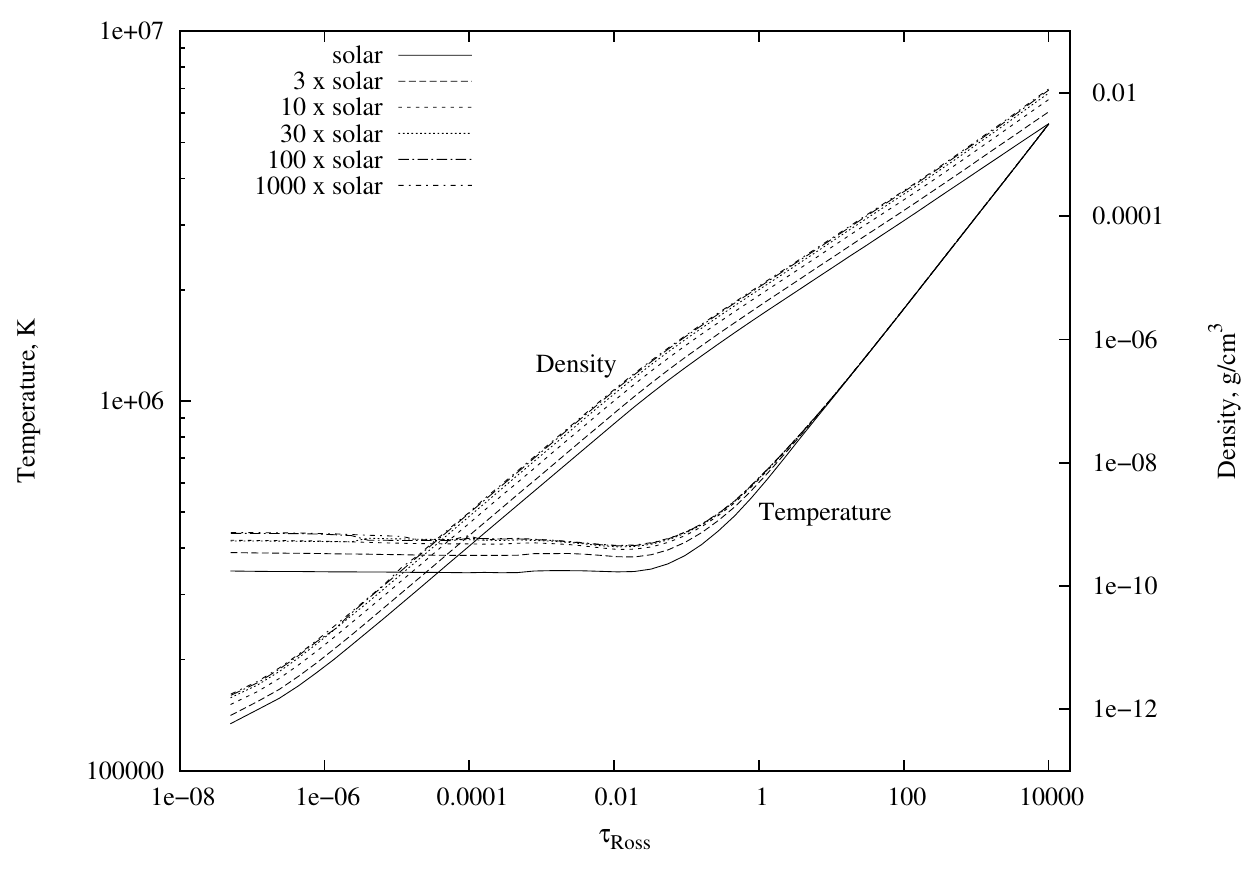}
\parbox{130mm}{\caption[Effects of metallicity on atmospheric structure.]
{Temperature and density as function of Rosseland optical-depth at solar and
  $3$, $10$, $30$, $100$ and $1000$ times solar abundances.
\label{fig:abnseq_td}}}
\end{center}\end{figure}

The temperature and density profiles for these atmospheres are shown in Figure
\ref{fig:abnseq_td}. Temperature structure does not change in the inner part
of the atmosphere where LTE is valid. At, and above the photosphere, however,
differences are more prominent. The strong opacity introduced by spectral
lines increases the temperature in the atmosphere by about $80,000$ Kelvins
when abundance was changed from solar to $10^3$ times the solar value.  

\subsection*{Radial extension}

\begin{figure}[htb]\begin{center}  
\includegraphics[width=13cm]{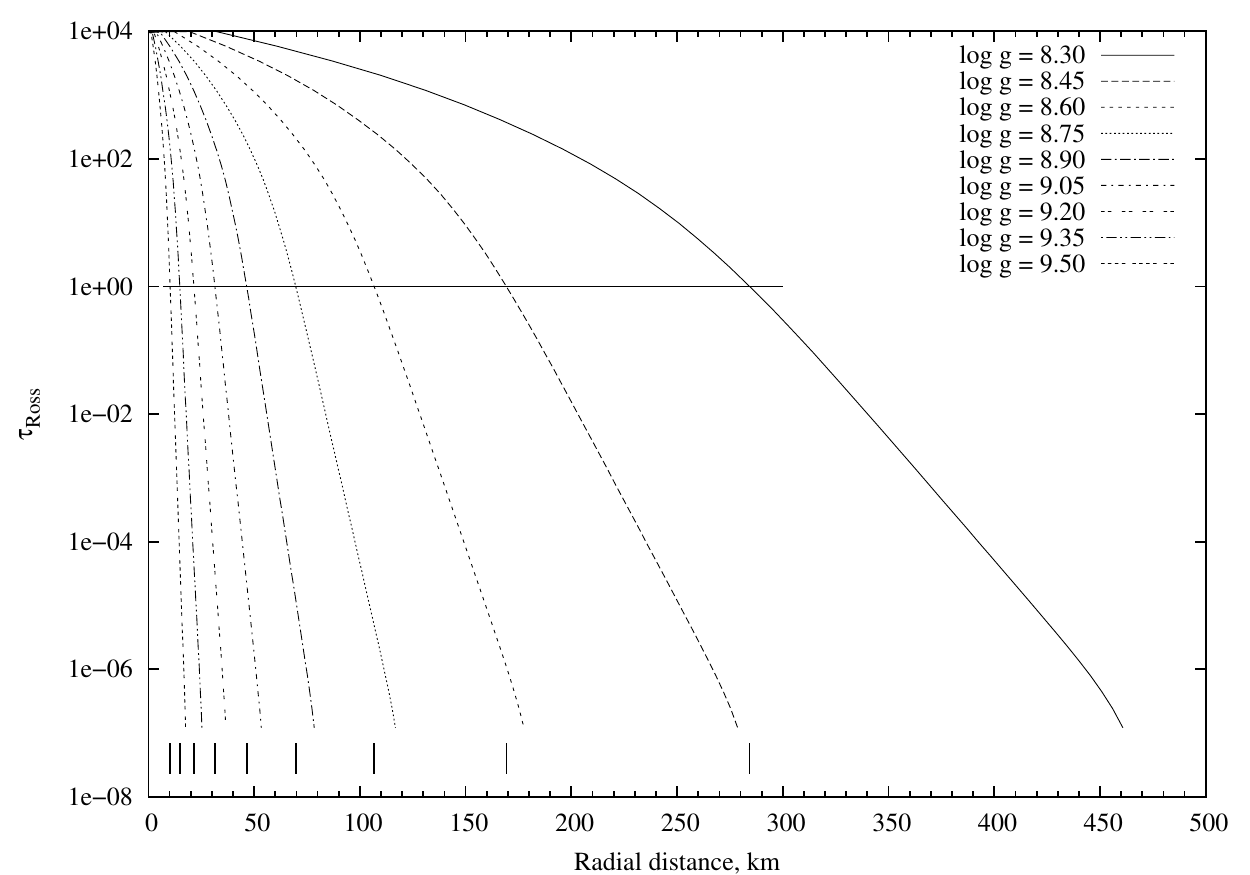}
\parbox{130mm}{\caption[Thickness of atmospheres.]
{Radial size of atmospheres as function of surface gravity (WD mass). Models
  were calculated at
  $T_{\rm eff}=600,000$ K and composition from Table \ref{tab:modelp} from
  $\log g = 8.3$ to $\log g=9.5$ cm/s$^2$. The 
photosphere was defined at $\tau_{\rm Ross}=1$,
  indicated by the horizontal line. Short vertical tics at the bottom show the
  distance of the photosphere from base of the atmosphere 
  ($\tau_{\rm Ross}=10^4$) for each model.  
\label{fig:radial}}}
\end{center}\end{figure}
The radial 
extension of model atmospheres can be derived from optical depth, opacity,
mass-density and column-mass. TLUSTY provides these quantities for fully
converged models at each depth point. The photosphere is defined at $\tau_{\rm
Ross}=1$ from where about 37\%\ of light escapes. Rosseland optical depths are
shown in Figure \ref{fig:radial} as function of radial geometrical distance in
the atmosphere. The horizontal line at $\tau_{\rm Ross}=1$ shows the location
of the photosphere. Short vertical lines on the bottom of the graph 
help to read the radial distance between $\tau_{\rm Ross}=10^4$ and $1$. The size
of the atmosphere grows progressively faster for lower surface
gravities. Assuming that at low effective temperatures ($T_{\rm eff}<50,000$ K) 
the radius of the atmosphere is negligible compared 
to the radius of the WD, one can
use the mass-radius relationship to find the relative size of the atmosphere
at higher temperatures. 
At $\log g=8.3$~cm/s$^2$ this is about $29$\%, at $\log g=8.5$~cm/s$^2$ about
$20$\%\ and at $\log g=9.0$~cm/s$^2$ about $7$\%\ of the WD radius.

\begin{figure}[!h]\begin{center}  
\includegraphics[width=13cm]{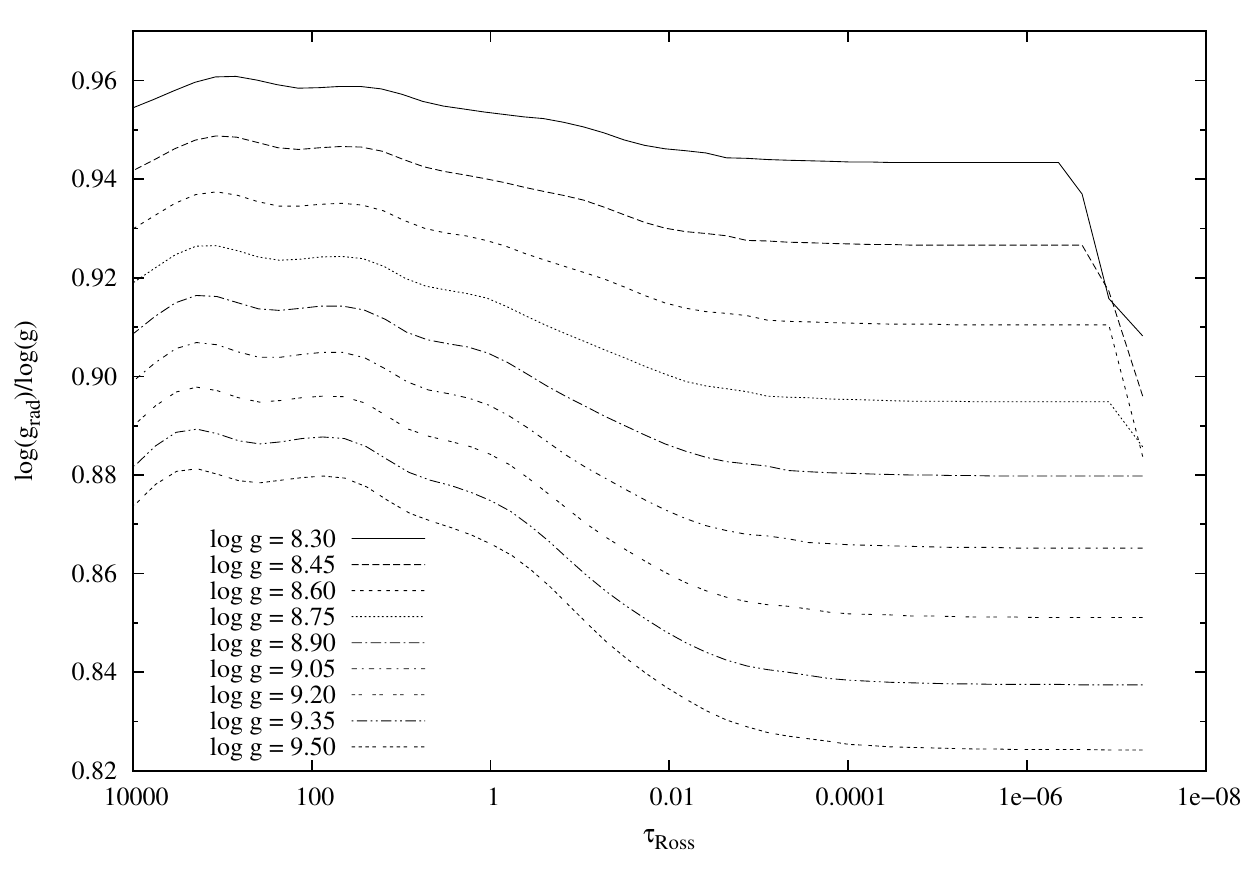}
\parbox{130mm}{\caption[Radiation pressure.]
{Radiation pressure as function of surface gravity. Atmospheric parameters are
  identical with those in Figure \ref{fig:radial}. Radiation pressure is
  expressed as fraction of the surface gravity for each model. 
\label{fig:radpres}}}
\end{center}\end{figure}
Figure \ref{fig:radpres} shows the radiation pressure versus optical depth 
at different surface gravities. The radiation pressure is
expressed as fraction of the surface gravity. At $\log g_{\rm Rad}/\log g=1$
the force by radiation balances gravity and the model cannot be considered
hydrostatic. This transition happens for $T_{\rm eff}= 600,000$ K at around
$\log g=8$~cm/s$^2$, however, due to numerical effects convergence breaks down
around
$\log g=8.3$~cm/s$^2$. This can be seen in Figure \ref{fig:radpres}, the
discontinuities at the top of the atmosphere grow larger for lower surface
gravities. Models below $\log g=8.3$~cm/s$^2$ failed to converge.  

\section{Comparison with TMAP}

\begin{figure}[t]
\begin{center}
\includegraphics[width=14.2cm]{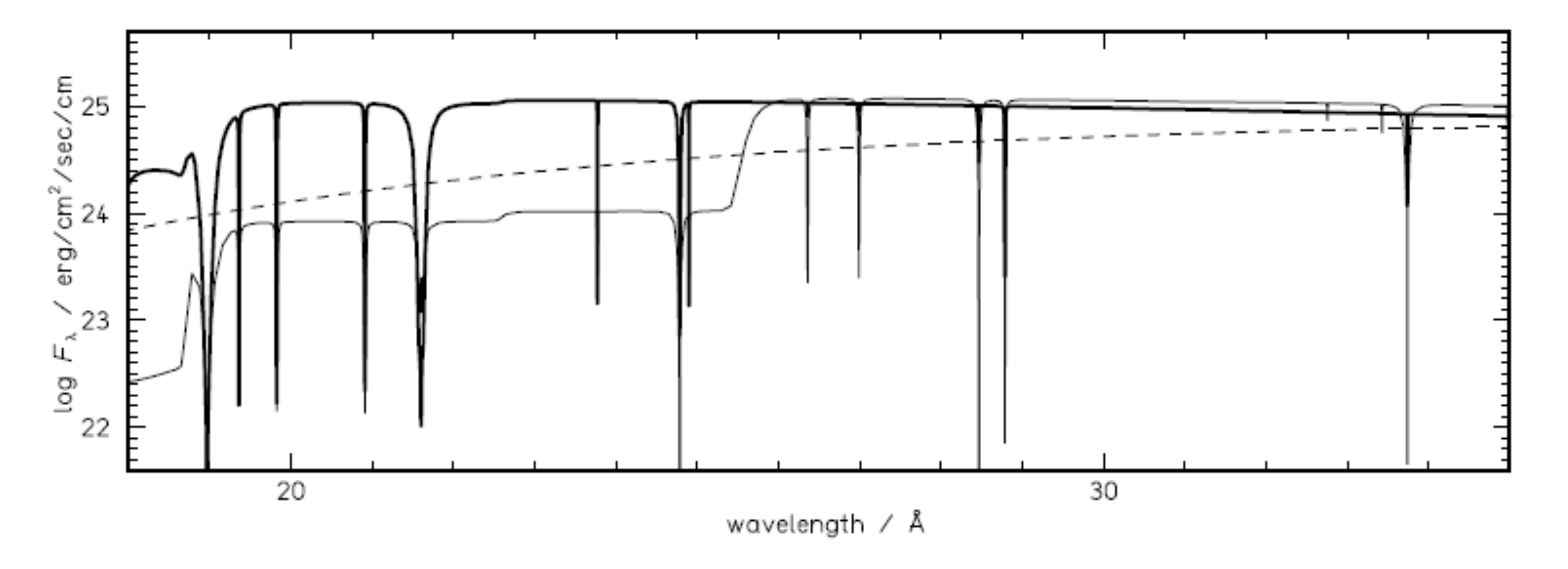}
\parbox{130mm}{\caption[TMAP models.]
{Comparison of fluxes for a model with $T_{\rm
      eff}=700,000$ K and $\log g=9.0$ cm/s$^2$ at different abundance
    ratios. The thin line shows solar composition and the thick line was
    calculated at H/H$_\odot=1$, He/He$_\odot=10.26$, C/C$_\odot=0.007$, 
    N/N$_\odot=0.544$ and O/O$_\odot=23.38$ by  number. Both models were
    calculated with TMAP \citep{rauch05}. Dashed line is a
    blackbody model. 
\label{fig:tmaprauch}}}
\end{center}
\end{figure}

\begin{figure}[b]
\begin{center}
\includegraphics[width=14.2cm]{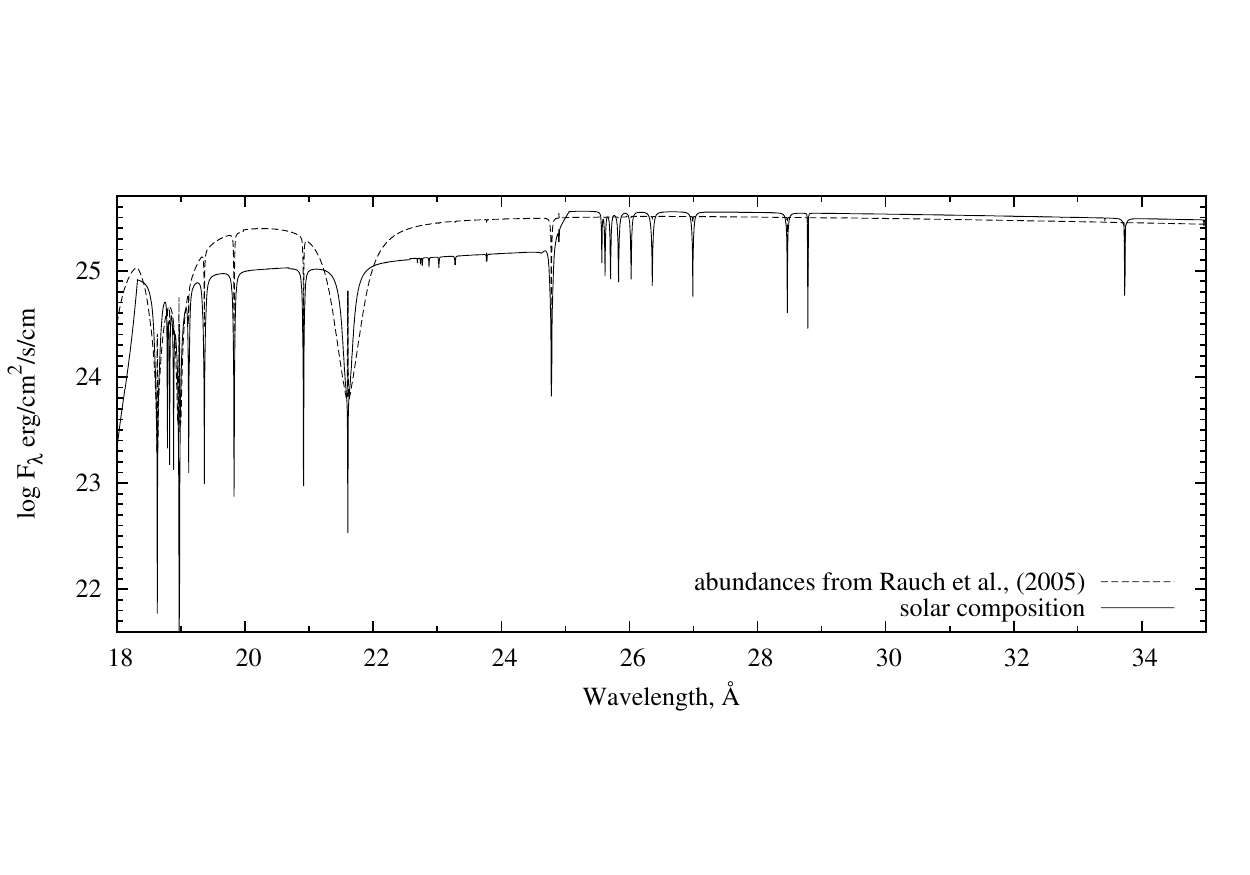}
\parbox{130mm}{\caption[TLUSTY models.]{Same models as in Figure
    \ref{fig:tmaprauch} calculated with TLUSTY. Thin line shows solar
    abundance, the other was calculated with abundance 
ratios used by \citet{rauch05}.   
\label{fig:tlustyme}}}
\end{center}
\end{figure}

Figures \ref{fig:tmaprauch} and \ref{fig:tlustyme} are comparisons between
TMAP\footnote{
\protect\url{http://astro.uni-tubingen.de/~rauch/TMAP/TMAP.html}} 
and TLUSTY models, Figure \ref{fig:tmaprauch} was taken from \citet{rauch05}. 
TMAP (T\"{u}bingen NLTE Model Atmosphere Package) calculates stellar
atmospheres in spherical or 
plane-parallel geometry in hydrostatic and radiative
equilibrium. The spectra shown here were calculated in plane-parallel
geometry. As both TLUSTY and TMAP share the same model assumptions, their
outputs are comparative. Beside the general agreement of the results
there are three important things to note: there are
more lines in the TLUSTY model, the super-solar oxygen abundance causes much
broader lines in the TLUSTY model, and the 
continuum is a factor of three higher than TMAP's
when the \citet{rauch05} abundance (H/H$_\odot=1$, He/He$_\odot=10.26$,
C/C$_\odot=0.007$, N/N$_\odot=0.544$ and O/O$_\odot=23.38$) was used. 
This can be seen at the flux levels of the solar spectra as well.

\chapter{Observations of Classical Novae}\label{observations}

 
X-ray measurements of celestial objects began with the era of space-borne 
observatories and the Einstein Observatory (HEAO-2) in
1978. During its three-year 
mission, Einstein discovered numerous X-ray sources in
the Andromeda Galaxy and the Magellanic Clouds, conducted deep X-ray surveys
and discovered that coronal emission in normal stars is stronger than
expected. Einstein's observations also helped the morphological studies of
supernova remnants. It was followed by EXOSAT from 1983 to 1986. EXOSAT
observed Low Mass X-ray Binaries, pulsars and AGN
variability. The first nova detected by EXOSAT in soft X-rays was GQ Mus 1983.

An important breakthrough came with the launch of ROSAT in 1990. Its higher
sensitivity, resolution and nine year mission significantly contributed
to the advancement of high energy astrophysics. 
The new class of Supersoft X-ray Sources (SSS)
was discovered with ROSAT. They were thought to be white dwarfs accreting matter
from a binary companion star at a rate sufficient to produce steady nuclear
burning on the white dwarf surface, or perhaps 
classical novae after outburst. However,
due to the relatively low resolution of ROSAT only the spectral energy
distribution (SED) was measurable and not line spectra. 
Unlike other accreting X-ray binaries
where the gravitational potential energy powers the X-ray source nuclear
burning is the primary energy source in SSS. That their SED  
can not be sufficiently modeled by black bodies, so detailed
white dwarf
atmosphere models are essential was realized by \citet{macdonald91}.
ROSAT observed 30 galactic
and nine LMC novae for up to ten years and found only three 
of them (V1974 Cyg 1992,
GQ Mus 1983 and N LMC 1995) active in soft
X-rays, \citep{orio01}.  

Detailed modeling requires even higher 
resolution spectra than ROSAT provided. Since 1999 
Chandra and XMM-Newton observatories have met this need. These are the data
sources that provide sufficiently high resolution observations for my
work. In the next sections I give more details on them.
Also worth mentioning are the RXTE, Swift, INTEGRAL, Suzaku, BeppoSAX X-ray 
satellites; their measurements also greatly 
contributed to the study of supersoft
sources, although this study does not use these data. 


\section{Chandra}\label{chandra}
Chandra was launched by Space Shuttle 
Columbia on July 23, 1999. It was the third of NASA's four Great
Observatories. Its original mission was planned for five years. Chandra has an
elliptical orbit with a 133,000 km apogee and a 16,000 km perigee with 
28.5$^\circ$
inclination to the ecliptic and a period of 64.2 hours. 
This allows integrations of up to two days.

The scientific instruments include the High-Resolution Mirror assembly (HRMA), 
the Low and High-Energy Transmission Gratings (LETG and HETG), and 
the Science Instruments Module (SIM) which 
has two focal plane instruments, the Advanced
CCD Imaging Spectrometer (ACIS) and the High Resolution Camera (HRC). Since it
is relevant to this subject the HRMA, LETG and the HRC are explained below
in detail. 
\begin{figure}[!h]
\begin{center}
\includegraphics[width=14.2cm]{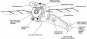}
\parbox{130mm}{\caption[The Chandra Observatory.]{The Chandra Observatory. 
Source: \protect\url{http://www.shuttlepresskit.com/STS-93/payload45.htm} }}
\label{fig:chandra}
\end{center}
\end{figure}

The HRMA consists of four pairs of mirrors with focal length of 10 m. Mirrors
are coated with iridium and the pair diameters range from 0.65~m to 1.23~m. 
Their total effective area is 1100~cm$^2$. 
\begin{figure}[!h]
\begin{center}
\includegraphics[width=14.2cm]{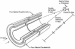}
\parbox{138mm}{\caption[The structure of HRMA.]{The structure of HRMA. 
Soruce: \protect\url{http://www2.jpl.nasa.gov/basics/axafmirror.gif}}}
\label{fig:hrma}
\end{center}
\end{figure}
The LETG is located behind the HRMA and can be rotated into the converging
beam to disperse X-rays to the focal plane where X-rays can be recorded by
either HRC or ACIS. Both cameras have spectroscopy and imaging modes. 
The LETG consists of $180$ 
segments having three modules each. These modules have gold wires with about
$991$~nm spacing and provide a high 
spectral resolution $E/\Delta{E}\approx10^3$ in the
$80$ -- $175$~\AA\ range and moderate resolution
$E/\Delta{E}\approx{20}$ between 3 and 50~\AA. LETG is most
commonly used for studies of on-axis observations of bright point sources. The
LETG/HRC-S combination is used for high resolution spectroscopy of soft
sources such as WDs and cataclysmic variables. 

The HRC is a 10~cm$^2$ 
microchannel plate instrument with two detectors: the HRC-I  is
optimized for imaging and the HRC-S is 
for spectroscopy. The HRC-S is well-suited
to serve as the readout detector for the LETG; it can also be used in a very
fast readout timing mode. The HRC provides a 30$\times$30~arcmin field of
view in imaging mode, and a 6$\times$99~arcmin field in spectroscopy mode.  

Chandra is the primary data source in this dissertation, 
since its spectral window covers the entire
supersoft range and has the neccessary resolution for spectral modeling.

\section{XMM-Newton}\label{xmm}
XMM-Newton was launched on Dec 10, 1999 by an Ariane 5 rocket from Guiana
Space Center at Kouron. Its orbit is a very eccentric 48 hour ellipse with
$40^\circ$ inclination. Its apogee is 114,000~km and perigee is only
7000~km. Similar to Chandra, the eccentric orbit makes long exposures
possible, here up to 40~hours. 

Scientific instruments related to my work are the Telescopes,
the two Reflection Grating Spectrometers (RGS) each consisting of a 
Reflection Grating Array (RGA) and an RGS Focal-plane Camera (RFC).
\begin{figure}[!h]
\begin{center}
\includegraphics[width=12cm]{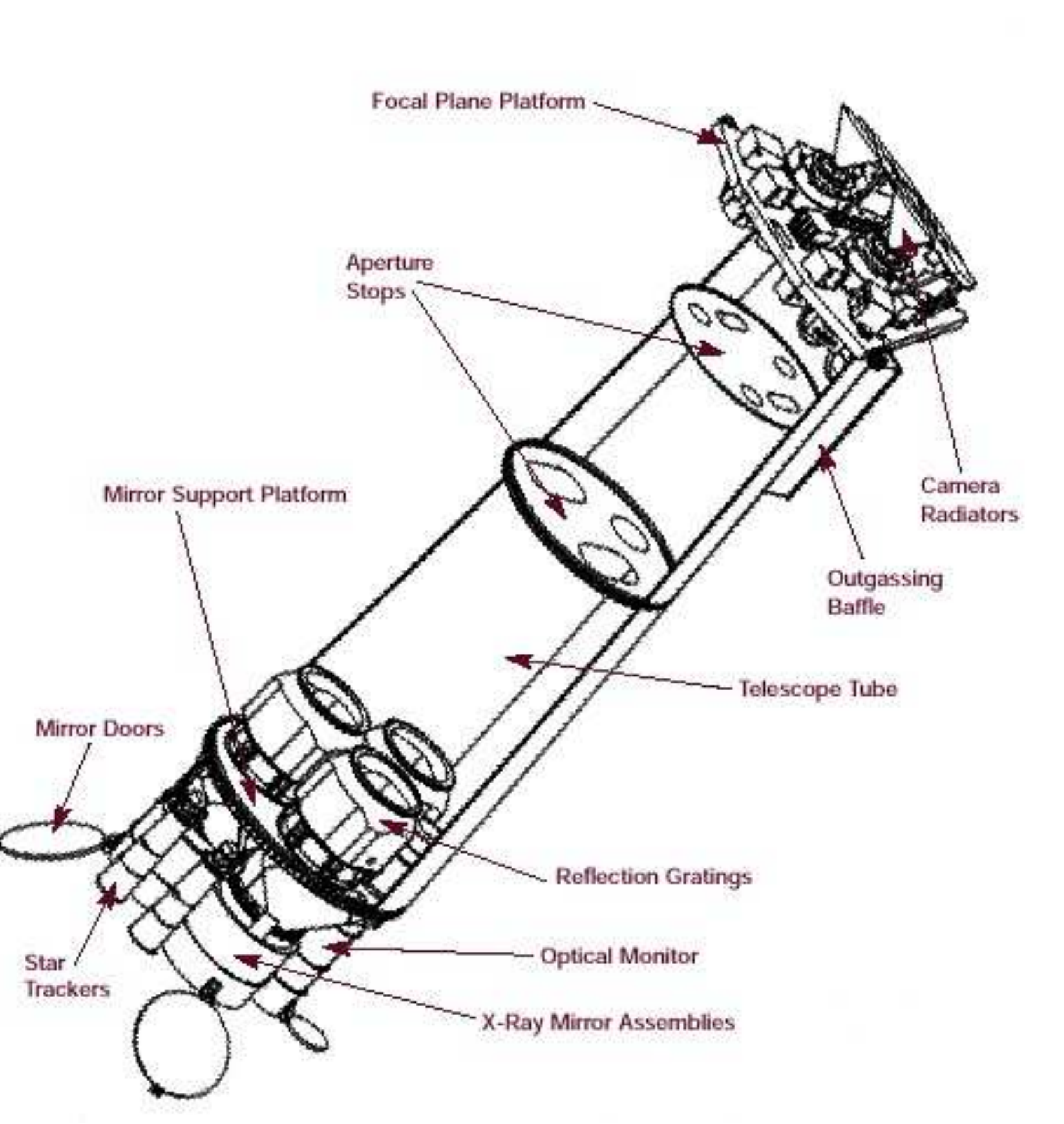}
\parbox{130mm}{\caption[The XMM-Newton Observatory.]{The XMM-Newton
    Observatory. 
Soruce:
\protect\url{http://xmm.esac.esa.int/
external/xmm_user_support/documentation/technical/Spacecraft/barre1.jpg} }}
\label{fig:chandra}
\end{center}
\end{figure}

XMM is equipped with three 
X-ray telescopes, each of them consists of 58 Wolter I
type grazing-incidence mirrors nested in a coaxial and confocal
configuration. The diameter of the largest mirror is 70~cm, which are made of
gold-coated nickel. 
Two telescopes are equipped with RGAs, which disperse about 58\% of
collected light at the RFCs in the 
sceondary focus. About 42\% of the light is directed
to the EPIC MOS cameras in the prime focus. The third 
is an open X-ray telescope for imaging .    
The total effective area of the three telescopes is 4650~cm$^{2}$ 
and their focal
length is 7.5~m with PSF (FWHM) 6~arcsec (HEW = 15~arcsec).

The grating plates in the RGAs have mean groove densities of about 645.6
lines/mm. Dispersion is a varying function of dispersion angle; 
at 15~\AA\ it is
about 8.3~mm/\AA\ in first order. The wavelegth accuracy is 7~m\AA\
(1 $\sigma$)
or better for both RGS1 and RGS2. 

RFCs consit of linear arrays of nine 
MOS CCD chips along the dispersion axis. The
chips are back-illuminated to maximize the soft energy response and aluminum
coated on the exposed side to suppress UV and optical light. Each detector has
1024x768 (27~$\mu$) pixels, half-exposed to the sky and half used as storage
area. At readout, pixels are binned to 81~$\mu{\rm m}$, 
which is sufficient to fully
sample the RGS line spread function. One bin corresponds to about 7, 10, and
14~m\AA\ in first order for wavelengths of 5, 15, and 38~\AA. 
Due to CCD failures
on orbit, effective areas are reduced by half between 20 and 24.1~\AA\ for RGS1
and between 10.6 and 13.8~\AA\ for RGS2. 

Although XMM-Newton has a narrower spectral window (5-38~\AA) than Chandra, 
there are many
K-shell transitions and He-like triplets of C, N, O, Ne, Mg and Si, and
L-shell transitions of heavier elements in its range. Furthermore, beyond 
40~\AA\ the strong interstellar absorption limits X-ray spectroscopy.  
Thanks to its higher
sensitivity and contribution to temporal coverage, XMM data 
complements the Chandra observations.

\section{Data Reduction}\label{chap:reduction}

\begin{figure}[!h]
\begin{center}
\includegraphics[width=12cm]{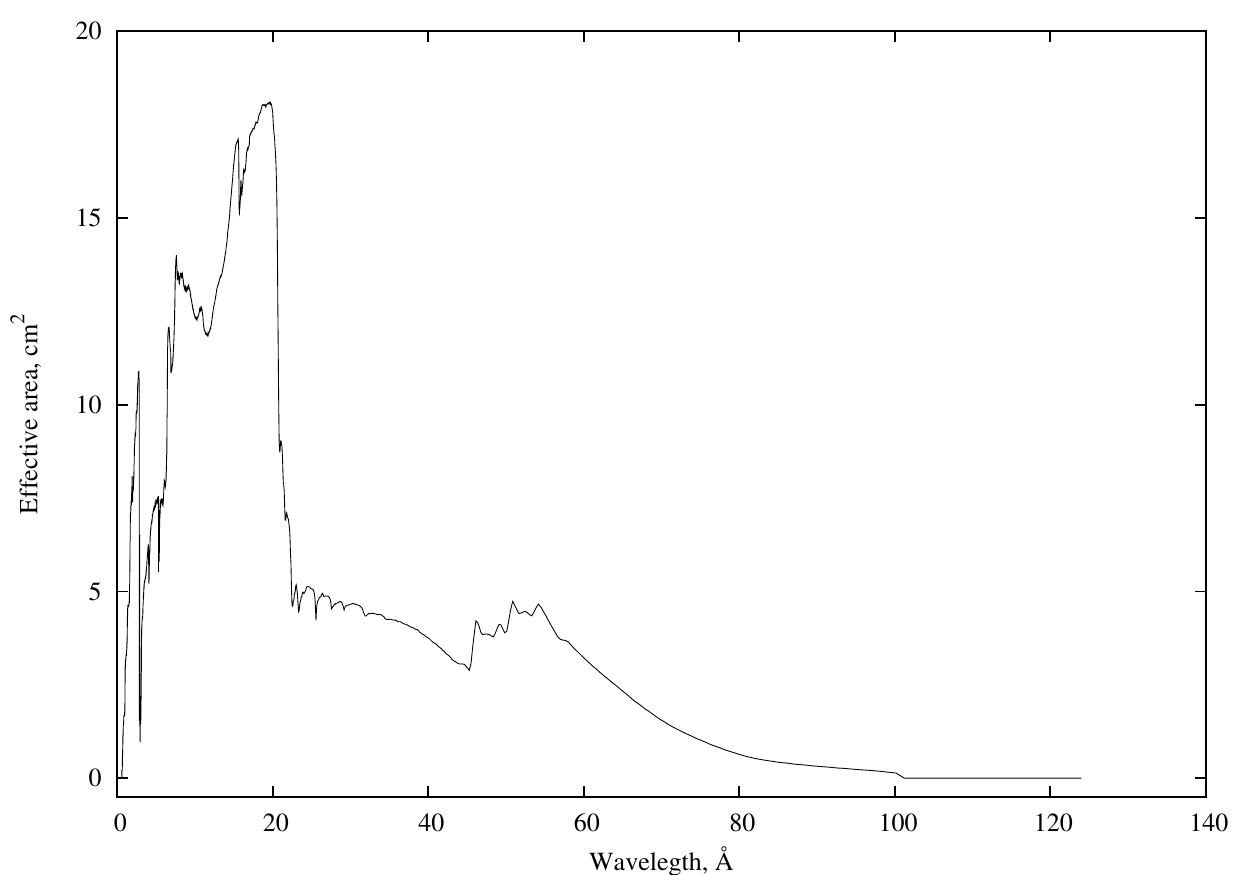}
\parbox{138mm}{\caption{LETG first order effective area for V4743 Sgr.
\label{fig:letg}}}
\end{center}
\end{figure}

Instead of
CIAO, the X-ray data analysis sofware of Chandra, 
a simpler procedure was applied. From the primary data products, the
spectra were extracted by using the tasks of
IRAF\footnote{\protect\url{http://iraf.noao.edu/}} and
STSDAS\footnote{\protect\url
{http://www.stsci.edu/resources/software_hardware/stsdas}} 
for binary tables. 
Positive and negative first-order 
spectra were processed separately and coadded to increase signal-to-noise
ratio. The effective area contains the properties of the telescope, the
gratings and the detector system. This function describes the total efficiency
of the instrument. With its help one can convert count rate to photon counts. 
The first order effective area for the negative arm of LETG can be seen
in Figure \ref{fig:letg}.

XMM-Newton data reduction was done in SAS version 8.0.1., using the standard
procedure presented for point-like sources on XMM's
page\footnote{\protect\url{http://xmm.esa.int/sas/current/documentation/threads/rgs\_thread.shtml}}
and in The XMM-Newton ABC
Guide\footnote{\protect\url{http://heasarc.nasa.gov/docs/xmm/abc/}}. 
Calibration
files were checked and updated automatically at startup. Data processing was
kept minimal, and 
only good time intervals (GTI) and background subtraction were
applied. Light curves were extracted and displayed to filter out
flares. During flares an abnormal high count rate can be identified over a
short period of time what is
not related to the source.  
During X-ray-on stages sources were bright enough and did not need rebinning. 

\section{Interstellar absorption}\label{sec:reddening}

Interstellar absorption is an important and
poorly understood factor in X-ray
modeling. The extinction curve is changing rapidly with wavelength. 
About 40\%\ of 
light extinction can be accounted for by interstellar matter
around 20~\AA\ and about 94\%\ at 50~\AA\ \citep{lanz05}. 
In my work I used the interstellar abundance and absorption
cross-sections for the 17 astrophysically important elements from 
\citet{morrison83} and 
\citet{balucinska92}\footnote{http://cdsweb.u-strasbg.fr/viz-bin/Cat?VI/62.}. 
Instead of their polynomial fits,
I used their functions to
calculate the total absorption cross-section at each data point
individually. 

I considered interstellar extinction as a sum of the standard
Galactic extinction plus the contribution of the circumstellar material. 
Galactic extinction was calculated from published hydrogen
column-densities ($N_{\rm H}$)
and relative interstellar abundances ($a_i^{\rm\tiny mm}$) from 
\citet{morrison83} using the formula:
\begin{equation}
\varphi(E,N_{\rm H},a_{\rm 1}^{{\rm\tiny mm}},...,a_{\rm 17}^{\rm\tiny mm}) =
e^{-N_{\rm H}\Sigma_{i}a_{i}^{\rm\tiny mm}\chi_{i}(E)},
\label{eq:galabs}
\end{equation}
where $\chi_{i}(E)$ is the photoelectric absorption cross-section of element
$i$ at energy
$E$. If there were no available data on column density
I estimated the standard extinction based on the
distance of the object \citep{binney98}:
\begin{equation}
N_{\rm H} \approx 3.1\times10^{21}\left(\frac{d}{\rm kpc}\right)\frac{1}{{\rm cm}^{2}}.
\label{eq:hcoldens}
\end{equation}
This provided a starting point for estimating the Galactic absorption along
the line of sight at each observed wavelength. However, Equation
\ref{eq:hcoldens} overestimates the column density for each nova. 
Their values were determined from model fits.

\begin{figure}[!h]
\begin{center}
\includegraphics[width=14.2cm]{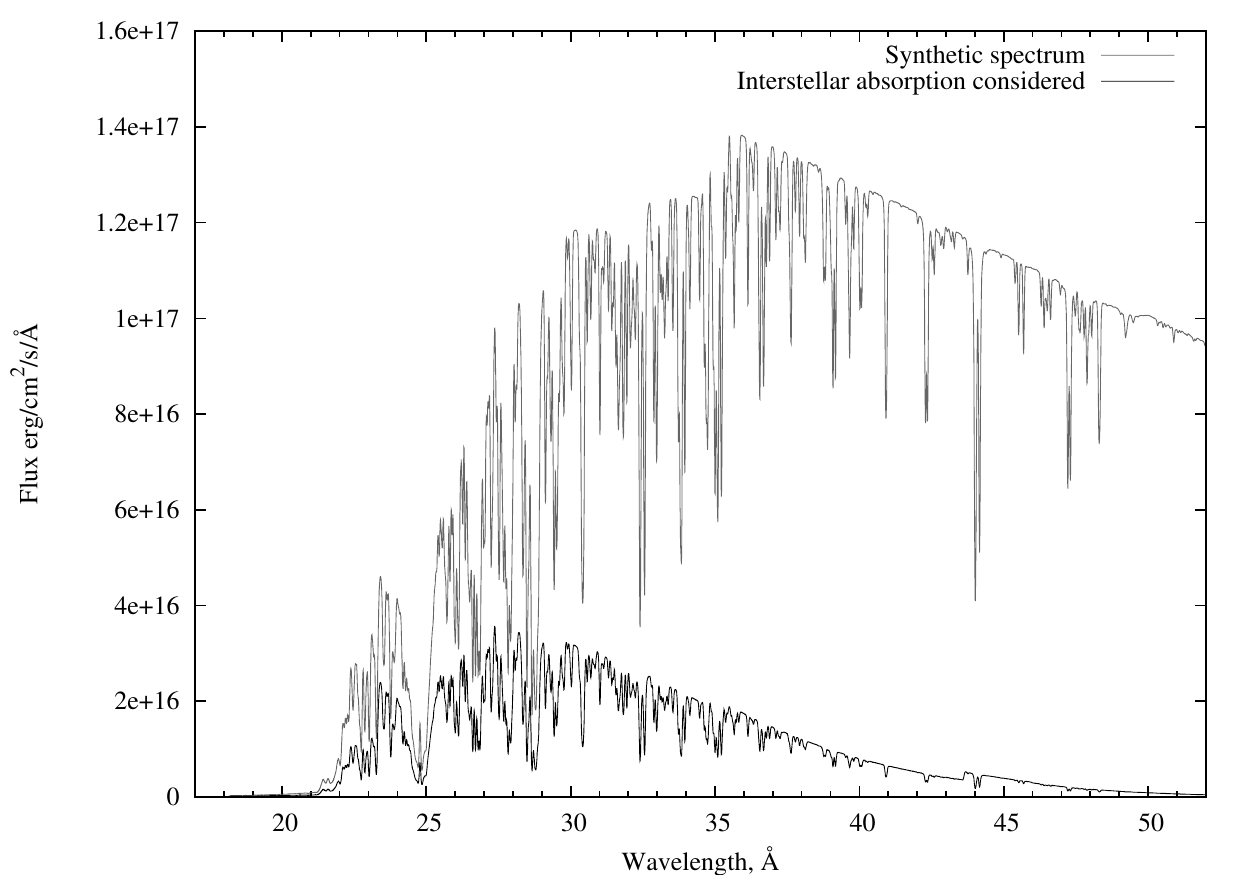}
\parbox{130mm}{\caption[Correction for interstellar absorption.]
{Correction for interstellar absorption. Galactic
    abundance was taken from \citet{morrison83} and a hydrogen column density 
N$_H=4\times10^{20}$ was used.\label{fig:abs}}}
\end{center}
\end{figure}

Also important is the extinction due to circumstellar material, which is
even more difficult to take into account for novae. The composition of the
shell, its clumpiness, geometry and evolution make this a 
very complex problem. The
column density of the shell can change significantly over time as well. 
Assuming that the composition of circumstellar material reflects the composition
of the nova shell, 
I modified the individual abundances in Equation 
\ref{eq:galabs}:
\begin{equation}
a_i^{total}=a_i^{\rm\tiny mm}+Ca_i^{\rm\tiny cs}
\label{eq:modab}
\end{equation}
where $a_i^{\rm\tiny cs}$ is the abundance of a given element relative 
to hydrogen
by number density in the shell and
$C$ is a scaling parameter which describes the amount of excess
hydrogen in the column relative to the standard column density. 

\section{Instrumental resolution}\label{sec:res}

\begin{figure}[!h]
\begin{center}
\includegraphics[width=12cm]{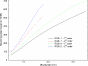}
\parbox{138mm}{\caption[RGS spectral resolution.]{Spectral resolution of XMM's RGS instruments, \citep{xmmuhb}
\label{fig:rgsres}}}
\end{center}
\end{figure}

The instrumental profiles of both Chandra and XMM was considered to be
Gaussian. The spectral resolution of XMM's RGS spectrometers can be seen in
Figure \ref{fig:rgsres}. Only first-order spectra were used in my work. These
were approximated with linear functions between 10 and 60 \AA. Here
$\lambda_s$ is the numeric value of the wavelength without unit, taken as
the slope:
\begin{eqnarray}
{\rm RGS1:}\hspace{5mm}& \frac{\lambda}{\Delta\lambda}=&17.66\lambda_s-33.33\\
{\rm RGS2:}\hspace{5mm}& \frac{\lambda}{\Delta\lambda}=&21\lambda_s-35
\label{eq:resrgs}
\end{eqnarray}
and for Chandra LETG/HRC-S:
\begin{eqnarray}
{\rm Chandra:}\hspace{5mm}& \frac{\lambda}{\Delta\lambda}=&20\lambda_s
\label{eq:reschan}.
\end{eqnarray}

\begin{figure}[!h]
\begin{center}
\includegraphics[width=12cm]{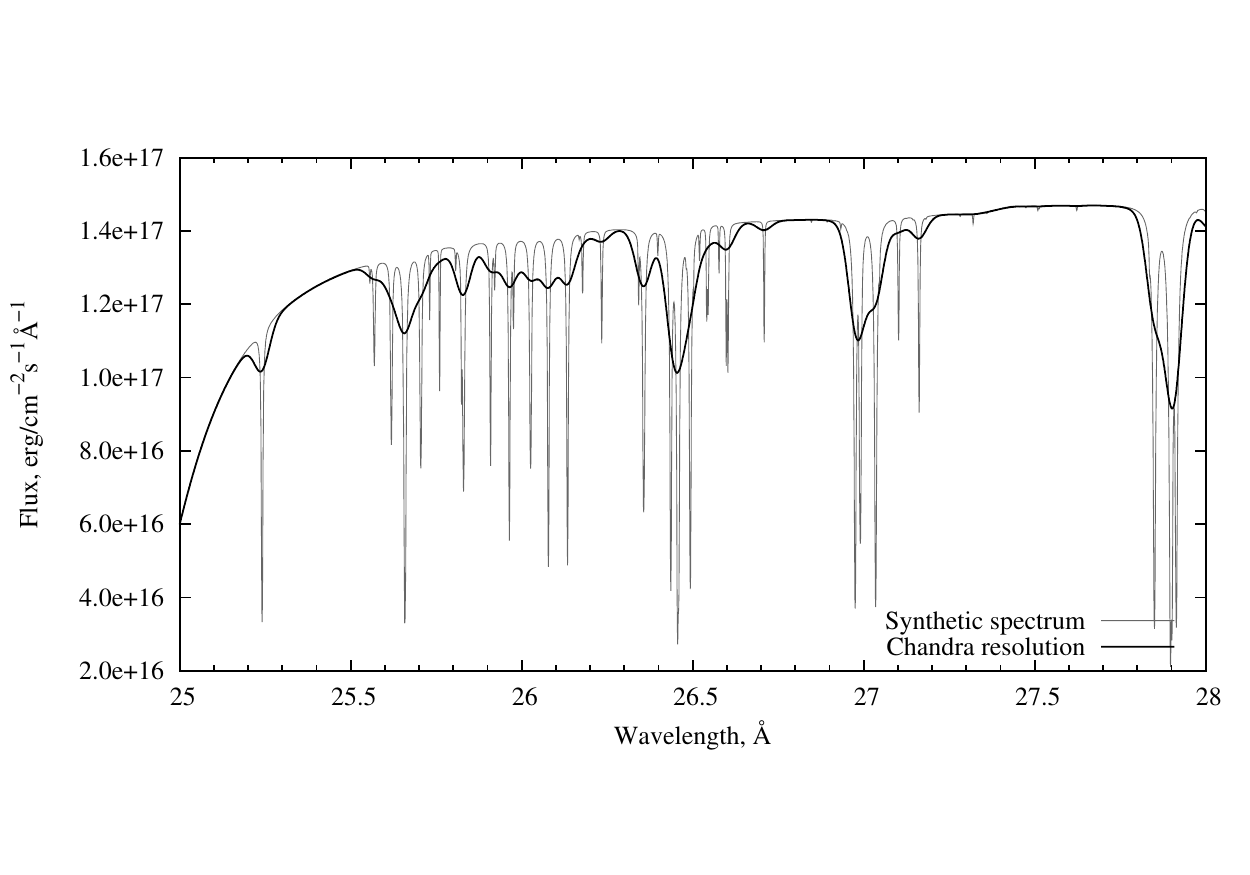}
\parbox{138mm}{\caption{Correction for instrumental resolution.
\label{fig:specres}}}
\end{center}
\end{figure}

\begin{figure}[!h]
\begin{center}
\includegraphics[width=12cm]{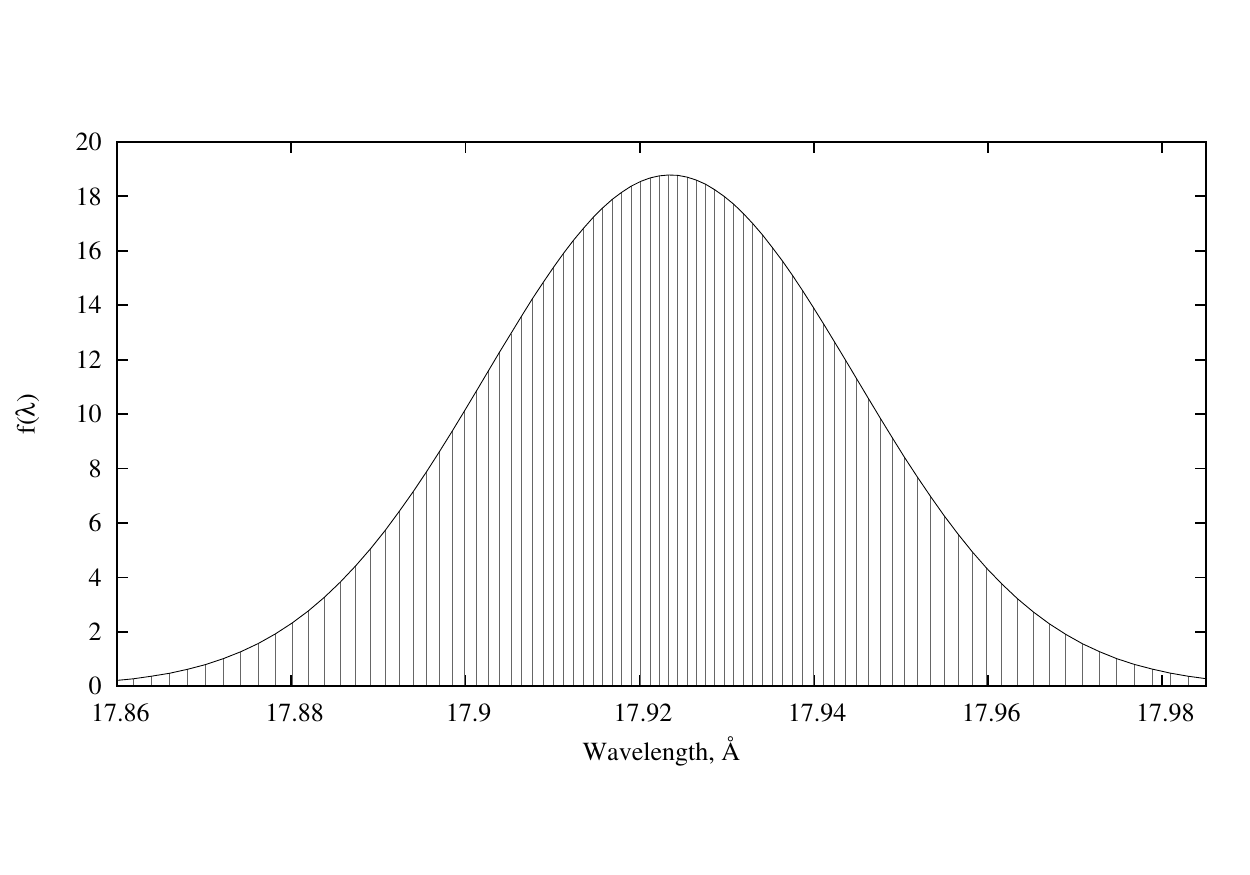}
\parbox{138mm}{\caption[Convolution function for instrumental resolution.]{Sampling of the convolution function for the
    calculation of the instrumental resolution.
\label{fig:convol}}}
\end{center}
\end{figure}

However, the difference in resolution between the instruments is practically
negligible. Figure \ref{fig:specres} is a plot of a synthetic and a
convolved model spectrum at the Chandra resolution.
The convolution function was calculated and normalized 
for the resolution of each data point. Next the spectrum 
was evaluated with a variable
step size to sample both the synthetic spectrum and the Gaussian well and 
the flux was then linearly interpolated for locations of the Gaussian sampling 
points. Figure \ref{fig:convol} shows the sampling of the kernel
function.

\newpage
\section{Application to Observations}\label{application}
In this chapter, I present public data on V4743 Sgr and V2491 Cyg 
retrieved from the Chandra and XMM data archives. The data on these two targets 
and the recurrent nova RS Oph are the
most comprehensive data sets available on CNe. However, both V2491
Cyg and RS Oph proved to be too luminous for static modeling. Their spectra
showed strong continuum emission below $18$ \AA\ what can be reproduced only at
relatively high effective temperatures and low surface gravities. This regime 
lies in the super Eddington area indicated in Figure \ref{fig:grid} and
cannot be modeled with TLUSTY. 
Both the Chandra and XMM
observations are released after one year of proprietary periods. As the
duration of the supersoft stage is relatively short, lasting only a 
few months, and novae appear randomly, scheduling observations is difficult and
mostly done as targets of opportunity.

\subsection{V4743 Sgr}
\begin{sidewaysfigure}[htbp]
\begin{center}
\begin{singlespace}
\includegraphics[width=20cm]{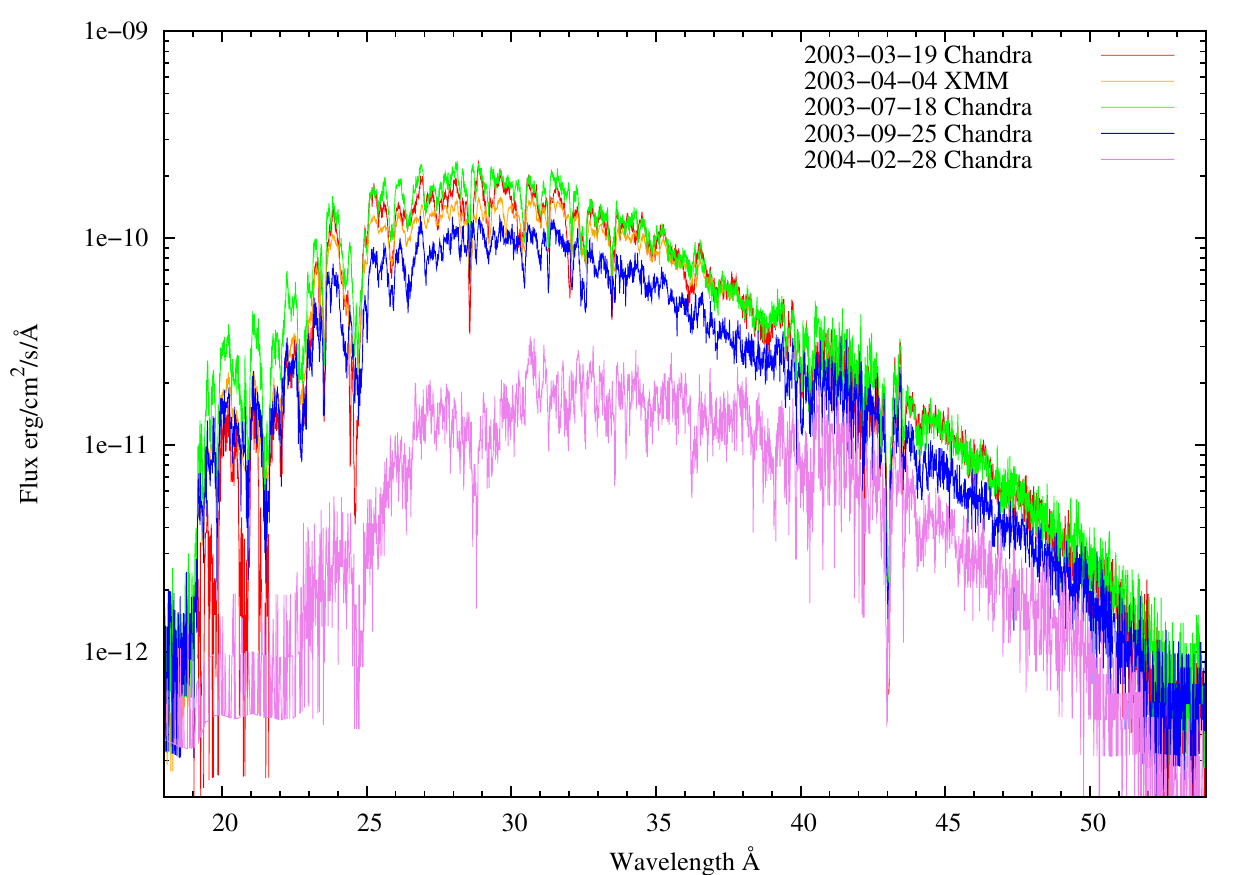}
\parbox{130mm}{\caption[Spectral evolution of V4743 Sgr.]{Spectral evolution of V4743 Sgr between March 2003 and February 2004.}}
\label{fig:v4743}
\end{singlespace}
\end{center}
\end{sidewaysfigure}

V4743 Sgr (Nova Sgr 2002c) was discovered on September 20, 2002 by
\citet{haseda02} at
its optical maximum of ${\rm V}=5^m$. It was a very fast nova (${\rm t}_3=15$ 
days)
and showed large expansion velocities. The FWHM of H$_{\alpha}$ reached 2400
km/s. The distance to the nova is uncertain, but estimated to be about 
$6.3$ kpc by
\citet{lyke02}, $3.9\pm0.3$ kpc by \citet{vanlandingham07} and $1.2\pm0.3$
kpc by \citet{nielbock03}. This makes direct radius and mass determinations
impossible. In March
2003 this nova was the brightest supersoft source in the sky. It was a
target of Chandra and XMM observatories over the following 
18 months. Follow up observations were made three years
later. The log of observations can be found in Table \ref{tab:v4743} and the
spectral evolution is shown in Figure \ref{fig:v4743}.  
\begin{table}[!h]
\begin{center}
\parbox{130mm}{\caption[Chandra and XMM observations of V4743 Sgr.]{Chandra
    LETG/HRC-S and XMM RGS X-ray observations of V4743 Sgr.}}
\vspace{2mm}
\begin{singlespace}
\begin{tabular}{r|c|c|l|l}
\hline
Observation ID & Date & Exposure (s)&Instrument& comment \\
\hline
3775       & 2003/03/19 & 24713& Chandra LETG/HRC-S&X-on\\
0127720501 & 2003/04/05 & 35214&XMM RGS    &X-on\\
3776       & 2003/07/18 & 11681&Chandra LETG/HRC-S&X-on\\
4435       & 2003/09/25 & 11485&Chandra LETG/HRC-S&X-on\\
5292       & 2004/02/28 & 10237&Chandra LETG/HRC-S&X-on\\
0204690101 & 2004/10/01 & 22317&XMM RGS    &X-off\\
0304720101 & 2006/03/29 & 33896&XMM RGS    &X-off\\
\hline
\end{tabular}
\label{tab:v4743}
\end{singlespace}
\end{center}
\end{table}

\begin{figure}[!h]
\begin{center}
\begin{singlespace}
\includegraphics[width=14.2cm]{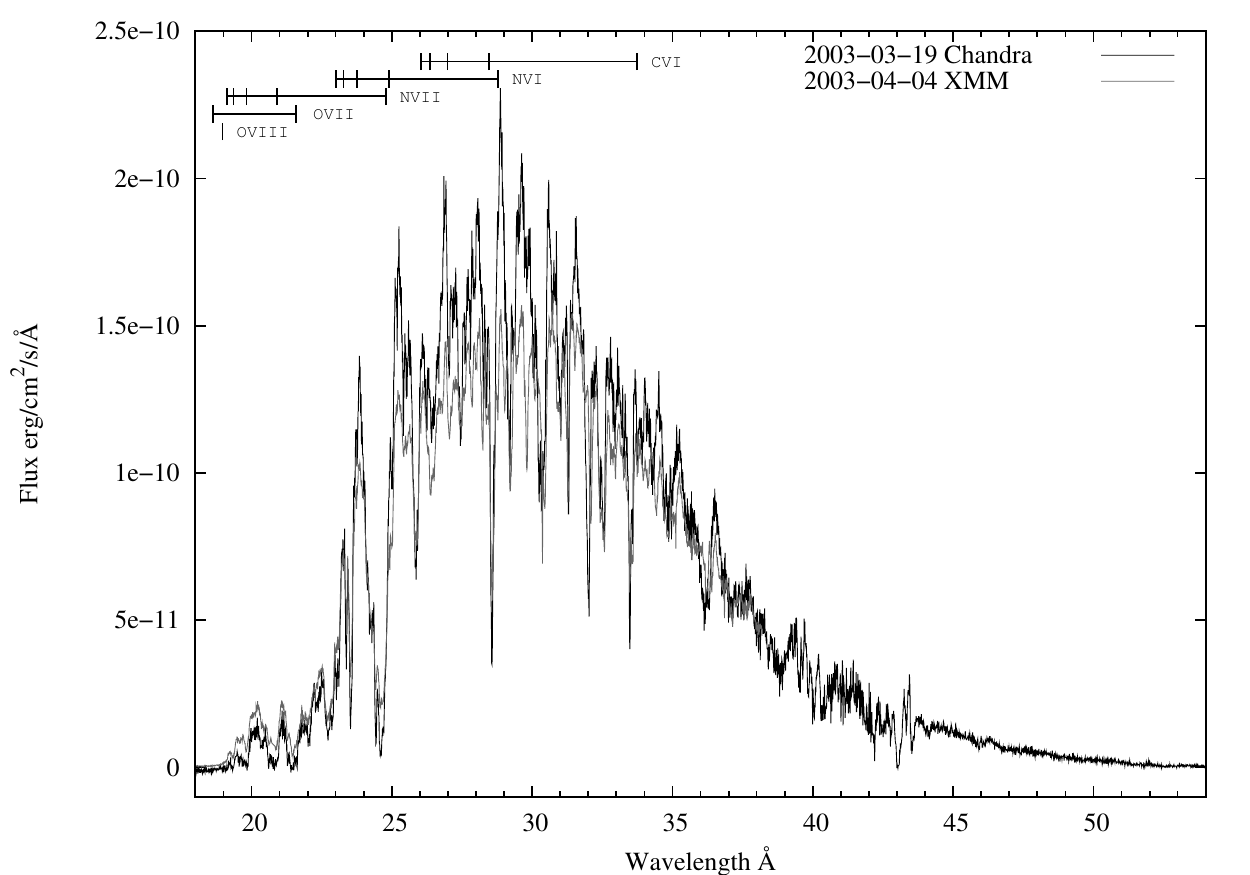}
\parbox{130mm}{\caption[Comparison 
of LETG and RGS spectra.]
{Comparison of Chandra LETG/HRC-S and XMM RGS observations.}\label{fig:v4743nc}}
\end{singlespace}
\end{center}
\end{figure}

V4743 Sgr was observed on March 19th and April 4th by Chandra and XMM,
respectively. A comparison of these spectra is shown in Figure
\ref{fig:v4743nc}. There were only 16 days between the two
observations, so the flux level did not change much; it was around $0.2$ cts/s 
in both observations, or at $1.5\times10^{-10}$ erg/cm$^2$/s/\AA. However the
peak flux decreased in April and line
strengths of both nitrogen and oxygen decreased compared to the earlier 
spectrum, while the high energy flux below 23 \AA\ increased. 
However, the line
strength variations are minor effects. 
By mid-July
the peak of the observed energy distribution 
reached $1.9\times10^{-10}$ erg/cm$^2$/s/\AA\, followed by a gradual
decrease to $1\times10^{-10}$ erg/cm$^2$/s/\AA\ in September 2003 and to
$2\times10^{-11}$ erg/cm$^2$/s/\AA\ in February 2004. V4743 Sgr was visited
later by XMM and no measurable X-ray spectra was found. It is 
important to note that
supersoft sources in general show rapid variability up to 40\% of the total
X-ray flux. In
novae this can be modulated further by the orbital period of the binary.  
\citet{leibowitz06} found six periodicities, ranging from a few hundred
seconds up to six hours. Two variations were particularly intense, with periods
1371 and 1310 seconds. The sum of these two frequencies was also
identified. These rapid variations may be related to the spin of the WD and/or
to non-radial pulsations. The relatively short timescale of these
variations relative to the length of spectroscopic observations were not
accounted for in this work. 
However, there was a dramatic change
during the March 2003 observation. At around 20,000 seconds into the
the observation the flux suddenly faded more than two
orders of magnitude. 
This variation might be related to the 6.6 hours orbital period
also observed in the optical or to other not yet identified mechanism. 
To account for 
this event the last 5000 seconds of
the March 2003 spectrum was deleted. 

The strong resonance lines of nitrogen and oxygen 
in the spectrum suggest overabundance of
these elements. A hydrogen deficiency and element ratio of N/C $\approx$ 1000
and 
N/O $\approx$ 50 is a signature of CNO-cycle processed material and typical
for novae
\citep{clayton68}. Such extreme compositions were not seen here, 
but the super
abundances of these elements relative to the Sun are obvious. The model 
composition was set up according to this over-abundance
of light metals, keeping all other elements at their solar value. However, 
models can be fit with an increased abundance 
of Ne, Mg, Al, and Si, as shown in Table
\ref{tab:modelp}. Individual element abundances are very uncertain. The most
reliable values could have been set up for nitrogen and oxygen; for them 
uncertainties are 
$\pm20$\%. Confining the error ranges of other element abundances are 
more difficult. 
Helium is about four to ten times more abundant than in the solar atmosphere.
It changes the spectral distribution homogeniously 
and does not have lines in this
range. However, a helium abundance over ten times the solar value raised
convergence problems. Carbon is somewhat different, as it regulates the flux at
the high energy end. Its spectral lines do not show up at the calculated
abundances and models failed to converge with an increased amount of
carbon. Based on its expected 
effects on the continuum the estimated abundance error is
$\pm40$\% for carbon. Neon and aluminum showed very similar behavior to
helium. Neon does not have spectral lines in the part of the spectrum where
both the signal-to-noise ratio is good and the interstellar absorption is
low. It does not make a significant contribution to the spectrum 
up to 100 times the solar
abundance. Aluminum has also a low effect on the spectra. Magnesium has
important lines in the supersoft spectra, but these were not found to be 
dominant, suggesting an upper limit on magnesium abundance at 50 times
solar. Silicon lines could not have been identified in the spectra suggesting
a low abundance close to the solar value. 
Heavier elements were not detectable, but they were considered to be near
solar values. It is important
to note that the simple model atoms of these elements do not allow detailed
modeling. Only some features of sulfur were identified in the spectra. 

By fitting the interstellar absorption for PHOENIX models \citet{petz05}
derived the hydrogen column density $N_H=4.0\times10^{21}$ cm$^{-2}$. In this
work it was found to be about a factor of ten lower, 
$N_H=(4.8\pm0.3)\times10^{20}$ cm$^{-2}$, much closer to
$N_H=3.47\times10^{20}$ cm$^{-2}$ found by \citet{rauch05} using TMAP.

From the model grid the possible temperature ($T_{\rm eff}$) range could
have been limited to $T_{\rm eff}=600,000$ -- $450,000$ 
K, at surface gravities ($\log g=8.2$ cm/s$^2$) indicating a
luminosity 
close to the Eddington limit at the beginning
and later decreasing ($\log g=8.0$ cm/s$^2$). 
The gravity did not change more than
$0.2$ dex between March 2003 and February 2004. The error in $\log g$
is $\pm0.1$ dex and the estimated error in effective temperature
is $\pm10,000$ K based on the resolution of the model grid. 
Considering the many free parameters affecting the
final spectrum and their complex interactions 
these values might not be accurate. 

\begin{sidewaysfigure}[p]
\begin{center}
\begin{singlespace}
\includegraphics[width=20cm]{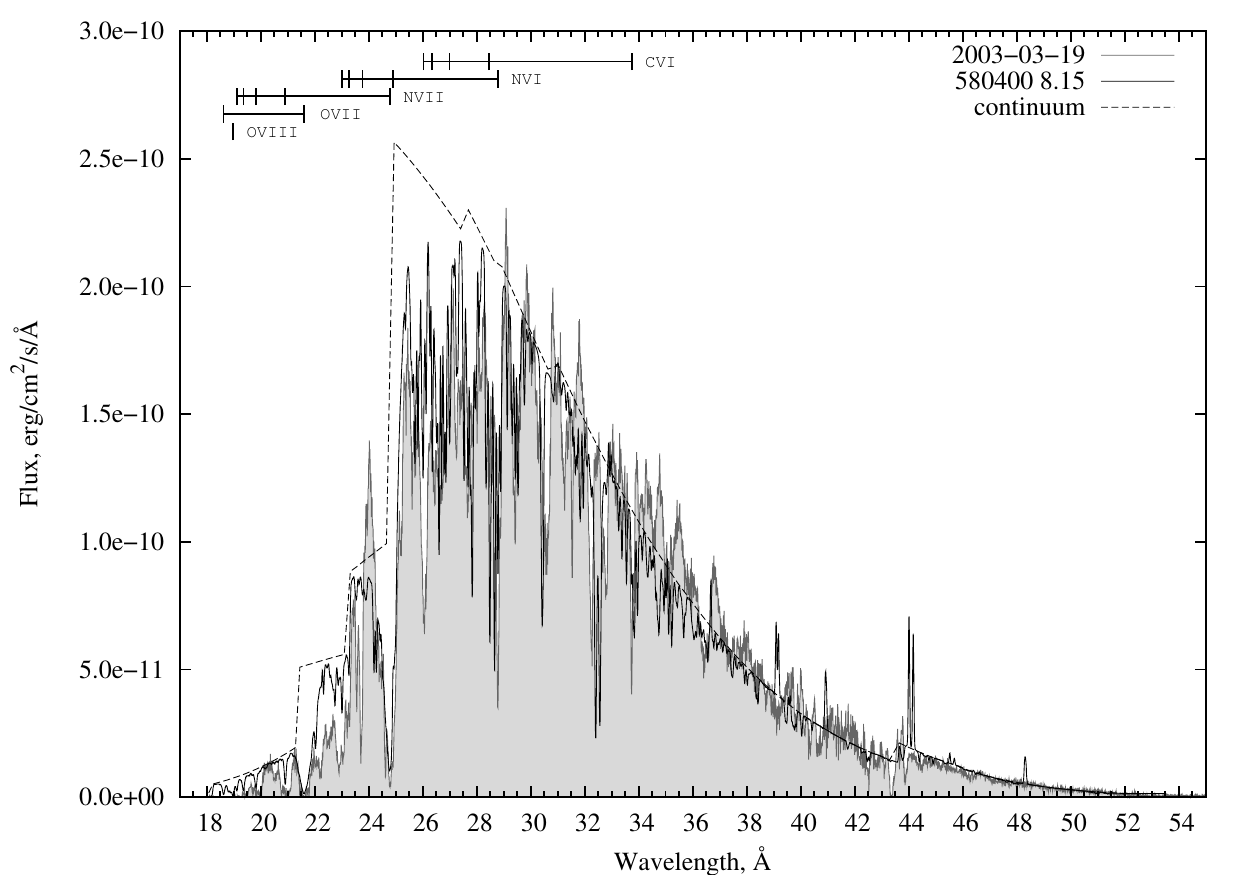}
\parbox{130mm}{\caption{Model fit for V4743 Sgr. March 19,
    2003}\label{fig:3775}}
\end{singlespace}
\end{center}
\end{sidewaysfigure}

\begin{sidewaysfigure}[p]
\begin{center}
\begin{singlespace}
\includegraphics[width=20cm]{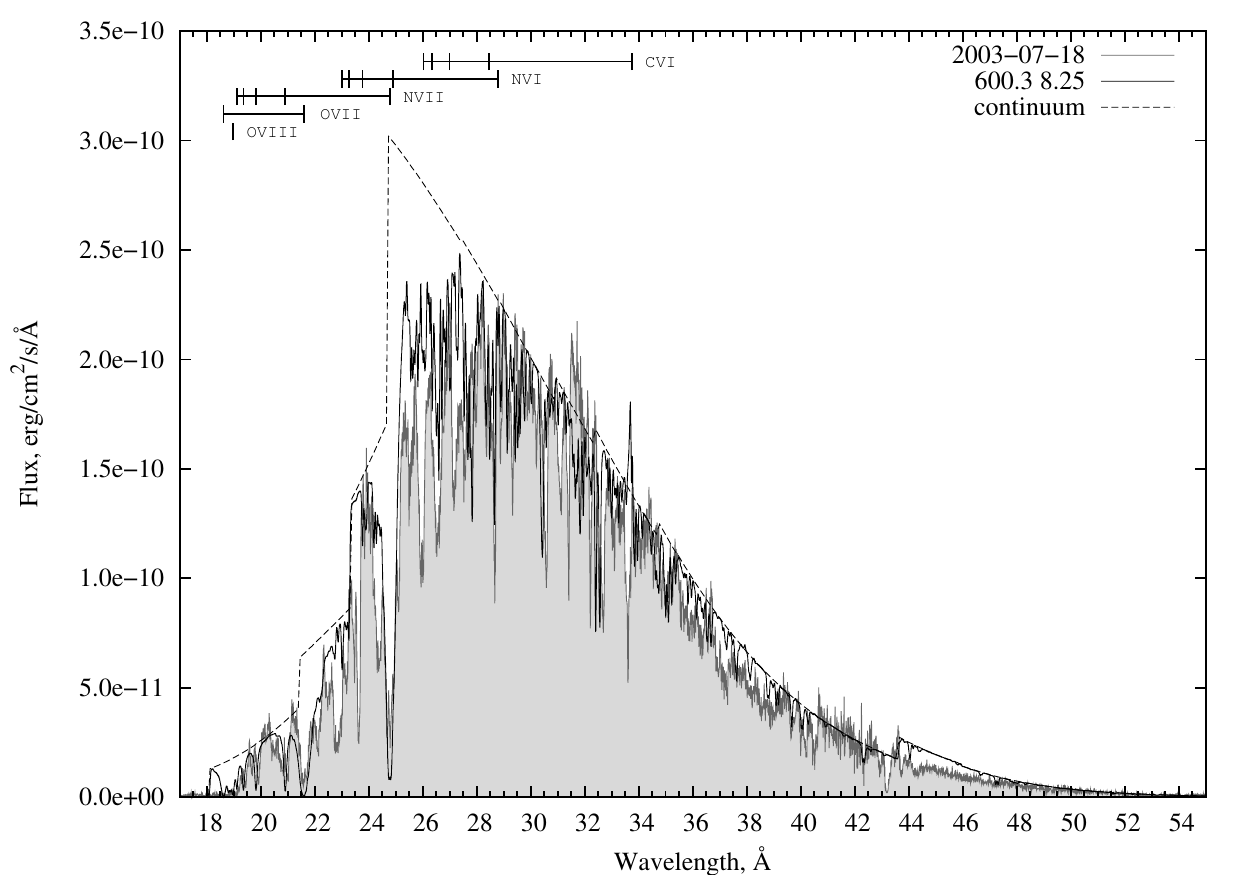}
\parbox{130mm}{\caption{Model fit for V4743 Sgr. July 18,
    2003.}\label{fig:3776}} 
\end{singlespace}
\end{center}
\end{sidewaysfigure}

\begin{sidewaysfigure}[p]
\begin{center}
\begin{singlespace}
\includegraphics[width=20cm]{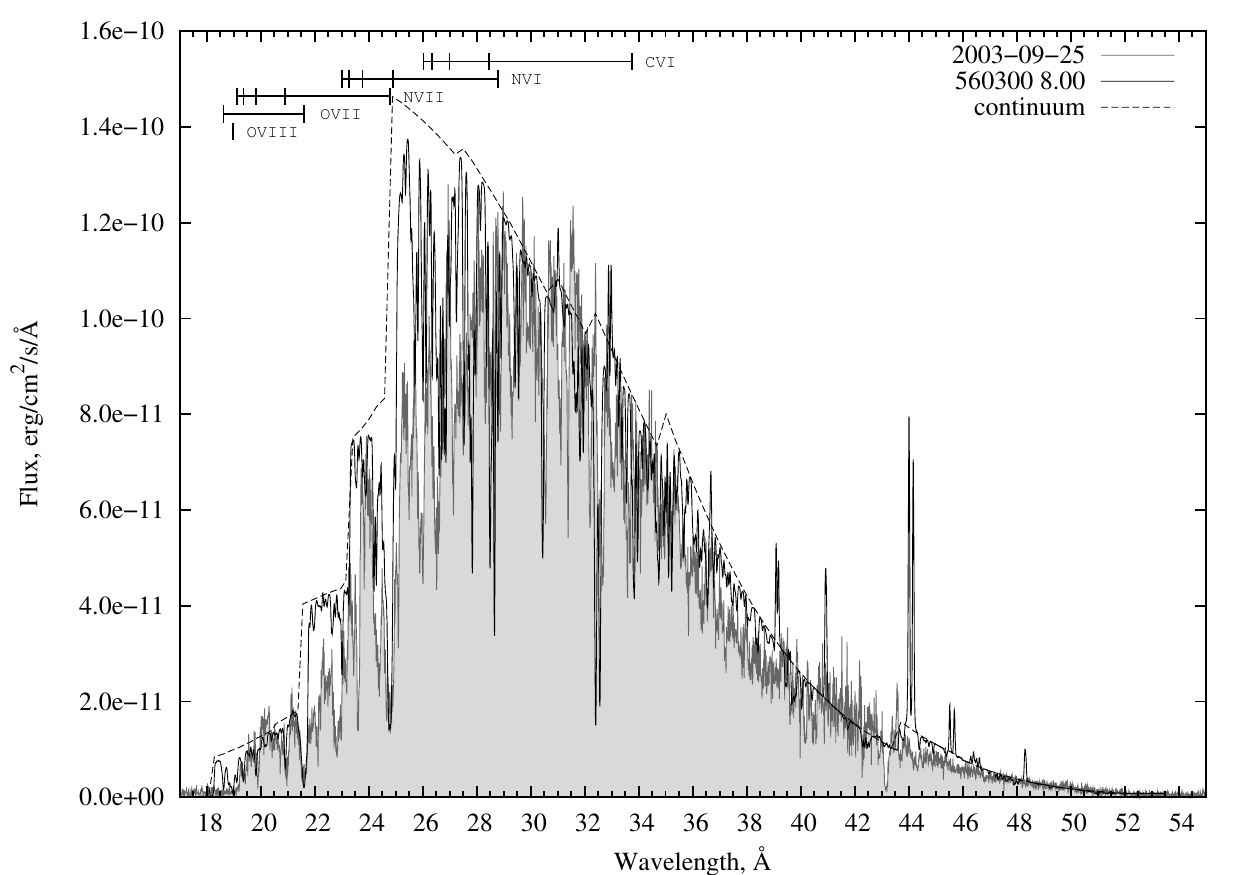}
\parbox{130mm}{\caption{Model fit for V4743 Sgr. September 25,
    2003}\label{fig:4435}} 
\end{singlespace}
\end{center}
\end{sidewaysfigure}

\begin{sidewaysfigure}[p]
\begin{center}
\begin{singlespace}
\includegraphics[width=20cm]{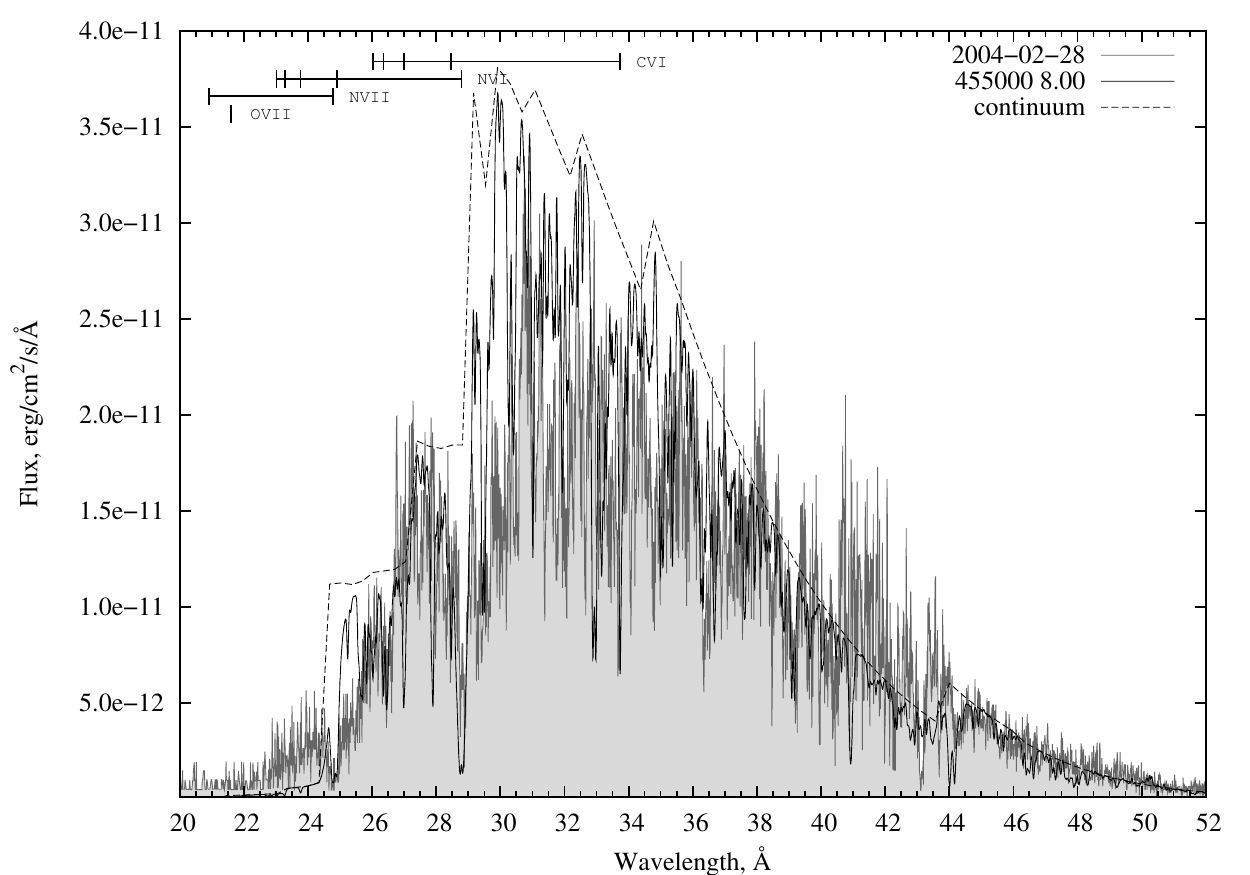}
\parbox{130mm}{\caption{Model fit for V4743 Sgr. February 28,
    2004}\label{fig:5292}} 
\end{singlespace}
\end{center}
\end{sidewaysfigure}


\begin{table}[htbp]
\begin{center}
\parbox{130mm}{\caption{Model parameters for V4743 Sgr.}\label{tab:v4743res}}
\vspace{2mm}
\begin{singlespace}
\begin{tabular}{c@{/}l|r|r|r|r}
\hline
\multicolumn{2}{l|}{Parameter}                        & 3775  & 3776  & 4435   & 5292 \\
\hline
\multicolumn{2}{l|}{T$_{\rm eff}$ $[\times$ 1000 K$]$} & $580.4$ & $600.0$ & $560.3$ & $455.0$ \\
\multicolumn{2}{l|}{$\log g$ $[$cm/s$^2]$}            & $8.15$  &  8.20   & 8.00   & 8.00 \\
\multicolumn{2}{l|}{${\rm N}_{\rm H}$ $[\times10^{21}$ cm$^{-2}]$}  & $0.46$  & $0.51$  &
$0.51$ & $0.51$ \\
\multicolumn{2}{l|}{${\rm H}_\nu$ $[10^{17}$ erg/cm$^2$/s/Hz$]$}& $5.122$     & $5.861$     & $4.448$
& $1.935$\\
\multicolumn{2}{l|}{$d^2/R^2$ [$\times 10^{26}$]}      & $1.5$     & $1.2$    & $1.2$ & $1.2$\\
\hline
He&He$_\odot$ &  7.5     &  5.0    & 3.0  & 4.0 \\
C&C$_\odot$   &  5.0     &  25.0   & 3.0  & 2.0  \\
N&N$_\odot$   &  140.0   & 70.0    & 26.0 & 60.0 \\
O&O$_\odot$   &  12.0    & 16.0    & 2.5  & 14.0 \\
\hline
Ne&Ne$_\odot$ & 200.0    & 250     & 50.0 & 10.0 \\
Mg&Mg$_\odot$ & 1.0      &  30     & 50.0 & 9.0  \\
Al&Al$_\odot$ & 100.0    &  30     & 100.0& 100.0\\
Si&Si$_\odot$ & 0.2      &  0.2    & 2.0  & 5.0  \\
\hline
S&S$_\odot$   & 1.0      & 1.0     & 1.0  & 1.0  \\
Ar&Ar$_\odot$ & 2.0      & 0.0     & 2.0  & 0.06 \\
Ca&Ca$_\odot$ & 0.05     & 0.0     & 0.05 & 0.01 \\
Fe&Fe$_\odot$ & 0.1      & 0.0     & 0.1  & 0.1  \\
\hline
\multicolumn{2}{l|}{$\Delta\lambda/\lambda$ N VI}& $7.6\times10^{-3}$
& $0.0$          & $0.0$     &  $0.0$    \\
\multicolumn{2}{l|}{$\Delta\lambda/\lambda$ N VII}& $6.1\times10^{-3}$       &
$4\times10^{-3}$  & $3.2\times10^{-3}$ & $4\times10^{-3}$  \\
\multicolumn{2}{l|}{$\Delta\lambda/\lambda$ O VII}& $6.9\times10^{-3}$      &
$3.2\times10^{-3}$  &  $3.2\times10^{-3}$ & --     \\
\hline
R&R$_\odot$   & $0.014$  & $0.016$ & $0.016$ & $0.016$\\
L&L$_\odot$   & $20300$  & $29000$ & $22100$ & $9600$\\
M&M$_\odot$   & $1.03$   & [$1.44$]& $0.91$  & $0.91$\\
\hline

\end{tabular}
\end{singlespace}
\end{center}
\end{table}

\begin{figure}[!h]
\begin{center}
\begin{singlespace}
\includegraphics[width=13cm]{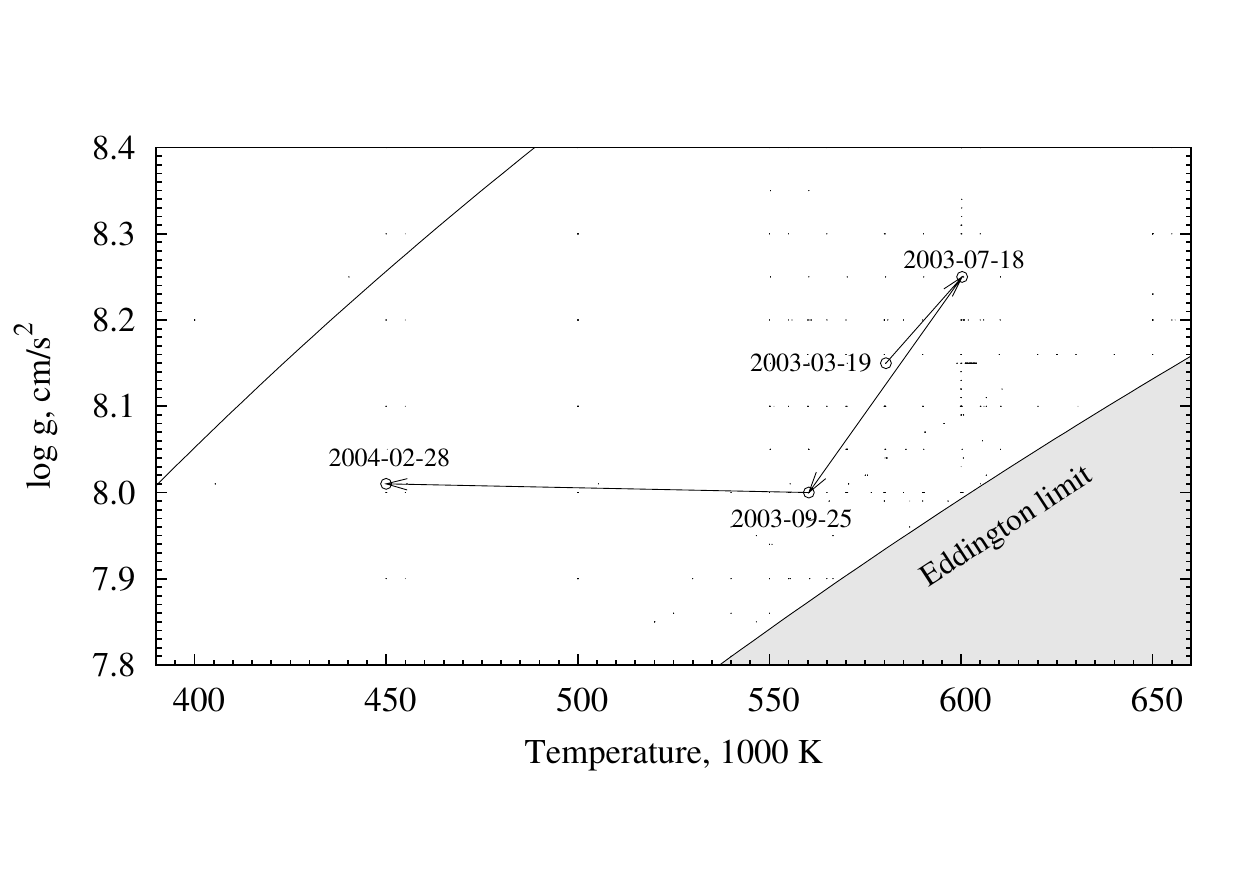}
\parbox{130mm}{\caption[Kiel-diagram for V4743 Sgr.]{Spectral evolution of
    V4743 Sgr between March 2003 and February 2004.}\label{fig:v4743ongrid}} 
\end{singlespace}
\end{center}
\end{figure}

Model fits for the 
Chandra data are shown in Figures \ref{fig:3775} -- \ref{fig:5292}. 
The parameters of model atmospheres 
are shown in Table \ref{tab:v4743res} for the 4 Chandra observations
of V4743 Sgr. The luminosity of the nova was close to the Eddington limit at
the beginning and decreased significantly to the end of the supersoft
phase. Figure \ref{fig:v4743ongrid} shows its evolution in the $\log
g$ -- $T_{\rm eff}$ diagram. The models of the first three observations fall on
a line near the Eddington luminosity and this 
indicates evolution at nearly constant
luminosity. The spectral evolution is primarily due to the change of effective
temperature, abundance evolution has lower order effects. 
The February 2004 spectrum shows a significant decrease in energy
production. Assuming that the mass of the nova was constant the change of 
surface gravity indicate only a few percent variation of the radius.  

The stellar radius derived from model fits is $R=0.015$ $R_\odot$, or
10500\ km. Adopting a WD mass of 1 $M_\odot$, which has a core radius of about
6000 km, the extension of the atmosphere is around 75\%. This result is 
somewhat lower, but it is still in agreement with earlier studies. 


\subsection{V2491 Cyg}

The classical nova V2491 Cyg was discovered on April 10, 2008 at visual
magnitude V$=7.7$,
\citep{nakano08}. The outburst occurred only 
six months after installation of the Ortega telescope 
at Florida Institute of Technology. Optical spectra were taken near optical
maximum, these data are shown in Figure \ref{fig:specevol2}. 
The nova reached its optical maximum at
V$=7.54^m$ on Apr 11th. Using the reddening $E(B-V)=0.43$ \citep{rudy08}
the distance to the nova was estimated by \citet{helton08} to be $d=10.5$ kpc.
V2491 Cyg was
a very fast nova with t$_2$=5.65 days. The broad spectral lines of He and N
shown in Figure \ref{fig:specevol2} indicated
extreme expansion velocities. The H$_\alpha$ FWHM was 161.2~\AA, implying 
an expansion velocity over 3600 km/s on May 17th (Nemeth et al., in prep.).

\begin{table}[!h]\begin{center}\begin{singlespace}
\caption{X-ray observations of V2491 Cyg.\label{tab:v2491}}
\begin{tabular}{c|c|c|c|l}
\multicolumn{5}{c}{}\\
\hline
Observation ID & Date & Exposure (s)&Instrument & Comment \\
\hline
0552270501 & 2008/05/20 &39339&XMM RGS& X-on\\
0552270601 & 2008/05/30 &29980&XMM RGS& X-on\\
\hline
\end{tabular}
\end{singlespace}\end{center}\end{table}

\begin{figure}[!h]\begin{center}\begin{singlespace}
\includegraphics[width=14.2cm]{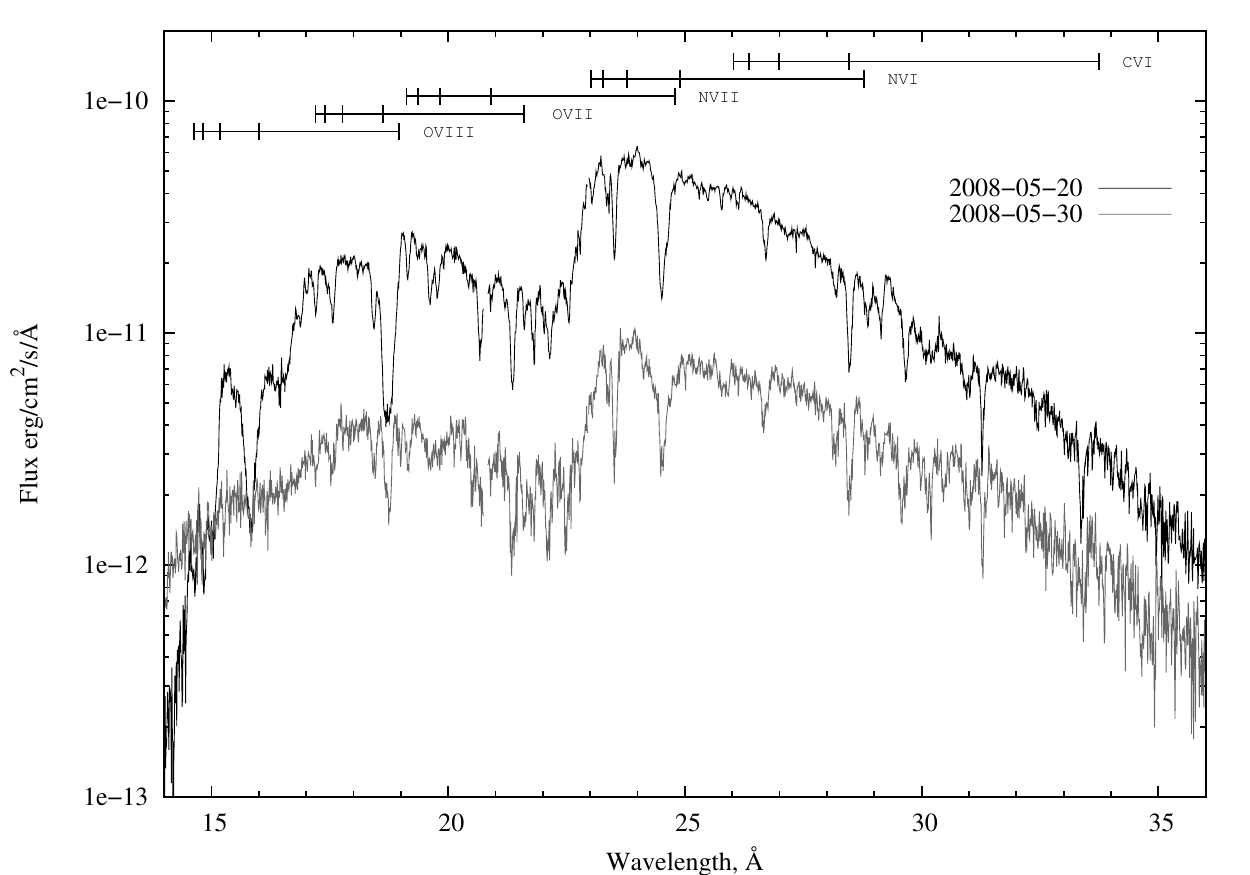}
\parbox{130mm}{\caption[Spectral evolution of V2491 Cyg.]{Spectral evolution of V2491 Cyg in May 2008. Both 
    spectra
    were taken with XMM/RGS spectrometers.}\label{fig:v2491}}
\end{singlespace}\end{center}\end{figure}

\begin{sidewaysfigure}[p]\begin{center}\begin{singlespace}
\includegraphics[width=20cm]{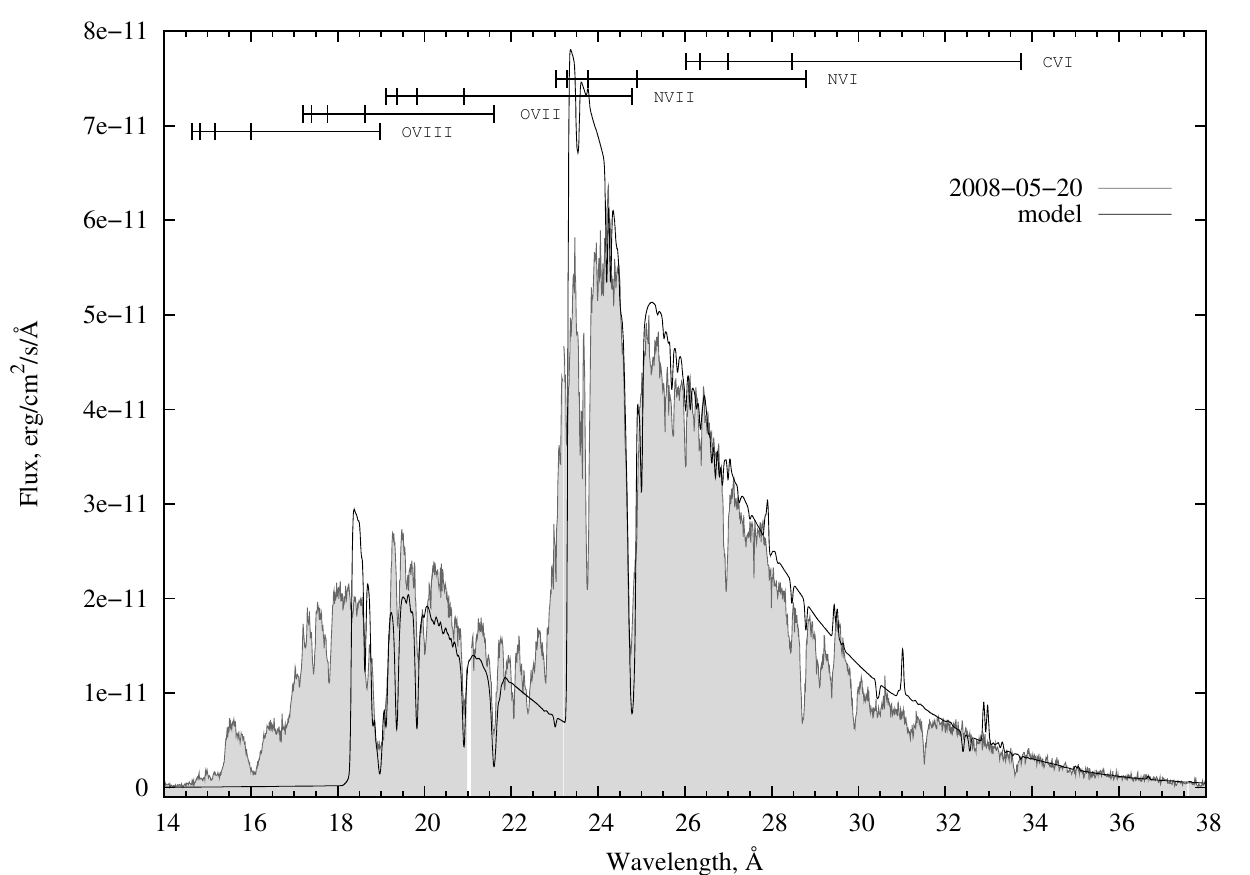}
\parbox{130mm}{\caption[Model fit for the May 20th spectrum.]{Best fit model for the May 20th spectrum. The strong
    discontinuity at 23 \AA\ is due to interstellar absorption, the other at
    18 \AA\ is due to static, plane-parallel modeling.}\label{fig:v2491fit1}}
\end{singlespace}\end{center}\end{sidewaysfigure}

\begin{sidewaysfigure}[p]\begin{center}\begin{singlespace}
\includegraphics[width=20cm]{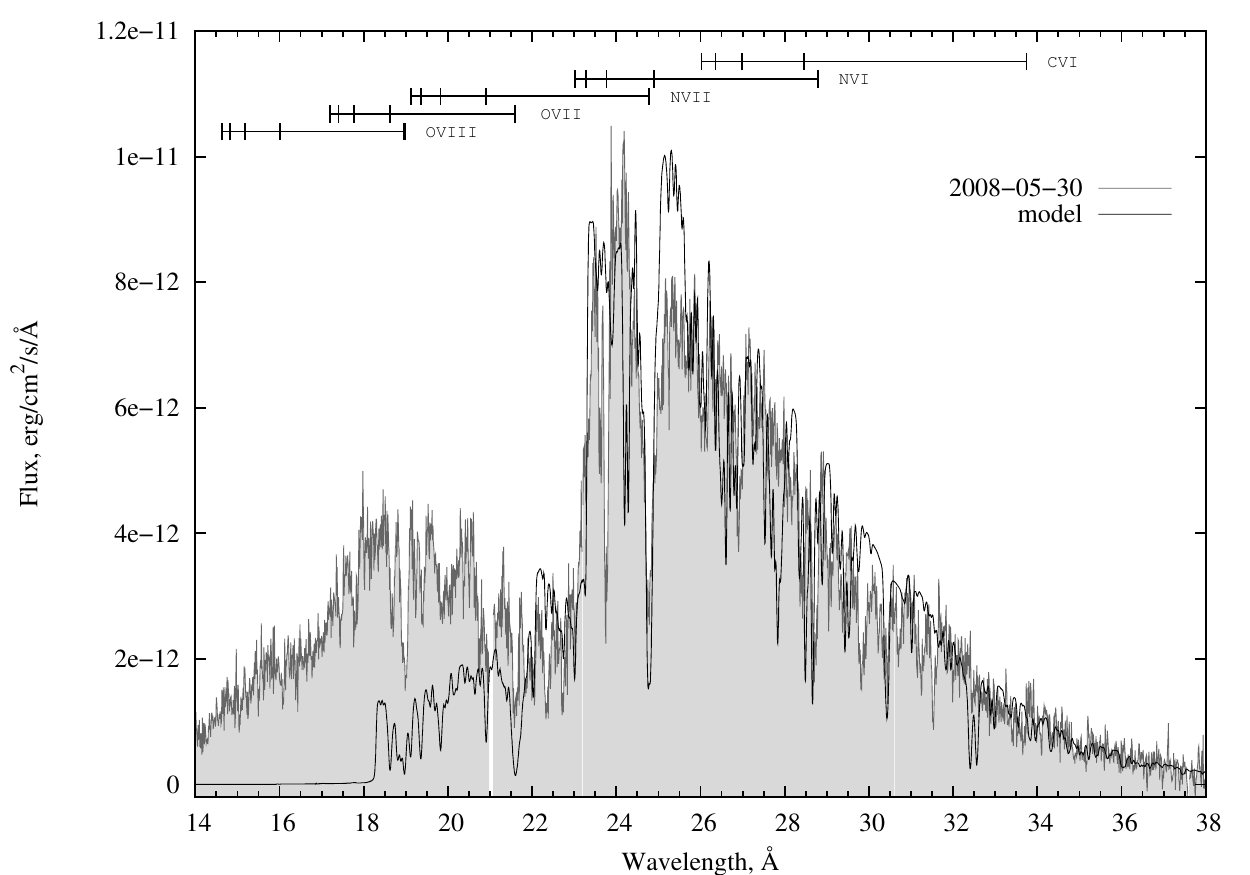}
\parbox{130mm}{\caption[Model fit for the May 30th spectrum.]{Best fit model for the May 30th spectrum.}\label{fig:v2491fit2}}
\end{singlespace}\end{center}\end{sidewaysfigure}

\begin{sidewaysfigure}[p]\begin{center}\begin{singlespace}
\includegraphics[width=20cm]{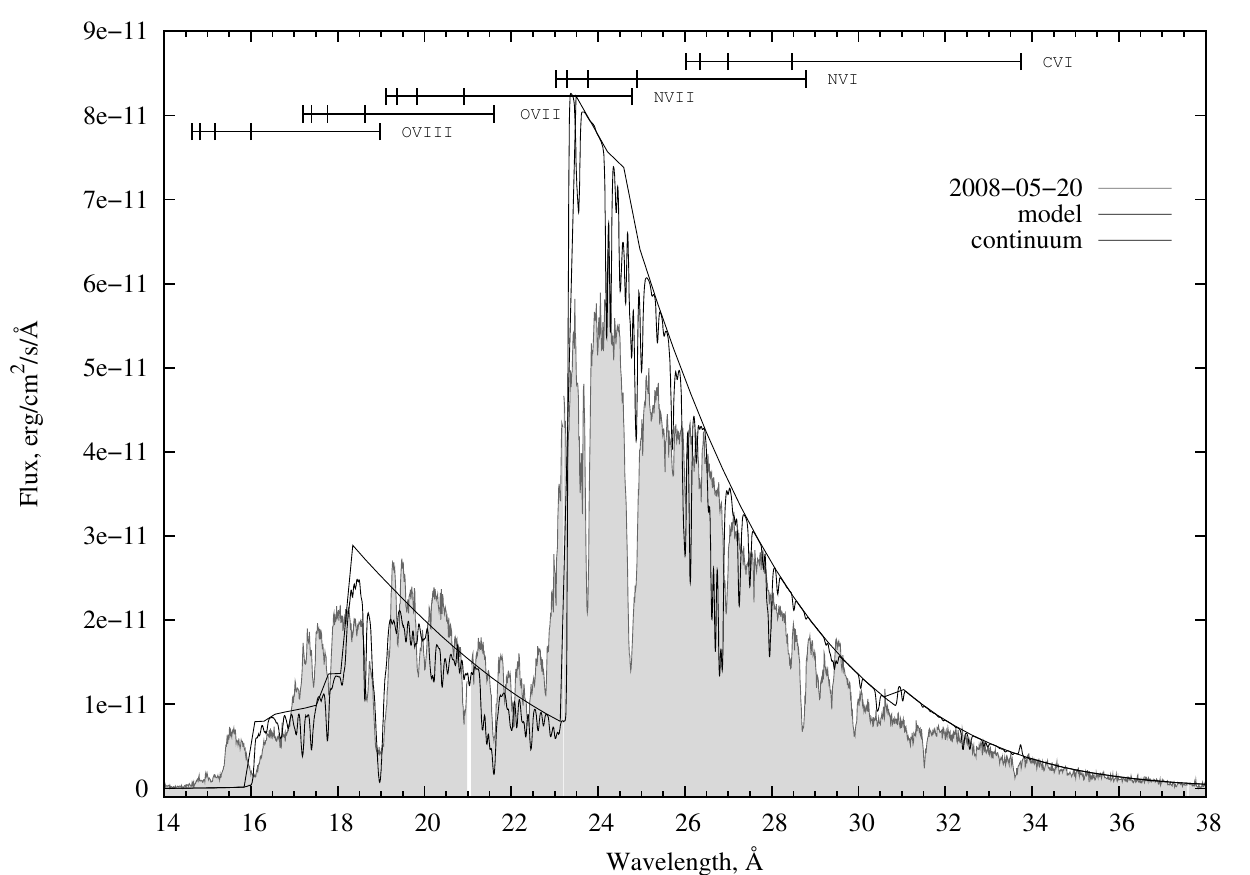}
\parbox{130mm}{\caption[Model fit with solar abundances.]{Model fit with solar
    abundances for the May 30th spectrum.}\label{fig:v2491fit3}}
\end{singlespace}\end{center}\end{sidewaysfigure}

V2491 Cyg was observed with XMM two times in May 2008 (Table \ref{tab:v2491}), 
and already has shown strong supersoft radiation. The spectra (Figure
\ref{fig:v2491}) indicated a hot continuum
with strong nitrogen and oxygen
absorption lines and very strong interstellar absorption. 
The first spectrum showed a peak flux at 
$6\times10^{-11}$ erg/cm$^2$/s/\AA, or 0.08 c/sec. The blue-shifted spectral
features suggest an expanding atmosphere as to be expected. 
The measured expansion velocity of O
VII absorption lines was v$_{\rm exp}=3330\pm140$ km/s and for O VIII
v$_{\rm exp}=3810\pm300$ km/s. The expansion velocity based on N VII lines was
v$_{\rm exp}=3280\pm240$ km/s and v$_{\rm exp}=3100\pm300$ km/s for N VI lines. 
Individual spectral features of heavier elements are very hard to distinguish.
Some blends suggest much lower velocities for these elements.  
The strong N VI and NVII lines as well as the differences in 
Doppler-shifts 
suggest an
extended line-forming region with different dominant ions and wind velocities. 
A similar trend was
found for oxygen lines in accordance with the results of \citet{ness10}. 
The line shifts indicate a strong wind and super Eddington luminosity.
Such atmospheres cannot be reliably modeled with TLUSTY 200. Figures
\ref{fig:v2491fit1} and \ref{fig:v2491fit2} show the best-fit models using the
parameters in Table \ref{tab:v2491res}. For comparison, 
a model with $T_{\rm eff}=600,100$ K, $\log g=8.1$ cm/s$^2$
and solar abundances is shown in Figure
\ref{fig:v2491fit3}. With solar composition there 
is more flux below $18$~\AA\ and the oxygen lines
show similar strength like the data. The computed 
nitrogen lines are much weaker than
observed and by increasing its abundance the high energy flux decreases
rapidly, giving rise to strong ionization edges. 
However, a closer look reveals that the model is not terribly off
compared to either the static WD or extended dynamical model presented in
\citet{rossum10}. The hydrogen column density was found to be
$N_{\rm H}=(2.5\pm0.4)\times10^{21}$ cm$^{-2}$ in the direction of V2491
Cyg. This is in good agreement with 
the column density $N_{\rm H}=2.8\times10^{21}$
cm$^{-2}$ used by \citet{rossum10}. However in this work interstellar
absoprtion was calculated with Galactic abundances, while they used solar
composition.  

\begin{table}[!h]\begin{center}
\parbox{130mm}{\caption{Model parameters for V2491 Cyg.}\label{tab:v2491res}}
\vspace{2mm}
\begin{singlespace}
\begin{tabular}{c@{/}l|r|r|r}
\hline
\multicolumn{2}{l|}{Parameter}
&0552270501&0552270601 & 0552270501\\
\hline
\multicolumn{2}{l|}{T$_{\rm eff}$ $[\times$ 1000 K$]$}
&$650.2$&$560.0$&$600.1$\\
\multicolumn{2}{l|}{$\log g$ $[$cm/s$^2]$}               &$8.2$&$8.2$&$8.1$\\
\multicolumn{2}{l|}{${\rm N}_{\rm H}$ $[\times10^{21}$
  cm$^{-2}]$}&$2.5$&$2.5$&$2.5$\\
\multicolumn{2}{l|}{${\rm H_{\nu}}$ $[10^{17}$ erg/cm$^2$/s/Hz$]$}&$8.064$&$4.444$&$5.852$ \\
\multicolumn{2}{l|}{$d^2/R^2$ [$\times 10^{26}$]}          &$2.5$&$7.5$&$2.5$\\

\hline
He&He$_\odot$ & 4.0     &  10.0   &1.0\\
C&C$_\odot$   & 1.0     &  2.0    &1.0\\
N&N$_\odot$   & 100.0   &  1000.0 &1.0\\
O&O$_\odot$   & 2.0     &  80.0   &1.0\\
\hline
Ne&Ne$_\odot$ & 50.0    &  50.0   &1.0\\
Mg&Mg$_\odot$ & 50.0    &  50.0   &1.0\\
Al&Al$_\odot$ & 100.0   &  1000.0 &1.0\\
Si&Si$_\odot$ & 10.0    &  2.0    &1.0\\
\hline
S&S$_\odot$   & 0.1     &  2.0    &1.0\\
Ar&Ar$_\odot$ & 0.1     &  10.0   &1.0\\
Ca&Ca$_\odot$ & 0.01    &  0.05   &1.0\\
Fe&Fe$_\odot$ & 0.1     &  0.1    &0.0\\
\hline
\multicolumn{2}{l|}{$\Delta\lambda/\lambda$ C VI}  & $1.05\times10^{-2}$ &  --     &        \\
\multicolumn{2}{l|}{$\Delta\lambda/\lambda$ N VI}  & $1.04\times10^{-2}$ &  --     &        \\
\multicolumn{2}{l|}{$\Delta\lambda/\lambda$ N VII} & $1.01\times10^{-2}$ &  $9.68\times10^{-3}$     &        \\
\multicolumn{2}{l|}{$\Delta\lambda/\lambda$ O VII} & $1.07\times10^{-2}$ &  $1.07\times10^{-2}$     &        \\
\multicolumn{2}{l|}{$\Delta\lambda/\lambda$ O VIII}& $1.26\times10^{-2}$ &  $1.05\times10^{-2}$     &        \\
\hline
\end{tabular}
\end{singlespace}\end{center}\end{table}

\chapter{Conclusions}\label{conclusions}

\begin{figure}[!h]
\begin{center}
\begin{singlespace}
\includegraphics[width=12cm]{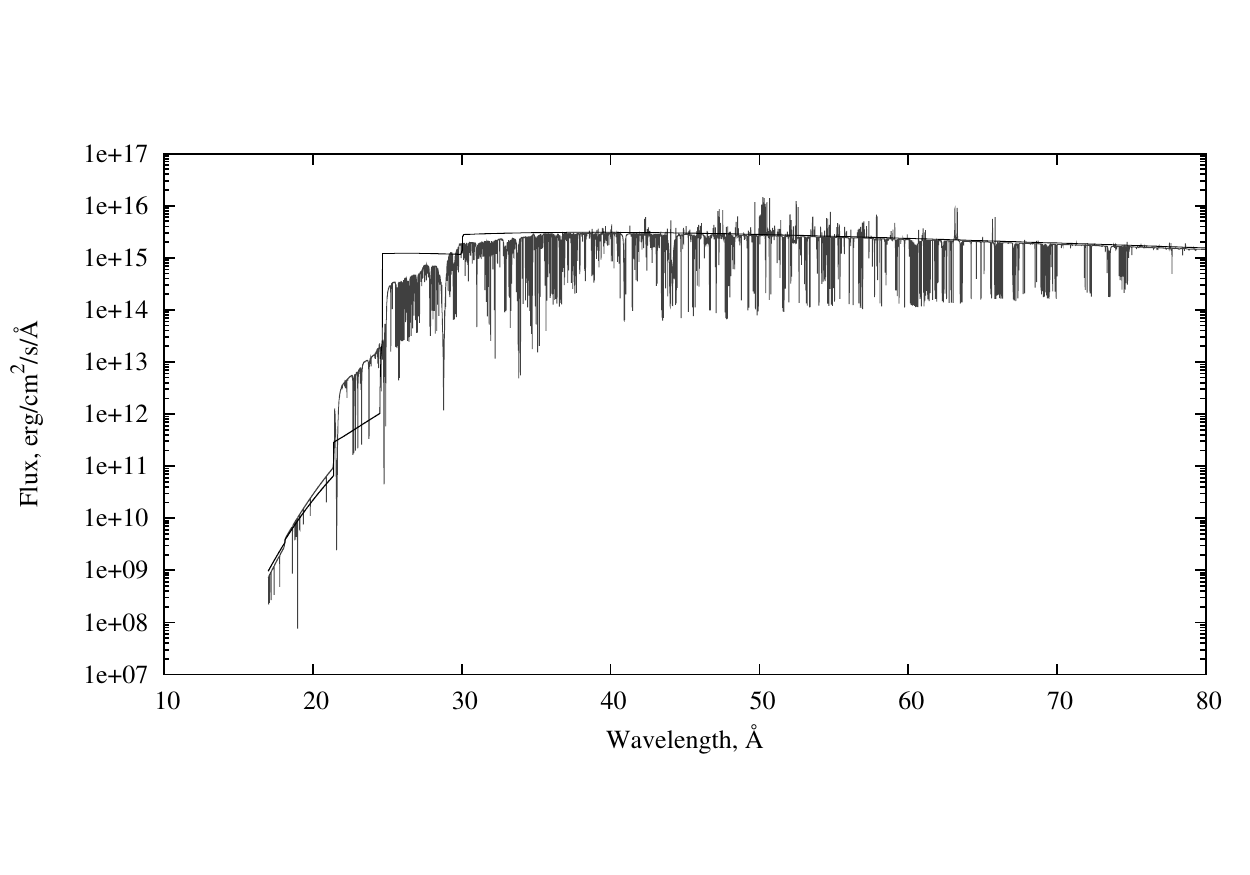}
\parbox{130mm}{\caption[Spectral lines added to the supersoft range.]{Comparison of a non-LTE CNO continuum spectrum (black
    line) and a full non-LTE spectrum (grey line) 
    including H, He, C, N, O, Ne, Mg, Al and
    Si with detailed model atoms. Energy levels of\ S, Ar and Ca were averaged
    into 9--10 superlevels, Fe was included only with its ground states.}\label{fig:line}}
\end{singlespace}
\end{center}
\end{figure}

Starting out from simple continuum models of light metals including 
only their ground states (black line in Figure \ref{fig:line}) 
I built multi-level model atoms from
NIST/ASD and TOPbase atomic data. I collected atomic data on
radiative transition strengths and natural broadening for about 
70,000 spectral lines
in the supersoft range from low ionization degrees of C, N, O, Ne, Mg, Al, Si,
S, Ar, Ca and Fe, up to hydrogenic ions. Altogether 21,148 energy levels
and 542,914 transitions were included. On bound-free cross sections,
data were extracted from TOPbase for all multiplets. 
With these opacity sources it is now possible
to calculate full non-LTE line blanketed spectra in the supersoft range
(the grey line shows in Figure \ref{fig:line}) with
TLUSTY and SYNSPEC. SYNSPEC also required slight
modifications to increase the highest ionization degrees to match the new
input files.

With an extensive grid I showed the effects of surface
temperature, 
gravity, abundance and the influence of model atoms on the final
spectrum. These relationships suggest that static plane-parallel models 
over-estimate the effective temperature and surface gravity for
classical novae.

My work also 
required numerous little programs to effectively manipulate TLUSTY and
SYNSPEC input and output files for spectral modeling. For example, 
model atoms have complex
structures and any change in the level structure must be propagated through
the entire model atom. By deleting one level, all higher level indices will
change. Bound-free transition of the given level must be removed and all
transitions originating from or ending on the given level must be removed.  
Such tasks can be done easily by hand 
for a few levels, but certainly not for hundreds of levels. Level
averaging requires similar automated methods as well. TLUSTY stores the
atmospheric structure and level populations in its own format. These data
blocks must be processed for graphing or to calculate ionization balance and
conditions in the atmosphere. Short Python scripts create Grotrian diagrams
from model atoms, convergence logs during model calculation, carrying out
Gaussian convolution of the spectra or calculating interstellar absorption.

With TOPAtom one can
build arbitrarily complex model atoms directly from the 
original TOPbase or NIST data files. Any other atomic data sets can be easily
incorporated by reformatting them to the original 
format of one of these primary atomic data tables. Model atoms with
superlevels and superlines (OS mode) are also available.

As part of this work a task-oriented pipeline (TGRID) was developed.
TGRID can do spectral analysis from grid calculations up to abundance
analysis. However, inaccuracies in the input atomic data, coupled with strong
blanketing and the enormous
interstellar reddening, made quantitative fits impossible. 
Steps of the spectral analysis are
in modular format, which makes them scriptable and 
they can also be used independently in other applications. 

Application of these techniques 
to some classical novae have shown the validity of the model atoms. 
I showed in the example of V4743 Sgr and V2491 Cyg 
how the new model atoms fit
observed data of Chandra and XMM-Newton satellites. V4743 Sgr was very close to
the Eddington limit in March -- September 2003. 
The blue-shifted nitrogen and oxygen 
spectral lines confirm
this. By February 2004 the luminosity decreased significantly and spectral
lines appeared at their laboratory wavelengths, which suggests that hydrogen
burning was turned off, the stellar wind stopped and the WD returned to a 
static state. Abundance analysis showed enhancement of helium, nitrogen and
oxygen compared to solar values. The abundance of nitrogen is about 10--20
times greater than was found by \citet{petz05}. Oxygen is over-abundant by a
factor of a few, while carbon is around solar values. Enhancement of neon,
magnesium and aluminum of about 50 times the solar values were found. A 
slight over-abundance of silicon and sulfur is possible (up to twice
the solar value). Argon and calcium are depleted compared to solar
abundance. However, the 
abundance analysis is still uncertain and these figures should be considered
only as approximations.   
These tests 
also showed the need for a comprehensive upgrade of TLUSTY/SYNSPEC to
model CNe. 
The case of V2491 Cyg was a good example showing the limitation of static,
plane-parallel modeling of luminous supersoft sources. The blue-shifted lines
and the extended blue continuum indicate a strong wind. 
The simultaneous strong lines of NVI and NVII suggest 
an extended line-forming region 
and these features together cannot be modeled by
static atmospheres.   
These objects need hydrodynamic calculations in spherical
geometry, treating both the extended line formation region and stellar
wind explicitly. This was beyond the scope of the current project. 
However, the technique used here work well 
for supersoft sources below the Eddington limit.
Finally, for qualitative analysis on abundance and
evolution of classical novae, many more spectroscopic observations are 
required. 


\section{Future directions}

My work pointed out the need of Opacity Sampling for all metals heavier than
Si. Their intermediate and high energy levels provide a lot of important
lines in the X-ray region and contribute to bound-free transitions, 
which affects both the atmospheric structure as well
as the final
spectra. However, the increased computational cost introduced by many
new frequency points (about 250,000 total) exceeds the capabilities of a single
computer. For detailed analysis, a large number of models must be generated,
and so it is preferable to use a computer grid.

Line broadening data can be updated with data for Stark broadening from
STARK-B\footnote{\protect\url{http://stark-b.obspm.fr/index.php}} maintained by
M.S. Dimitrijevi\'{c} and S. Sahal-Br\'{e}chot. This extension 
requires further updates in SYNSPEC, which currently accepts broadening
parameters as given in \citet{griem74} for temperatures ranging 
up to 50,000 K.  

The example of V2491 Cyg emphasized the importance of spherical
geometry. Novae in their supersoft stage 
are luminous, usually well over the Eddington
limit. Their modeling must be done in spherical geometry, and line formation 
considered in an extended and expanding non-LTE atmosphere. The most 
important update would be to reformulate TLUSTY for spherical geometry and
incorporate stellar wind and hydrodynamics with a $\beta$-type velocity profile:
\begin{equation}
v = v_{\infty}\left( 1-\frac{R_0}{R}  \right)^\beta
\label{eq:beta}
\end{equation}
where, $v_\infty$ is the terminal velocity of the stellar wind, $R$ is the
radial distance from the stellar surface and $R_0$ is the radius of the
star. The empirical exponent $\beta$ defines the 
velocity profile in the wind. 

It is also important to update and incorporate new and more reliable atomic 
data in the model atoms and the line list as they become available. For higher
energy levels a further extension can be done using calculated data from 
The Atomic Line
List\footnote{\protect\url{http://www.pa.uky.edu/~peter/newpage/}} 
V2.05B12 by van Hoof. This work could be done with TOPAtom.  
Helium-like and lithium-like ions of metals are 
important in modeling and their energy levels and transitions are not
calculated in TOPbase over azimuthal quantum number $l>4$. These missing 
levels and
transitions can be calculated with Cowan's Atomic Structure 
Code\footnote{\protect\url{http://tcd.ie/Physics/People/Cormac.McGuinness/
Cowan/}}. It is also important to update collision
strengths for bound-free and bound-bound transitions. 
Although data are limited to the lowest levels up to $N\le5$, 
Chianti \citep{dere09} 
provides spline fits for electron collision strengths for a
wide range of temperatures and species.

At the current level of modeling and atomic databases 
all these updates are considered 
secondary compared to geometrical and dynamical considerations.

\bibliographystyle{elsart-harv} 
\bibliography{mybib}

\begin{thebibliography}{66}
\expandafter\ifx\csname natexlab\endcsname\relax\def\natexlab#1{#1}\fi
\expandafter\ifx\csname url\endcsname\relax
  \def\url#1{\texttt{#1}}\fi
\expandafter\ifx\csname urlprefix\endcsname\relax\def\urlprefix{URL }\fi

\bibitem[{Andre\"{a} et~al.(1994)Andre\"{a}, Drechsel, and Starrfield}]{ads94}
Andre\"{a}, J., Drechsel, H., Starrfield, S., 1994. Astronomy and Astrophysics
  291, 869.

\bibitem[{Ba{\l}uci\'nska-Church and McCammon(1992)}]{balucinska92}
Ba{\l}uci\'nska-Church, M., McCammon, D., 1992. Astrophysical Journal 400, 699.

\bibitem[{Bautista et~al.(1998)Bautista, Romano, and Pradhan}]{bautista98}
Bautista, M.~A., Romano, P., Pradhan, A.~K., 1998. Astrophysical Journal
  Supplement Series 118, 259.

\bibitem[{Binney and Merrifield(1998)}]{binney98}
Binney, J., Merrifield, M., 1998. Galactic Astronomy. Princeton University
  Press.

\bibitem[{Burke and Seaton(1971)}]{burke71}
Burke, P.~G., Seaton, M.~J., 1971. Methods in Computational Physics 10, 1.

\bibitem[{Cassatella et~al.(2004)Cassatella, Lamers, and Rossi}]{cassatella04}
Cassatella, A., Lamers, H., Rossi, C., 2004. Astronomy and Astrophysics 420,
  571.

\bibitem[{Chandrasekhar(1935)}]{chandrasekhar35}
Chandrasekhar, S., 1935. Monthly Notices of the Royal Astronomical Society 95,
  207.

\bibitem[{Clayton(1968)}]{clayton68}
Clayton, D., 1968. Principles of Stellar Evolution and Nucleosynthesis.
  McGraw-Hill Book Company.

\bibitem[{Cunto et~al.(1993)Cunto, Mendoza, Ochsenbein, and Zeippen}]{cunto93}
Cunto, W., Mendoza, C., Ochsenbein, F., Zeippen, C., 1993. Astronomy and
  Astrophysics 275, L5.

\bibitem[{Dahlem(1999)}]{xmmuhb}
Dahlem, M., January 1999. XMM User's Handbook. 1st Edition.

\bibitem[{{Della Valle} and Livio(1995)}]{dellavalle95}
{Della Valle}, M., Livio, M., 1995. Astrophysical Journal 452, 704.

\bibitem[{Dere et~al.(1997)Dere, Landi, Mason, Monsignori~Fossi, and
  Yound}]{dere97}
Dere, K.~P., Landi, E., Mason, H.~E., Monsignori~Fossi, B.~C., Yound, P.~R.,
  1997. Astronomy and Astrophysics Supplement Series 125, 149.

\bibitem[{Dere et~al.(2009)Dere, Landi, Young, Del~Zanna, Landini, and
  Mason}]{dere09}
Dere, K.~P., Landi, E., Young, P.~R., Del~Zanna, G., Landini, M., Mason, H.~E.,
  2009. Astronomy and Astrophysics 498, 915.

\bibitem[{Gehrz et~al.(1998)Gehrz, Truran, Williams, and Starrfield}]{gehrz98}
Gehrz, R., Truran, J.~W., Williams, R.~E., Starrfield, S., 1998. Publications
  of the Astronomical Society of the Pacific 110, 3.

\bibitem[{Gilmozzi and {Della Valle}(2003)}]{stellarcandles}
Gilmozzi, R., {Della Valle}, M., 2003. Lecture Notes in Physics: Stellar
  Candles for the Extragalactic Distance Scale. Springer, Ch.~12.

\bibitem[{Greiner(2000)}]{greiner00}
Greiner, J., 2000. New Astronomy 5, 137.

\bibitem[{Grevesse and Sauval(1998)}]{grevesse98}
Grevesse, N., Sauval, A.~J., 1998. Space Science Reviews 85, 161.

\bibitem[{Griem(1974)}]{griem74}
Griem, H., 1974. Spectral Line Broadening by Plasmas. Academic Press, New York.

\bibitem[{Hartmann and Heise(1997)}]{hartmann97}
Hartmann, H.~W., Heise, J., 1997. Astronomy and Astrophysics 322, 591.

\bibitem[{Haseda et~al.(2002)Haseda, West, Yamaoka, and Masi}]{haseda02}
Haseda, K., West, D., Yamaoka, H., Masi, G., 2002. International Astronomical
  Union Circulars 7975.

\bibitem[{Helton and Woodward(2008)}]{helton08}
Helton, L.~A., Woodward, C.~E., 2008. The Astronomer's Telegram 1379.

\bibitem[{Hubeny(1988)}]{hubeny88}
Hubeny, I., 1988. Computer Physics Comm. 52, 103.

\bibitem[{Hubeny and Lanz(1992)}]{hubeny92}
Hubeny, I., Lanz, T., 1992. Astronomy and Astrophysics 262, 501.

\bibitem[{Hubeny and Lanz(1995)}]{hl95}
Hubeny, I., Lanz, T., 1995. Astrophysical Journal 439, 875.

\bibitem[{Hubeny and Lanz(2000)}]{synspec43}
Hubeny, I., Lanz, T., April 2000. SYNSPEC - A User's Guide, Version 43.

\bibitem[{Hubeny and Lanz(2003)}]{tlusty200}
Hubeny, I., Lanz, T., February 2003. TLUSTY - A User's Guide, Version 200.

\bibitem[{Kantorovich(1949)}]{kantorovich49}
Kantorovich, L.~V., 1949. Trudy Mat. Inst. Steklov. , 104.

\bibitem[{Kato(2010)}]{kato10}
Kato, M., 2010. Astronomishce Nachrichten 331, 140.

\bibitem[{Kohn(1996)}]{kohn96}
Kohn, W., 1996. Physical Review Letters 76, 3168.

\bibitem[{Kolb and Politano(1997)}]{kolb97}
Kolb, U., Politano, M., 1997. Astronomy and Astrophysics 319, 909.

\bibitem[{Kunasz and Olson(1988)}]{kunasz88}
Kunasz, P.~B., Olson, G.~L., 1988 39, 1.

\bibitem[{Lada(2006)}]{lada06}
Lada, C., 2006. Astrophysical Journal 640, L63.

\bibitem[{Lanz and Hubeny(2003)}]{lanz03}
Lanz, T., Hubeny, I., 2003. Astrophysical Journal Supplement Series 146, 417.

\bibitem[{Lanz and Hubeny(2007)}]{lanz07}
Lanz, T., Hubeny, I., 2007. Astrophysical Journal Supplement Series 169, 83.

\bibitem[{Lanz et~al.(2005)Lanz, Telis, Audard, Paerels, P., and
  Hubeny}]{lanz05}
Lanz, T., Telis, G.~A., Audard, M., Paerels, F., P., R.~A., Hubeny, I., 2005.
  Astrophysical Journal 619, 517.

\bibitem[{Leibowitz et~al.(2006)Leibowitz, Orio, Gonzalez-Riestra, Lipkin,
  Ness, Starrfield, Still, and Tepedelenlioglu}]{leibowitz06}
Leibowitz, E., Orio, M., Gonzalez-Riestra, R., Lipkin, Y., Ness, J.-U.,
  Starrfield, S., Still, M., Tepedelenlioglu, E., 2006. Monthly Notices of the
  Royal Astronomical Society 371, 424.

\bibitem[{Livio(1992)}]{livio92}
Livio, M., 1992. Astrophysical Journal 393, 516.

\bibitem[{Lundmark(1921)}]{lundmark21}
Lundmark, K., 1921. Publications of the Astronomical Society of the Pacific 33,
  255.

\bibitem[{Lyke et~al.(2002)Lyke, Kelly, Gehrz, and Woodward}]{lyke02}
Lyke, J.~E., Kelly, M.~S., Gehrz, R.~D., Woodward, C.~E., 2002. Bulletin of the
  American Astronomical Society 34, 1161.

\bibitem[{MacDonald et~al.(1985)MacDonald, Fujimoto, and Truran}]{macdonald85}
MacDonald, J., Fujimoto, M., Truran, J., 1985. Astrophysical Journal 294, 263.

\bibitem[{MacDonald and Vennes(1991)}]{macdonald91}
MacDonald, J., Vennes, S., 1991. Astrophysical Journal 373, L51.

\bibitem[{McLaughlin(1943)}]{mclaughlin43}
McLaughlin, D.~B., 1943. Publications of the Observatory of the University of
  Michigan 8, 149.

\bibitem[{McLaughlin(1945)}]{mclaughlin45}
McLaughlin, D.~B., 1945. Publications of the Astronomical Society of the
  Pacific 57, 69.

\bibitem[{Mihalas(1970)}]{mihalas70}
Mihalas, D., 1970. Stellar Atmospheres. W. H. Freeman and Company.

\bibitem[{Mihalas(1978)}]{mihalas78}
Mihalas, D., 1978. Stellar Atmospheres. W. H. Freeman and Company.

\bibitem[{Morrison and McCammon(1983)}]{morrison83}
Morrison, R., McCammon, D., 1983. Astrophysical Journal 270, 119.

\bibitem[{Nakano et~al.(2008)Nakano, Beize, Jin, Gao, Yamaoka, Haseda, Giudo,
  Sostero, Klingenberg, and Kadoka}]{nakano08}
Nakano, S., Beize, J., Jin, Z.-W., Gao, X., Yamaoka, H., Haseda, K., Giudo, E.,
  Sostero, G., Klingenberg, G., Kadoka, K., 2008. International Astronomical
  Union Circulars 8934.

\bibitem[{Ness(2010)}]{ness10}
Ness, J.-U., 2010. Astronomishce Nachrichten 331, 179.

\bibitem[{Ng(1974)}]{ng74}
Ng, K.~C., 1974. Journal of Quantitative Spectroscopy and Radiative Transfer
  61, 2680.

\bibitem[{Nielbock and Schmidtobreick(1997)}]{nielbock03}
Nielbock, M., Schmidtobreick, L., 1997. Astronomy and Astrophysics Supplement
  Series 125, 149.

\bibitem[{Ninkovi\'c and Trajkovska(2006)}]{ninkovic06}
Ninkovi\'c, S., Trajkovska, V., 2006. Serb. Astron. J. 172, 17.

\bibitem[{{\"O}gelman et~al.(1984){\"O}gelman, Beuermann, and
  Krautter}]{ogelman83}
{\"O}gelman, H., Beuermann, K., Krautter, J., 1984. Astrophysical Journal 287,
  L31.

\bibitem[{Orio et~al.(2001)Orio, Covington, and \"{O}gelman}]{orio01}
Orio, M., Covington, J., \"{O}gelman, 2001. Astronomy and Astrophysics 373,
  542.

\bibitem[{Petz(2005)}]{apetz05}
Petz, A., 2005. Modeling atmospheres of classical novae in x-rays with phoenix.
  Ph.D. thesis, University of Hamburg.

\bibitem[{Petz et~al.(2005)Petz, Hauschildt, Ness, and Starrfield}]{petz05}
Petz, A., Hauschildt, P.~H., Ness, J.-U., Starrfield, S., 2005. Astronomy and
  Astrophysics 431, 321.

\bibitem[{Rauch et~al.(2005)Rauch, Orio, Gonzales-Riestra, and Still}]{rauch05}
Rauch, T., Orio, M., Gonzales-Riestra, R., Still, M., A. A. G.-R.~R., 2005 334,
  423R.

\bibitem[{Rudy et~al.(2008)Rudy, Lynch, and Russell}]{rudy08}
Rudy, R.~J., Lynch, D.~K., Russell, R.~W., 2008. International Astronomical
  Union Circulars 8938.

\bibitem[{Rybicki and Hummer(1991)}]{rybicki91}
Rybicki, G.~B., Hummer, D.~G., 1991. Astronomy and Astrophysics 245, 171.

\bibitem[{Seaton(1985)}]{seaton85}
Seaton, M., 1985. Journal of Physics B 18, 2111.

\bibitem[{Seaton(1987)}]{seaton87}
Seaton, M., 1987. Journal of Physics 20, 6363.

\bibitem[{Shafter(2002)}]{shafter02}
Shafter, A.~W., 2002. AIP Conference Proceedings 637, 462.

\bibitem[{van Rossum and Ness(2010)}]{rossum10}
van Rossum, D., Ness, J.-U., 2010. Astronomishce Nachrichten 331, 175.

\bibitem[{Vanlandingham et~al.(2007)Vanlandingham, Schwarz, Starrfield,
  Woodward, Wagner, Ness, and Helton}]{vanlandingham07}
Vanlandingham, K.~M., Schwarz, G., Starrfield, S., Woodward, C., Wagner, M.,
  Ness, J.-U., Helton, A., 2007. Bulletin of the American Astronomical Society
  38, 99.

\bibitem[{Vennes(1999)}]{vennes99}
Vennes, S., 1999. Astrophysical Journal 525, 995.

\bibitem[{Verner et~al.(1996)Verner, Ferland, Korista, and Yakovlev}]{verner96}
Verner, D.~A., Ferland, G.~J., Korista, K.~T., Yakovlev, D.~G., 1996.
  Astrophysical Journal Supplement Series 103, 467.

\bibitem[{Williams(1992)}]{williams92}
Williams, R.~E., 1992. Astronomical Journal 104, 725.

\end{thebibliography}

\appendix
\addcontentsline{toc}{chapter}{Appendices}
\chapter{O VI model atom}\label{appendixa}

The detailed output for O VI has more than 1600 rows, the averaged model
atom file still has 340 rows, what cannot be presented here. 
Instead of the whole model atom, I show only the first ten
levels, their bound-free cross sections and transitions. The line
list consists of 471 transitions total; 
out of those I show are shown the first 
64 lines, up to 200 \AA. Although O VI is a low
ionization stage to be significantly populated in the hot atmospheres of CNe,
I generated this model atom to validate the procedure.  
This model atom file can be found in good agreement with the O VI data 
file built by \citet{lanz03}, at TLUSTY's web page:
\newline{\protect\url{http://nova.astro.umd.edu/Tlusty2002/database/atom/o6.dat}}

%

\newpage
\section*{Simplified output of TOPAtom}

\begin{singlespace}
\begin{tiny}
\begin{verbatim}
****** Levels
3.339711965517e+16      2.      2       '____1s2.2s'    0       0.      0           ! 1 2Se 0.000000 
3.050786984119e+16      2.      2       '____1s2.2p'    0       0.      0           ! 2 2Po 96.375000 
3.049190589280e+16      4.      2       '____1s2.2p'    0       0.      0           ! 3 2Po 96.907500 
1.420920916918e+16      2.      3       '____1s2.3s'    0       0.      -103        ! 4 2Se 640.039800 
1.342441846848e+16      6.      3       '___2lev___'    0       0.      -103        ! 5 2Po 666.217600 
1.317143080819e+16      10.     3       '___2lev___'    0       0.      -103        ! 6 2De 674.656360 
7.833936678490e+15      2.      4       '____1s2.4s'    0       0.      -104        ! 7 2Se 852.696000 
7.513746341647e+15      6.      4       '___2lev___'    0       0.      -104        ! 8 2Po 863.376400 
7.408323124622e+15      10.     4       '___2lev___'    0       0.      -104        ! 9 2De 866.892940 
7.402624840890e+15      14.     4       '___2lev___'    0       0.      -104        ! 10 2Fo 867.083014 
****** Continuum transitions
1 11            1 174 0 0 0 9.2359e-20 0.000e+00 
 0.019 0.039 0.059 0.078 0.097 0.116 0.135 0.153 0.171 0.190 0.208 0.226 0.279 
 0.372 0.448 0.513 0.569 0.619 0.664 0.704 0.705 0.706 0.708 0.709 0.725 0.740 
 0.756 0.771 0.787 0.802 0.818 0.833 0.849 0.865 0.880 0.896 0.911 0.927 0.942 
 0.958 0.973 0.989 1.004 1.020 1.035 1.051 1.067 1.082 1.098 1.113 1.129 1.144 
 1.160 1.175 1.191 1.206 1.222 1.237 1.253 1.269 1.284 1.300 1.315 1.331 1.346 
 1.362 1.377 1.393 1.408 1.424 1.440 1.455 1.471 1.489 
 -0.512 -0.547 -0.583 -0.620 -0.656 -0.692 -0.728 -0.764 -0.799 -0.836 -0.873 
 -0.910 -1.032 -1.219 -1.380 -1.521 -1.647 -1.761 -1.865 -0.955 -0.819 -0.730 
 -0.676 -0.639 -0.599 -0.637 -0.676 -0.714 -0.753 -0.792 -0.832 -0.872 -0.912 
 -0.952 -0.992 -1.032 -1.073 -1.114 -1.155 -1.197 -1.238 -1.280 -1.322 -1.364 
 -1.406 -1.448 -1.491 -1.533 -1.576 -1.619 -1.663 -1.706 -1.750 -1.793 -1.837 
 -1.881 -1.925 -1.970 -2.014 -2.059 -2.103 -2.148 -2.193 -2.238 -2.284 -2.329 
 -2.375 -2.420 -2.466 -2.512 -2.558 -2.604 -2.649 -2.661 
2 11            1 168 0 0 0 1.1173e-19 0.000e+00 
 0.012 0.025 0.037 0.050 0.062 0.075 0.087 0.100 0.112 0.124 0.137 0.149 0.161 
 0.174 0.186 0.198 0.210 0.222 0.234 0.246 0.258 0.270 0.282 0.294 0.306 0.318 
 0.330 0.341 0.353 0.365 0.376 0.388 0.400 0.411 0.423 0.434 0.439 0.455 0.471 
 0.488 0.504 0.520 0.536 0.552 0.568 0.584 0.600 0.616 0.631 0.647 0.663 0.678 
 0.694 0.709 0.725 0.740 0.756 0.771 0.786 0.801 0.816 0.831 0.846 0.861 0.876 
 0.891 0.906 0.923 
 -0.429 -0.467 -0.504 -0.542 -0.580 -0.616 -0.655 -0.692 -0.730 -0.767 -0.807 
 -0.844 -0.879 -0.917 -0.958 -0.995 -1.032 -1.070 -1.108 -1.146 -1.184 -1.222 
 -1.260 -1.298 -1.336 -1.374 -1.412 -1.451 -1.489 -1.527 -1.565 -1.604 -1.642 
 -1.680 -1.719 -1.757 -1.772 -1.826 -1.880 -1.935 -1.990 -2.044 -2.099 -2.154 
 -2.209 -2.264 -2.318 -2.373 -2.428 -2.483 -2.538 -2.593 -2.648 -2.703 -2.758 
 -2.813 -2.868 -2.923 -2.978 -3.033 -3.088 -3.143 -3.198 -3.254 -3.309 -3.364 
 -3.419 -3.442 
3 11            1 168 0 0 0 1.1173e-19 0.000e+00 
 0.012 0.025 0.037 0.050 0.062 0.075 0.087 0.100 0.112 0.124 0.137 0.149 0.161 
 0.174 0.186 0.198 0.210 0.222 0.234 0.246 0.258 0.270 0.282 0.294 0.306 0.318 
 0.330 0.341 0.353 0.365 0.376 0.388 0.400 0.411 0.423 0.434 0.439 0.455 0.471 
 0.488 0.504 0.520 0.536 0.552 0.568 0.584 0.600 0.616 0.631 0.647 0.663 0.678 
 0.694 0.709 0.725 0.740 0.756 0.771 0.786 0.801 0.816 0.831 0.846 0.861 0.876 
 0.891 0.906 0.923 
 -0.429 -0.467 -0.504 -0.542 -0.580 -0.616 -0.655 -0.692 -0.730 -0.767 -0.807 
 -0.844 -0.879 -0.917 -0.958 -0.995 -1.032 -1.070 -1.108 -1.146 -1.184 -1.222 
 -1.260 -1.298 -1.336 -1.374 -1.412 -1.451 -1.489 -1.527 -1.565 -1.604 -1.642 
 -1.680 -1.719 -1.757 -1.772 -1.826 -1.880 -1.935 -1.990 -2.044 -2.099 -2.154 
 -2.209 -2.264 -2.318 -2.373 -2.428 -2.483 -2.538 -2.593 -2.648 -2.703 -2.758 
 -2.813 -2.868 -2.923 -2.978 -3.033 -3.088 -3.143 -3.198 -3.254 -3.309 -3.364 
 -3.419 -3.442 
4 11            1 174 0 0 0 1.5288e-19 0.000e+00 
 0.019 0.038 0.057 0.076 0.095 0.114 0.132 0.151 0.169 0.188 0.206 0.224 0.242 
 0.260 0.278 0.296 0.314 0.332 0.349 0.367 0.384 0.402 0.419 0.436 0.454 0.471 
 0.488 0.505 0.522 0.539 0.555 0.572 0.589 0.605 0.622 0.638 0.655 0.671 0.688 
 0.704 0.720 0.736 0.752 0.768 0.784 0.800 0.807 0.830 0.853 0.875 0.898 0.920 
 0.943 0.965 0.987 1.009 1.031 1.052 1.074 1.096 1.117 1.138 1.160 1.181 1.202 
 1.223 1.244 1.265 1.285 1.306 1.327 1.347 1.368 1.385 
 -0.293 -0.328 -0.363 -0.399 -0.434 -0.470 -0.504 -0.540 -0.575 -0.611 -0.646 
 -0.682 -0.717 -0.752 -0.788 -0.824 -0.860 -0.896 -0.932 -0.966 -1.002 -1.037 
 -1.073 -1.109 -1.144 -1.180 -1.216 -1.252 -1.287 -1.323 -1.359 -1.396 -1.431 
 -1.467 -1.503 -1.539 -1.575 -1.611 -1.648 -1.684 -1.719 -1.754 -1.790 -1.827 
 -1.863 -1.900 -1.917 -1.968 -2.021 -2.073 -2.125 -2.178 -2.230 -2.282 -2.335 
 -2.387 -2.440 -2.493 -2.545 -2.598 -2.650 -2.703 -2.756 -2.809 -2.861 -2.914 
 -2.967 -3.020 -3.073 -3.126 -3.178 -3.231 -3.284 -3.300 
5 11            1 173 0 0 0 2.0972e-19 0.000e+00 
 0.015 0.031 0.046 0.062 0.077 0.092 0.107 0.122 0.137 0.152 0.166 0.181 0.196 
 0.210 0.225 0.239 0.253 0.268 0.282 0.296 0.310 0.324 0.338 0.351 0.365 0.379 
 0.393 0.406 0.420 0.433 0.446 0.460 0.473 0.486 0.499 0.513 0.526 0.539 0.552 
 0.564 0.577 0.590 0.603 0.615 0.628 0.641 0.653 0.666 0.678 0.691 0.696 0.714 
 0.731 0.749 0.766 0.784 0.801 0.818 0.835 0.852 0.869 0.886 0.903 0.919 0.936 
 0.952 0.969 0.985 1.001 1.018 1.034 1.050 1.067 
 -0.155 -0.193 -0.230 -0.267 -0.304 -0.341 -0.379 -0.415 -0.452 -0.489 -0.527 
 -0.564 -0.602 -0.638 -0.675 -0.712 -0.749 -0.788 -0.824 -0.863 -0.900 -0.939 
 -0.975 -1.012 -1.049 -1.087 -1.125 -1.162 -1.200 -1.237 -1.275 -1.312 -1.350 
 -1.388 -1.426 -1.463 -1.500 -1.539 -1.577 -1.614 -1.652 -1.690 -1.728 -1.764 
 -1.804 -1.841 -1.879 -1.917 -1.954 -1.992 -2.010 -2.064 -2.119 -2.173 -2.228 
 -2.282 -2.337 -2.392 -2.446 -2.501 -2.556 -2.610 -2.665 -2.720 -2.775 -2.829 
 -2.884 -2.939 -2.994 -3.049 -3.104 -3.158 -3.179 
6 11            1 177 0 0 0 1.3941e-19 0.000e+00 
 0.011 0.022 0.033 0.044 0.055 0.066 0.077 0.088 0.099 0.110 0.121 0.132 0.142 
 0.153 0.164 0.174 0.185 0.196 0.206 0.217 0.227 0.238 0.248 0.259 0.269 0.280 
 0.290 0.300 0.311 0.321 0.331 0.341 0.351 0.362 0.372 0.382 0.392 0.402 0.412 
 0.422 0.432 0.442 0.452 0.462 0.471 0.481 0.491 0.501 0.511 0.520 0.530 0.540 
 0.549 0.559 0.569 0.578 0.588 0.597 0.607 0.616 0.626 0.635 0.644 0.654 0.663 
 0.672 0.682 0.691 0.700 0.703 0.716 0.729 0.742 0.756 0.769 0.782 0.799 
 -0.333 -0.371 -0.410 -0.448 -0.487 -0.525 -0.564 -0.602 -0.640 -0.679 -0.717 
 -0.757 -0.795 -0.833 -0.872 -0.910 -0.950 -0.987 -1.026 -1.065 -1.103 -1.142 
 -1.181 -1.219 -1.258 -1.296 -1.335 -1.374 -1.413 -1.452 -1.490 -1.529 -1.568 
 -1.607 -1.646 -1.684 -1.723 -1.762 -1.801 -1.841 -1.879 -1.917 -1.955 -1.995 
 -2.035 -2.073 -2.112 -2.151 -2.190 -2.230 -2.268 -2.307 -2.346 -2.385 -2.424 
 -2.463 -2.503 -2.542 -2.581 -2.620 -2.659 -2.699 -2.737 -2.777 -2.815 -2.854 
 -2.893 -2.932 -2.973 -2.984 -3.038 -3.094 -3.150 -3.206 -3.262 -3.317 -3.343 
7 11            1 179 0 0 0 2.2219e-19 0.000e+00 
 0.018 0.037 0.055 0.074 0.092 0.111 0.129 0.147 0.165 0.184 0.202 0.220 0.238 
 0.256 0.274 0.291 0.309 0.327 0.345 0.362 0.380 0.397 0.415 0.432 0.450 0.467 
 0.484 0.502 0.519 0.536 0.553 0.570 0.587 0.604 0.621 0.638 0.655 0.672 0.689 
 0.705 0.722 0.739 0.755 0.772 0.788 0.805 0.821 0.837 0.854 0.870 0.886 0.903 
 0.919 0.935 0.951 0.967 0.983 0.999 1.015 1.031 1.034 1.057 1.080 1.103 1.126 
 1.148 1.171 1.193 1.215 1.238 1.260 1.282 1.304 1.326 1.348 1.370 1.392 1.414 
 1.431 
 -0.130 -0.165 -0.201 -0.236 -0.271 -0.305 -0.341 -0.376 -0.411 -0.446 -0.481 
 -0.516 -0.551 -0.587 -0.622 -0.657 -0.692 -0.728 -0.762 -0.798 -0.833 -0.869 
 -0.903 -0.939 -0.975 -1.010 -1.045 -1.080 -1.116 -1.152 -1.187 -1.222 -1.258 
 -1.293 -1.329 -1.364 -1.400 -1.436 -1.471 -1.507 -1.543 -1.578 -1.614 -1.650 
 -1.686 -1.721 -1.757 -1.793 -1.829 -1.863 -1.899 -1.935 -1.971 -2.008 -2.044 
 -2.080 -2.116 -2.152 -2.188 -2.224 -2.232 -2.283 -2.335 -2.386 -2.438 -2.490 
 -2.542 -2.594 -2.646 -2.698 -2.750 -2.802 -2.855 -2.907 -2.959 -3.011 -3.063 
 -3.115 -3.130 
8 11            1 177 0 0 0 3.1248e-19 0.000e+00 
 0.016 0.033 0.049 0.065 0.082 0.098 0.114 0.130 0.146 0.162 0.177 0.193 0.209 
 0.224 0.240 0.255 0.271 0.286 0.301 0.316 0.332 0.347 0.362 0.376 0.391 0.406 
 0.421 0.435 0.450 0.465 0.479 0.493 0.508 0.522 0.536 0.550 0.565 0.579 0.593 
 0.606 0.620 0.634 0.648 0.662 0.675 0.689 0.702 0.716 0.729 0.742 0.756 0.769 
 0.782 0.795 0.808 0.821 0.834 0.847 0.860 0.873 0.886 0.898 0.908 0.926 0.944 
 0.962 0.980 0.998 1.016 1.034 1.051 1.069 1.086 1.104 1.121 1.138 1.155 
 0.018 -0.018 -0.054 -0.090 -0.127 -0.163 -0.199 -0.236 -0.272 -0.309 -0.346 
 -0.382 -0.418 -0.455 -0.492 -0.529 -0.565 -0.602 -0.638 -0.676 -0.712 -0.749 
 -0.785 -0.824 -0.860 -0.896 -0.932 -0.970 -1.008 -1.045 -1.082 -1.119 -1.156 
 -1.194 -1.231 -1.268 -1.305 -1.343 -1.380 -1.418 -1.455 -1.492 -1.530 -1.567 
 -1.604 -1.642 -1.680 -1.717 -1.754 -1.793 -1.830 -1.866 -1.906 -1.943 -1.979 
 -2.018 -2.056 -2.093 -2.131 -2.169 -2.207 -2.245 -2.274 -2.326 -2.380 -2.434 
 -2.488 -2.542 -2.597 -2.651 -2.705 -2.760 -2.814 -2.868 -2.923 -2.977 -2.995 
9 11            1 182 0 0 0 2.7680e-19 0.000e+00 
 0.013 0.027 0.040 0.054 0.067 0.080 0.093 0.106 0.119 0.132 0.145 0.158 0.170 
 0.183 0.195 0.208 0.220 0.233 0.245 0.257 0.270 0.282 0.294 0.306 0.318 0.330 
 0.341 0.353 0.365 0.377 0.388 0.400 0.411 0.423 0.434 0.446 0.457 0.468 0.480 
 0.491 0.502 0.513 0.524 0.535 0.546 0.557 0.568 0.579 0.590 0.600 0.611 0.622 
 0.633 0.643 0.654 0.664 0.675 0.685 0.696 0.706 0.716 0.727 0.737 0.747 0.757 
 0.767 0.777 0.788 0.798 0.808 0.818 0.828 0.837 0.847 0.857 0.867 0.877 0.887 
 0.896 0.906 0.913 0.930 
 -0.035 -0.073 -0.111 -0.149 -0.187 -0.225 -0.263 -0.301 -0.339 -0.378 -0.415 
 -0.453 -0.492 -0.530 -0.567 -0.605 -0.644 -0.682 -0.721 -0.759 -0.796 -0.835 
 -0.873 -0.910 -0.950 -0.987 -1.027 -1.065 -1.104 -1.142 -1.181 -1.219 -1.257 
 -1.296 -1.334 -1.373 -1.412 -1.450 -1.489 -1.527 -1.565 -1.605 -1.644 -1.682 
 -1.721 -1.759 -1.798 -1.836 -1.876 -1.914 -1.954 -1.991 -2.030 -2.069 -2.108 
 -2.147 -2.186 -2.224 -2.263 -2.302 -2.341 -2.380 -2.419 -2.458 -2.497 -2.536 
 -2.575 -2.614 -2.653 -2.692 -2.730 -2.769 -2.809 -2.847 -2.886 -2.928 -2.966 
 -3.004 -3.044 -3.083 -3.113 -3.135 
10 11           1 182 0 0 0 2.7680e-19 0.000e+00 
 0.013 0.027 0.040 0.054 0.067 0.080 0.093 0.106 0.119 0.132 0.145 0.158 0.170 
 0.183 0.195 0.208 0.220 0.233 0.245 0.257 0.270 0.282 0.294 0.306 0.318 0.330 
 0.341 0.353 0.365 0.377 0.388 0.400 0.411 0.423 0.434 0.446 0.457 0.468 0.480 
 0.491 0.502 0.513 0.524 0.535 0.546 0.557 0.568 0.579 0.590 0.600 0.611 0.622 
 0.633 0.643 0.654 0.664 0.675 0.685 0.696 0.706 0.716 0.727 0.737 0.747 0.757 
 0.767 0.777 0.788 0.798 0.808 0.818 0.828 0.837 0.847 0.857 0.867 0.877 0.887 
 0.896 0.906 0.913 0.930 
 -0.035 -0.073 -0.111 -0.149 -0.187 -0.225 -0.263 -0.301 -0.339 -0.378 -0.415 
 -0.453 -0.492 -0.530 -0.567 -0.605 -0.644 -0.682 -0.721 -0.759 -0.796 -0.835 
 -0.873 -0.910 -0.950 -0.987 -1.027 -1.065 -1.104 -1.142 -1.181 -1.219 -1.257 
 -1.296 -1.334 -1.373 -1.412 -1.450 -1.489 -1.527 -1.565 -1.605 -1.644 -1.682 
 -1.721 -1.759 -1.798 -1.836 -1.876 -1.914 -1.954 -1.991 -2.030 -2.069 -2.108 
 -2.147 -2.186 -2.224 -2.263 -2.302 -2.341 -2.380 -2.419 -2.458 -2.497 -2.536 
 -2.575 -2.614 -2.653 -2.692 -2.730 -2.769 -2.809 -2.847 -2.886 -2.928 -2.966 
 -3.004 -3.044 -3.083 -3.113 -3.135 
*** Line transitions
1 2            -1 0 1 0 0 6.6000e-02    7.000e-01 
   F  1  7  0.  0. 
1 3            -1 0 1 0 0 1.3300e-01    7.000e-01 
   F  1  7  0.  0.
1 4             0 0 4 0 0 0.0000e+00    5.000e-01 
1 5            -1 0 1 0 0 2.6550e-01    2.000e-01 
   F  1  7  0.  0.
1 6             0 0 4 0 0 0.0000e+00    5.000e-01 
1 7             0 0 4 0 0 0.0000e+00    5.000e-01 
1 8            -1 0 1 0 0 7.4200e-02    2.000e-01 
   F  1  7  0.  0.
1 9             0 0 4 0 0 0.0000e+00    5.000e-01 
1 10            0 0 4 0 0 0.0000e+00    5.000e-01 
2 3             0 0 4 0 0 0.0000e+00    5.000e-01 
2 4            -1 0 1 0 0 2.8900e-02    2.000e-01 
   F  1  7  0.  0.
2 5             0 0 4 0 0 0.0000e+00    5.000e-01 
2 6            -1 0 1 0 0 6.5669e-01    2.000e-01 
   F  1  7  0.  0.
2 7            -1 0 1 0 0 5.7100e-03    2.000e-01 
   F  1  7  0.  0.
2 8             0 0 4 0 0 0.0000e+00    5.000e-01 
2 9            -1 0 1 0 0 1.2250e-01    2.000e-01 
   F  1  7  0.  0.
2 10            0 0 4 0 0 0.0000e+00    5.000e-01 
3 4            -1 0 1 0 0 2.9000e-02    2.000e-01 
   F  1  7  0.  0.
3 5             0 0 4 0 0 0.0000e+00    5.000e-01 
3 6            -1 0 1 0 0 6.5662e-01    2.000e-01 
   F  1  7  0.  0.
3 7            -1 0 1 0 0 5.7000e-03    2.000e-01 
   F  1  7  0.  0.
3 8             0 0 4 0 0 0.0000e+00    5.000e-01 
3 9            -1 0 1 0 0 1.2230e-01    2.000e-01 
   F  1  7  0.  0.
3 10            0 0 4 0 0 0.0000e+00    5.000e-01 
4 5            -1 0 1 0 0 3.3500e-01    7.000e-01 
   F  1  7  0.  0.
4 6             0 0 4 0 0 0.0000e+00    5.000e-01 
4 7             0 0 4 0 0 0.0000e+00    5.000e-01 
4 8            -1 0 1 0 0 2.7730e-01    2.000e-01 
   F  1  7  0.  0.
4 9             0 0 4 0 0 0.0000e+00    5.000e-01 
4 10            0 0 4 0 0 0.0000e+00    5.000e-01 
5 6            -1 0 1 0 0 4.8033e-02    7.000e-01 
   F  1  7  0.  0.
5 7            -1 0 1 0 0 9.5650e-02    2.000e-01 
   F  1  7  0.  0.
5 8             0 0 4 0 0 0.0000e+00    5.000e-01 
5 9            -1 0 1 0 0 5.5698e-01    2.000e-01 
   F  1  7  0.  0.
5 10            0 0 4 0 0 0.0000e+00    5.000e-01 
6 7             0 0 4 0 0 0.0000e+00    5.000e-01 
6 8            -1 0 1 0 0 1.5295e-02    2.000e-01 
   F  1  7  0.  0.
6 9             0 0 4 0 0 0.0000e+00    5.000e-01 
6 10           -1 0 1 0 0 1.6875e+00    2.000e-01 
   F  1  7  0.  0.
7 8            -1 0 1 0 0 4.6300e-01    7.000e-01 
   F  1  7  0.  0.
7 9             0 0 4 0 0 0.0000e+00    5.000e-01 
7 10            0 0 4 0 0 0.0000e+00    5.000e-01 
8 9            -1 0 1 0 0 8.5391e-02    7.000e-01 
   F  1  7  0.  0.
8 10            0 0 4 0 0 0.0000e+00    5.000e-01 
9 10           -1 0 1 0 0 3.9660e-03    7.000e-01 
   F  1  7  0.  0.
\end{verbatim}
\end{tiny}
\end{singlespace}

\newpage
\section*{Line list}

\begin{singlespace}
\begin{tiny}
\begin{verbatim}
    9.3110 08.05  -2.192          0.000 0.5    1074000.000 0.5 9.39  0.00 0.00 0 !TBtr
    9.3110 08.05  -2.192          0.000 0.5    1074000.000 1.5 9.39  0.00 0.00 0 !TBtr
    9.3915 08.05  -2.047          0.000 0.5    1064790.000 0.5 9.53  0.00 0.00 0 !TBtr
    9.3915 08.05  -2.047          0.000 0.5    1064790.000 1.5 9.53  0.00 0.00 0 !TBtr
    9.5082 08.05  -1.883          0.000 0.5    1051724.000 0.5 9.68  0.00 0.00 0 !TBtr
    9.5082 08.05  -1.883          0.000 0.5    1051724.000 1.5 9.68  0.00 0.00 0 !TBtr
    9.6840 08.05  -1.695          0.000 0.5    1032631.000 0.5 9.85  0.00 0.00 0 !TBtr
    9.6840 08.05  -1.695          0.000 0.5    1032631.000 1.5 9.85  0.00 0.00 0 !TBtr
    9.9688 08.05  -1.471          0.000 0.5    1003130.000 0.5 10.05 0.00 0.00 0 !TBtr
    9.9688 08.05  -1.471          0.000 0.5    1003130.000 1.5 10.05 0.00 0.00 0 !TBtr
   10.2247 08.05  -2.110      96375.000 0.5    1074400.000 1.5 9.39  0.00 0.00 0 !TBtr
   10.2247 08.05  -2.110      96375.000 0.5    1074400.000 2.5 9.39  0.00 0.00 0 !TBtr
   10.2303 08.05  -1.809      96907.500 1.5    1074400.000 1.5 9.39  0.00 0.00 0 !TBtr
   10.2303 08.05  -1.809      96907.500 1.5    1074400.000 2.5 9.39  0.00 0.00 0 !TBtr
   10.2456 08.05  -3.413      96375.000 0.5    1072406.849 0.5 8.09  0.00 0.00 0 !TBtr
   10.2512 08.05  -3.112      96907.500 1.5    1072406.849 0.5 8.09  0.00 0.00 0 !TBtr
   10.3203 08.05  -1.960      96375.000 0.5    1065337.000 2.5 9.53  0.00 0.00 0 !TBtr
   10.3206 08.05  -1.956      96375.000 0.5    1065311.000 1.5 9.24  0.00 0.00 0 !NISTtr
   10.3260 08.05  -1.702      96907.500 1.5    1065337.000 2.5 9.32  0.00 0.00 0 !NISTtr
   10.3263 08.05  -2.655      96907.500 1.5    1065311.000 1.5 8.54  0.00 0.00 0 !NISTtr
   10.3473 08.05  -3.262      96375.000 0.5    1062811.418 0.5 8.23  0.00 0.00 0 !TBtr
   10.3530 08.05  -2.961      96907.500 1.5    1062811.418 0.5 8.23  0.00 0.00 0 !TBtr
   10.4611 08.05  -1.789      96375.000 0.5    1052301.000 2.5 9.69  0.00 0.00 0 !TBtr
   10.4612 08.05  -1.788      96375.000 0.5    1052288.000 1.5 9.39  0.00 0.00 0 !NISTtr
   10.4669 08.05  -1.533      96907.500 1.5    1052301.000 2.5 9.47  0.00 0.00 0 !NISTtr
   10.4670 08.05  -2.488      96907.500 1.5    1052288.000 1.5 8.69  0.00 0.00 0 !NISTtr
   10.4803 08.05  -3.103      96375.000 0.5    1050543.000 0.5 8.38  0.00 0.00 0 !NISTtr
   10.4813 08.05  -1.194          0.000 0.5     954080.000 0.5 10.29 0.00 0.00 0 !TBtr
   10.4813 08.05  -1.194          0.000 0.5     954080.000 1.5 10.29 0.00 0.00 0 !TBtr
   10.4862 08.05  -2.804      96907.500 1.5    1050543.000 0.5 8.68  0.00 0.00 0 !NISTtr
   10.6728 08.05  -1.589      96375.000 0.5    1033334.000 2.5 9.88  0.00 0.00 0 !TBtr
   10.6731 08.05  -1.588      96375.000 0.5    1033310.000 1.5 9.58  0.00 0.00 0 !NISTtr
   10.6789 08.05  -1.333      96907.500 1.5    1033334.000 2.5 9.66  0.00 0.00 0 !NISTtr
   10.6792 08.05  -2.287      96907.500 1.5    1033310.000 1.5 8.88  0.00 0.00 0 !NISTtr
   10.7020 08.05  -2.905      96375.000 0.5    1030780.000 0.5 8.56  0.00 0.00 0 !NISTtr
   10.7081 08.05  -2.605      96907.500 1.5    1030780.000 0.5 8.86  0.00 0.00 0 !NISTtr
   11.0155 08.05  -1.347      96375.000 0.5    1004184.000 2.5 10.09 0.00 0.00 0 !TBtr
   11.0157 08.05  -1.347      96375.000 0.5    1004170.000 1.5 9.79  0.00 0.00 0 !NISTtr
   11.0220 08.05  -1.093      96907.500 1.5    1004184.000 2.5 9.87  0.00 0.00 0 !NISTtr
   11.0222 08.05  -2.048      96907.500 1.5    1004170.000 1.5 9.09  0.00 0.00 0 !NISTtr
   11.0656 08.05  -2.670      96375.000 0.5    1000080.000 0.5 8.77  0.00 0.00 0 !NISTtr
   11.0721 08.05  -2.365      96907.500 1.5    1000080.000 0.5 9.07  0.00 0.00 0 !NISTtr
   11.5821 08.05  -1.004          0.000 0.5     863397.700 1.5 10.09 0.00 0.00 0 !NISTtr
   11.5830 08.05  -1.306          0.000 0.5     863333.800 0.5 10.09 0.00 0.00 0 !NISTtr
   11.6349 08.05  -1.039      96375.000 0.5     955860.000 2.5 10.35 0.00 0.00 0 !TBtr
   11.6350 08.05  -1.041      96375.000 0.5     955851.000 1.5 10.05 0.00 0.00 0 !NISTtr
   11.6421 08.05  -0.784      96907.500 1.5     955860.000 2.5 10.13 0.00 0.00 0 !NISTtr
   11.6422 08.05  -1.740      96907.500 1.5     955851.000 1.5 9.35  0.00 0.00 0 !NISTtr
   11.7328 08.05  -2.355      96375.000 0.5     948690.000 0.5 9.03  0.00 0.00 0 !NISTtr
   11.7401 08.05  -2.056      96907.500 1.5     948690.000 0.5 9.33  0.00 0.00 0 !NISTtr
   12.9781 08.05  -0.611      96375.000 0.5     866901.500 2.5 10.68 0.00 0.00 0 !TBtr
   12.9785 08.05  -0.609      96375.000 0.5     866880.100 1.5 10.39 0.00 0.00 0 !NISTtr
   12.9871 08.05  -0.357      96907.500 1.5     866901.500 2.5 10.46 0.00 0.00 0 !NISTtr
   12.9875 08.05  -1.308      96907.500 1.5     866880.100 1.5 9.69  0.00 0.00 0 !NISTtr
   13.2219 08.05  -1.942      96375.000 0.5     852696.000 0.5 9.34  0.00 0.00 0 !NISTtr
   13.2312 08.05  -1.642      96907.500 1.5     852696.000 0.5 9.64  0.00 0.00 0 !NISTtr
   15.0089 08.05  -0.451          0.000 0.5     666269.800 1.5 10.42 0.00 0.00 0 !NISTtr
   15.0125 08.05  -0.752          0.000 0.5     666113.200 0.5 10.42 0.00 0.00 0 !NISTtr
   17.2920 08.05   0.118      96375.000 0.5     674676.800 2.5 11.16 0.00 0.00 0 !TBtr
   17.2935 08.05   0.119      96375.000 0.5     674625.700 1.5 10.87 0.00 0.00 0 !NISTtr
   17.3079 08.05   0.374      96907.500 1.5     674676.800 2.5 10.94 0.00 0.00 0 !NISTtr
   17.3095 08.05  -0.581      96907.500 1.5     674625.700 1.5 10.16 0.00 0.00 0 !NISTtr
   18.3937 08.05  -1.238      96375.000 0.5     640039.800 0.5 9.76  0.00 0.00 0 !NISTtr
   18.4117 08.05  -0.936      96907.500 1.5     640039.800 0.5 10.06 0.00 0.00 0 !NISTtr
\end{verbatim}
\end{tiny}
\end{singlespace}

\newpage
\section*{Grotrian diagram}

\begin{figure}[!h]
\begin{center}
\includegraphics[width=12cm]{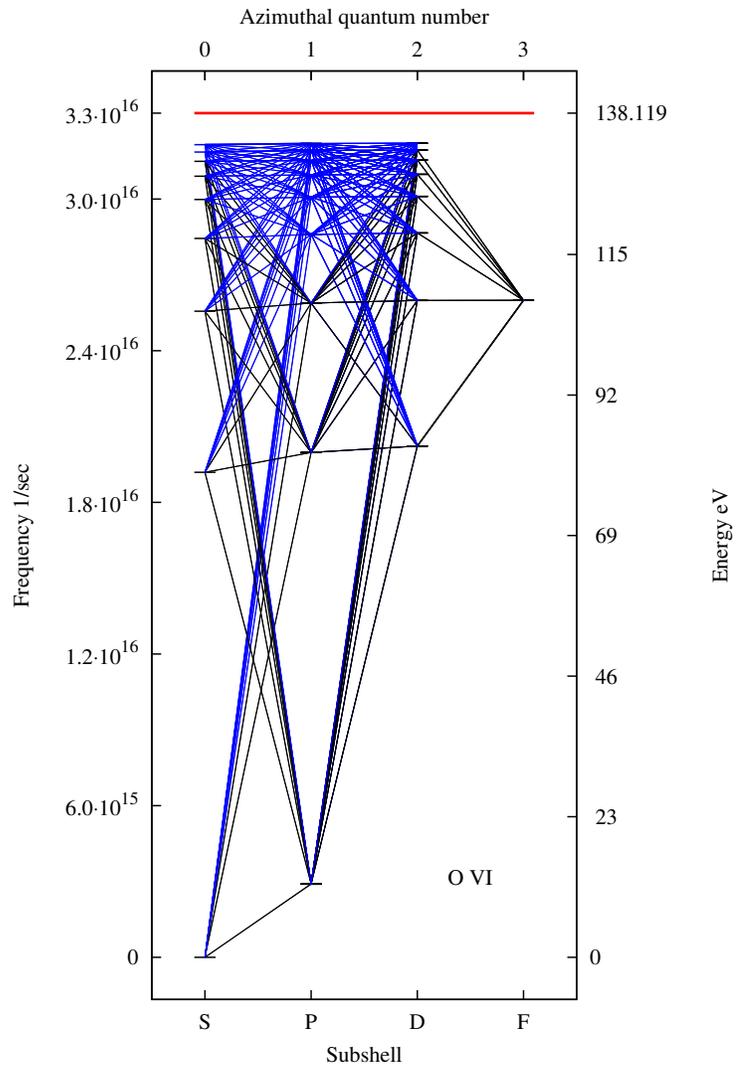}
\parbox{130mm}{\caption{Grotrian diagram for O VI.\label{fig:Fe26gro}}}
\end{center}
\end{figure}

\chapter{Catalog of Galactic CNe with Supersoft Phase}

This catalog lists the most recent galactic CNe which were observed to have
supersoft phase. It can be considered as an update of the
online\footnote{\protect\url{http://www.mpe.mpg.de/~jcg/sss/ssscat.html}} 
catalog of 
\citet{greiner00}.
For each novae equatorial coordiantes for year 2000 and
finder charts were collected from the
SIMBAD\footnote{\protect\url{http://simbad.u-strasbg.fr/simbad/}}
database. Other data on each novae were collected from the scientific
literature. The time of visual maximum ($t_0$) is given in UT, where available
decline rates in the optical band ($t_2$, $t_3$) are also given. The object
type CN refers to classical novae, RN stands for recurrent novae. The general
data table has data on maximum optical brightness in V band, 
interstellar reddening (either with (B-V) color index or E(B-V) color excess),
measured expansion or terminal velocity, distance to the object, 
absolute brightness in V band and
orbital period. The times X-on and X-off refers to the dates when the nova was
first and last observed in the supersoft range, respectively. For some novae
only the estimated length of the supersoft phase were available, in those cases
this is given as $t_{\rm BOL}$ in years. Model data is
listed if either black-body (BB) or WD model fits were available. From
modeling only the most important parameters are shown, the effective
temperature ($T_{\rm eff}$), the date after $t_0$ when the spectrum was taken, 
the derived luminosity and hydrogen column-density. 
Data are labeled with superscripts and their sources can be found 
in the References.

\noindent\begin{table*}[!h]
\vspace{1mm}
\begin{tabular}{p{1.8cm}p{10.0cm}}
\multicolumn{2}{l}{Journal Abbreviations:}\\
\\
New Astr.&New Astronomy\\
RMxAC2   &Revista Mexicana de Astronomia y Astrof\'{i}sica\\
ApJ      &Astrophysical Journal\\
AN       &Astronomische Nachrichten\\ 
ATel     &The Astronomer's Telegram\\
IAUC     &International Astronomical Union Circulars\\
MNRAS    &Monthly Notices of the Royal Astronomical Society\\
A\&A     &Astronomy \& Astrophysics\\
AAS      &American Astronomical Society\\
AJ       &The Astronomical Journal\\
AB       &Astrophysical Bulletin\\
\end{tabular}
\end{table*}



\newpage
\begin{table*}[t]\begin{center}
\begin{tabular}{p{6cm}r}
{\Huge {\bf GQ Mus}}& \multirow{6}{*}{
\includegraphics[width=7cm]{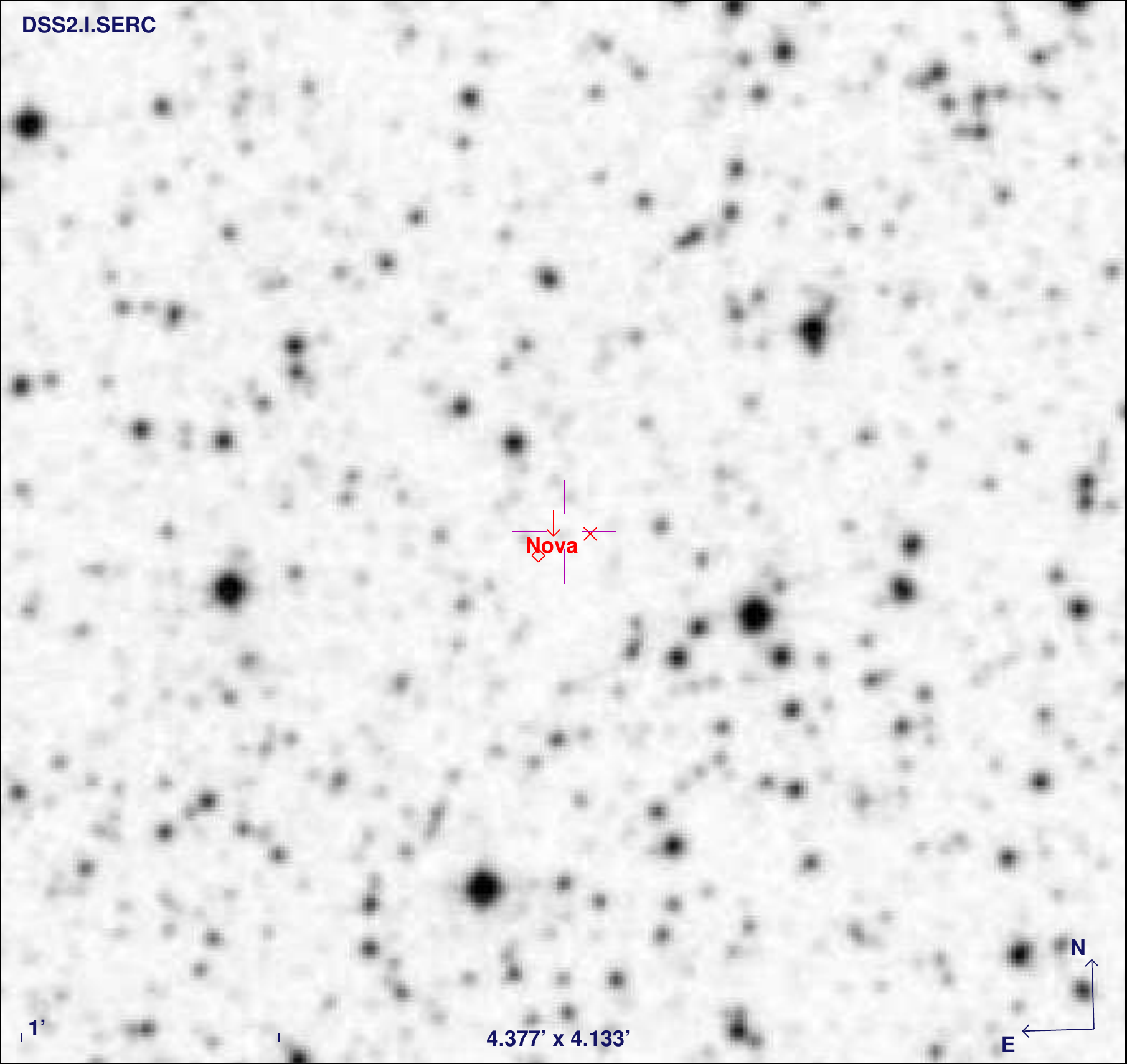}}\\
&\\
\verb,RA (2000) =  11 52 02.5,&\\ 
\verb,DEC(2000) = -67 12 24,&\\
&\\
t$_0=$ 1983/01/18$^{[2]}$ &\\
t$_3=40^{[2]}$ [d] &\\
&\\
Type: CN&\\
&\\
&\\
&\\
&\\
\end{tabular}\end{center}\end{table*}
\begin{table*}[h]\small\begin{center}
\begin{tabular}{p{4cm}|p{4cm}|p{4cm}}
\multicolumn{3}{c}{General Data}\\
\hline
m$_B$ [mag] = 7.21$^{[1]}$&E(B-V) = 0.45$\pm$0.15$^{[2]}$&v$_{\infty}$ [km/s]
= 550 -- 800$^{[1]}$\\
D [kpc] = 4.7$\pm$1.5$^{[1]}$   &M$_V$ = -7.75$^{[2]}$& P$_{\rm orb}$ [d] = 0.0588$^{[1]}$ \\
X-ray on =       &t$_{\rm Bol}$ [yr] = 9 -- 10$^{[3]}$& \\ 
\hline
\end{tabular}\end{center}\end{table*}
\begin{table*}[h]\small\begin{center}
\begin{tabular}{p{1.1cm}|p{2.1cm}|p{3cm}|p{2cm}|p{3cm}}
\multicolumn{5}{c}{Model Data}\\
\hline
Model&T$_{\rm eff}$[$\times$10$^3$ K]&t after t$_0$ [d]&L [erg/s]&N$_{H}$
[10$^{21}$ cm$^{-2}$]\\
\hline
BB$^{[1]}$   &290 -- 400     &                 &                   & 1 -- 3.4\\
WD   &           &                 &                   &     \\
\hline
\end{tabular}\end{center}\end{table*}
\begin{table*}[!h]\small\begin{center}
\begin{tabular}{p{.8cm}p{11.7cm}}
\multicolumn{2}{c}{References}\\
\hline
1&J. Greiner; 2000, New Astr., 5, 137\\
2&J. Krautter et al.; 1984, A\&A, 137, 307\\
3&M. Orio; 2004, RMxAC, 20, 182\\
\hline
\end{tabular}\end{center}\end{table*}

 \newpage 
\begin{table*}[t]\begin{center}\begin{tabular}{p{6cm}r}
{\Huge {\bf V1974 Cyg}}& \multirow{6}{*}{\includegraphics[width=7cm]
{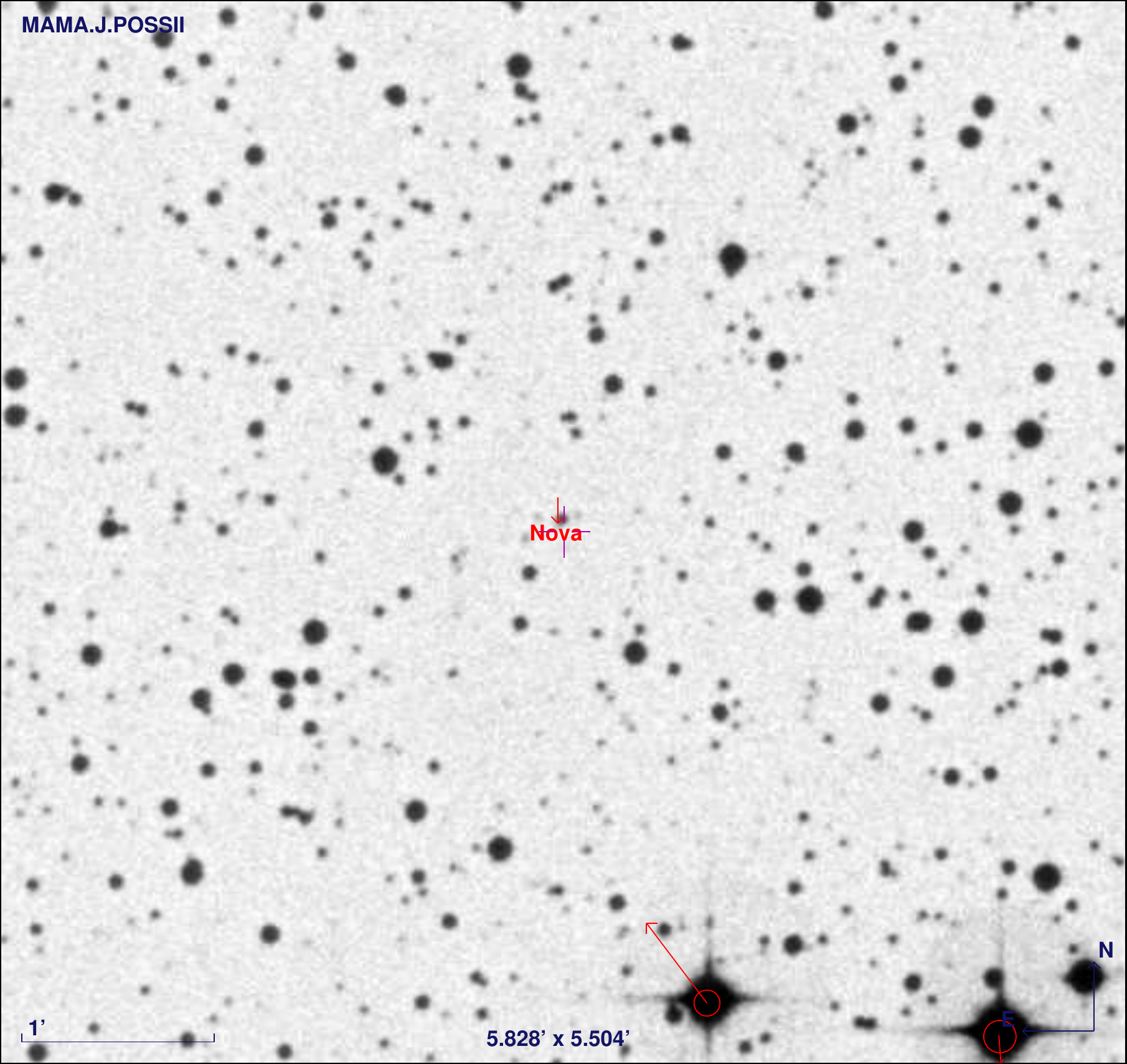}}\\
&\\
\verb,RA (2000) =  20 30 31.8,&\\ 
\verb,DEC(2000) = +52 37 53,&\\
&\\
t$_0=$ 1992/02/22$^{[3]}$ &\\
t$_2= 17^{[1]}$ d   &\\
&\\
Type: CN&\\
&\\&\\&\\&\\
\end{tabular}\end{center}\end{table*}\begin{table*}[h]\small\begin{center}
\begin{tabular}{p{3.8cm}|p{4.2cm}|p{4cm}}
\multicolumn{3}{c}{General Data}\\
\hline
m$_V$ [mag] = $4.3^{[3]}$&(B-V) = -$0.6\pm0.45^{[1]}$&v$_{\infty}$ [km/s] = $2000^{[1]}$\\
D [kpc] = $1.8$ -- $3.2^{[1]}$        &M$_V$ [mag] = $-7.78\pm0.12^{[1]}$ &
P$_{\rm orb}$ [d] = $0.081263^{[1]}$ \\
X-ray on = t$_0+434^{[2]}$      &t$_{\rm Bol}$ [yr] = $2$ -- $3^{[2]}$ & \\ 
\hline
\end{tabular}\end{center}\end{table*}\begin{table*}[h]\small\begin{center}
\begin{tabular}{p{1.1cm}|p{2.1cm}|p{3cm}|p{2cm}|p{3cm}}
\multicolumn{5}{c}{Model Data}\\
\hline
Model&T$_{\rm eff}$[$\times$10$^3$ K]&t after t$_0$ [d]&L [erg/s]&N$_{H}$
[10$^{21}$ cm$^{-2}$]\\
\hline
BB$^{[1]}$   &$220$ -- $300$ &                 & &      \\
WD$^{[1]}$   &$350$ -- $600$ &                 & & $1.9$ -- $2.5$ \\
\hline
\end{tabular}\end{center}\end{table*}
\begin{table*}[!h]\small\begin{center}
\begin{tabular}{p{.8cm}p{11.7cm}}
\multicolumn{2}{c}{References}\\
\hline
1&J. Greiner; 2000, New Astr., 5, 137\\
2&M. Orio; 2004, RMxAC, 20, 182\\
3&J.A. DeYoung et al.; 1994, ApJ, 431, L47\\
\hline
\end{tabular}\end{center}\end{table*}

\newpage 
\begin{table*}[t]\begin{center}
\begin{tabular}{p{6cm}r}
{\Huge {\bf V2491 Cyg}}& \multirow{6}{*}{\includegraphics[width=7cm]
{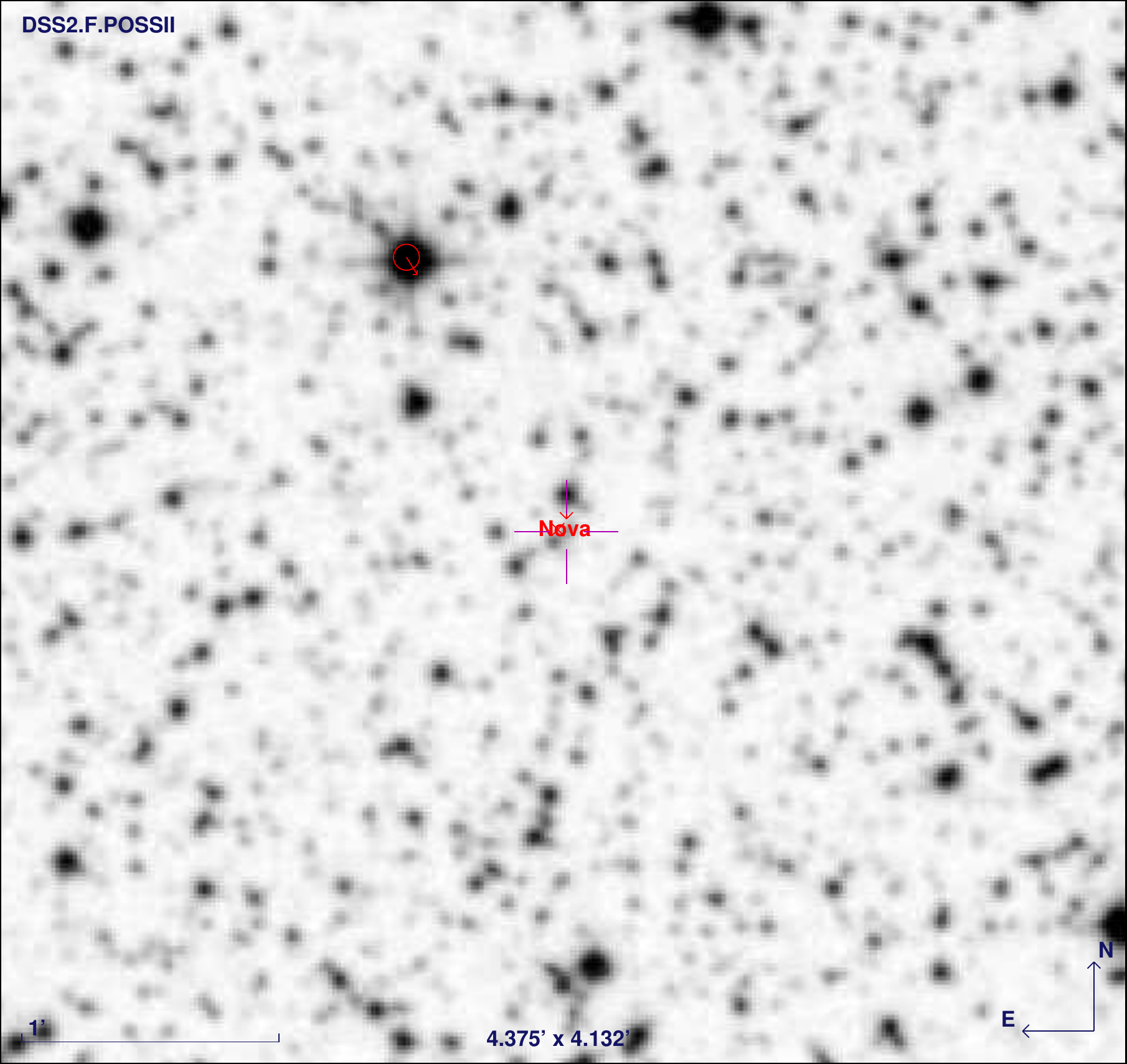}}\\
&\\
\verb,RA (2000) =  19 43 01.96,&\\ 
\verb,DEC(2000) = +32 19 13.8,&\\
&\\
t$_0=$ 2008/04/10.7$^{[2]}$ &\\
t$_2=$ 5.65$\pm$0.1$^{[3]}$ d  &\\
&\\
Type: CN&\\
&\\
&\\
&\\
&\\
\end{tabular}\end{center}\end{table*}
\begin{table*}[h]\small\begin{center}
\begin{tabular}{p{4cm}|p{4cm}|p{4cm}}
\multicolumn{3}{c}{General Data}\\
\hline
m$_V$ [mag] = 7.5$^{[4]}$&E(B-V) = 0.43$^{[5]}$&v$_{\infty}$ [km/s] = 4800\\
D [kpc] = $10.5^{[2]}$   &M$_V$ = & P$_{\rm orb?}$ = 0.09580$^{[4]}$\\
X-ray on =       &t$_{\rm Bol}$ [month] $<3^{[3]}$ & \\ 
\hline
\end{tabular}\end{center}\end{table*}
\begin{table*}[h]\small\begin{center}
\begin{tabular}{p{1.1cm}|p{2.1cm}|p{3cm}|p{2cm}|p{3cm}}
\multicolumn{5}{c}{Model Data}\\
\hline
Model&T$_{\rm eff}$[$\times$10$^3$ K]&t after t$_0$ [d]&L [erg/s]&N$_{H}$
[10$^{21}$ cm$^{-2}$]\\
\hline
BB$^{[2,3]}$   &$520$        & 39.9     & 1$\times$10$^{40}$ & $5^{[2]}$\\
BB$^{[2,3]}$   &$570$        & 49.7     & 8.7$\times$10$^{38}$& $4.7^{[2]}$\\
WD$^{[1]}$      &$600$        &          &        &$2.8$ -- $5$\\
\hline
\end{tabular}\end{center}\end{table*}
\begin{table*}[!h]\small\begin{center}
\begin{tabular}{p{.8cm}p{11.7cm}}
\multicolumn{2}{c}{References}\\
\hline
1&D.R. van Rossum et al.; 2010, AN, 331 175 \\
2&J.-U. Ness; 2010, AN, 331, 179\\
3&this work\\
4&A. Baklanov et al.; 2008, ATel, 1514\\ 
5&R.J. Rudy et al.; 2008, IAUC, 8938\\
\hline
\end{tabular}\end{center}\end{table*}

\newpage              
\begin{table*}[!h]\begin{center}
\begin{tabular}{p{6cm}r}
{\Huge {\bf V4743 Sgr}}& \multirow{6}{*}{\includegraphics[width=7cm]
{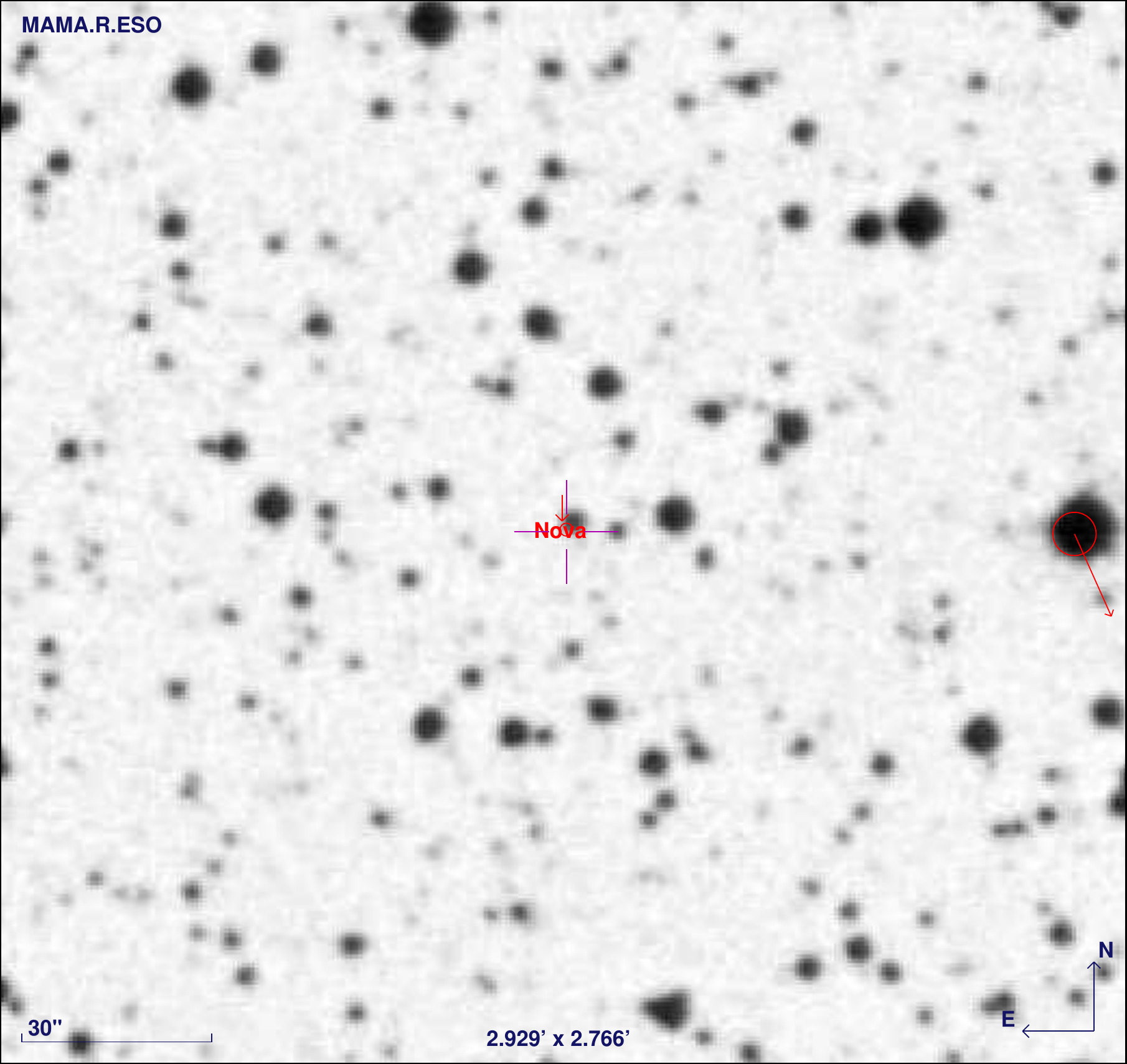}}\\
&\\
\verb,RA (2000) =  19 01 09.38,&\\ 
\verb,DEC(2000) = -22 00 05.9,&\\
&\\
t$_0=$ 2002/09/18.3$^{[1]}$ &\\
t$_2=$ 9 d$^{[1]}$  &\\
t$_3=$ 16 d$^{[1]}$  &\\
&\\
Type: CN&\\
&\\
&\\
&\\
&\\
\end{tabular}\end{center}\end{table*}
\begin{table*}[!h]\small\begin{center}
\begin{tabular}{p{4cm}|p{4cm}|p{4cm}}
\multicolumn{3}{c}{General Data}\\
\hline
m$_V$ [mag] = $5^{[1]}$&E(B-V) = 0.25$^{[6]}$&v$_{\infty}$ [km/s] =
2200-2700$^{[1]}$\\
D [kpc] = $3.9\pm0.3^{[6]}$       &M$_V$ = & P$_{\rm orb}$ [d] = 0.2799$^{[5]}$\\
X-ray on = t$_0+180^{[4]}$ d   &t$_{\rm Bol}$ [yr] $\approx1.5^{[4]}$& \\ 
\hline
\end{tabular}\end{center}\end{table*}
\begin{table*}[!h]\small\begin{center}
\begin{tabular}{p{1.1cm}|p{2.1cm}|p{3cm}|p{2cm}|p{3cm}}
\multicolumn{5}{c}{Model Data}\\
\hline
Model&T$_{\rm eff}$[$\times$10$^3$ K]&t after t$_0$
[d]&L[erg/s]&N$_{H}$[10$^{21}$ cm$^{-2}$]\\
\hline
BB$^{[2]}$ & $390$      & 180.4                & 1.6$\times$10$^{40}$ & $2.1$  \\
BB$^{[2]}$ & $460$      & 371                  & 1.8$\times$10$^{39}$ & $1.8$  \\
WD$^{[3]}$ & $600$      & 182                &        & $4.6$ \\
WD$^{[3]}$ & $580$      & 303                &        & $4.4$ \\
WD$^{[3]}$ & $580$      & 372                &        & $4.2$ \\
WD$^{[3]}$ & $540$      & 528                &        & $3.3$  \\
WD$^{[4]}$ & 580             & 182             & 7.8$\times$10$^{37}$ & 0.46\\
WD$^{[4]}$ & 600             & 303             & 1.1$\times$10$^{38}$ & 0.51\\
WD$^{[4]}$ & 560             & 372             & 8.5$\times$10$^{37}$ & 0.51\\
WD$^{[4]}$ & 455             & 528             & 3.7$\times$10$^{37}$ & 0.51\\
\hline
\end{tabular}\end{center}\end{table*}
\begin{table*}[!h]\scriptsize\begin{center}
\begin{tabular}{p{.8cm}p{11.7cm}}
\multicolumn{2}{c}{References}\\
\hline
1&G.E. Morgan et al.; 2003, MNRAS, 344 521\\
2&J.-U. Ness; 2010, AN, 331, 179\\
3&A. Petz et al.; 2005, A\&A, 431, 321\\
4&this work\\
5&T.W. Kang et al.; 2006, ApJ, 132, 608\\
6&K.M. Vanlandingham et al.; 2007, AAS, 210\\
\hline
\end{tabular}\end{center}\end{table*}

\newpage 
\begin{table*}[t]\begin{center}
\begin{tabular}{p{6cm}r}
{\Huge {\bf RS Oph}}& \multirow{6}{*}{\includegraphics[width=7cm]
{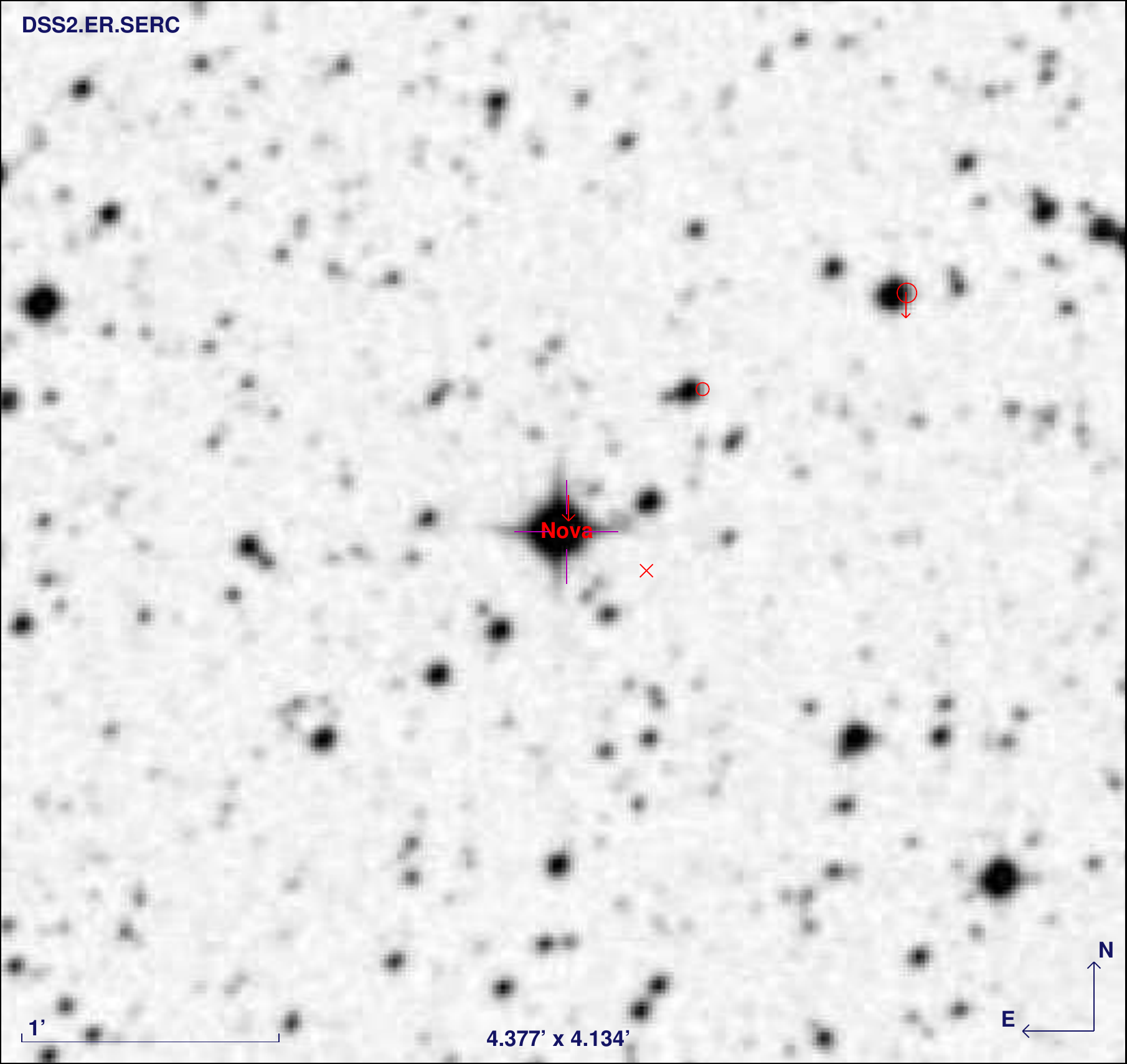}}\\
&\\
\verb,RA (2000) =  17 50 13.20 ,&\\ 
\verb,DEC(2000) = -06 42 28.48 ,&\\
&\\
t$_0=$ 2006/02/12.85$^{[2]}$ &\\
t$_2=$ $6.8^{[1]}$ d  &\\
t$_3=$ $14.0^{[1]}$ d  &\\
&\\
Type: RN&\\
&\\
&\\
&\\
&\\
\end{tabular}\end{center}\end{table*}
\begin{table*}[h]\small\begin{center}
\begin{tabular}{p{4cm}|p{4cm}|p{4cm}}
\multicolumn{3}{c}{General Data}\\
\hline
m$_V$ [mag] = $4.5^{[2]}$&E(B-V) = $0.73^{[1]}$ &v$_{\infty}$ [km/s] = $5600\pm1100^{[2]}$\\
D [kpc] = $4.3\pm0.7^{[1]}$, \newline\verb,         ,\mbox{$1.6\pm0.3^{[3]}$}
&M$_{\rm Bol}$ = $-3.66^{[1]}$ & P$_{\rm orb}$ $[$d$]$ = $453.6\pm0.4^{[6]}$ \\
X-ray on =       &t$_{\rm Bol}$ [yr] = & \\ 
\hline
\end{tabular}\end{center}\end{table*}
\begin{table*}[h]\small\begin{center}
\begin{tabular}{p{1.1cm}|p{2.1cm}|p{3cm}|p{2cm}|p{3cm}}
\multicolumn{5}{c}{Model Data}\\
\hline
Model&T$_{\rm eff}$[$\times$10$^3$ K]&t after t$_0$
[d]&L[erg/s]&N$_{H}$[10$^{21}$ cm$^{-2}$]\\
\hline
BB$^{[4]}$   & 620    & 39.7            &  $2.7\times10^{38}$ & 5.1      \\
BB$^{[4]}$   & 500    & 66.9            &  $7.2\times10^{39}$ & 6.9      \\
WD$^{[5]}$   & 550    & 39.7            &                   & 4.8         \\
\hline
\end{tabular}\end{center}\end{table*}
\begin{table*}[!h]\small\begin{center}
\begin{tabular}{p{.8cm}p{11.7cm}}
\multicolumn{2}{c}{References}\\
\hline
1&B. E. Schaefer; 2009, ApJ, 697, 721\\
2&K. Hirosava; 2006, IAUC, 8671\\
3&M.F. Bode; 2009, AN, 331, 160\\
4&J.U. Ness; 2010, AN, 331, 179\\
5&D.R. van Rossum et al.; AN, 331, 175\\
6&E. Brandi et al.; 2009, A\&A, 497, 815\\
\hline
\end{tabular}\end{center}\end{table*}

\newpage 
\begin{table*}[t]\begin{center}
\begin{tabular}{p{6cm}r}
{\Huge {\bf KT Eri}}& \multirow{6}{*}{\epsfig{file=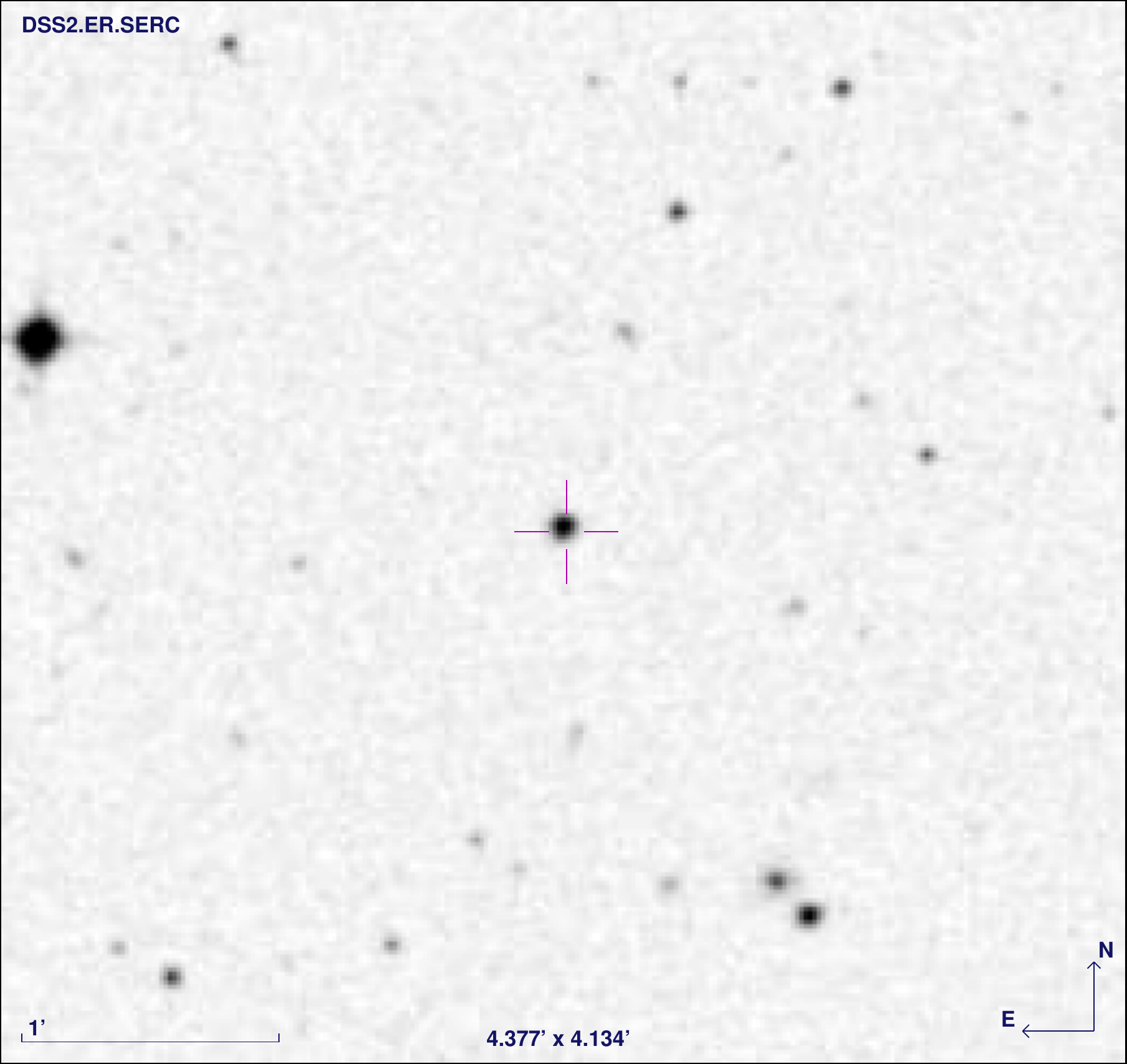,
width=7cm}}\\
&\\
\verb,RA (2000) =  04 47 54.21 ,&\\ 
\verb,DEC(2000) = -10 10 43.1,&\\
&\\
t$_0=$ 2009/11/14.572$^{[1]}$ &\\
t$_0=$ 2009/11/19.241$^{[3]}$ &\\
t$_2=$ 8$^{[3]}$  &\\
t$_3=$ 15$^{[3]}$ &\\
&\\
Type: CN&\\
&\\
&\\
\end{tabular}\end{center}\end{table*}
\begin{table*}[h]\small\begin{center}
\begin{tabular}{p{4cm}|p{4cm}|p{4cm}}
\multicolumn{3}{c}{General Data}\\
\hline
m$_V$ [mag] = $5.6^{[3]}$   &E(B-V) = $0.08^{[3]}$&v$_{\infty}$ [km/s] =
$5800^{[1]}$\\
D [kpc] = $6.5^{[3]}$, $10.5^{[]}$      &M$_V$ = $-8.7^{[3]}$& P$_{\rm orb}$ = \\
X-ray on = 2010/01/19$^{[1]}$&t$_{\rm Bol}$ [yr] = & \\ 
\hline
\end{tabular}\end{center}\end{table*}
\begin{table*}[h]\small\begin{center}
\begin{tabular}{p{1.1cm}|p{2.1cm}|p{3cm}|p{2cm}|p{3cm}}
\multicolumn{5}{c}{Model Data}\\
\hline
Model&T$_{\rm eff}$[$\times$10$^3$ K]&$t$ after $t_0$ [day]&L [erg/s]&N$_{H}$
[10$^{21}$ cm$^{-2}$]\\
\hline
BB$^{[1]}$   &  $520$      & $61.4$        & & $2.8\pm0.1$     \\
BB$^{[1]}$   &  $210$      & $65.5$        & & $2.8\pm0.1$     \\
BB$^{[2]}$   &  $440$      & $70.4$        & & $1.2$           \\
WD   &               &                 &                   &     \\
\hline
\end{tabular}\end{center}\end{table*}
\begin{table*}[!h]\small\begin{center}
\begin{tabular}{p{.8cm}p{11.7cm}}
\multicolumn{2}{c}{References}\\
\hline
1&M.F. Bode et al.; 2010, ATel, 2392 \\
2&J.-U. Ness et al.; 2010, ATel, 2418\\
3&E. Ragan et al.; 2009, ATel, 2327\\
\hline
\end{tabular}\end{center}\end{table*}

\newpage   
\begin{table*}[t]\begin{center}
\begin{tabular}{p{6cm}r}
{\Huge {\bf V458 Vul}}& \multirow{6}{*}{\epsfig{file=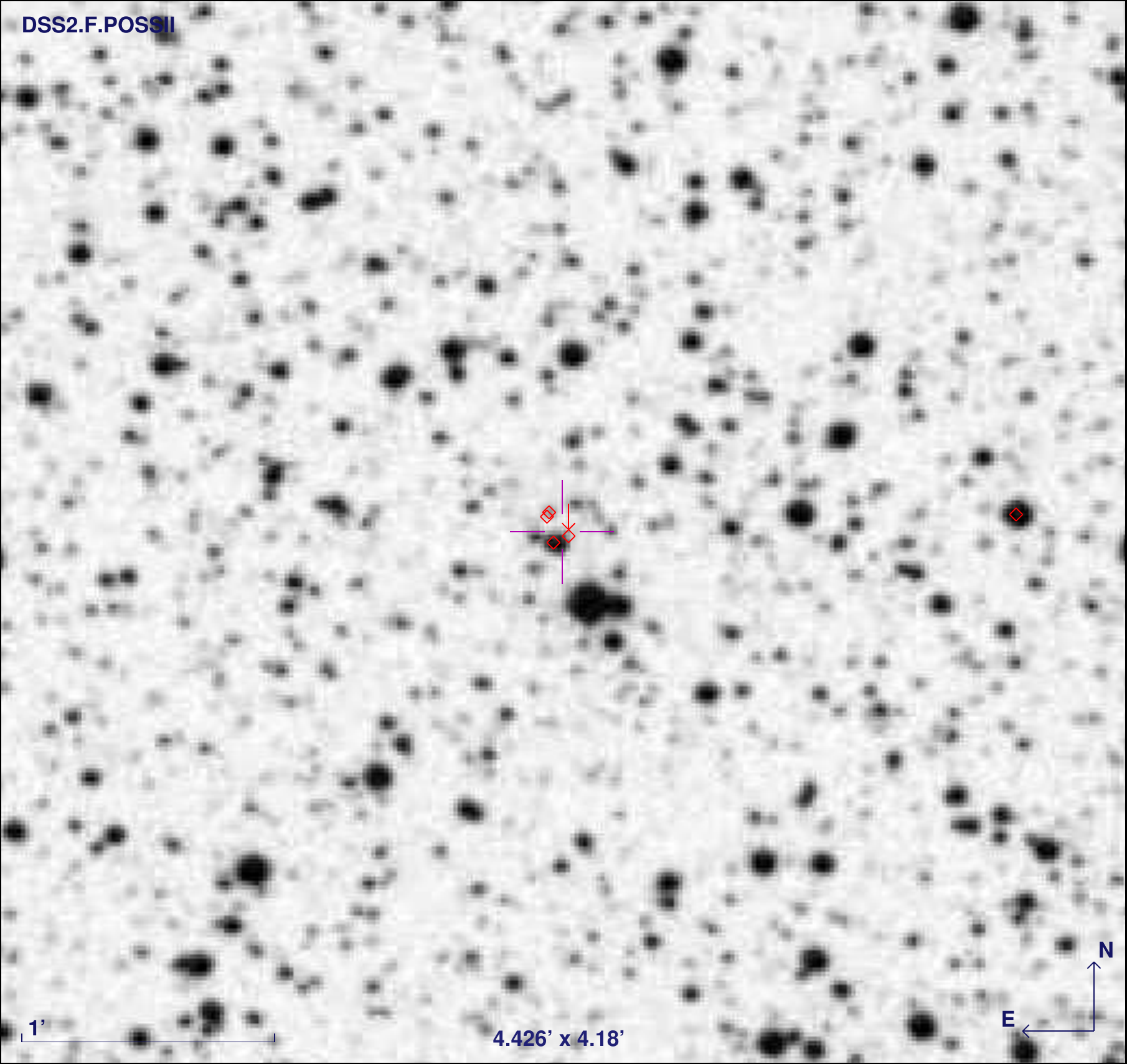,
    width=7cm}}\\
&\\
\verb,RA (2000) =  19 54 24.3,&\\ 
\verb,DEC(2000) = +20 52 47,&\\
&\\
t$_0=$ 2007/08/8.54$^{[1]}$ &\\
t$_2=$ 7$^{[1]}$ d &\\
&\\
Type: CN&\\
&\\
&\\
&\\
&\\
\end{tabular}\end{center}\end{table*}
\begin{table*}[h]\small\begin{center}
\begin{tabular}{p{4cm}|p{4cm}|p{4cm}}
\multicolumn{3}{c}{General Data}\\
\hline
m$_V$ [mag] =    &E(B-V) = 0.6$^{[1]}$&v$_{\infty}$ [km/s] = \\
D [kpc] = $8.8\pm0.8^{[1]}$ &M$_V$ = & P$_{\rm orb}$ [d] = 0.58946 \\
X-ray on = t$_0 + 315^{[1]} $ d    &t$_{\rm Bol}$ [yr] $>2^{[1]}$& \\ 
\hline
\end{tabular}\end{center}\end{table*}
\begin{table*}[h]\small\begin{center}
\begin{tabular}{p{1.1cm}|p{2.1cm}|p{3cm}|p{2cm}|p{3cm}}
\multicolumn{5}{c}{Model Data}\\
\hline
Model&T$_{\rm eff}$[$\times$10$^3$ K]&t after t$_0$ [d]&L [erg/s]&N$_{H}$
[10$^{21}$ cm$^{-2}$]\\
\hline
BB$^{[2]}$   & 333     & 37              &   & 3.3     \\
WD   &               &                 &                   &     \\
\hline
\end{tabular}\end{center}\end{table*}
\begin{table*}[!h]\small\begin{center}
\begin{tabular}{p{.8cm}p{11.7cm}}
\multicolumn{2}{c}{References}\\
\hline
1&J.-U. Ness et al.; 2010, AJ, 137, 4160\\
2&J.J. Drake et al.; 2008, ATel, 1721\\
\hline
\end{tabular}\end{center}\end{table*}

\newpage     
\begin{table*}[t]\begin{center}
\begin{tabular}{p{6cm}r}
{\Huge {\bf V5116 Sgr}}& \multirow{6}{*}{\epsfig{file=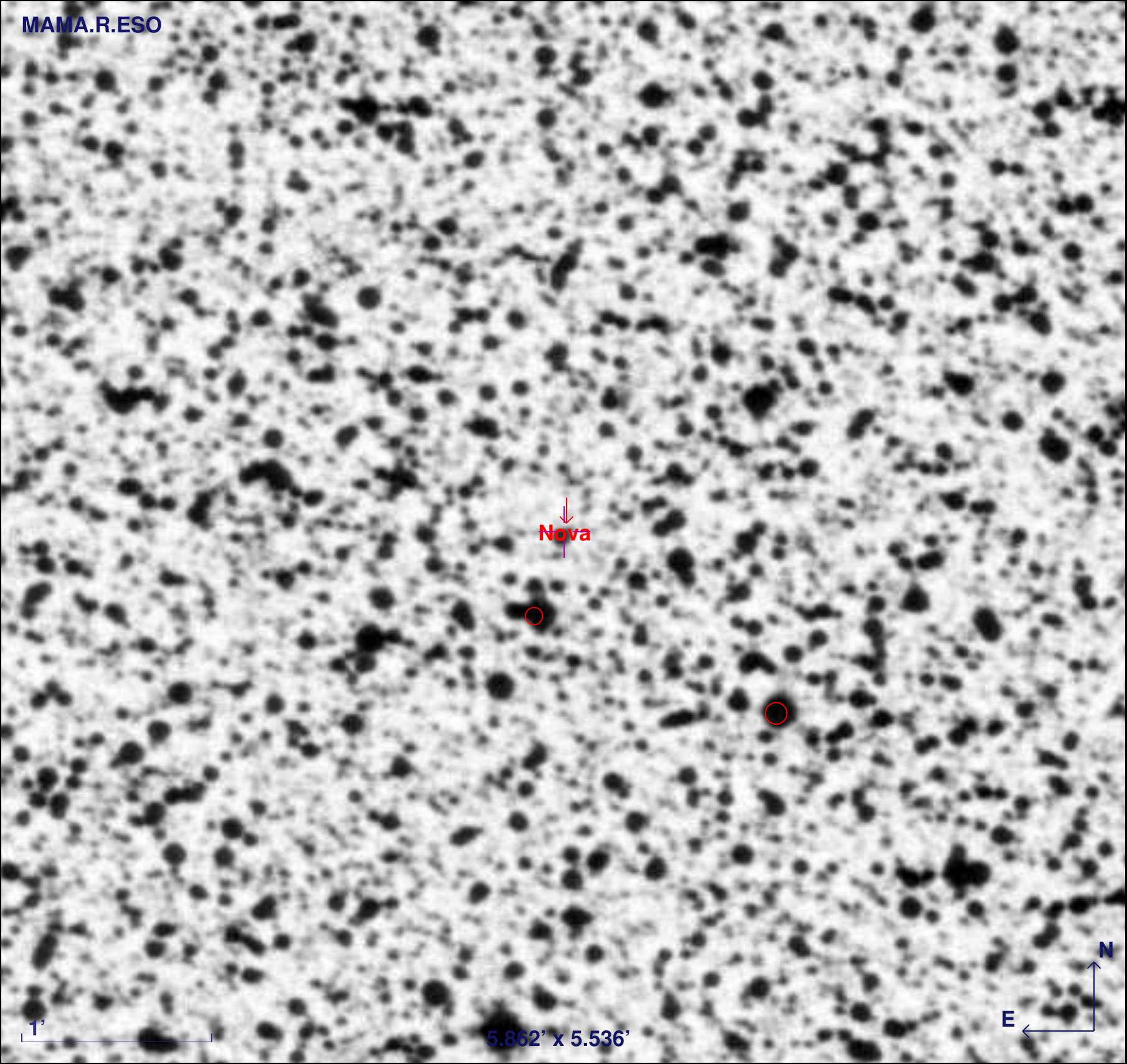,
    width=7cm}}\\
&\\
\verb,RA (2000) =  18 17 50.77,&\\ 
\verb,DEC(2000) = -30 26 31.2,&\\
&\\
t$_0=$ 2005/07/5.085$^{[2]}$  &\\
t$_2=6.5\pm1^{[5]}$ days   &\\
t$_3=39^{[3]}$ days   &\\
&\\
Type: CN&\\
&\\
&\\
&\\
\end{tabular}\end{center}\end{table*}
\begin{table*}[h]\small\begin{center}
\begin{tabular}{p{4cm}|p{4cm}|p{4cm}}
\multicolumn{3}{c}{General Data}\\
\hline
m$_V$ [mag] = $7.2^{[2]}$   &(B-V) = $0.48^{[8]}$&v$_{\infty}$ [km/s] = $2200^{[4]}$\\
D [kpc] = $11.3\pm3^{[6]}$       &M$_V$ [mag]= $-8.85\pm0.04^{[5]}$& P$_{\rm orb}$ $[{\rm h}]=$\newline
\verb,    ,$2.9712\pm0.0024^{[5]}$ \\
X-ray on $\approx{t_0+700}^{[9]}$ d     &t$_{\rm Bol}$ [yr] $\approx{1}^{[9]}$& \\ 
\hline
\end{tabular}\end{center}\end{table*}
\begin{table*}[h]\small\begin{center}
\begin{tabular}{p{1.1cm}|p{2.1cm}|p{2.5cm}|p{2.7cm}|p{2.8cm}}
\multicolumn{5}{c}{Model Data}\\
\hline
Model&T$_{\rm eff}$[$\times$10$^3$ K]&t after t$_0$ [d]&L [erg/s]&N$_H$ [$\times10^{21}$ cm$^{-2}$]\\
\hline
BB   &                              &                 &         &  \\
WD$^{[1]}$   &$610\pm0.6$ & &{\scriptsize $2.6\pm0.7\times(d/10{\rm kpc})$}         &$1.3\pm0.1$ \\
WD$^{[7]}$   & 465                         &                 &         &  \\       
\hline
\end{tabular}\end{center}\end{table*}
\begin{table*}[!h]\small\begin{center}
\begin{tabular}{p{.1cm}p{5.7cm}|p{.1cm}p{5.7cm}}
\multicolumn{4}{c}{References}\\
\hline
1&G. Sala et al.; 2007, ATel, 1184&6&G. Sala et al.; 2008b, A\&A, 473, 61\\
2&W. Liller; 2005, IAUC, 8559 &7&T. Nelson et al.; 2007, ATel, 1202\\
3&R. Williams et al.; 2008, ApJ, 685, 451&8&A.C. Gilmore et al.; 2005, IAUC, 8559\\
4&R. Russell et al.; 2005, IAUC 8579&9&G. Sala et al.; 2010, AN, 331, 201\\ 
5&A. Dobrotka et al.; 2008, A\&A, 478, 815&&\\
\hline
\end{tabular}\end{center}\end{table*}

\newpage 
\begin{table*}[t]\begin{center}
\begin{tabular}{p{6cm}r}
{\Huge {\bf V723 Cas}}& \multirow{6}{*}{\epsfig{file=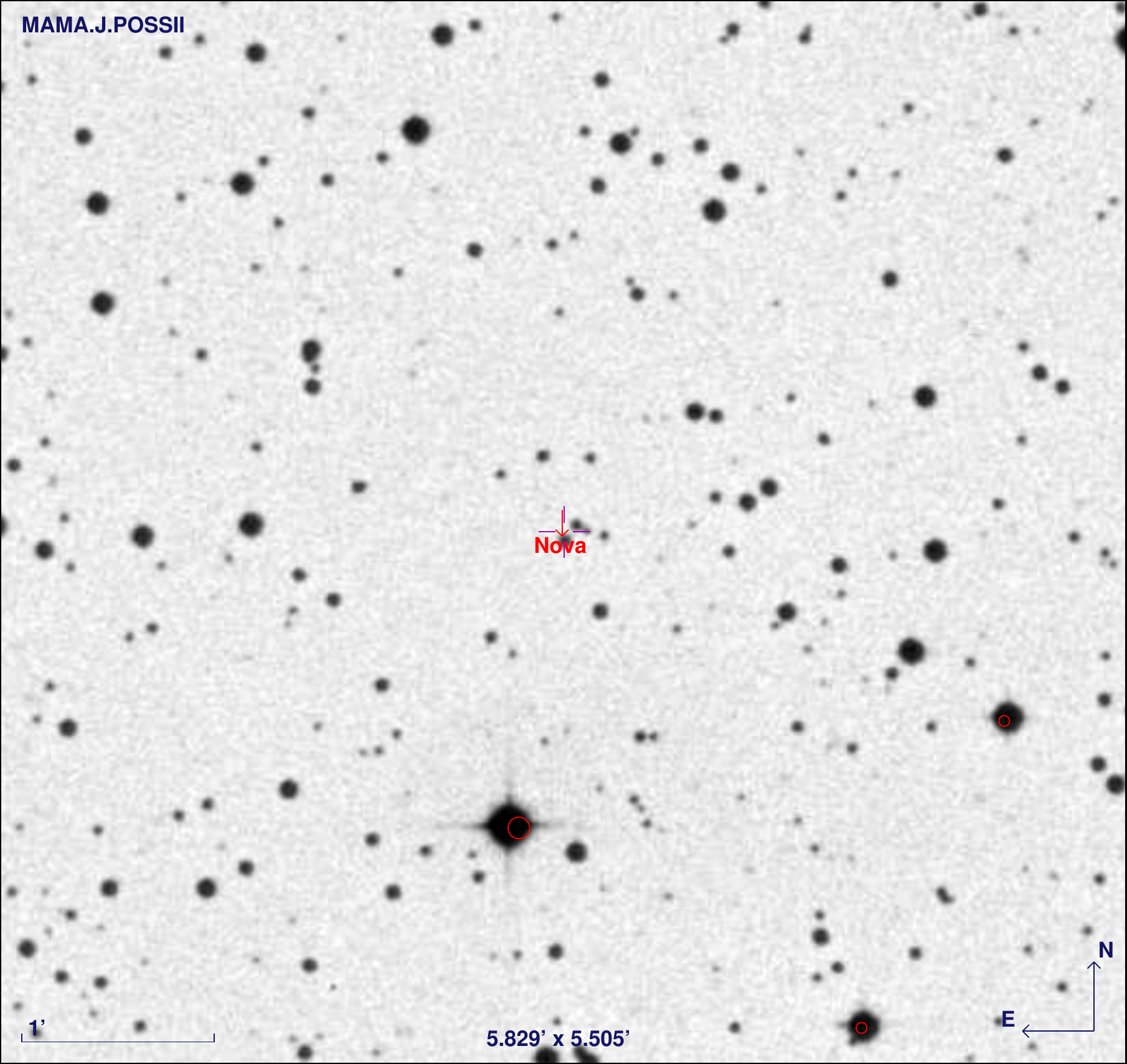,
    width=7cm}}\\
&\\
\verb,RA (2000) =  01 05 05.37,&\\ 
\verb,DEC(2000) = +54 00 40.5,&\\
&\\
t$_0=$ 1995/12/17$^{[3]}$  &\\
t$_2=$ $340^{[4]}$  &\\
t$_3=$ $515^{[4]}$  &\\
&\\
Type: CN&\\
&\\
&\\
&\\
\end{tabular}\end{center}\end{table*}
\begin{table*}[h]\small\begin{center}
\begin{tabular}{p{4cm}|p{4cm}|p{4cm}}
\multicolumn{3}{c}{General Data}\\
\hline
m$_V$ [mag] = $7.1^{[3]}$    &E(B-V) = $0.57\pm0.05^{[4]}$ & v$_{\infty}$ [km/s] = \\
D [kpc] = ${3.85^{+0.23}_{-0.21}}^{[1]}$       &M$_V$ = & P$_{\rm orb}$ [d] = 
0.693265$^{[4]}$\\
X-ray on =       &t$_{\rm Bol}$ [yr] = & \\ 
\hline
\end{tabular}\end{center}\end{table*}
\begin{table*}[h]\small\begin{center}
\begin{tabular}{p{1.1cm}|p{2.1cm}|p{2.7cm}|p{2.3cm}|p{3cm}}
\multicolumn{5}{c}{Model Data}\\
\hline
Model&T$_{\rm eff}$[$\times$10$^3$ K]&Peak flux [cts/s]&L [erg/s]&N$_{H}$
[$10^{21}$ cm$^{-2}$]\\
\hline
BB$^{[2]}$   & $280$ -- $380$& $0.022\pm0.01$&$>0.5\times10^{37}$ & $1.6$ -- $4.8$     \\
WD   &               &                 &                   &     \\
\hline
\end{tabular}\end{center}\end{table*}
\begin{table*}[!h]\small\begin{center}
\begin{tabular}{p{.1cm}p{5.7cm}|p{.1cm}p{5.7cm}}
\multicolumn{4}{c}{References}\\
\hline
1&J.E. Lyke et al.; 2009, ApJ, 138, 1090&4&V.P. Goranskij et al.; 2007, AB,
62, 2\\
2&J.-U. Ness et al.; 2008, AJ, 135, 1328&&\\
3&Munari et al.; 1996, A\&A, 315, 166&&\\
\hline
\end{tabular}\end{center}\end{table*}

\newpage 
\begin{table*}[!t]
\begin{center}
\begin{tabular}{p{6cm}r}
{\Huge {\bf V1494 Aql}}& \multirow{6}{*}{\epsfig{file=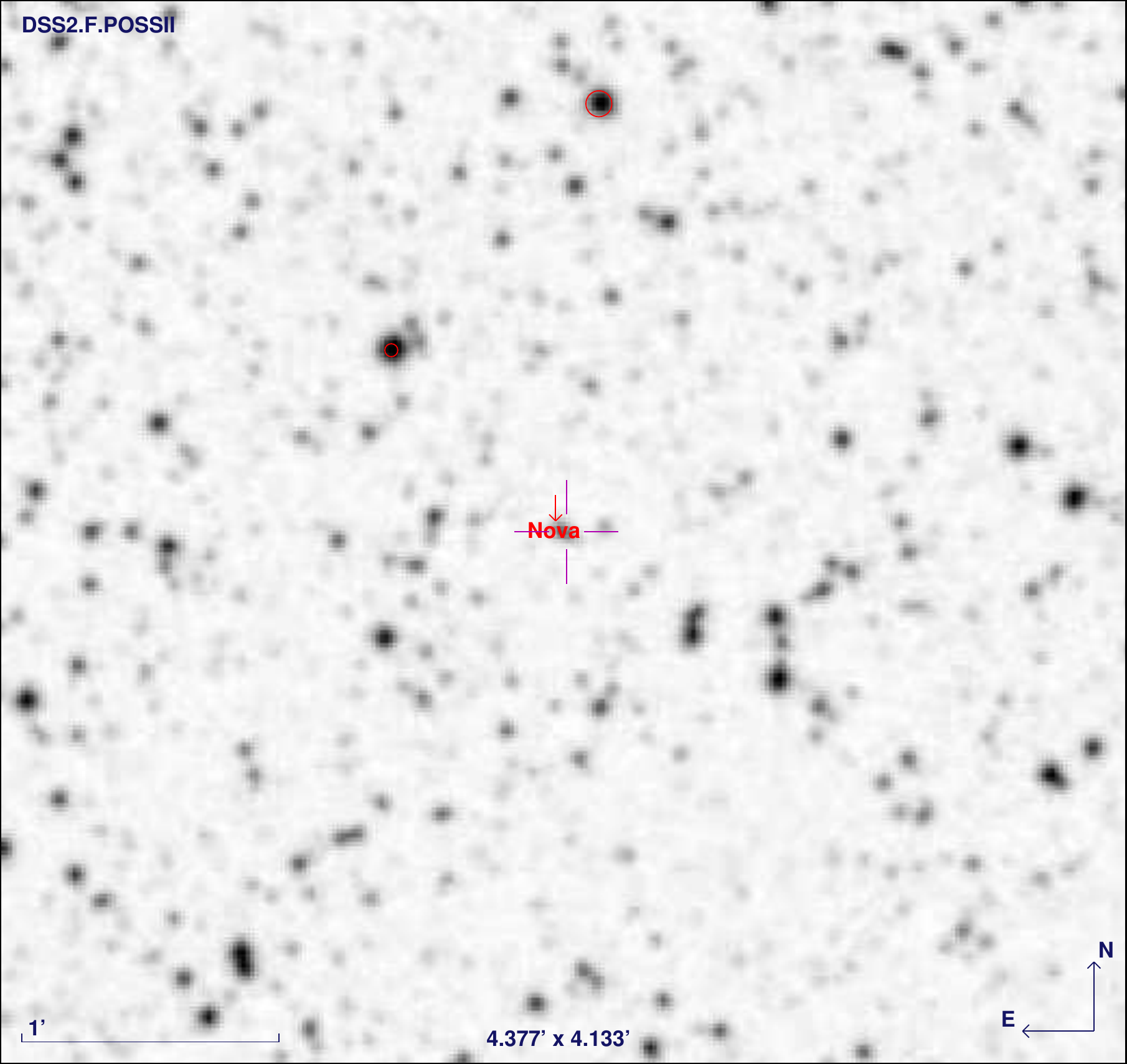,
    width=7cm}}\\
&\\
\verb,RA (2000) =  19 23 05.30 ,&\\ 
\verb,DEC(2000) = +04 57 19.1,&\\
&\\
t$_0=$ 1999/12/3.4$^{[2]}$  &\\
t$_2=6.6\pm0.5^{[2]}$ d  &\\
t$_3=16\pm0.5^{[2]}$ d  &\\
&\\
Type: CN&\\
&\\&\\&\\
\end{tabular}\end{center}\end{table*}\begin{table*}[!h]\small\begin{center}
\begin{tabular}{p{4cm}|p{4cm}|p{4cm}}
\multicolumn{3}{c}{General Data}\\
\hline
m$_V$ [mag] = $4.0$   &E(B-V) =$0.6\pm0.1^{[1]}$ &v$_{\infty}$ [km/s] =
$2900^{[1]}$\\
D [kpc] = \mbox{$1.6\pm0.2^{[1]}$}, \newline\verb,         ,\mbox{$3.6\pm0.3^{[2]}$}       &M$_V$ =$-8.8\pm0.2^{[2]}$ & P$_{\rm orb}$ = \\

X-ray on = t$_0+180^{[3]}$ d      &t$_{\rm Bol}$ [year] $=2.5$ -- $3^{[3]}$ & \\ 
\hline
\end{tabular}\end{center}\end{table*}\begin{table*}[h]\small\begin{center}
\begin{tabular}{p{1.1cm}|p{2.1cm}|p{3cm}|p{2cm}|p{3cm}}
\multicolumn{5}{c}{Model Data}\\
\hline
Model&T$_{\rm eff}$[$\times$10$^3$ K]&t after t$_0$ [d]&L [erg/s]&N$_{H}$
[10$^{21}$ cm$^{-2}$]\\
\hline
BB   &                &                 & &      \\
WD   &               &                 &                   &     \\
\hline
\end{tabular}\end{center}\end{table*}\begin{table*}[!h]\small\begin{center}
\begin{tabular}{p{.8cm}p{11.7cm}}
\multicolumn{2}{c}{References}\\
\hline
1&T. Iijima et al.; 2003, A\&A, 404, 997\\
2&L.L. Kiss et al.; 2000, A\&A, 355, L9\\
3&M. Orio; 2000, RMxAC, 20, 182\\
\hline
\end{tabular}\end{center}\end{table*}

\end{document}